\documentclass[12pt,preprint]{aastex}

\usepackage{apjfonts}
\usepackage{natbib}

\usepackage{subfigure}
\usepackage{verbatim}
\usepackage{graphics}
\usepackage{graphicx}
\usepackage{amssymb}
\usepackage{cite}
\usepackage[toc,page]{appendix}
\usepackage{float}
\usepackage{amssymb,amsmath}
\usepackage{color}

\begin{document}

\title{Star Formation in Orion's L1630 Cloud: an Infrared and Multi-epoch X-ray Study}

\author{\vspace{0.25in}David A. Principe,\altaffilmark{1} J. H. Kastner,\altaffilmark{1} Nicolas Grosso,\altaffilmark{2} Kenji Hamaguchi,\altaffilmark{3}\altaffilmark{4} Michael Richmond,\altaffilmark{1} William K. Teets,\altaffilmark{5} David A. Weintraub\altaffilmark{5}  }

\email{daveprincipe@astro.rit.edu}

\altaffiltext{1}{Chester F. Carlson Center for Imaging Science and the Laboratory for Multiwavelength Astronomy (LAMA), 
Rochester Institute of Technology, 54 Lomb Memorial Drive, Rochester, NY 14623}

\altaffiltext{2}{Observatoire Astronomique de Strasbourg, Universit\'{e} de Strasbourg, CNRS, UMR 7550, 11 rue de l'Universit\'{e}, F-67000 Strasbourg, France}

\altaffiltext{3}{CRESST and X-ray Astrophysics Laboratory NASA/GSFC, Greenbelt, MD 20771}

\altaffiltext{4}{Department of Physics, University of Maryland, Baltimore County, 1000 Hilltop Circle, Baltimore, MD 21250}

\altaffiltext{5}{Department of Physics \& Astronomy, Vanderbilt University, Nashville TN 37235, USA}

\begin{abstract}

 X-ray emission is characteristic of young stellar objects (YSOs) and is known to be highly variable. We investigate, via an infrared and multi-epoch X-ray study of the L1630 dark cloud, whether and how X-ray variability in young stellar objects is related to protostellar evolutionary state.  We have analyzed 11 Chandra X-ray Observatory observations, obtained over the course of four years and totaling $\sim$240 ks exposure time, targeting the eruptive Class I YSO V1647 Ori in L1630. We used 2MASS and Spitzer data to identify and classify IR counterparts to L1630 X-ray sources and  identified a total of 52 X-ray emitting YSOs with IR counterparts, including 4 Class I sources and 1 Class 0/I source.  We have detected cool ($<$ $3$ $MK$) plasma, possibly indicative of accretion shocks, in three classical T Tauri stars.  A subsample of 27 X-ray-emitting YSOs were covered by 9 of the 11 Chandra observations targeting V1647 Ori and vicinity.  For these 27 YSOs, we have constructed X-ray light curves spanning approximately four years.  These light curves highlight the variable nature of pre-main sequence X-ray emitting young stars; many of the L1630 YSOs vary by orders of magnitude in count rate between observations.  We discuss possible scenarios to explain apparent trends between various X-ray spectral properties, X-ray variance and YSO classification.

\vspace{30mm}

\end{abstract}

\section{Introduction}

 Multiwavelength observations of star-forming regions are essential in understanding pre-main sequence (pre-MS) stellar evolution, from shortly after the collapse of a molecular cloud to the dissipation of circumstellar disks.  YSOs such as those in L1630 are usually subdivided into four stages of evolution based on their spectral energy distributions (SED).   To classify YSOs of low masses ($\leq$ 2 $M_{\odot}$), the slope of the spectral energy distribution is measured using the IR spectral slope from 2 $\mu$m to 10-25 $\mu$m \citep[$\alpha$;][]{Lada1987, Wilking1989, Greene1994}.  Class 0 objects represent an early stage of stellar evolution preceding Class I, during which the mass in the envelope is significantly higher than that of the central protostar \citep{Andre1993}. Such objects are characterized by a lack of detectable near-IR counterparts and a steady increase of flux from the mid-IR to the submillimeter regime.  Objects with a rising infrared continuum ($\alpha$ > 0.3) are designated as Class I sources and are thought to be composed of a remnant infalling envelope surrounding the protostar-disk system.  The flat spectrum phase represents an intermediate phase between Class I and Class II ($0.3 > \alpha > -0.3$; Greene et al. 1994).  Class II objects have decreasing IR SEDs (-1.5 < $\alpha$ < -0.3) and hence most likely have dissipated their outer envelopes, i.e., they consist of a pre-MS star and disk; Finally, Class III objects  \citep[$\alpha$ < -1.5;][]{Andre1994} have little or no detectable IR excess and hence likely have almost completely dissipated their disks.

 Numerous Chandra X-ray Observatory and XMM-Newton studies of star-forming regions  have resulted in detections of X-ray emission from Class I, Class II, and Class III YSOs \citep[e.g.,][]{ Getman2005,Ozawa2005, Guedel2007, Townsley2011, Pillitteri2013}.  This X-ray emission is generated in hot, circumstellar plasmas.  Magnetic reconnection events at YSOs heat plasma to temperatures $>>10^6$ K, such that magnetic activity at YSOs can be effectively probed via X-rays.  Such magnetic reconnection may be confined to a YSO's (proto-)corona \citep{Preibisch2005} or may trace star-disk interactions \citep{Kastner2006}.  Furthermore, coronal emission and/or star-disk interactions are known to produce X-ray variability \citep{Favata2005}. Bright X-ray flares have been observed in YSOs, and are thought to result from large magnetic loop structures, some of which extend to distances large enough to interact with disks, if present \citep{Favata2005}.  Although T Tauri stars can be bright X-ray sources ($L_{X} \sim 10^{-3}$  $L_\mathrm{bol}$), the X-ray detection of {\sl bona fide} Class 0 objects is rare and controversial \citep[and refs. therein]{Hamaguchi2005, Prisinzano2008}. The paucity of X-ray-detected Class 0 YSOs may be due to the large (molecular cloud and circumstellar) absorbing column densities toward such objects, or may indicate that the high-energy emission does not occur during the transition from primordial (cloud) to protostellar (stellar dynamo) fields.

Obviously in the case of star-disk interactions -- e.g., magnetically-funneled accretion, disk instabilities resulting in accretion outbursts, or magnetically confined plasma suddenly released in star-disk magnetic-reconnection events-- a circumstellar disk must be present.  Hence, X-ray activity due to such interactions should only be present during Class I, Class II and transition disk stages of evolution. Accretion shocks in YSOs are associated with lower temperature ($\sim 3 \times 10^{6}$ K), high-density X-ray emitting plasma \citep[e.g.,][]{Kastner2002, Brickhouse2010}. The accretion process can be time dependent on several timescales, due to a variety of mechanisms \citep[as discussed in][]{Bouvier2007}: non-steady accretion \citep[hours;][]{stempels2002}, coronal loop flaring \citep[days;][]{Favata2005}, rotational modulation \citep[weeks; ][]{Smith1999}, global magnetospheric instabilities \citep[months;][]{Bouvier2003}, and major accretion-driven eruptions \citep[years;][]{Herbig1989}.
This suggests that the X-ray production and/or variability produced by star-disk interactions should depend on the stage of young stellar evolution.  

The Chandra Orion Ultradeep Project (COUP), a 10-day-long ($\sim$840 ks) X-ray observation of the Orion Nebula Cluster (ONC; D$\sim$400 pc), yielded one of the nearest and richest samples of X-ray emitting pre-main sequence stars \citep[$\sim$1300 X-ray detected YSOs;][]{Getman2005}.  Many different characteristics of X-ray emitting YSOs were investigated, including the evolution of X-ray emission in YSOs \citep{Preibisch2005}, rotational modulation of X-ray emission \citep{Flaccomio2005}, X-ray flaring \citep{Favata2005} and X-ray emission from "proplyds," i.e., YSOs with circumstellar disks, and in some cases, jets \citep{Kastner2005}. The most powerful $\approx$1$\%$ of flares reported in the COUP survey had measured decay times in the range 10-400 ks \citep{Favata2005}.  In a follow up variability study of the COUP X-ray data \citet{Flaccomio2012} concluded that the Class II YSOs in the COUP sample were more variable than the Class III YSOs at all timescales (within the $\sim$ 10 day observation), suggesting that time-variable absorption from circumstellar structures (e.g., inner disk warps and/or accretion streams) is responsible for the variable X-ray emission in Class II YSOs.

Several studies have compared the IR and X-ray properties of YSOs in nearby (D $<$ 400 pc) star-forming regions \citep{Winston2007, Getman2005, Forbrich2007}.  Correlations between X-ray emission and YSO classification have been difficult to establish, however,  due in large part to the highly variable nature of YSO sources.  Winston et al. (2007; henceforth W07), studied an embedded stellar cluster in the Serpens cloud (D $\sim$ 260 pc).  They discussed the mid-IR spectral energy distributions of protostellar objects detected in the submillimeter, the spatial distribution of cluster members, and the X-ray properties of classified YSOs. Included in their classification were flat-spectrum objects.    A total of 138 YSOs were identified, 60 of which had X-ray counterparts and 7 of which had submillimeter counterparts.  W07 correlated the X-ray properties of these Serpens YSO classes by calculating plasma temperatures, foreground absorbing column density, and reddening for each YSO.  Although there were large uncertainties due to scatter in their data, they conclude that there is no clear observational motivation for invoking different mechanisms for generating X-ray emission in the Class I, II, and III phases.

Forbrich \& Preibisch (2007; henceforth FP07), investigated the infrared and X-ray properties of young stellar objects in the Coronet Cluster in the R CrA star-forming region.  Although Coronet is not nearly as rich in YSOs as the ONC, given the effective exposure time of the FP07 Chandra observations ($\sim$160 ks) and the proximity of the Coronet Cluster ($\sim$130 pc), this observation represents one of the most sensitive X-ray studies of a star-forming region, rivaling the COUP ONC survey.  FP07 examined the X-ray properties of 23 YSOs.  Their spectral fitting and light curve analysis provided plasma temperatures, column densities, and insight into the variable nature of young stellar objects.   FP07 found that  Class I and Class II objects tend to show hotter plasma temperatures than Class III YSOs.  FP07 also asserted that Class I YSOs tend to be more variable than Class II and Class III YSOs, possibly as a consequence of strong magnetic reconnection events in either stellar coronae or star-disk magnetic fields.

The Lynds 1630 (L1630) dark cloud \citep[D $\sim$ 400 pc;][]{Anthony-Twarog1982,Gibb2008} is part of the Orion B molecular cloud and is a well-studied star-forming region at radio, infrared and X-ray wavelengths, all of which are useful regimes for probing different aspects of pre-MS stellar evolution.   Radio and submillimeter observations trace the earliest stages of star formation, when a cold molecular cloud collapses to form a young stellar object (YSO).  Many such deeply embedded YSOs have been detected in L1630: dense cores of molecular gas harboring high-velocity outflows \citep{Snell1982, Gibb1993, Gibb1995}, radio continuum sources \citep{Bontemps1995, Verdes-Montenegro1996, Reipurth2002}, and luminous far-infrared and sub-mm sources \citep{Cohen1984, Cohen1987, Gibb1993, Mitchell2001}.  Infrared observations, useful for probing dust in infalling envelopes and circumstellar disks around YSOs, have also resulted in the detection of many low-mass, pre-MS (T Tauri) stars in L1630 \citep{Herbig1963, Wiramihardja1989, Fang2009, Megeath2012}.  A combination of X-ray and infrared observations has also been used to study embedded young stars in L1630  \citep{Simon2004}.

During the course of a multi-year campaign to monitor the optical/IR/X-ray outbursts of the YSO V1647 Ori \citep{Kastner2004, Grosso2005b, Teets2011, Hamaguchi2012}, we have obtained a series of Chandra X-ray observations, totaling $\sim$240 ks exposure, of a region of the L1630 dark cloud in the Orion molecular cloud.  In this paper we use these Chandra V1647 Ori monitoring data, combined with archival Spitzer data, to perform an infrared and multi-epoch X-ray study of the population of YSOs in L1630.  We present the reduction and analysis of the {\it Spitzer} and {\it Chandra} data in Section 2, the results obtained for YSO classification and interesting individual sources in Section 3, and a discussion of these results in Section 4.  The last section describes X-ray detection fractions, variability, and spectral characteristics with respect to YSO classification, and includes a comparison of our results to the aforementioned IR/X-ray studies of star-forming regions.


\section{Observations and Data Analysis}

\subsection{Spitzer and 2MASS IR Source Classification}

\begin{figure}
\centering
\includegraphics[scale=.31]{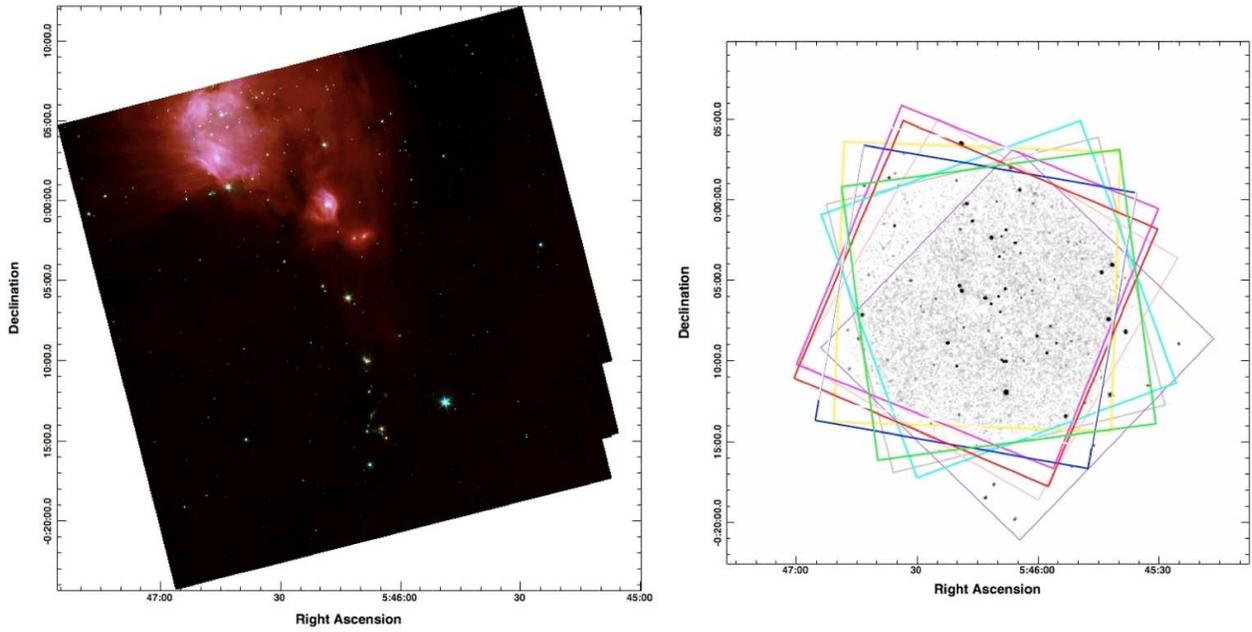}
\caption{\label{chan_spitz_fov22} {\it Left:} Spitzer RGB IRAC mosaic of L1630 constructed with images at 3.6, 5.8 and 8.0 $\mu$m.   {\it Right:} 11 merged Chandra X-ray observations (color inverted) totaling $\sim$240 ks of effective integration time.  Colored squares represent each Chandra FOV; 9 of the 11 are centered on the eruptive YSO V1647 Ori.}
\end{figure}

The Spitzer InfraRed Array Camera (IRAC) archival data for L1630 analyzed here comprises a subset of observations carried out in October 2004  \citep{Fazio2006,Fang2009}.  The observations were obtained using IRAC's High Dynamic Range (HDR) mode, yielding images with 10.4 second and 0.4 second exposure times.  The MOsaicker and Point source EXtractor (MOPEX) and the Astronomical Point source EXtractor (APEX)\footnote{http://irsa.ipac.caltech.edu/data/SPITZER/docs/dataanalysistools/tools/mopex/ Version 18.4.9} were used, respectively,  to create mosaics of L1630 using archival Spitzer IRAC observations and for performing aperture photometry on the resulting mosaics.    Two mosaics were made, one for the long (10.4 s) exposures (Fig. \ref{chan_spitz_fov22}, {\it left}) and one for the short (0.4 s)  exposures, so as to detect faint sources as well as bright sources that might have been saturated in the long exposure.  Before mosaicking the images, overlap correction was performed via the MOPEX overlap correction module.  The MOPEX sliding box routine was employed for background subtraction.   Aperture photometry was performed using an aperture radius of 10 pixels.  

\begin{figure}
\centering
\includegraphics[scale=0.65]{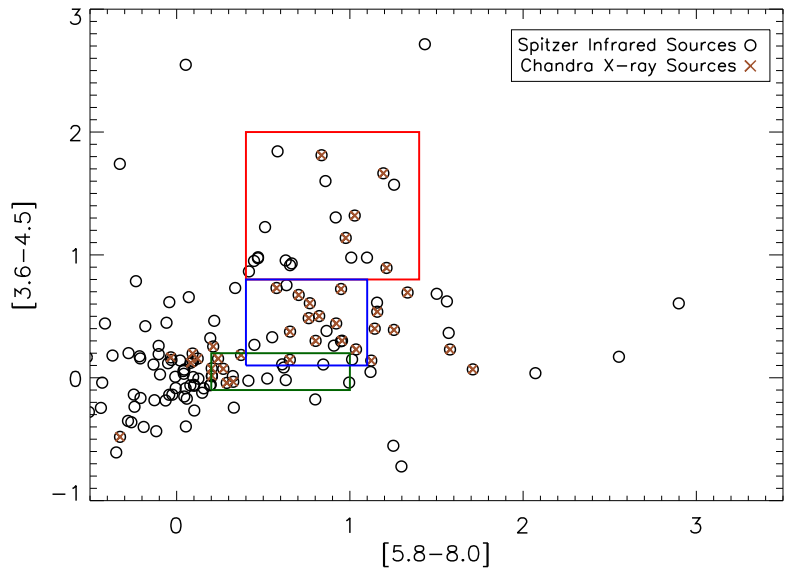}
\caption{\label{spitz_cc} Spitzer IRAC color-color diagram of L1630 for sources that lie within the Chandra field of view  (see Fig. \ref{chan_spitz_fov22}, right).  Colored boxes represent YSO classifications with green, blue and red boxes representing transition disks, Class II and Class 0/I YSOs respectively \citep{Fang2009}.  X-ray detected YSOs (Table \ref{xray_IR_22}) are overlaid with a brown X.  }
\end{figure}

A [3.6-4.5] vs. [5.8-8.0] color-color diagram (Figure \ref{spitz_cc}) was created to classify YSOs as Class 0/I, Class II, or "transition disks".  The last category represents objects whose spectral energy distributions (SEDs) indicate that they are in transition between Class II (disk-bearing) and Class III (apparently diskless) YSOs  \citep{Fang2009, Allen2004}.  In general, this classification scheme is useful for determining the evolutionary state of a YSO because disk dissipation leads to a sharp decrease in $\sim$ 1-10 $\mu$m emission. The Class 0/I, Class II, and transition disk YSOs were classified according to the Spitzer-based color-color criteria originally used in \citet{Fang2009}.  These classification criteria are as follows: (a)  0.4 $\leq$ $[5.8] - [8.0]$ $\leq$ 1.4 and 0.8 $\leq$ $[3.6] - [4.5]$ $\leq$ 2.0 (Class 0/I); (b) 0.4 $\leq$ $[5.8] - [8.0]$ $\leq$ 1.1 and 0.1 $\leq$ $[3.6] - [4.5]$ $\leq$ 0.8 (Class II);  (c) 0.2$\leq$ $[5.8] - [8.0]$ $\leq$ 1.0 and -0.1 $\leq$ $[3.6] - [4.5]$ $\leq$ 0.2 (transition disk).  Class III sources (defined as having little or no detectable IR excess above photospheric emission) were classified using X-ray observations and a 2MASS color-color diagram (see below). Only sources with a signal-to-noise ratio $>$ 3 were considered.  

Some sources display excess 8 $\mu$m emission in Figure \ref{spitz_cc}, placing them outside the standard YSO classification regions in color-color space.  After individual inspection, four of these infrared sources were found to be positionally correlated with diffuse, bow-shock-like infrared emission (Fig. \ref{bow}).  All four of these sources are also X-ray emitters.  Since the bow-shock-like extended infrared emission in the Spitzer observations contaminates the point-source flux at 8 $\mu$m, they have been tentatively classified assuming a smaller amount of intrinsic 8 $\mu$m emission (i.e., by reducing their values of [$5.8 - 8.0$] in Figure \ref{spitz_cc}).  

Chandra and Spitzer sources were positionally matched with the Tool for OPerations on Catalogues And Tables (TOPCAT), version 3.9.  X-ray and infrared sources were considered coincident if separated by $<$ 3 $''$ (to account for the PSF of off-axis Chandra detections and the pointing accuracy of Spitzer and Chandra) and if they met the aforementioned requirements for both infrared and X-ray detection.  Infrared colors for X-ray sources with infrared counterparts are listed in Table \ref{xray_IR_22}.  The infrared luminosity $L_{IR}$ was determined by integrating the 2MASS and Spitzer flux from 1.23 $\mu$m to 8.0 $\mu$m.  Infrared SEDs created using 2MASS, Spitzer and WISE photometry for all X-ray sources with infrared counterparts are presented in the Appendix.  Some SEDs in the Appendix show WISE excesses $>$ 8 $\mu$m that could potentially be signatures of a cool dust component not accounted for in our (2MASS/Spitzer-based) classifications.  However, many of these sources have been flagged by WISE pipeline reduction procedures and are likely either spurious detections, contaminated by a diffraction spike or scattered light halo from a nearby bright source, or are highly infrared-variable.

\begin{figure}
\centering
\includegraphics[scale=0.65]{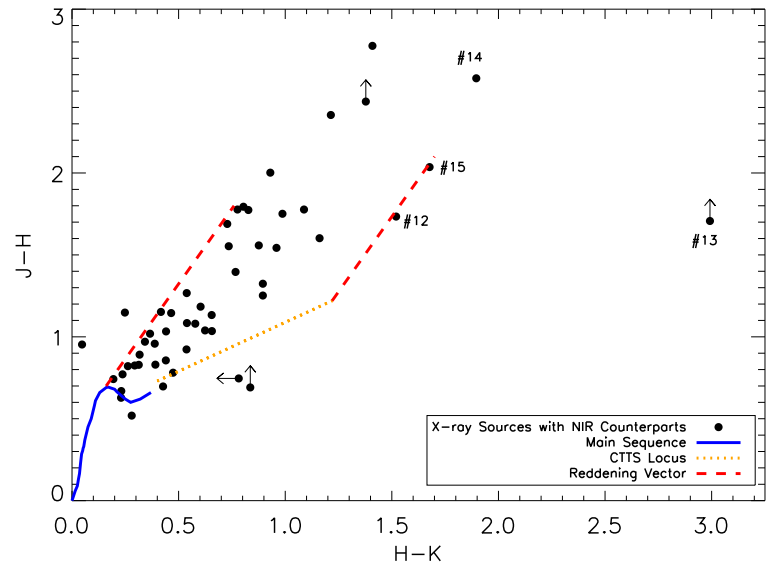}
\caption{\label{2mass_cc} 2MASS color-color diagram of near-infrared sources with X-ray counterparts in L1630. The blue line indicates where unreddened main sequence stars would lie on this plot. The cTTS locus is taken from \citet{Meyer1997}. The reddening vectors represent $A_{V} = 10$.}
\end{figure}

In order to ensure classification of very red (potential Class 0 or Class I) sources, we required detection at 5.8 $\mu$m for all YSOs classified in this study other than Class III.  This requirement led to the inclusion of some sources that were not included in the photometric study of \citet{Fang2009}, who required detection in both 3.6 $\mu$m and 4.5 $\mu$m IRAC bands.  This condition also potentially leads to the exclusion of many Class III YSOs from Fig. \ref{spitz_cc}.  To include these X-ray luminous objects, we compiled a list of 2MASS counterparts to X-ray sources and extracted their photometry in the J, H and K bands using the NASA/IPAC Infrared Science Archive (Table \ref{xray_IR_22}).  We then constructed a near-infrared color-color diagram (Fig. \ref{2mass_cc}) to identify Class III YSOs.  The main sequence line and reddening vectors are taken from \citet{Bessell1988} and \citet{Cohen1981} respectively.  The classical T Tauri star (cTTS) locus is the area in the plot where cTTSs suffering little or no reddening are located \citep{Meyer1997}.  Given that pre-MS stars are typically $\sim 10^{3}$ brighter in X-rays than main sequence stars \citep{Feigelson1999}, we conclude that all X-ray sources in L1630 that appear as reddened main-sequence stars are likely Class III pre-MS stars.

\subsection{Chandra X-ray Analysis}

	The analysis is focused on 11 Chandra X-Ray Observatory observations obtained with the Advanced CCD Imaging Spectrometer array (ACIS-I), spanning four years, of the star forming region L1630.  Individual exposure times were 18-30 ks.  V1647 Ori, a well-studied  highly variable YSO, was the target in 9 of these 11 observations.  Details concerning results for V1647 Ori itself can be found in \citet{Teets2011}, \citet{Hamaguchi2012} and \citet{Kastner2006}.  The date and exposure time of each observation included in the present study are listed in Table \ref{chan_obsid}.    X-ray data reduction was performed using Chandra's Interactive Analysis and Observation (CIAO)  \footnote{http://cxc.harvard.edu/ciao/index.html CIAO version 4.3}  version 4.3.  The images were cropped to contain only data from the ACIS-I imaging array.  All 11 observations were merged with CIAO's \verb+dmmerge+ and \verb+reproject-events+ to generate the master merged event file and merged exposure map used for source detection.  The merged image is shown in Figure \ref{chan_spitz_fov22} {\it (right)} with the respective fields of view of the 11 component exposures represented by colored boxes.



  All Chandra X-ray observations, including the $\sim$240ks merged observation, were filtered into soft (0.5-2.0 keV), hard (2.0-8.0 keV) and broad (0.5-8.0 keV) energy bands.  We removed photons with energies $<$ 0.5 keV and $>$ 8.0 keV so as to avoid spurious and background events.  The CIAO process \verb+fluximage+  was used to create exposure maps for each observation.  All exposure maps were created with an effective energy of 1.6 keV.  Tests of exposure maps constructed with effective energies larger than 1.6 keV do not significantly affect the results.  The effective exposure time is calculated for each source using a merged exposure map that accounts for instrumental effects such as effective area, quantum efficiency, and telescope pointing motion.  To determine count rates and other source properties, CIAO's \verb+wavdetect+ source detection algorithm was employed in each band in all observations, including the final merged observation, with the detection probability threshold set to P=$10^{-6}$.  The wavdetect scales (i.e., wavelet radii in pixels) used were 1, 2, 4, 8, and 16 in order to optimize detection of point sources well off-axis.  The wavdetect source extraction region is based on the Chandra PSF and chip position and is calculated considering a PSF energy of 1.4967 keV and an encircled energy fraction of 40\%. The faintest source detected by wavdetect in any of our observations had four events; therefore, at least four events are needed to be considered a reliable on-axis detection in this study. 
  
   Source count rates were determined using exposure maps to account for the effective exposure time corresponding to the position of each source. Hardness ratios, defined as $(H-S)/(H+S)$, were calculated using the hard (H) and soft (S) count rates determined for each source.  Median X-ray energies (Table \ref{xray_IR_22}) were determined for the X-ray detected L1630 YSOs using {the wavdetect position for each source and} the CIAO tool \verb+dmstat+.  Errors on these median energies were calculated using the standard error on the mean as a proxy for standard error on the median. Values of $N_{H}$ for all sources  were estimated using the relationship between median energy and absorption in YSOs determined by \citet{Feigelson2005}; for sources with more than 100 counts, $N_{H}$ was calculated using the best-fit model flux.  Assuming a distance to L1630 of 400 pc, intrinsic X-ray luminosity was calculated from the best-fit flux model.  For sources with fewer than 100 counts, X-ray luminosity was not estimated.  
  
\begin{figure}
\centering
\includegraphics[scale=0.65]{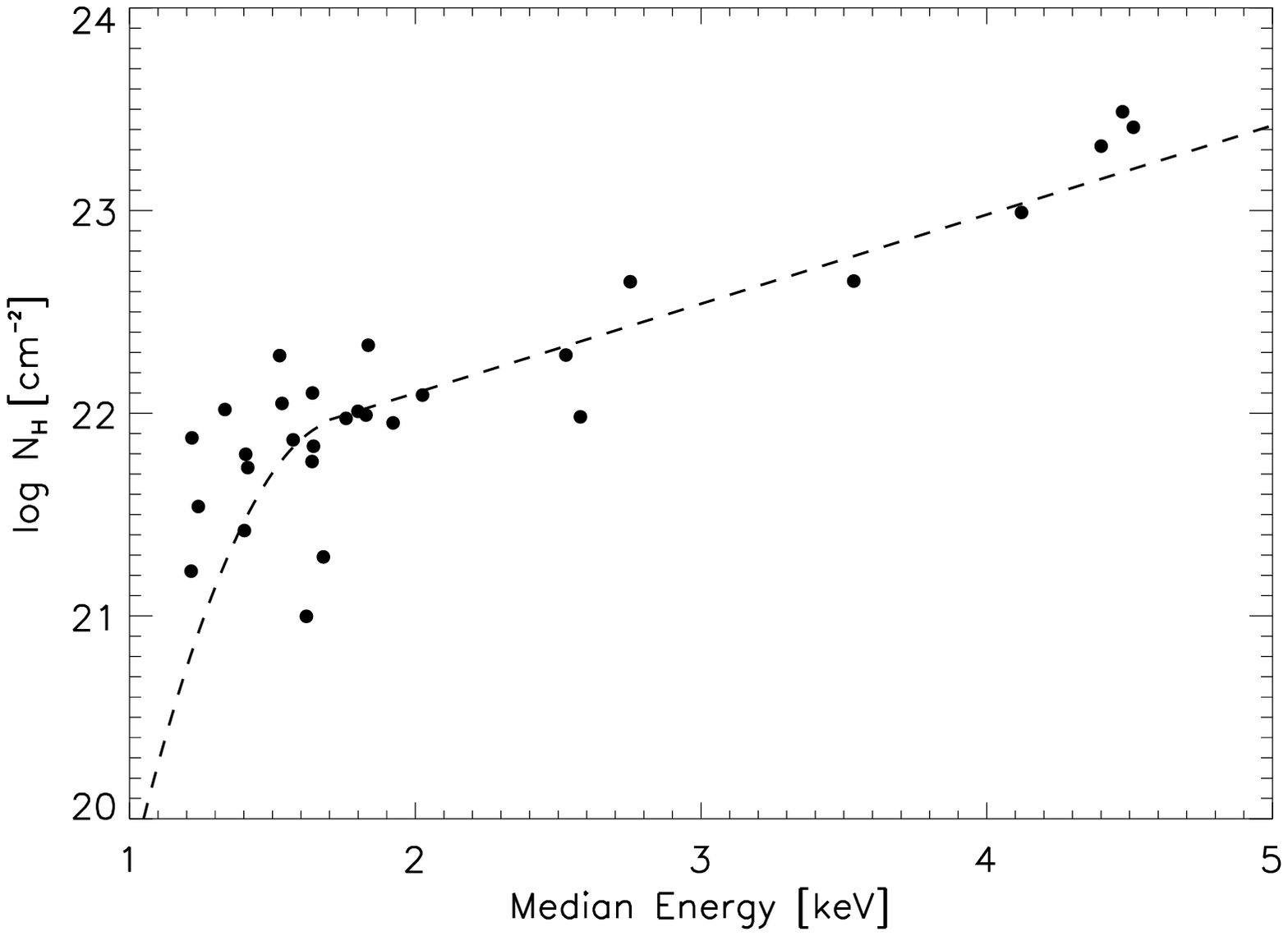}
\caption{ \label{nh_medE} Spectrally derived absorption for X-ray emitting YSOs with > 100 counts.  Dashed curve is a best-fit relationship taken from \citet{Feigelson2005} representing the relationship between median energy and column density of YSOs in the COUP survey. }
\end{figure}

The flux detection limit of the Chandra data-set was determined using the Portable, Interactive Multi-Mission Simulator (PIMMS)\footnote{http://heasarc.gsfc.nasa.gov/docs/software/tools/} software developed by the NASA High Energy Astrophysics Science Archive Science Center.  In a 234 ks observation, one X-ray photon corresponds to an energy flux of $2.4$ $\times$ $10^{-17}$ erg s$^{-1}$cm$^{-2}$, assuming an intrinsic source spectrum consisting of a 10 MK thermal plasma with a metal abundance of 0.4 times solar \citep{Getman2005, Forbrich2007}, a representative (Figure \ref{nh_medE}) intervening column density of $7.5$ x $10^{21}$ cm$^{-2}$, and adopting an energy band of 0.5-8.0 keV.  At our assumed distance of 400 pc to L1630, an on-axis four event source detection therefore corresponds to an intrinsic luminosity of 1.8 $\times$ $10^{27}$ erg s$^{-1}$.

\begin{figure}
\centering

\includegraphics[scale=0.46]{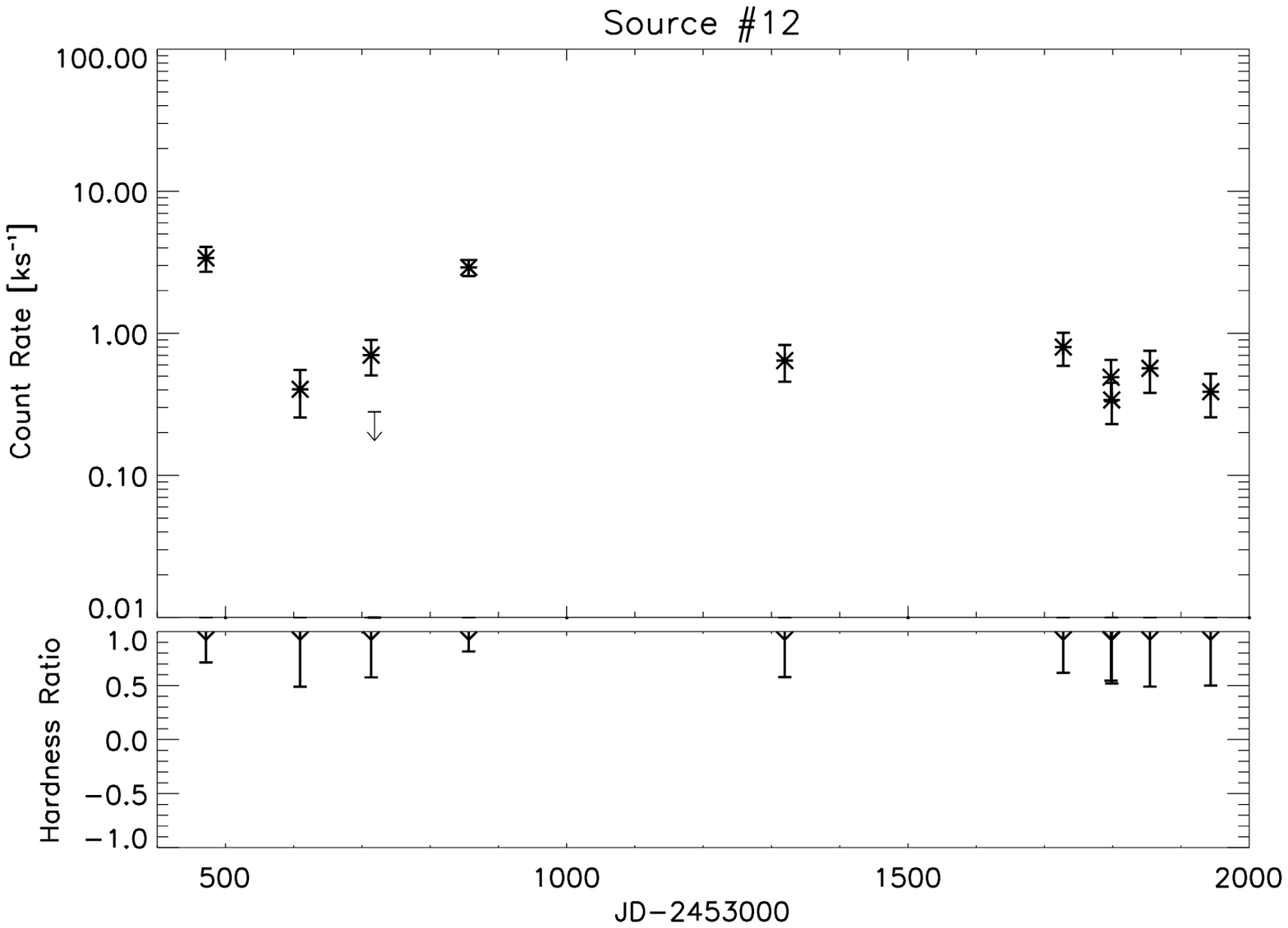} 
\includegraphics[scale=0.46]{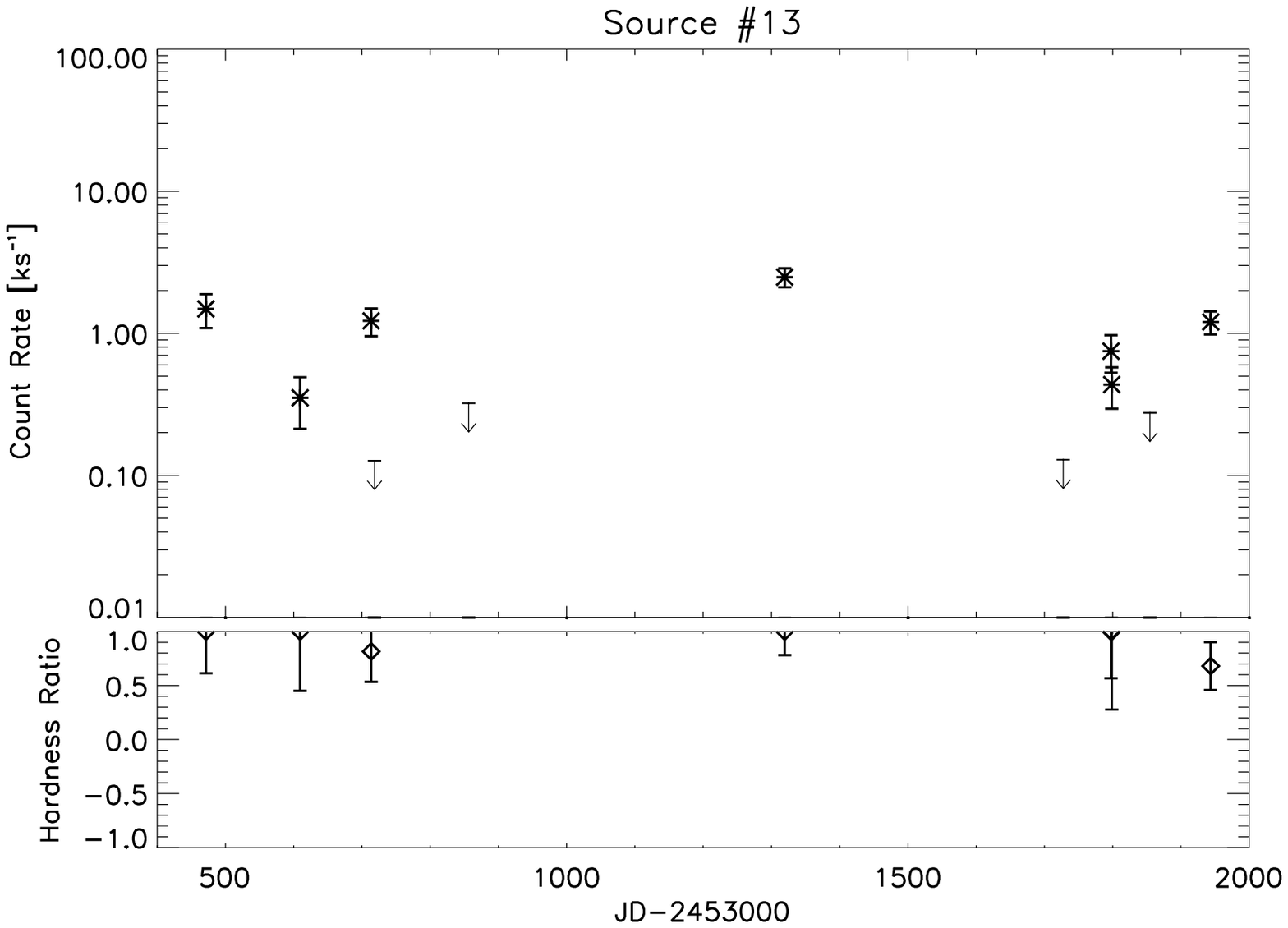} 
\includegraphics[scale=0.46]{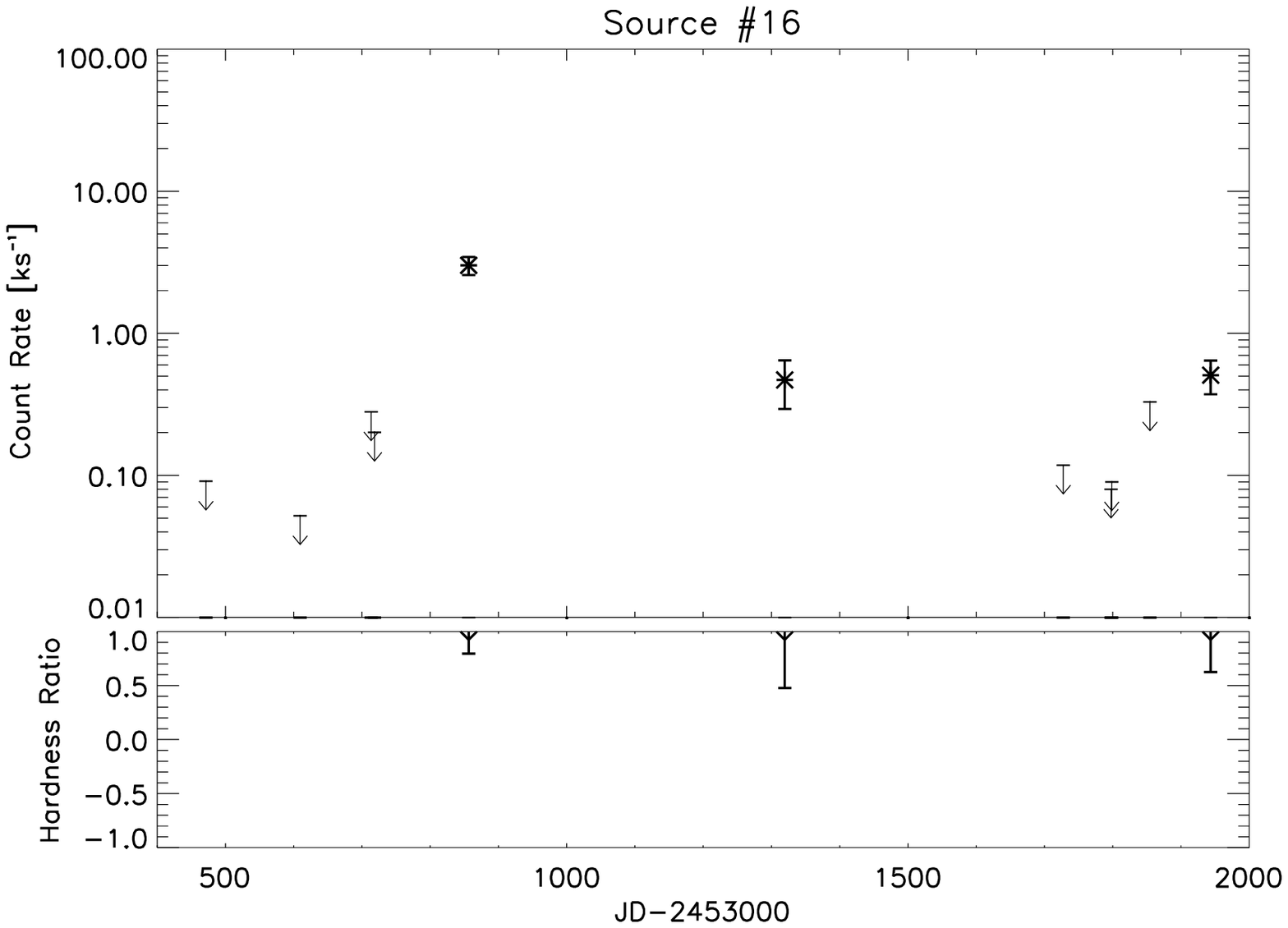} 
\includegraphics[scale=0.46]{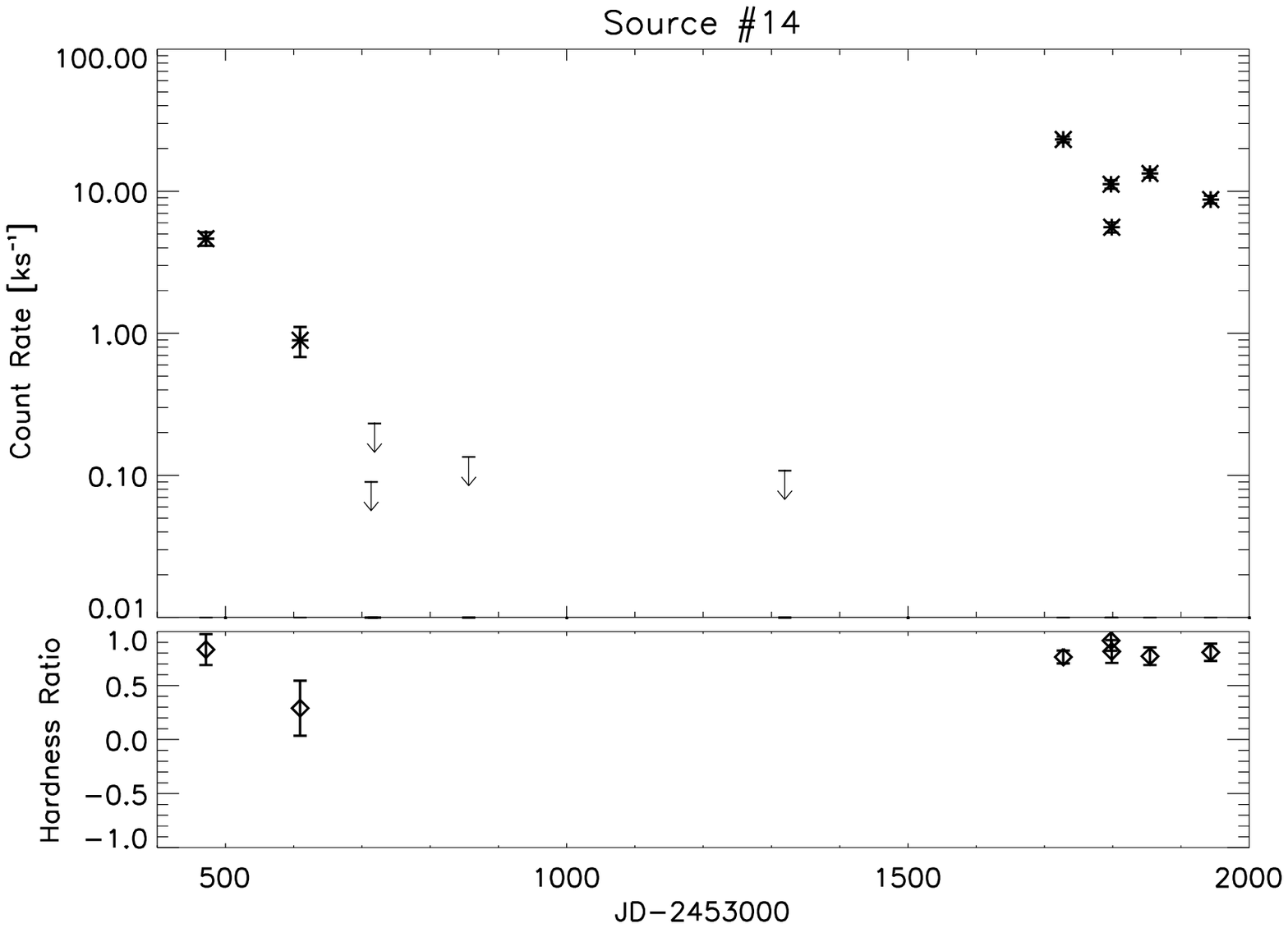} 

\caption{\label{lc_1} Count rate and hardness ratio variability over $\sim$ 4 years for Class 0/I YSOs in L1630.  }
\end{figure}
\begin{figure}
\centering

\includegraphics[scale=0.46]{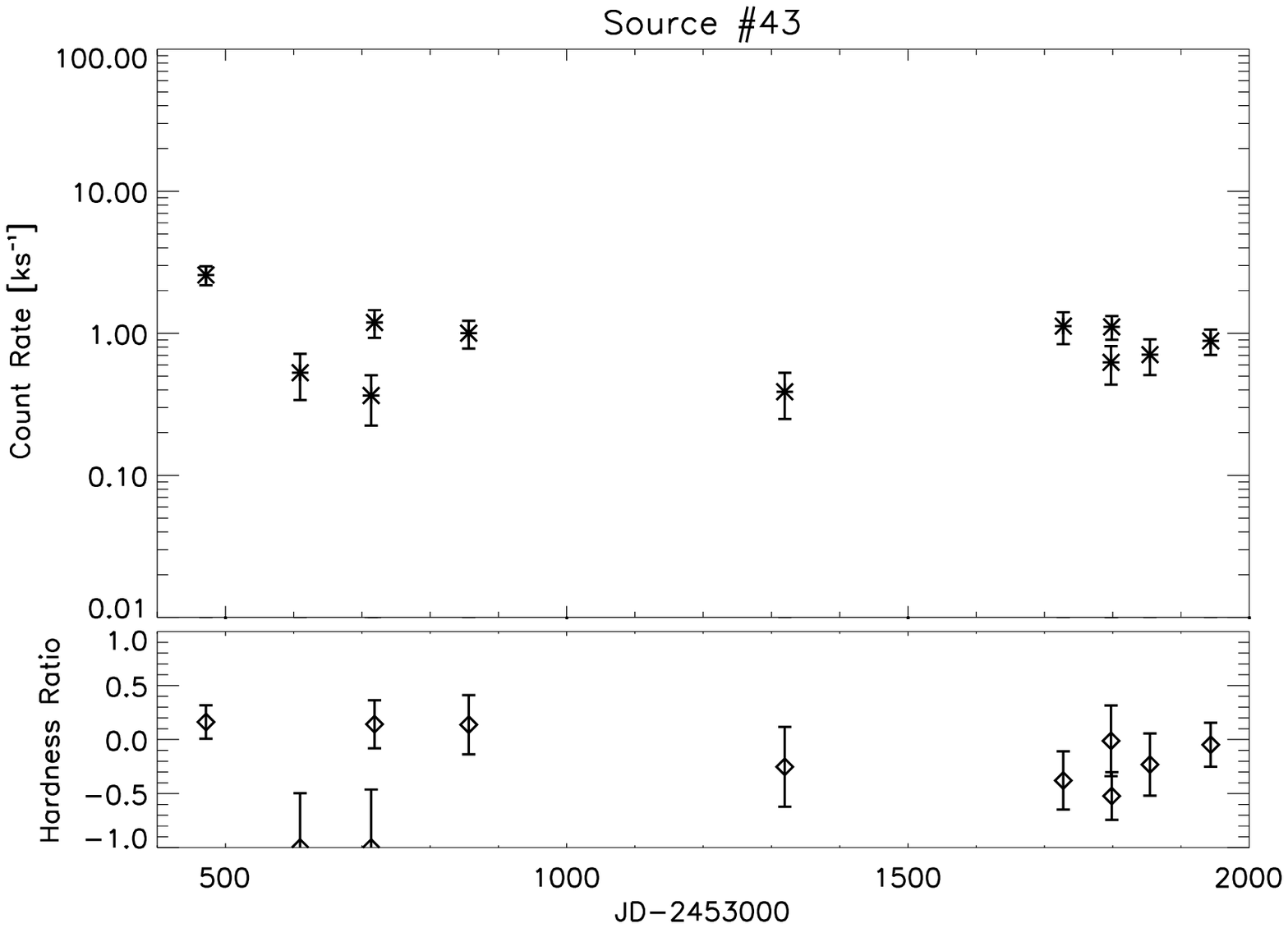} 
\includegraphics[scale=0.46]{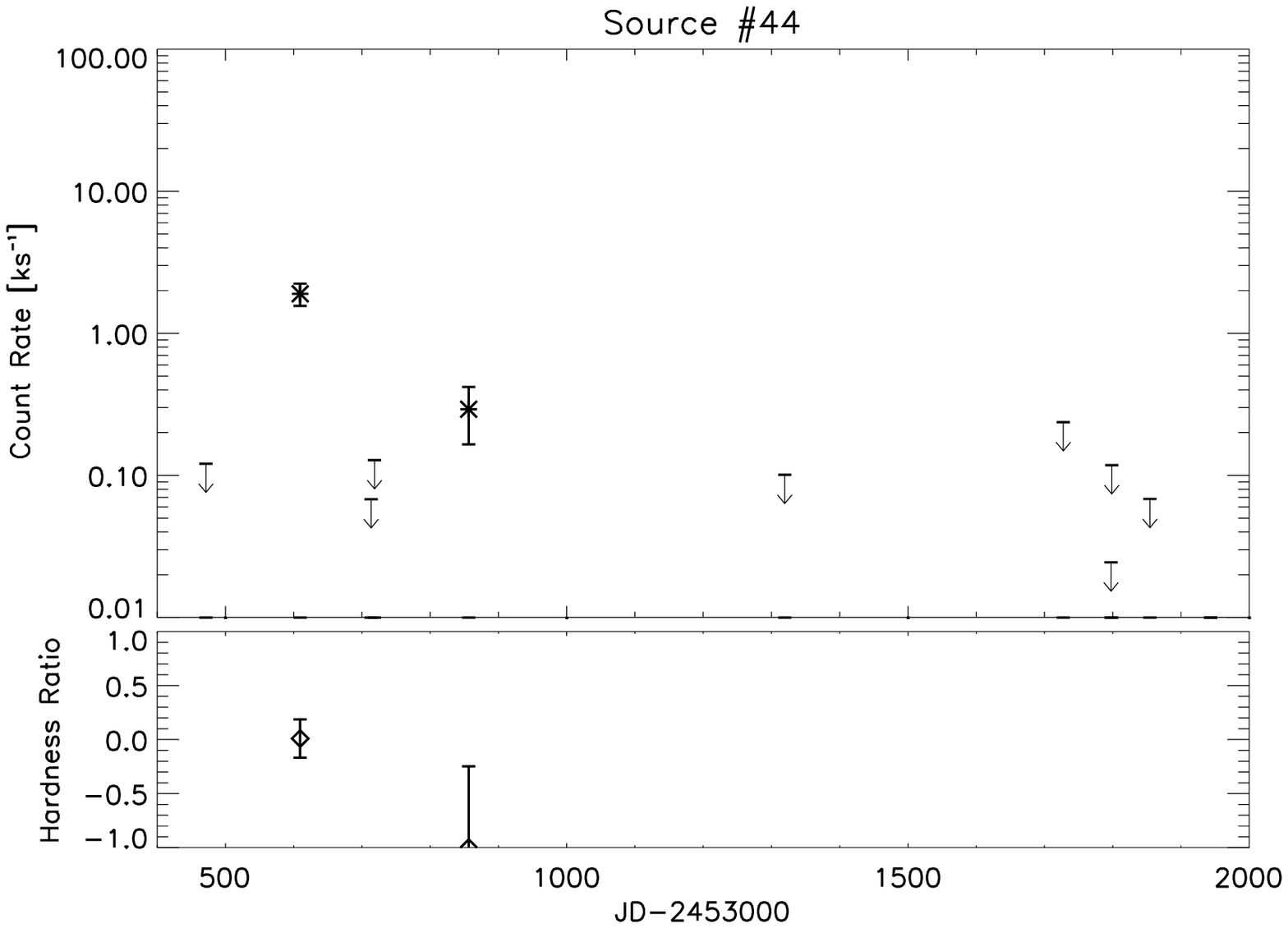} 
\includegraphics[scale=0.46]{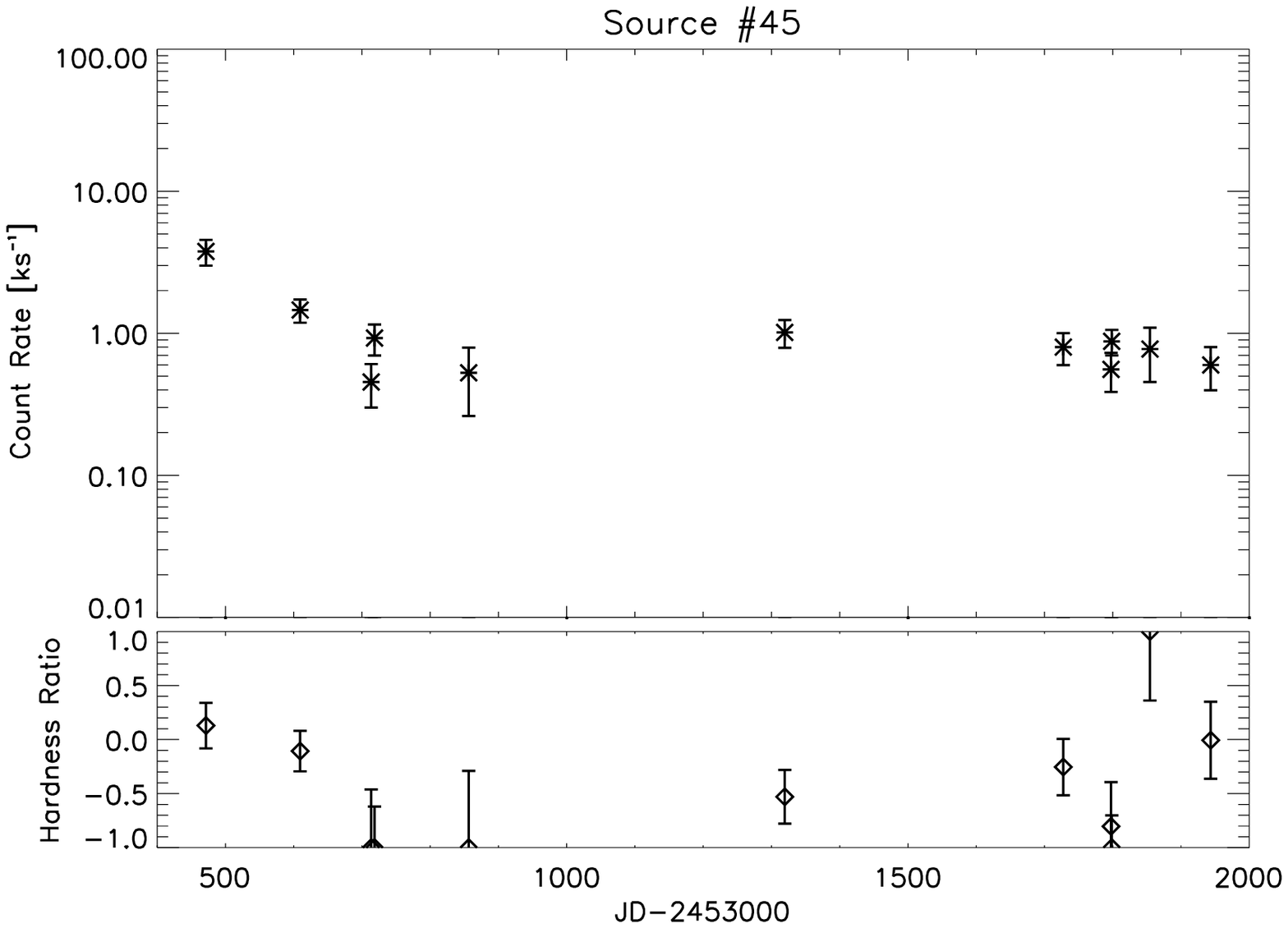} 
\includegraphics[scale=0.46]{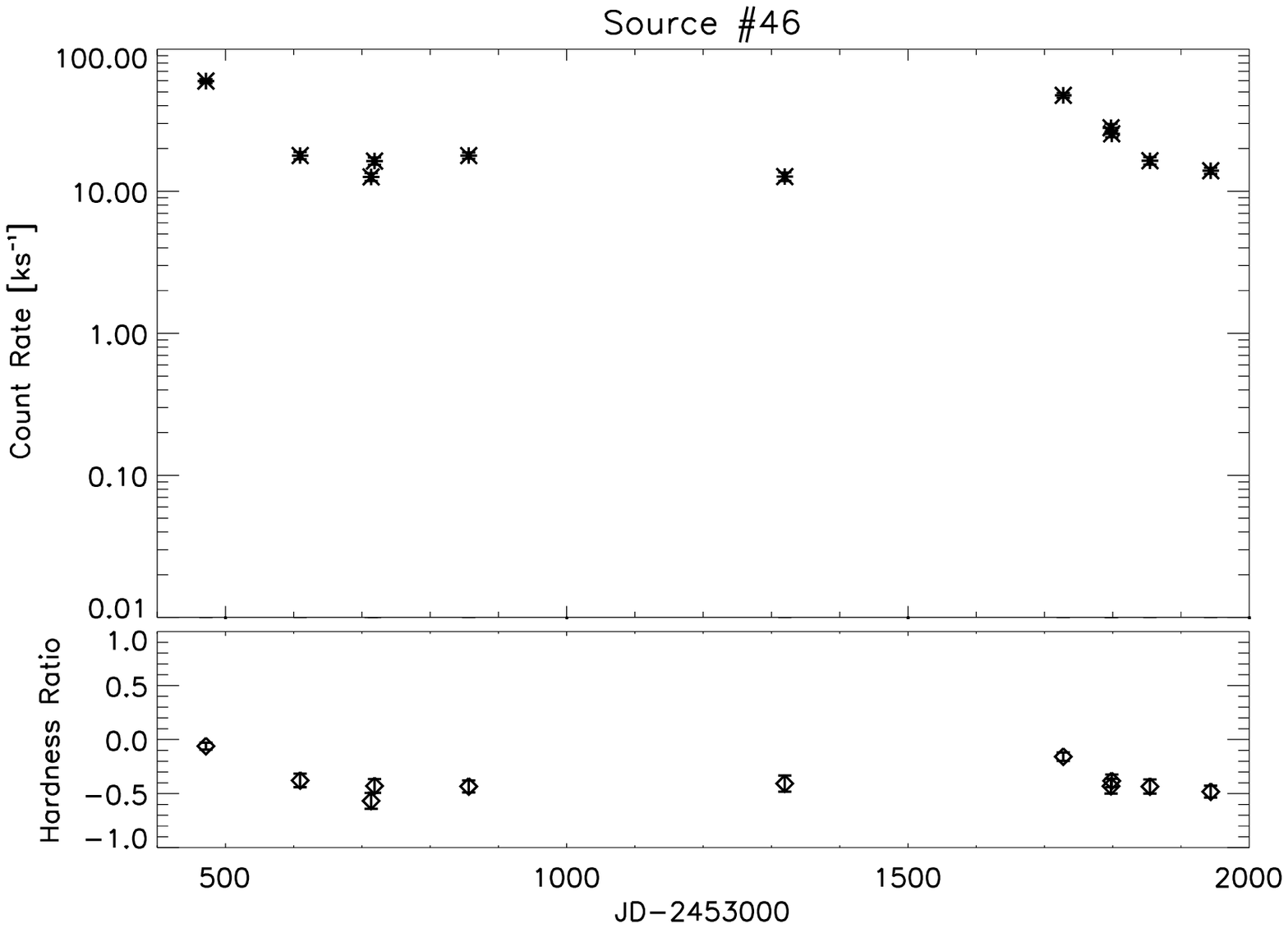} 
\includegraphics[scale=0.46]{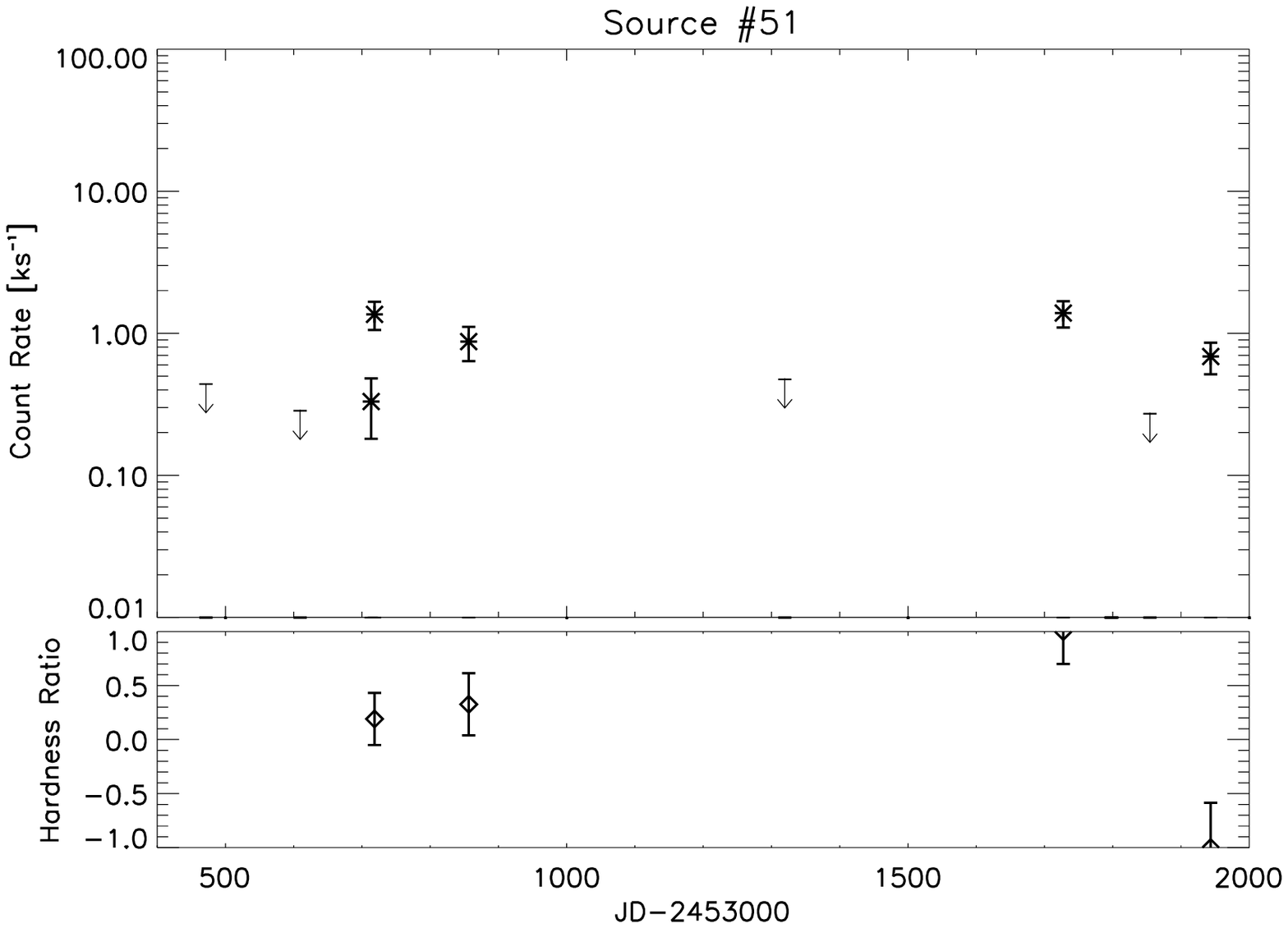} 
\includegraphics[scale=0.46]{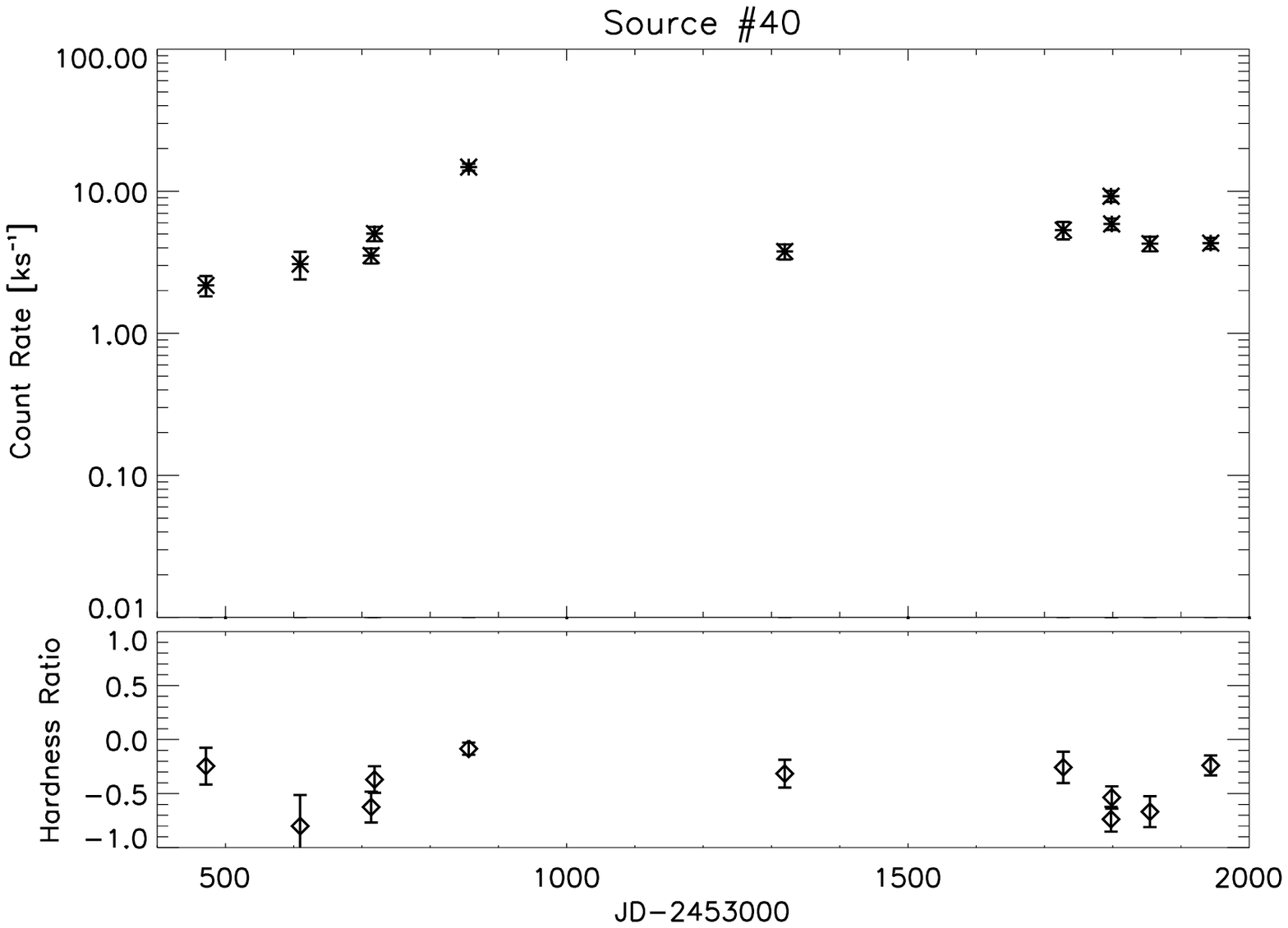}

\caption{\label{lc_2}Count rate and hardness ratio variability over $\sim$ 4 years for Class II YSOs in L1630.   }
\end{figure}

X-ray spectral analysis was performed on all sources with $\gtrsim$ 100 counts in the merged observation. CIAO's \verb+specextract+ routine was run for each source to generate/extract spectra and associated spectral response files for each individual observation.  If more than 10 events were detected from a source in any particular observation, a merged spectrum for that source was created using the \verb+specextract+ "combine" parameter.   It is important to note that the resulting merged spectrum from multiple observations could be strongly influenced by source variability (see, e.g., Figures \ref{lc_1}, \ref{lc_2}, and \ref{xray_var}).  Many of the sources did not have enough counts in an individual observation to constrain the spectral fitting; therefore, we only performed spectral analysis on the combined observations.

  Models of absorbed one- or two-temperature thermal plasmas were fit to the merged spectra to estimate plasma temperature, intervening hydrogen absorbing column density, and flux. We performed our spectral analysis with HEASARC's X-ray Spectral Fitting Package (XSPEC) using the \verb+wabs+ absorption model and the \verb+apec+ plasma model, assuming a uniform density plasma with a global pattern of abundances fixed to the value $Z=0.8$ $Z_{\odot}$. If the reduced $\chi^2$ of a two-temperature fit was smaller than for a one-temperature fit, an F-test was performed to ensure the fit was improved statistically (i.e., that the F-Test P $<$ 0.05).  

The X-ray spectral properties for each source are listed in Table \ref{spec_tbl}, and mean values for each class are listed in Table \ref{avg_prop}.  The count rate of each source listed in Table \ref{spec_tbl} was calculated from the merged observation and the exposure map corrected effective exposure time.  The remaining parameters listed in Table \ref{spec_tbl} are the best-fit model parameters from spectral fitting. The listed values for $F_{X}$ are observed (i.e., "absorbed") X-ray fluxes, while the values of $L_{X}$ are intrinsic (i.e., "unabsorbed") X-ray luminosities. 


  To be considered for our variability analysis (Figures \ref{lc_1}-\ref{xray_var}; Sect. 4.2),  X-ray emitting YSOs had to meet the following criteria:  (1) be detected with Spitzer 5.8 $\mu$m emission or have little near-infrared excess (to preserve a Class III sample), (2) lie in the field of view of at least 9 of the 11 Chandra observations, (3) be detected in at least two of the 11 Chandra observations, and (4) have a signal to noise ratio $\geq$ 3 in both Chandra and Spitzer photometry.   Each source's broad count rate was integrated over the entire exposure and accounts for effective exposure time.  Upper limits for non-detected sources in individual exposures were calculated with the CIAO tool  \verb+aprates+.  The location and shape of the source, as detected by \verb+wavdetect+ in the merged Chandra observation, was used to calculate upper limits on the count rate at the location of any source not detected in an individual observation. 

 The variance of each X-ray emitting YSO in our variability analysis ($s^{2}$) was calculated over the course of all 11 observations and normalized to its mean broad band count rate 
 \begin{equation*}
{\rm \mathit{s^{2}}}=\frac{\sum (x-\bar{x})^{2}}{\bar{x}^{2}(N-1)}
\label{var_eq}
\end{equation*}
where $x$ is the broad (0.5-8.0 keV) count rate per observation, $\bar{x}$ is the mean count rate, and N is the number of observations. The count rate upper-limit was used for non-detected sources in individual observations. If a source was detected in fewer than 6 of the 11 observations, an upper limit was used to determine the count rate ($x$) and the mean count rate ($\bar{x}$) for the ($s^{2}$) lower-limit calculation.  The variance values of these objects are hence considered to be lower limits.


\section{Results }

\subsection{X-ray Characteristics and YSO Class}
The sources in the upper-right of the 2MASS color-color plot in Figure \ref{2mass_cc} ($[H-K] > 1.5$) have large infrared excesses and are heavily reddened, and therefore most likely represent highly embedded Class I YSOs (e.g., sources \#12, \#13, \#14 and \#15).  Source \#16 was not detected at J, H or K wavelengths and therefore we consider it Class 0/I candidate.  These inferences are supported by classifications of the same five embedded YSOs based on Spitzer IRAC colors.  These five (Class I and candidate Class 0/I) are hereafter referred to, collectively, as the L1630 Class 0/I sample.

\begin{figure}
\centering
\includegraphics[scale=0.60]{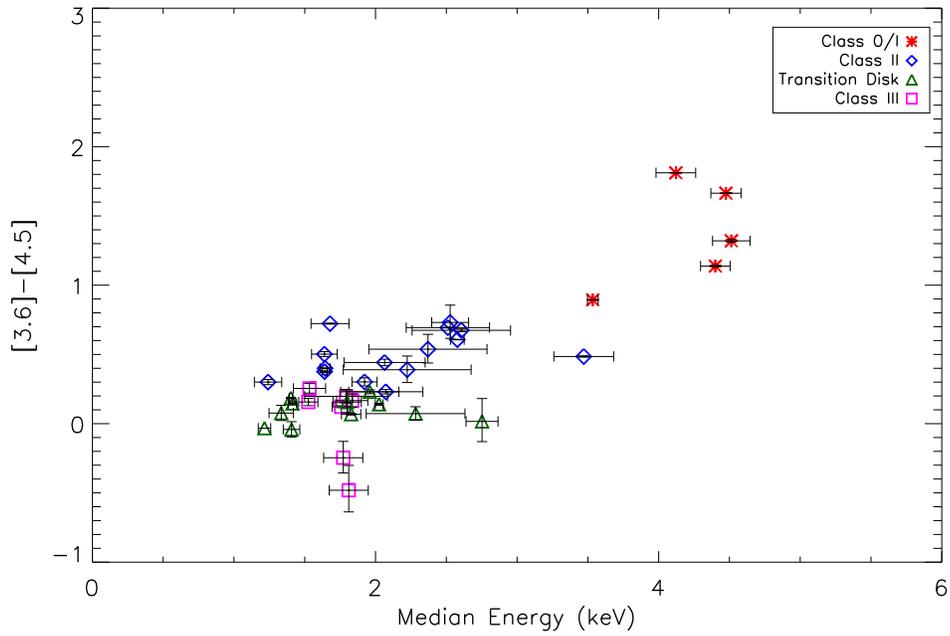}
\caption{\label{medE_cc} IRAC [3.6]-[4.5] color vs. Chandra median X-ray photon energy for YSOs in L1630.  Median X-ray photon energies were calculated for source photons extracted from the merged Chandra observation.  Error bars for the median energy were calculated using the standard error of the mean. Color error bars are over plotted and in most cases are smaller than the plot symbol.} 
\end{figure}

Infrared colors can be used in conjunction with median X-ray photon energy to distinguish between highly embedded Class 0/I YSOs and other sources.  Soft X-rays ( $\lesssim 1$ keV)  are susceptible to absorption by intervening cloud and/or disk material.  This allows estimates of the line-of-sight absorbing column solely on the basis of the median energy of the X-ray source \citep{Feigelson2005, Getman2010}.  Figure \ref{medE_cc} makes apparent that, for L1630 YSOs, median energy is correlated with infrared colors; i.e., that redder infrared sources have larger median energies.  The relationship from \citet{Feigelson2005} was used to estimate the absorbing columns for X-ray sources in L1630 listed in Table \ref{xray_IR_22}.  The resulting typical absorbing columns determined for L1630 YSOs lie in the determined range N$_{H}$ $\sim$ 10$^{19}$-10  $^{22}$ cm$^{-2}$ for Class II, Class III and transition disk YSOs, whereas highly embedded Class 0/I sources have estimated N$_{H}$ values as high as 10$^{23}$ cm$^{-2}$. The value of N$_{H}$ has also been determined via X-ray spectral modeling for strong ($>$ 100 count) X-ray sources.  The relationship between best-fit column density and median energy for these X-ray-bright YSOs is shown in Figure \ref{nh_medE} and is evidently well described by the best-fit curve relating $N_{H}$ and median energy for ONC YSOs \citep{Feigelson2005}.  The spectrally derived $N_{H}$ for L1630 YSOs are in good agreement with those measured from the COUP survey.

\begin{figure}
\centering

{\includegraphics[ scale=0.46]{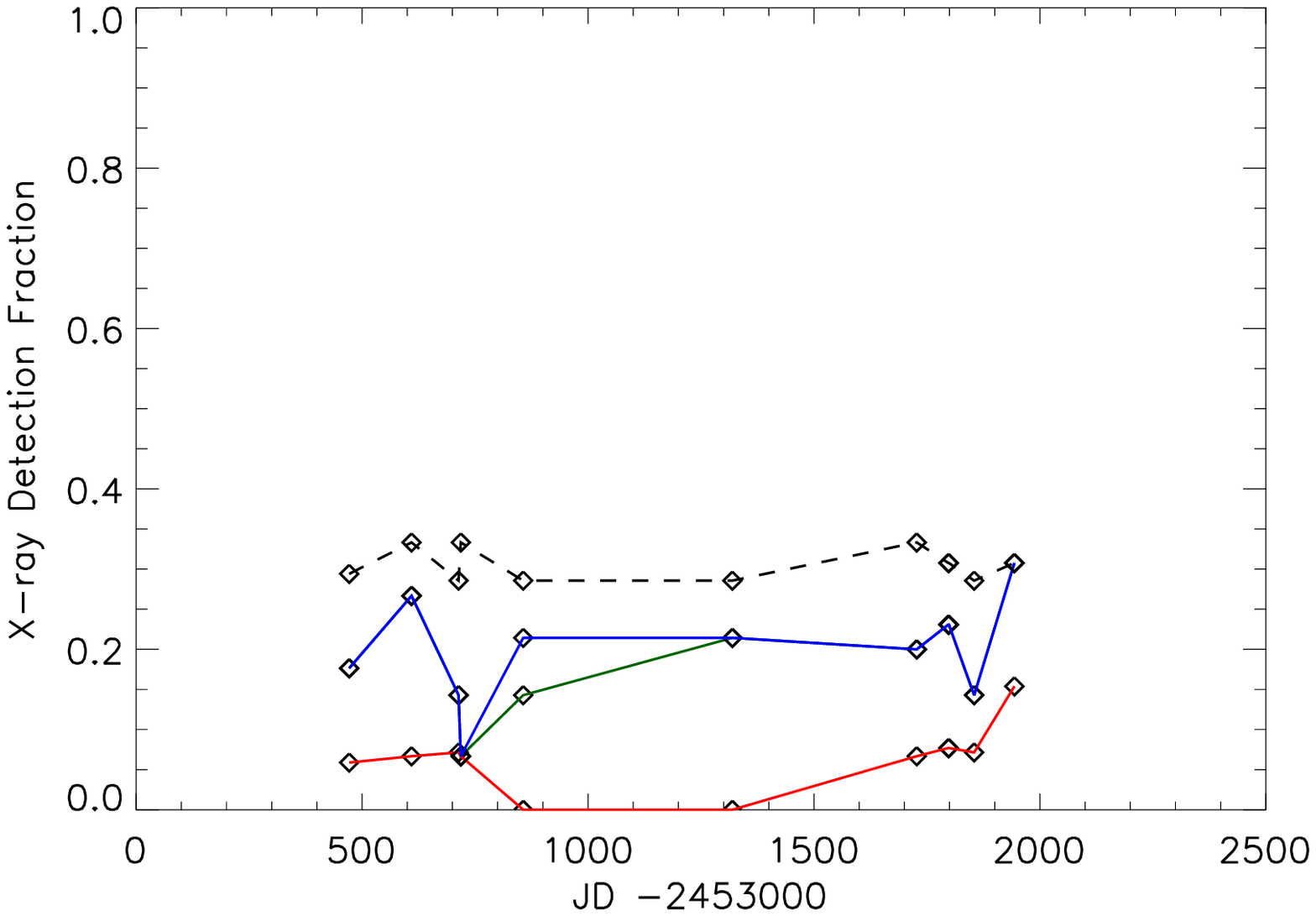}}
{\includegraphics[ scale=0.46]{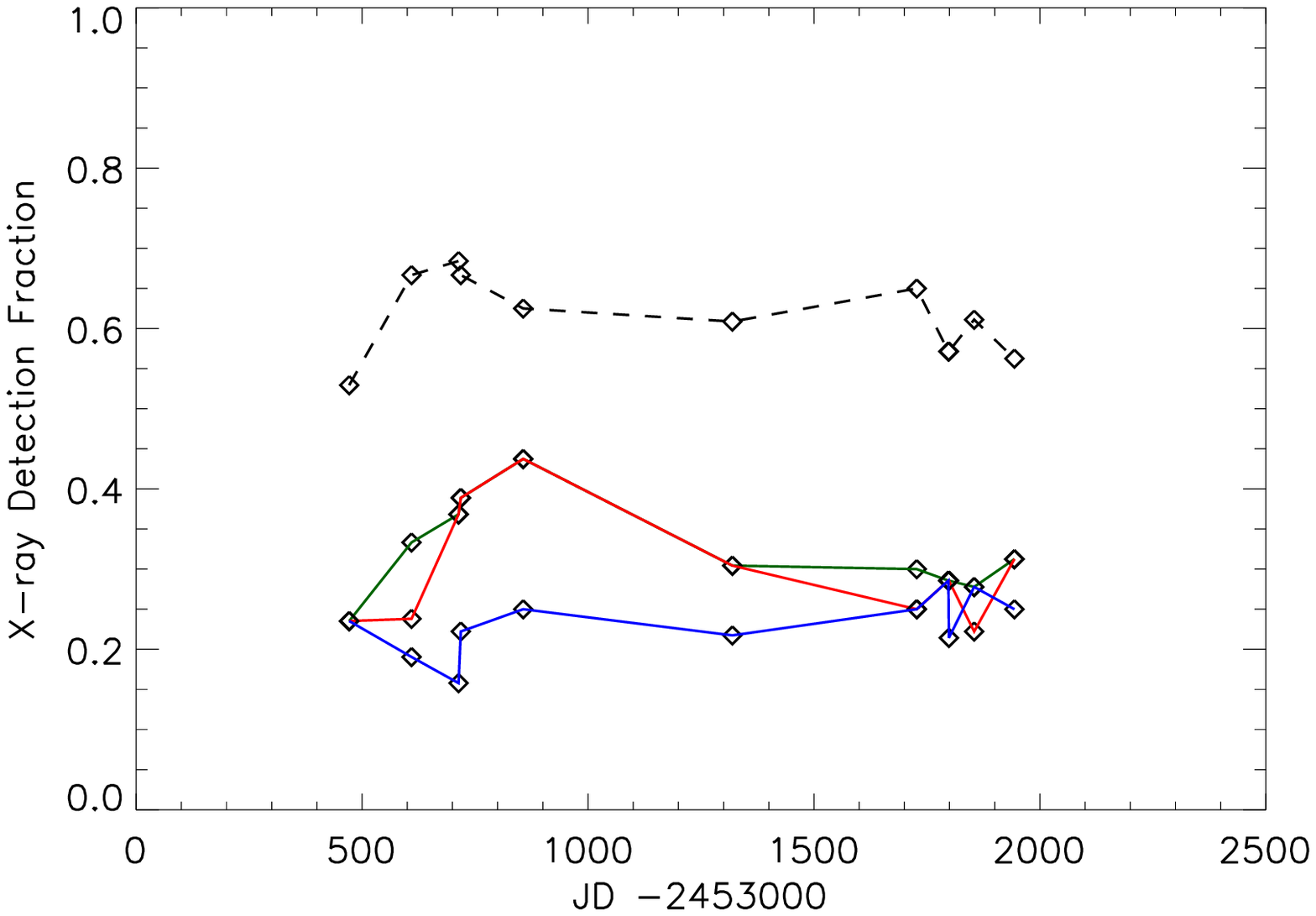}}
{\includegraphics[scale=0.46]{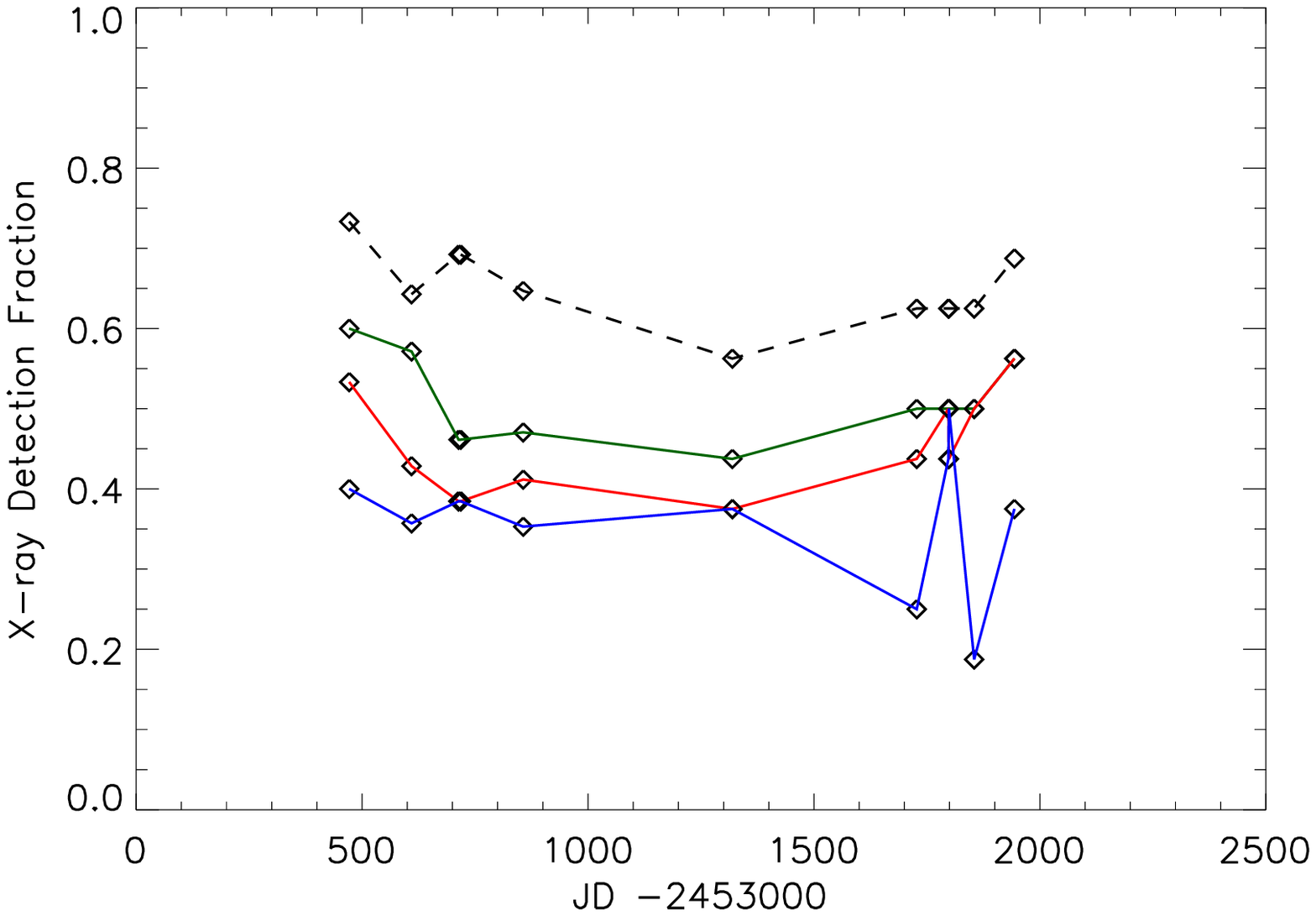}}
{\includegraphics[scale=0.46]{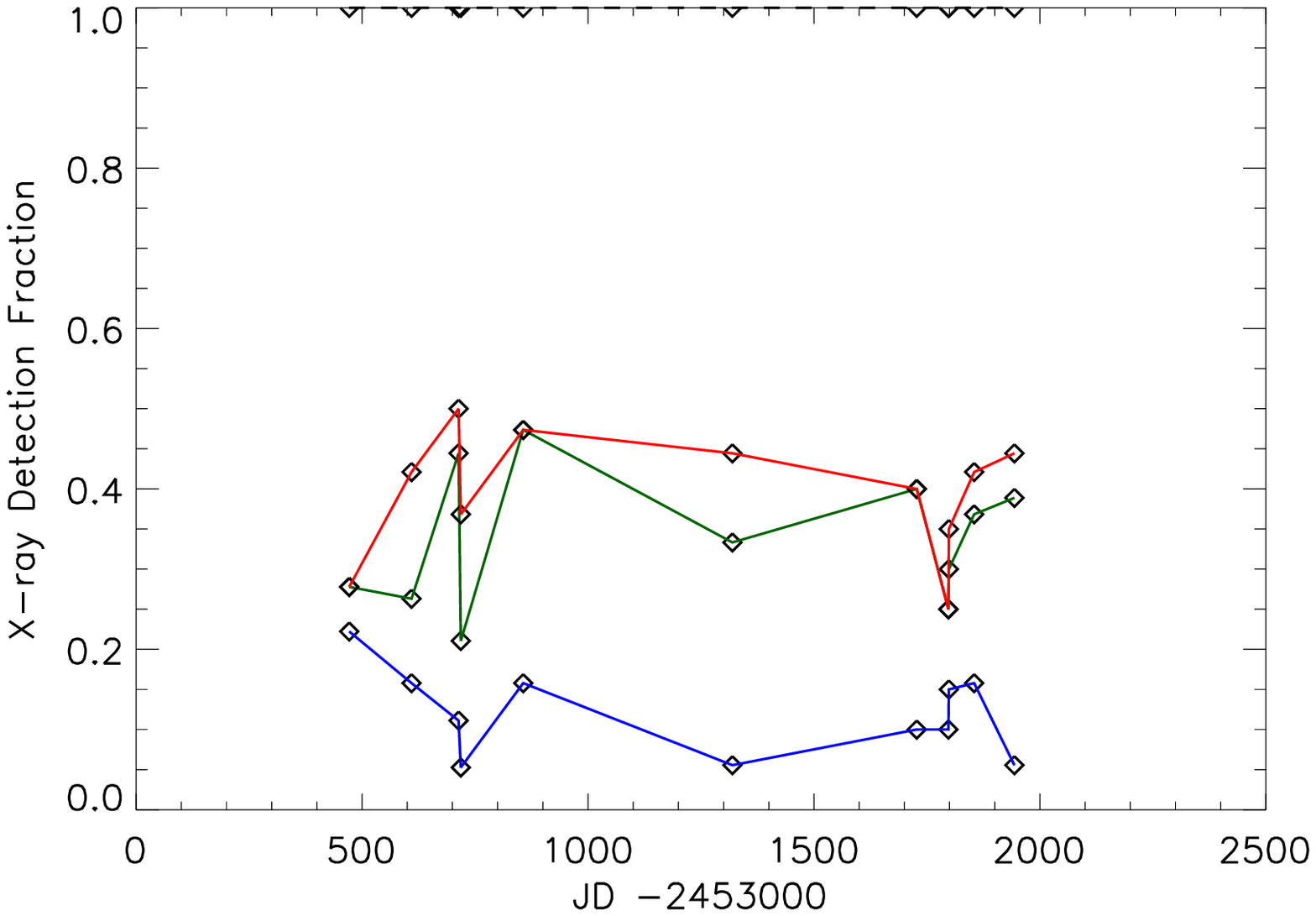}}

\caption{\label{yso_frac} X-ray detection fraction by YSO classification in L1630; Class 0/I (top-left), Class II (top-right), Transition Disk (bottom-left), and Class III (bottom-right).  Solid red, blue and green lines represent soft (0.5-2.0 kev), hard (2.0-8.0 keV) and broad (0.5-8.0 keV) energy filters respectively. The black dashed line represents the maximum possible X-ray detection fraction of YSOs using the merged $\sim$240 ks image and, thus, indicates the largest possible X-ray detection fraction for each observation. Since X-ray emission is a required condition for Class III YSO candidacy, the total X-ray detection fraction of Class III YSOs is 1.0 by definition.  Observation-specific fields of view are partly responsible for differing total X-ray detection fractions from exposure to exposure.}

\end{figure}

  Of the 59 X-ray sources detected in the merged broad band filter, 40 were detected in the hard band filter and all 59 were detected in the soft band filter. The total number of X-ray-emitting YSOs in each observation is listed in Table \ref{yso_fov} where S, H, and B correspond to soft, hard, and broad energy filters, respectively.  In a few cases, low count rate sources were detected in the broad energy filter but not in the soft or hard filters.  In Figure \ref{yso_frac}, the detection fraction, i.e., the ratio of X-ray-emitting YSOs to non-X-ray emitting YSOs, for each observation is plotted by YSO classification. Class 0/I sources have the lowest detection fraction, whereas transition disks and Class II YSOs are more readily detected (See Sect. 4.1).  Figures and tables in this paper combine Class 0 and Class I YSOs (i.e., Class 0/I) because we seek to compare the X-ray detectability, variability and spectral characteristics of deeply embedded YSO phases with X-ray emission characteristic at later stages of YSO evolution.  The black dashed line in Figure \ref{yso_frac} represents the maximum possible detection fraction based on the merged image and accounts for sources included and excluded by the ACIS field-of-view for each observation.  The colored solid lines indicate the detection fraction within specific energy bands.

With observations of $\sim$50 L1630 YSOs at up to 11 epochs, we can examine X-ray variability as it relates to YSO infrared class.  Of the 52 X-ray emitting YSOs with infrared classifications, 27 meet our variability requirements (see Sect. 2.2).   X-ray light curves and hardness ratios spanning $\sim$4 years were constructed for each YSO meeting these criteria, as shown in Figures \ref{lc_1}, \ref{lc_2}, \ref{lc_3} and \ref{lc_4}. All YSO classes show significant X-ray variability with at least one X-ray source from each infrared class undergoing an order of magnitude change in count rate. Furthermore, many YSOs show significant variability even on timescales of days (ObsIDs 5384-6413 and 10763-8585; Table\ref{chan_obsid}). X-ray variability with respect to YSO classification is compared to X-ray and IR spectral parameters in Figures \ref{xray_var} and \ref{var_lum} a-e (See sect. 4.2).  Not all spectrally analyzed sources met the X-ray variability selection requirements (described in Sect. 2.2) and thus not all X-ray sources are shown in Figure \ref{var_lum}.  We find that L1630 Class 0/I YSOs tend to be more variable than other YSO classifications.  We find that four of the five Class II YSOs that met our variability requirements have very similar levels of variability ($s^{2} \approx$ 0.4; Figure \ref{var_lum}) yet display differing spectral characteristics.




  Examples of model fits to X-ray spectra of representative objects from each YSO class are shown in Figure \ref{spec}.  Various X-ray ($L_{X}$, $T_{X}$, $N_{H}$) and infrared ($L_{IR}$) properties of L1630 YSOs are plotted in Figures \ref{temp_lum} a-e. Such parameters are used to investigate trends based on YSO classification (see Sect. 4).  We find that Class III YSOs in L1630 tend to have lower intrinsic $L_{X}$ than other YSO classifications yet have similar plasma temperatures.  We find no clear correlation between YSO classification and plasma temperature and we find three (one Class II and two transition disk) YSOs  that display emission from low temperature ($\sim$3 MK) plasma that could be produced via accretion shocks.

\subsection{Individual Sources}

   Only 5 of the 19 Class 0/I (candidate Class 0/I or Class I) YSOs in L1630 were detected in X-rays, possibly due to factors such as large obscuration, low quiescent X-ray luminosity and, perhaps, less frequent events than Class 0/I YSOs in Serpens or Coronet or Class II/III YSOs.  SSV 63NE (ID\# 16) has previously been discussed as an X-ray emitting Class 0/I candidate in \citet{Simon2004} due to its spatially correlated X-ray and radio emission and lack of near-infrared counterpart. Three of the five X-ray emitting YSOs (ID$\#$s 12, 13 and 14) are designated as Class I YSOs in \citet{Cohen1984}, \citet{Simon2004} and \citet{Muzerolle2005}.   All five are classified as protostars in \citet{Megeath2012}.  Furthermore, the detection of ID$\#$s 12, 13 and 14 at J, H and K wavelengths suggest they are indeed Class I (as opposed to Class 0) sources (Fig. \ref{2mass_cc}).

\begin{figure}
\centering
\includegraphics[scale=0.35]{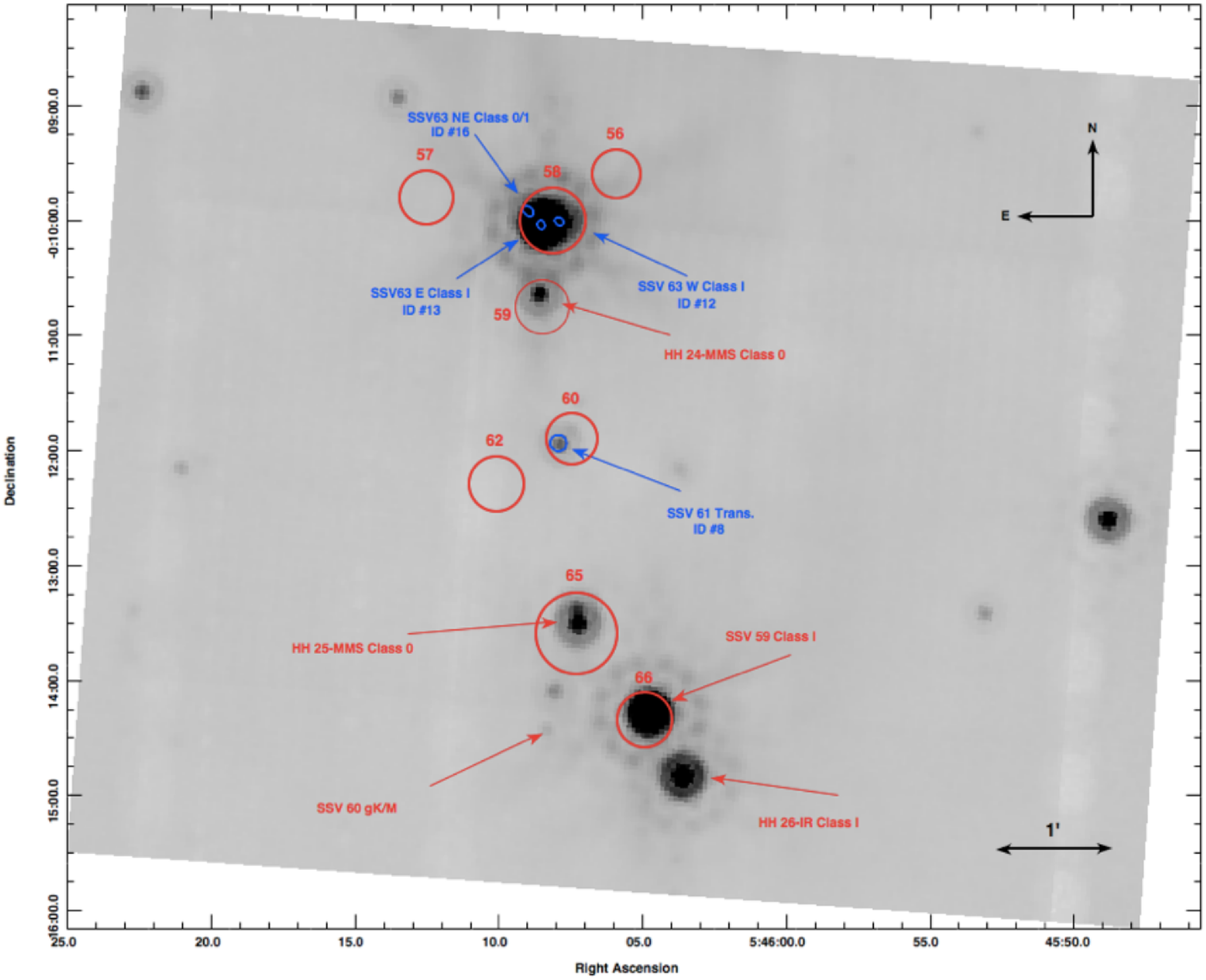}
\caption{ \label{ssv} Spitzer 24 $\mu$m inverted-color image of a southern region in L1630.  Red circles indicate 850 $\mu$m continuum emission from JCMT SCUBA (Mitchell et al. 2001) and blue circles indicate Chandra X-ray emitting YSOs. }

\end{figure}

\begin{figure}[H]
\centering

\includegraphics[scale=0.45]{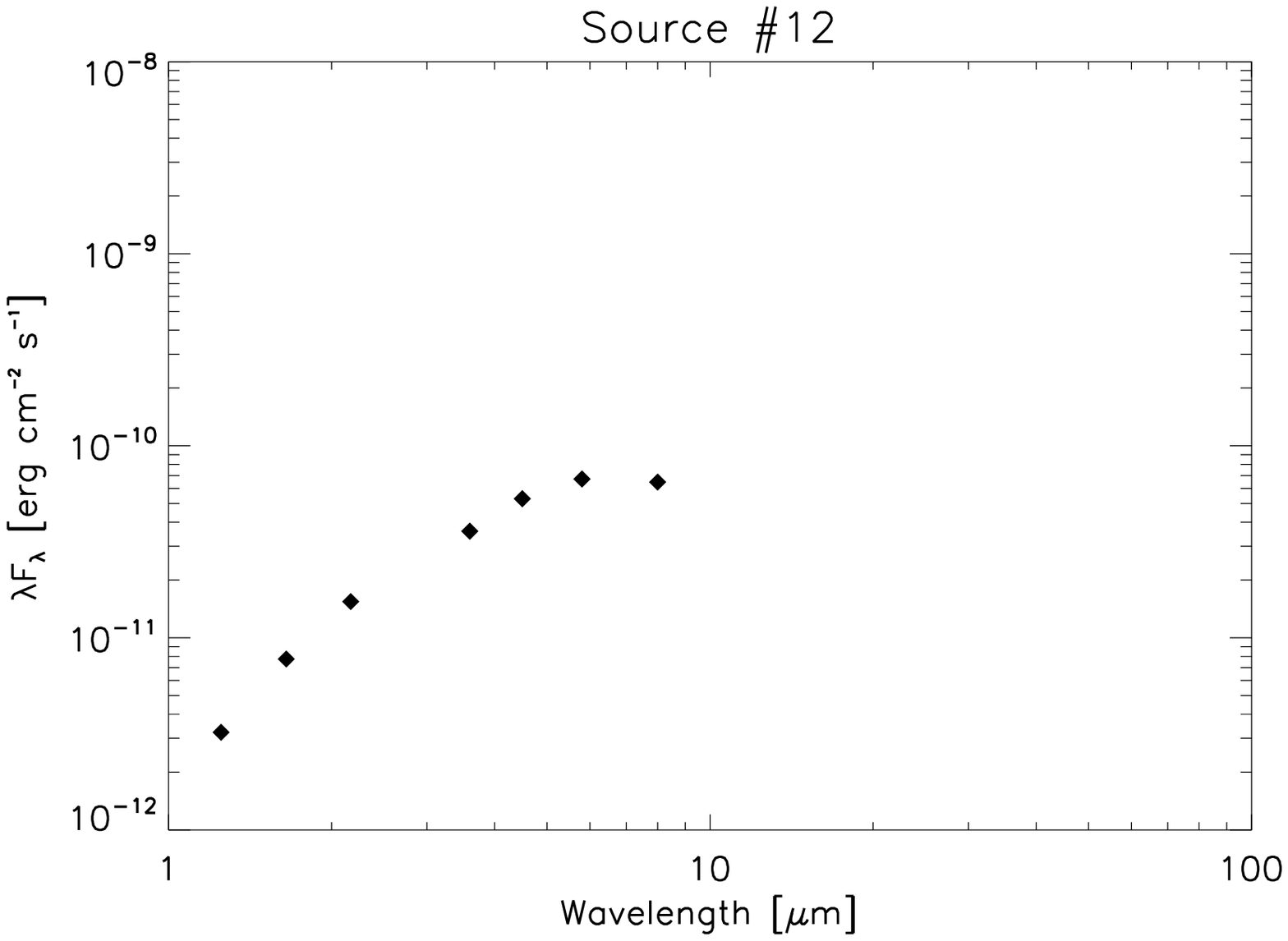}
\includegraphics[scale=0.45]{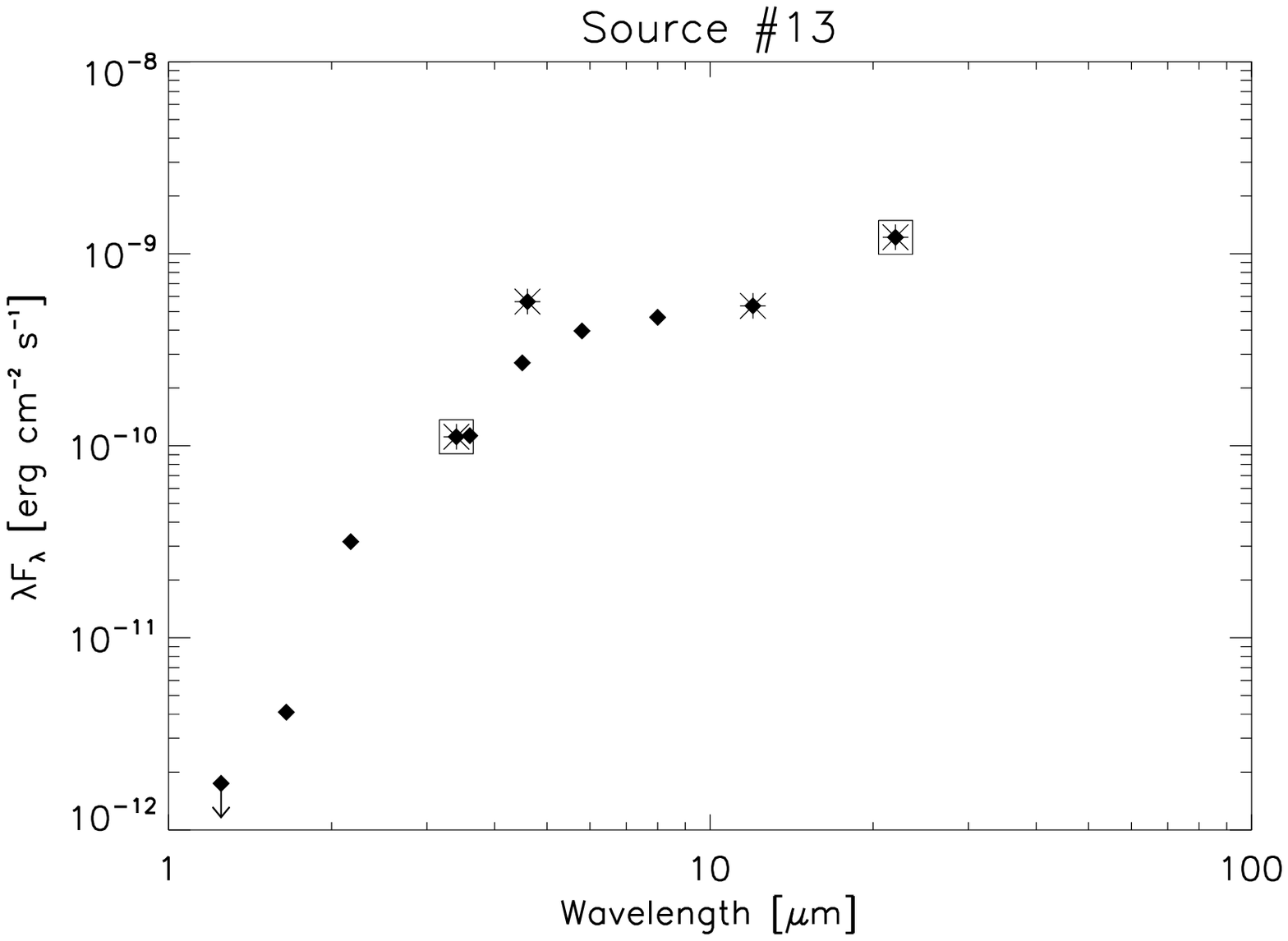}
\includegraphics[scale=0.45]{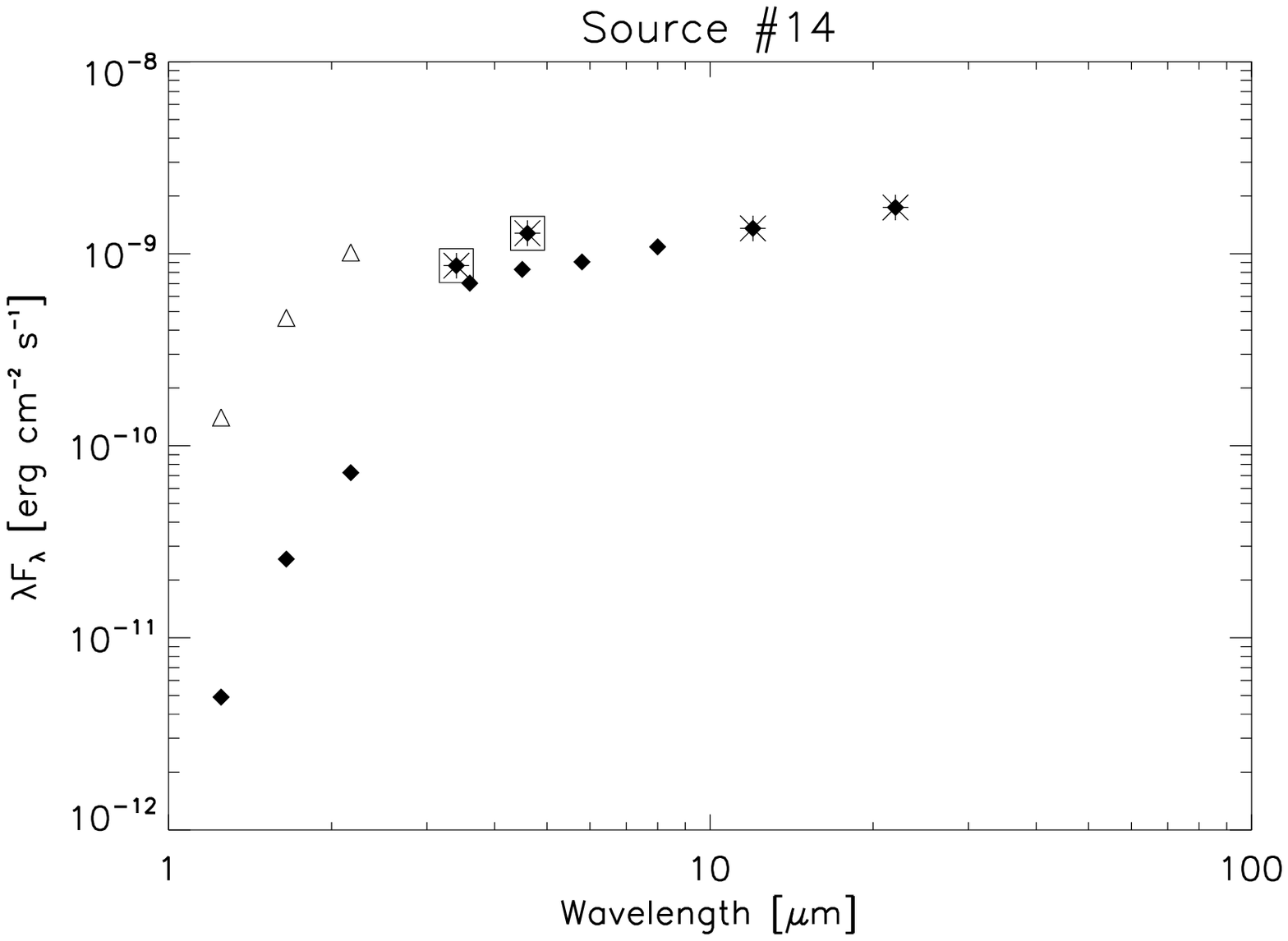}
\includegraphics[scale=0.45]{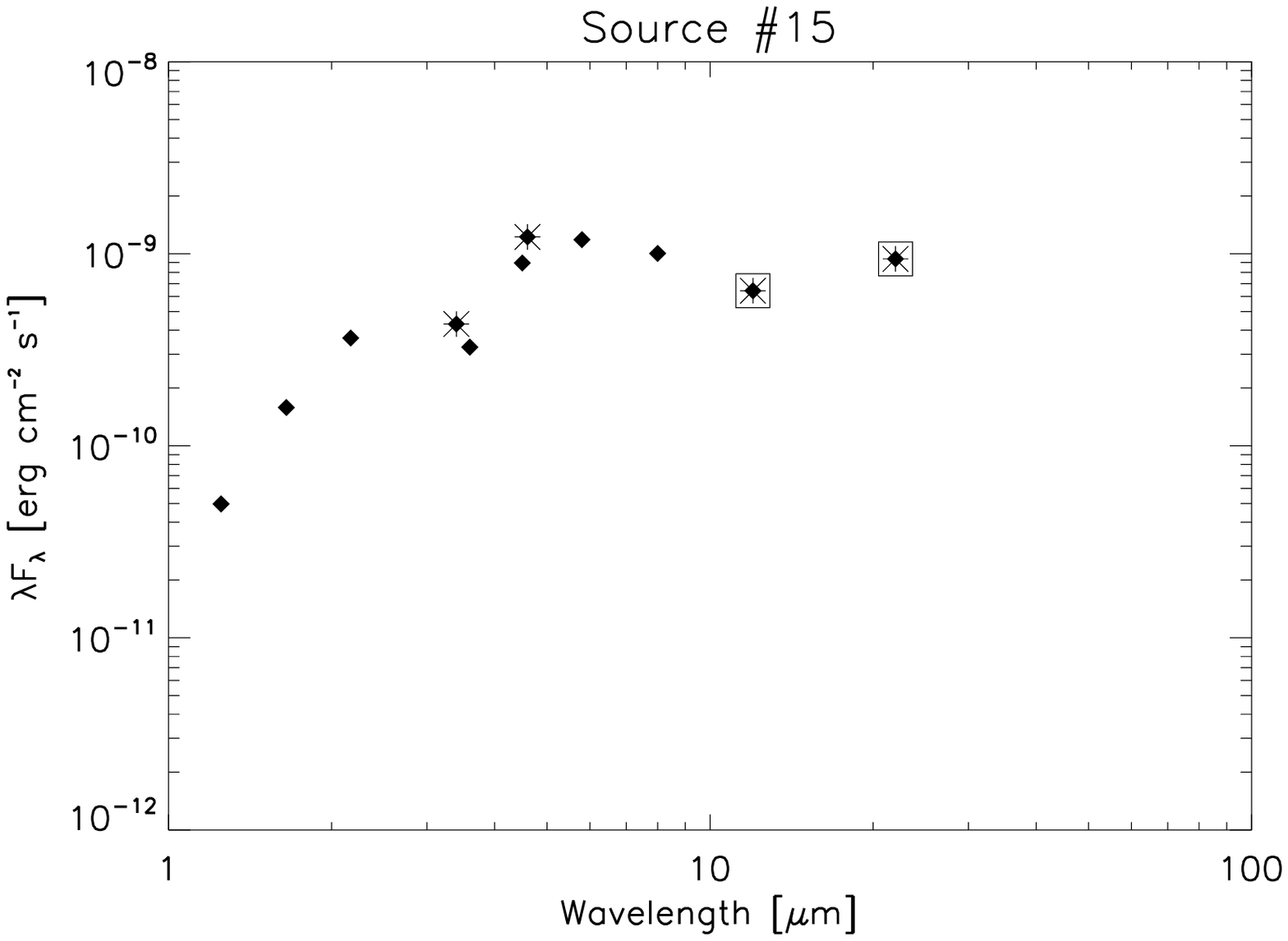}
\includegraphics[scale=0.45]{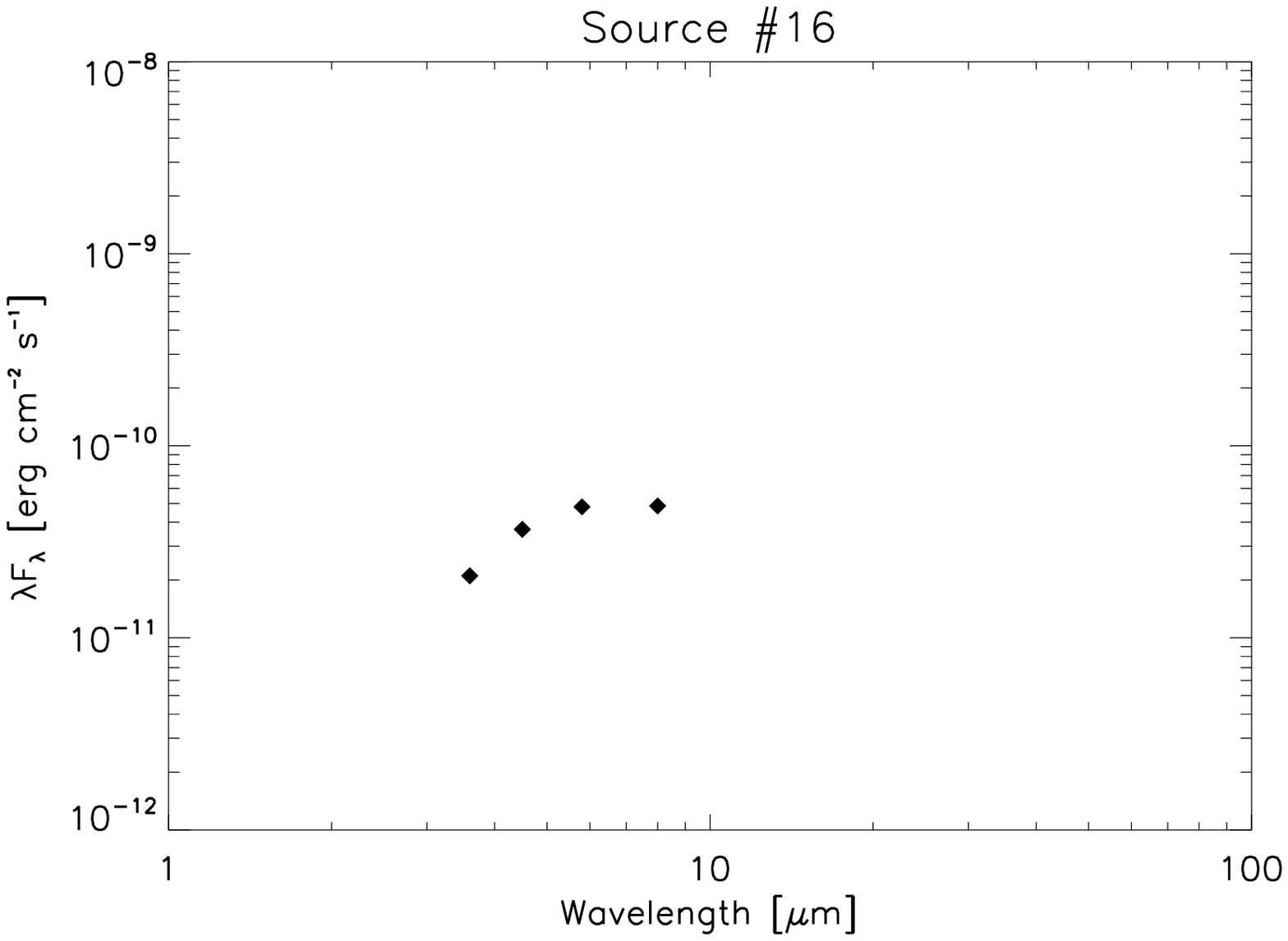}

\caption{\label{sed_01} SEDs of Class 0/I X-ray detected YSOs constructed from 2MASS (diamond), Spitzer (diamond), and WISE (asterisk) data when available.  In the case of V1647 Ori (Source $\#$14), post-outburst JHK photometry \citep[triangle;][]{Aspin2009}. Boxes plotted over WISE data signify high variability and/or possible contamination or confusion of photometry.} 
\end{figure}

 The X-ray spectrum of SSV 63NE, is shown in Figure \ref{spec} and its SED is shown in Figure \ref{sed_01}.  Model fitting confirms a highly absorbed ($N_H \sim 10^{23}$ cm$^{-2}$) source and yields a plasma temperature of $T_X \sim 7-27$ $MK$ (Table \ref{spec_tbl}).  Higher temperatures for Class 0 candidates have been found previously:  $T_X \sim50$ $MK$ and $N_H=2.47\times10^{23}$ cm$^{-2}$ for source X$_{E}$, a sub-region of IRS 7 in the R Corona Australis (R CrA) star-forming core \citep{Hamaguchi2005} and $T_X =79.4\pm20$ $MK$ and $N_H=1.48\times10^{23}$ cm$^{-2}$ for IRS7E in the Coronet Cluster (FP07).  Submillimeter emission has been observed in the region of SSV 63NE with JCMT-SCUBA \citep{Mitchell2001}; however, due to the close proximity of two other X-ray emitting protostars (Class I YSOs SSN 63E and SSV 63W; Figure \ref{ssv}), it is unclear from which of the three (or a combination thereof)  the unresolved submillimeter emission is originating.  The large value of $N_{H}$ we derived from SSV 63NE is consistent with Class 0/I status and --combined with strong submillimeter emission in its vicinity and its lack of a near-IR counterpart-- support Class 0 candidacy. However, higher-resolution submillimeter and far-infrared imaging are needed to confirm this source's Class 0 status.

The strongest X-ray emitting YSO in our survey is the transition disk/wTTS SSV 61 (ID $\#$8; Figures \ref{spec}, \ref{ssv} and  \ref{sed_tran} ).  \citet{Simon2004} had previously classified SSV 61 as a wTTS based on H$\alpha$ emission and Li I $\lambda$6708 absorption equivalent widths; however, our analysis of IRAC photometry placed SSV 61 on the border between Class III (wTTS) and transition disk YSOs.  Our 234 ks merged spectrum of SSV 61 (Figure \ref{spec}b) contains $\sim$ 7000 counts, which corresponds to an effective count rate of 32.3 ks$^{-1}$.    \citet{Simon2004} previously published an effective count rate of 251.92 ks$^{-1}$ for a 62.8 ks observation in November 2002.  Evidently the count rate of SSV 61 had declined during the epochs of our Chandra observations and, with a (relatively small) normalized variance of 0.07 (Fig. \ref{xray_var}), we conclude that \citet{Simon2004} must have observed SSV 61 in outburst.  It is interesting to note that SSV 61 lies close to a submillimeter source (Fig. \ref{ssv}) originally reported by \citet{Mitchell2001}.  Such an association is highly unusual for a wTTS but could be consistent with transition disk status.

\begin{figure}
\centering
\includegraphics[scale=0.38]{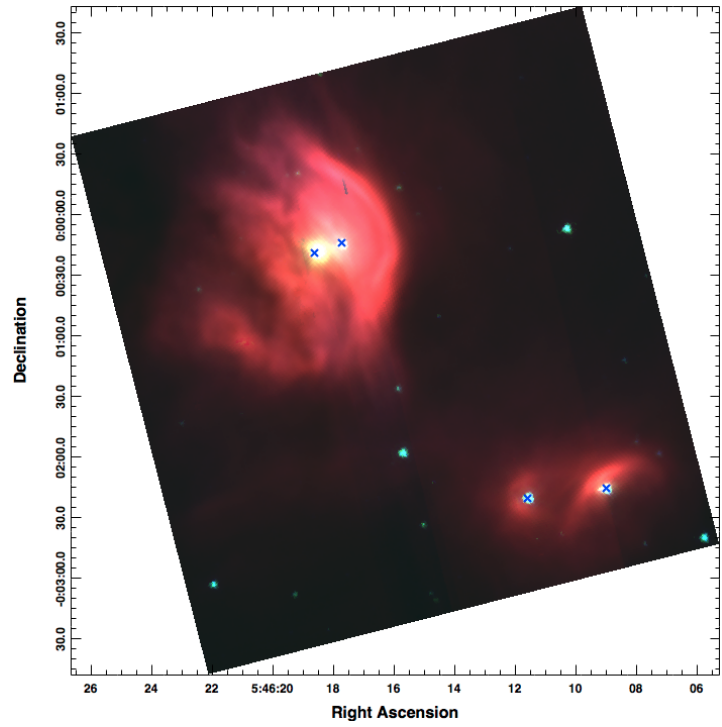}
\caption{ \label{bow} A Spitzer IRAC 3.6, 4.5 and 8.0 $\mu$m three-color image of IDs $\#$4, $\#$38, $\#$3 and $\#$1 showing spatially correlated bow-shaped emission likely indicative of a wind-collision front.  Chandra X-ray detections are indicated with overlaid blue crosses.}

\end{figure}

Other interesting X-ray/infrared sources are IDs \# 1, 3, 4 and 38 (Figures \ref{bow}, \ref{sed_2} and \ref{sed_tran}). \citet{Fang2009} classified ID \# 4 as a K3.5 cTTS with an optically thick disk.  It was detected as an X-ray source in our observations with an effective count rate of 1.94 $ks^{-1}$. The other three sources were not included in \citet{Fang2009}; however IDs \# 1 and 38 were classified as pre-MS stars with disks in \citet{Megeath2012}. If the bow-shock-like nebulae seen in Spitzer imaging (Figure \ref{bow}) are attributed to these YSOs then these structures may be wind-collision fronts similar to those associated with other YSOs in the Orion region \citep{Bally2000}.

\section{Discussion}

\subsection{L1630 YSO X-ray Detection Fraction}

 When observing young stellar objects, X-ray detection fraction depends primarily on sensitivity, spectral source hardness, and the degree of intervening obscuration toward the source.  Our results further illustrate that, in the case of YSOs characterized by strong variability, the number and/or duration of observation epochs plays a significant role in detection statistics  when compared to studies with similar sensitivity.  The X-ray detection fractions as a function of time in Table \ref{yso_fov} and Figure \ref{yso_frac} provide insight into the variable nature of YSOs.  Class 0/I YSOs have the lowest overall detection fraction in all bands.  These YSOs are rarely detected in the soft energy band in our data, most likely due to their highly embedded environments; whereas in the hard band, photons are more likely to penetrate, thus leading to a higher hard-band detection fraction.  The detection fractions for Class II sources are slightly larger than for Class 0/I sources; furthermore, it appears that for several observations, Class II sources are more readily detected in the soft band than in the hard band.  This could be an indication either of the relatively steady nature of soft accretion shock emission or of the generally smaller degree of intervening absorption towards Class II sources.  The transition disk detection fraction tends to be similar in hard and soft bands, indicative of relatively steady, non-variable emission.  Class III YSOs, whose X-ray detection fraction in a given observation is 1.0 by definition, are consistently more detectable in the soft band than the hard band, indicating lower levels of intervening absorption with respect to the other YSO classes.  The relatively low detection fractions in individual exposures evident in Fig. \ref{yso_frac} may reflect the highly variable nature of Class III YSOs.

Our X-ray detection fraction results are compared with other infrared/X-ray studies of star-forming regions in Table \ref{yso_compare}.   X-ray detection fractions for individual observations were not available in the other studies so we restrict the comparison to the respective merged exposures.  The respective sensitivities between 0.5-8.0 keV are  L$_{x, min}\approx4 \times 10^{26}$ erg/s, L$_{x, min}\approx3 \times 10^{27}$ erg/s, and L$_{x, min}\approx2 \times 10^{27}$ erg/s for  FP07, W07 and our L1630 observations. This comparison of X-ray luminosity sensitivities is only approximate, given the different source detection thresholds and angular size scales used in the three studies.  When comparing YSO detection fraction, it may be important to consider not only the total exposure time and source distances, but also how the component exposures were distributed in time. FP07 used eight exposures of the Coronet Cluster over the course of $\sim$5 years; however, five of the exposures were obtained only a day apart. The Serpens Core (W07) was observed for one 90 ks period.  L1630 was observed 11 times over the course of four years with only two sets of two exposures obtained less than 2.5 months apart (Table \ref{chan_obsid}).

For Class 0/I YSOs, FP07 and W07 found detection fractions of 90\% and 41\% in R CrA and Serpens, respectively, whereas for L1630 we obtain a Class 0/I detection fraction of only 26\%. The Class II detection fraction for L1630 (58\%) is significantly larger than the detection fraction for Class II YSOs in the Serpens Cloud (32\%) but again smaller than in the Coronet Cluster (82\%).  It appears the high detection fraction in R CrA can most likely be attributed to the less deeply embedded nature of its Class 0/I YSOs and the high sensitivity of the R CrA observations.  If Class II YSOs were characterized by relatively steady X-ray emission over timescales longer than a typical X-ray observation, punctuated by relatively rare, strong outbursts, then the probability of detecting X-ray emitting Class II YSOs should increase with the number of X-ray observations.  It would therefore appear that, for studies with similar sensitivities -- as for W07's Serpens study and our L1630 study -- observations spanning longer time scales may yield a higher X-ray detection fraction of Class II sources.

\subsection{L1630 YSO X-ray Variability by Class}

 Both coronal activity and star-disk interactions -- e.g., magnetically-funneled accretion,  time-variable absorption from inner disk warps and/or accretion streams, disk instabilities resulting in accretion outbursts, or magnetically confined-plasma suddenly released in star-disk magnetic-reconnection events -- are phenomena that can plausibly cause significant X-ray variability.  In the case of star-disk interactions, obviously, a circumstellar disk must be present.  Class III YSOs have little or no detectable IR excesses and hence their disks have likely already dispersed.  Ergo, if star-disk interactions are are at least partially responsible for the highly X-ray variable nature in observed YSOs with disks, then Class I, Class II, and transition disk sources should show X-ray variability that is distinct from that of Class III sources. In particular YSOs with disks might be expected to display larger levels of variability than Class III YSOs over a timespan of several years and multiple observations \citep[][and references therein]{Bouvier2007}.  Transition disk YSOs, representative of an evolutionary stage between Class II and Class III, are characterized by disks with large ($\sim$1-10 A.U.) inner dust clearings \citep{Calvet2005}.  If magnetic field lines interacting with gas in the disk are required to provide some mechanism for variability, then transition disks whose "holes" are large enough might extend beyond the immediate influence of the protostellar magnetic field, thus halting such variability.  However, even though there is a large gap in the dusty component of a transition disk, this does not necessarily imply a large gap in the gaseous component \citep{Chiang2007, Sacco2012}, and hence star-disk interactions could still be an important driver of YSO X-ray variability.

\begin{figure}
\centering
\includegraphics[scale=0.60]{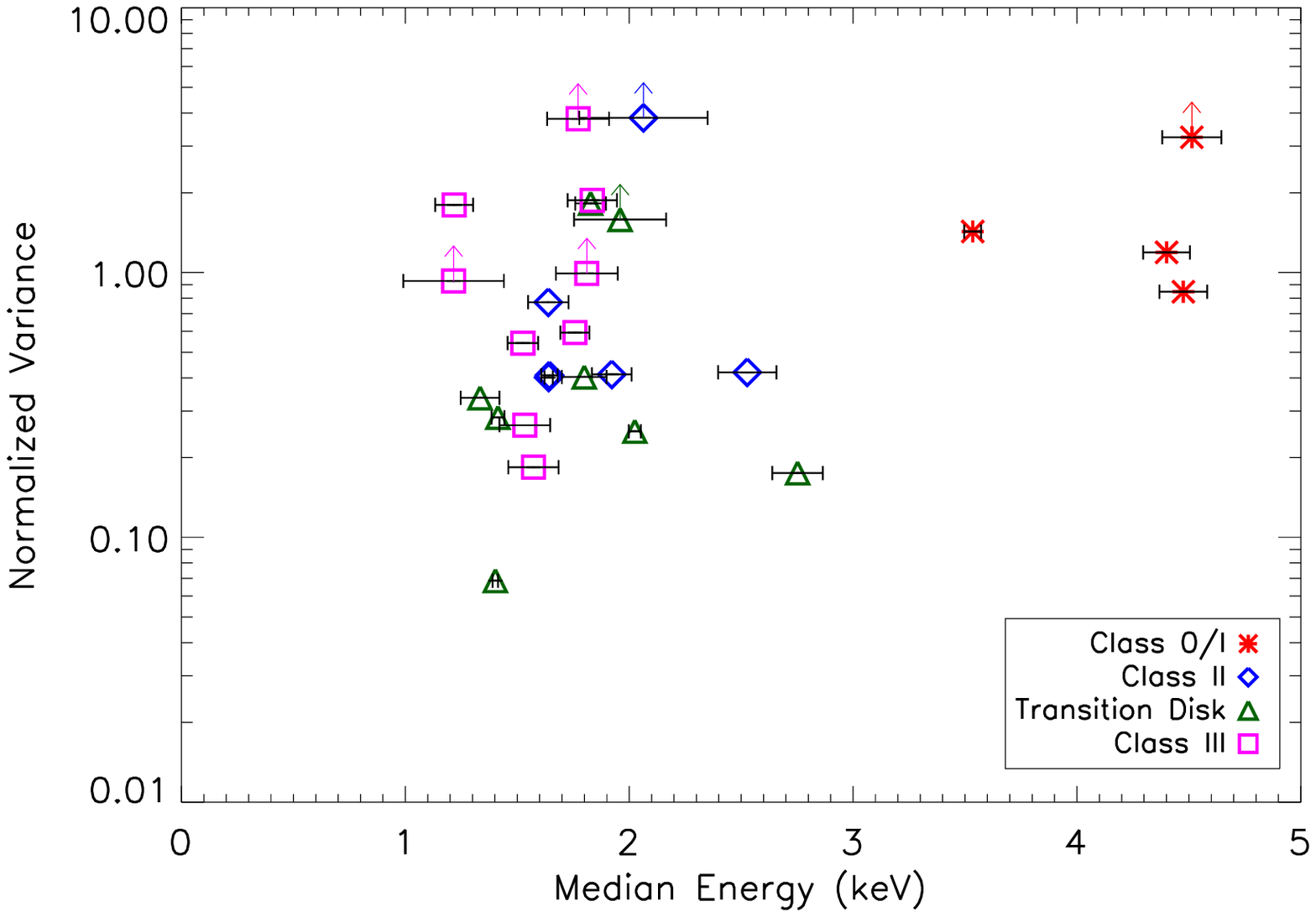}
\caption{\label{xray_var} X-ray variability vs. median X-ray photon energy for YSOs in L1630. Sources with lower limits were detected in only 2-5 observations.  Error bars for the median energy were calculated using the standard error of the mean.}   
\end{figure}

Regardless of the mechanisms underlying variability, Figures \ref{lc_1}, \ref{lc_2}, \ref{lc_3} and \ref{lc_4} show that each class in our variability sample has at least one YSO that underwent an order of magnitude increase in count rate over the $\sim$4 year course of our study.  Several YSOs also showed significant variability on timescales of days, which is suggestive of magnetic loop flaring \citep{Favata2005}.   FP07 displayed similar X-ray duration light curves for YSOs in the Coronet Cluster and suggested that these light curves indicated that Class I YSOs displayed stronger levels of variability than Class II YSOs.  Likewise, Figures \ref{lc_1}, \ref{lc_2} and \ref{xray_var} suggest that L1630 Class 0/I YSOs are more variable than Class II YSOs, though these figures also suggest that Class 0/I and Class III objects are similarly variable.  Similar distinctions in variability are apparent for YSOs in the $\rho$ Ophiuchi dark cloud \citep{Imanishi2001a, Imanishi2001b}.  

\begin{figure}

\centering

\subfigure[]{\includegraphics[scale=0.46]{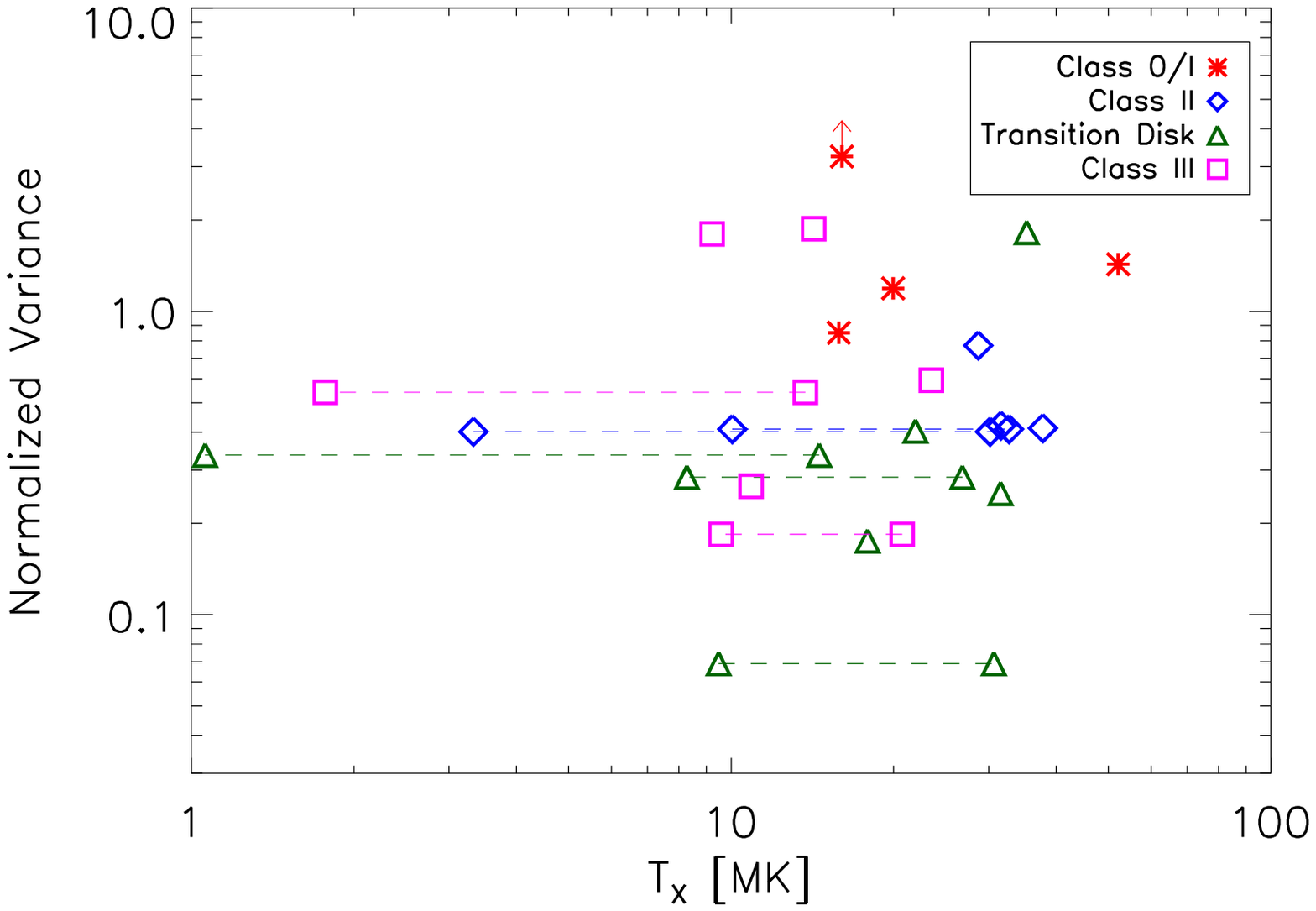}}
\subfigure[]{\includegraphics[scale=0.46]{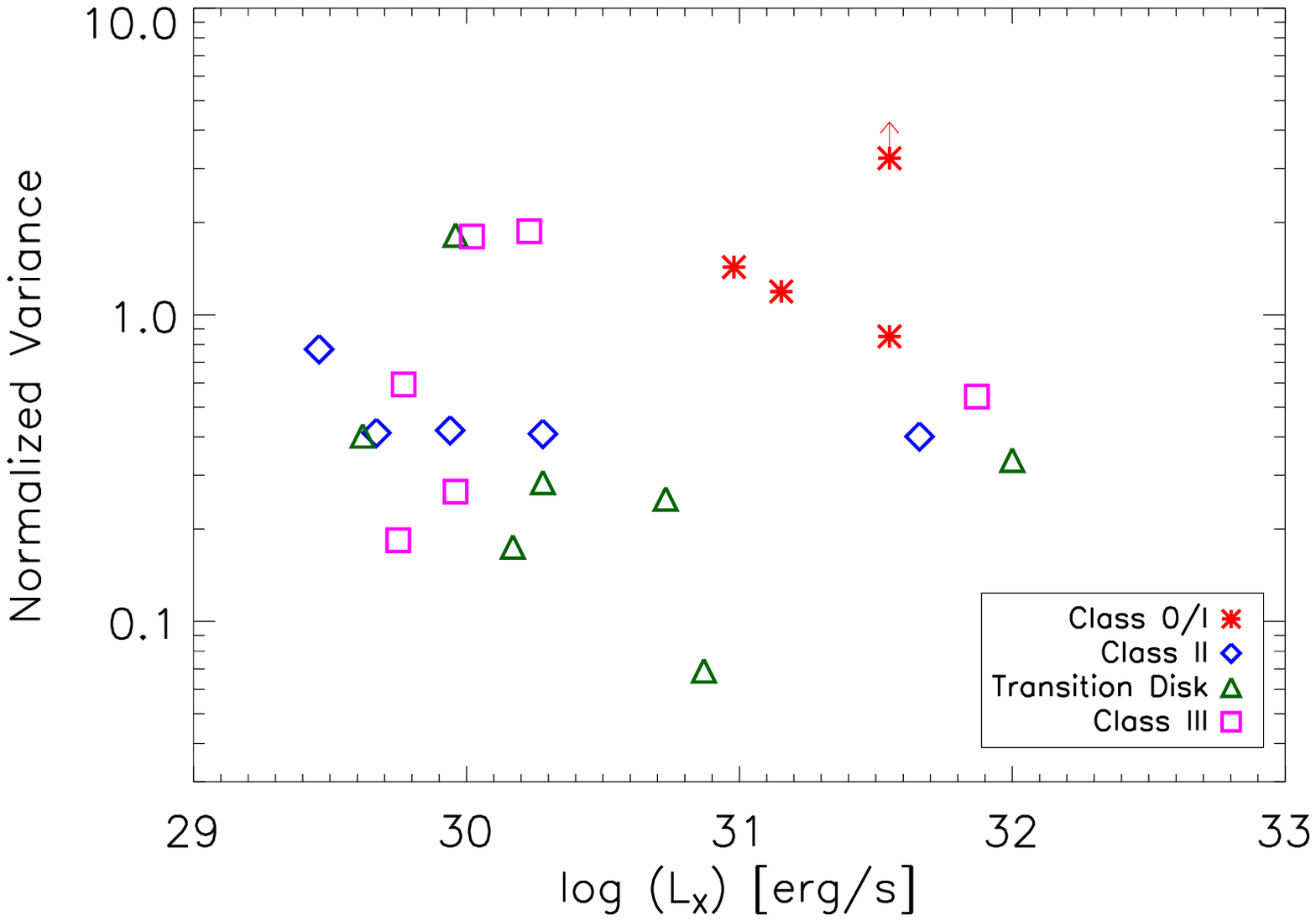}}
\subfigure[]{\includegraphics[scale=0.46]{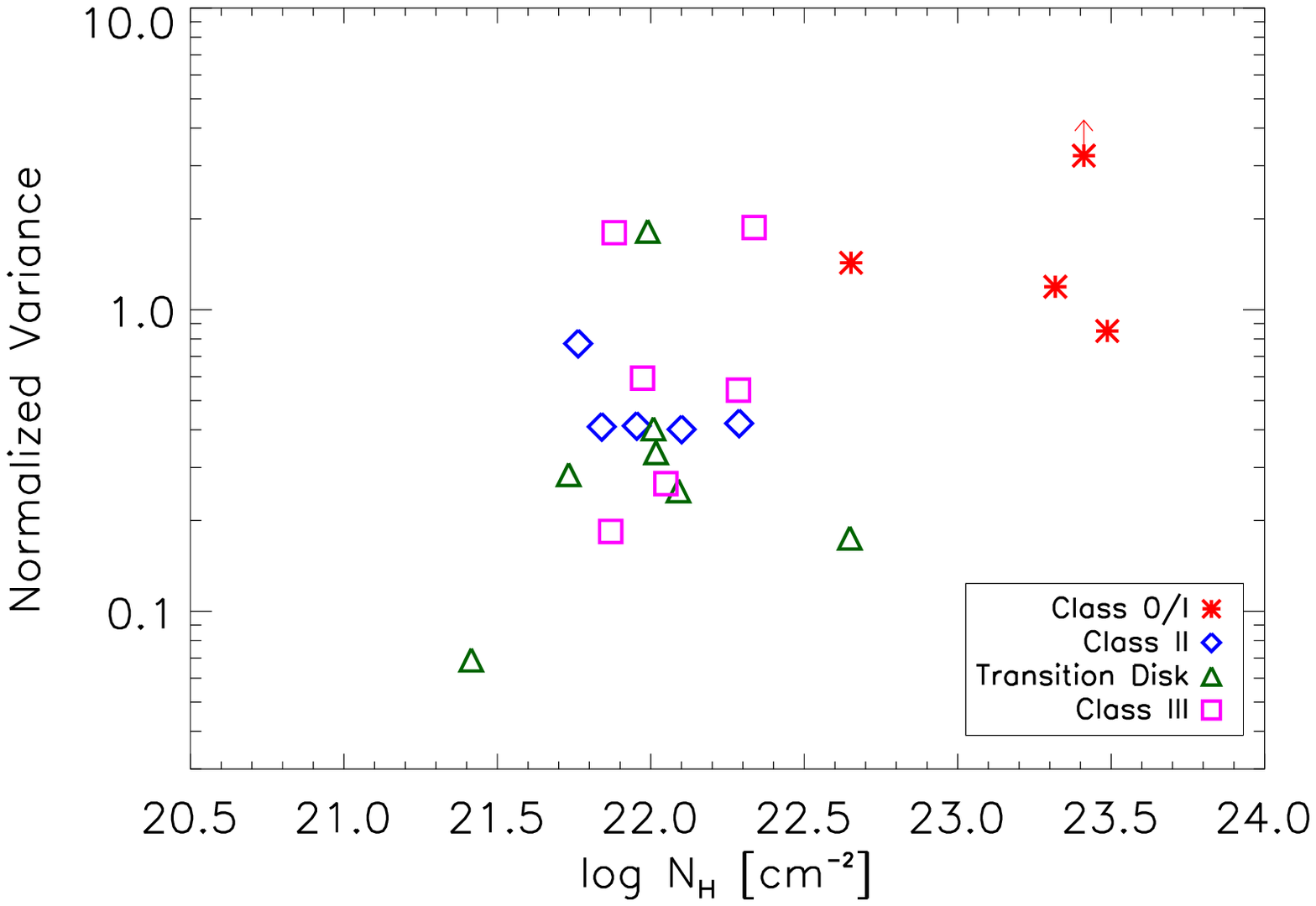}}
\subfigure[]{\includegraphics[scale=0.46]{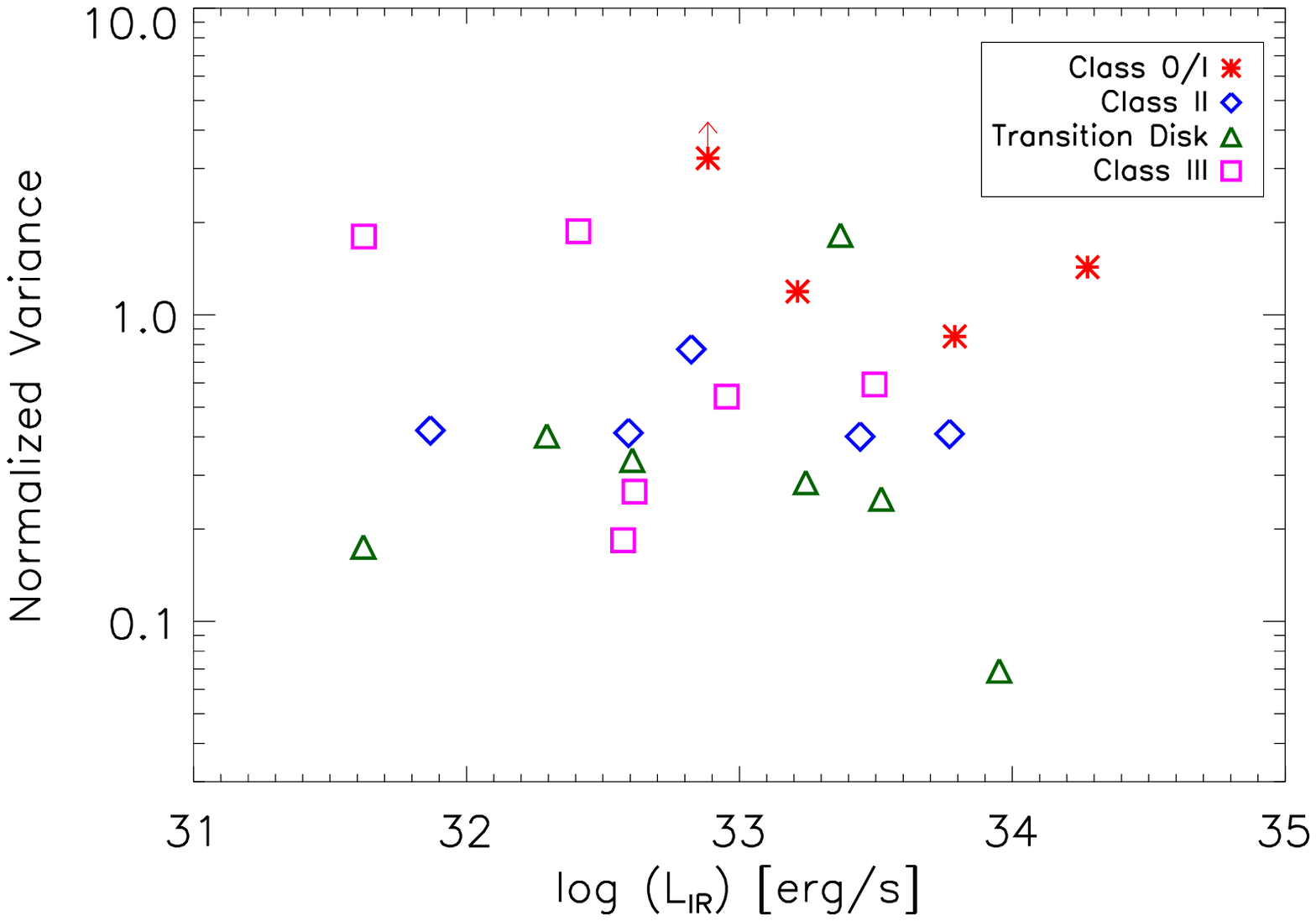}}
\subfigure[]{\includegraphics[scale=0.46]{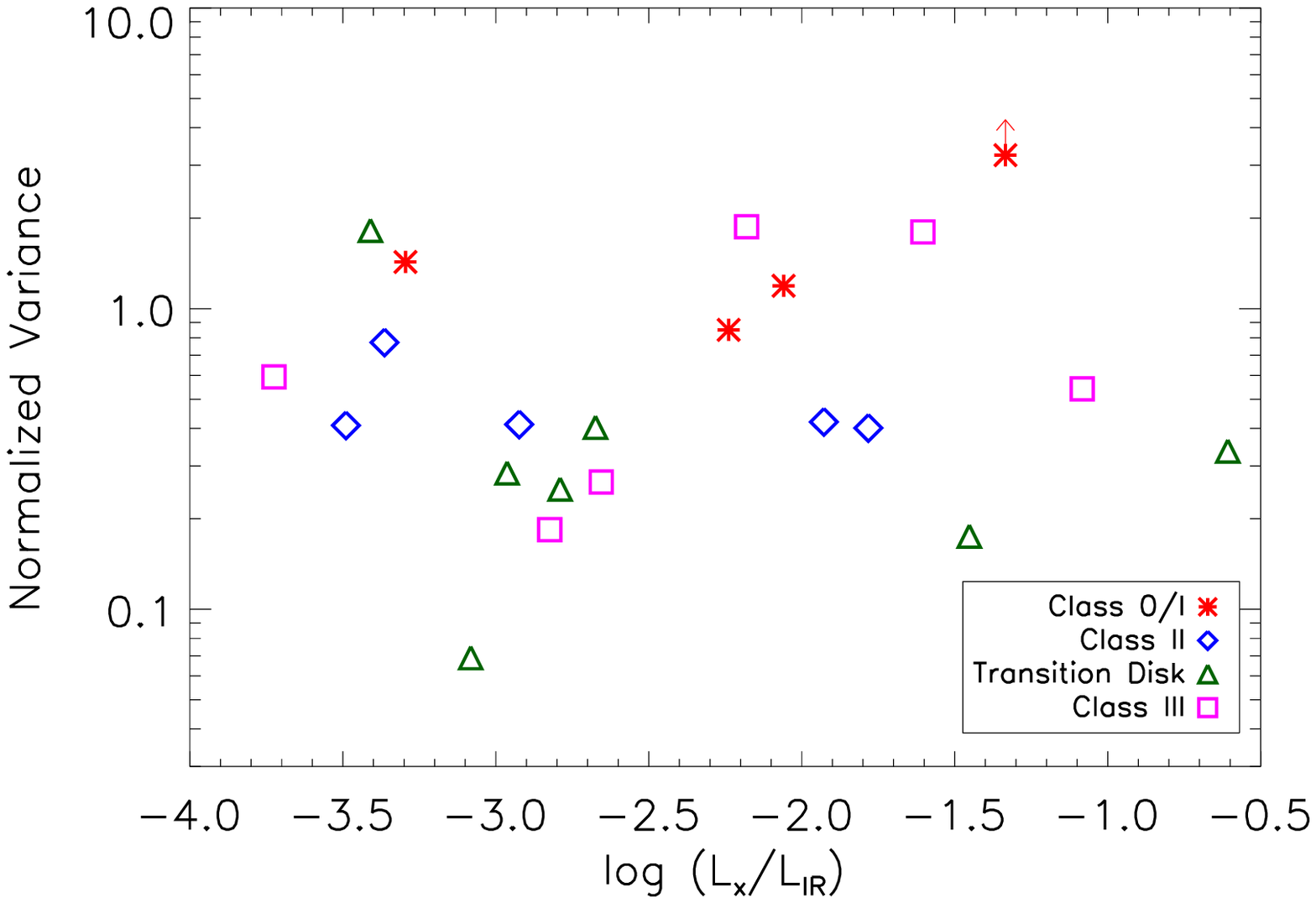}}

\caption{Normalized variance versus derived spectral properties from model fitting for YSOs in L1630.   Dashed horizontal lines represent two-temperature model fits.  }

\label{var_lum}

\end{figure}

Figure \ref{xray_var} illustrates the overall variability and spectral hardness (median photon energy) for YSOs of various classes in L1630 over the course of $\sim$4 years.   Figures \ref{medE_cc} and \ref{xray_var}  make clear that Class 0/I sources are significantly harder than Class II and III YSOs.  This is likely mainly due to their highly embedded nature (i.e., absorption of $<$1 keV photons by circumstellar material) given that their inferred plasma temperatures are similar to those of Class II and Class III YSOs (Fig. \ref{var_lum}a and Sect. 4.3). Large ($\gtrsim$ 3 keV) median energies for similarly deeply embedded objects have been inferred previously, including V1647 Ori (ID\#14) \citep{Grosso2005a, Grosso2005b}.

 Figures  \ref{xray_var} and \ref{var_lum}b illustrate the highly variable nature of Class 0/I YSOs in L1630.  It is important to note that when comparing these two figures, one must account for the lower limits of sources that were only detected in 2-5 observations.  Many sources in Figure \ref{xray_var} did not have enough counts for spectral fitting and, therefore, were not included in Figure \ref{var_lum}.  Furthermore, for these sources with non-detections, count rate upper limits (i.e., background count rate at the source position) were included in calculations of the normalized variances (see Sect. 2.2).  Figure \ref{xray_var} suggests both Class 0/I and Class III YSOs are highly variable.  However, Figure \ref{var_lum} implies that of the YSOs detected in more than 5 observations, the Class 0/I YSOs are marginally more variable than all other YSO classes, including Class III YSOs.  Transition disks in L1630 tend to be less variable than other classes (Figs. \ref{xray_var} and \ref{var_lum}a).  
 
 Of the five Class II YSOs that met our variability and spectral fitting requirements, four of them have very similar levels of variance ($s^{2} \approx$ 0.4; Figure \ref{var_lum}). It is interesting to note that all four of these Class II YSOs also show similar plasma temperatures ($T_{X} \sim30MK$), yet their other spectral characteristics differ.  \citet{Flaccomio2012} found that COUP Class II YSOs were more variable than Class III YSOs and suggested that rotational modulation was the mechanism responsible for high levels of variability.  We do not find such a trend for YSOs in L1630.  This could be due to the fact that the quasi-continuous 10-day-long observation of the ONC study was more sensitive to variability due to rotational modulation but was not as sensitive to rare, energetic flaring events as our L1630 study, in which the X-ray observations span four years.


 \subsection{L1630 YSOs X-ray Spectral Characteristics}

Before considering the X-ray spectral properties of L1630 YSOs obtained from model fitting, it is important to note that the absorbed one-or two-component thermal plasma models typically used to fit X-ray spectra of YSOs are not complete representations of the intrinsic properties of YSOs.  Detailed studies of magnetically and accretionally active YSOs show it is likely that their plasmas span a wide range of temperatures \citep{Brickhouse2010}.  However, one- or two-component thermal plasma models are sufficient to ascertain global plasma properties, and have been used successfully in numerous previous studies \citep[e.g.,][]{Feigelson2005,Guedel2007,Forbrich2007,Winston2007}.

\begin{figure}

\centering

\subfigure[]{\includegraphics[scale=0.46]{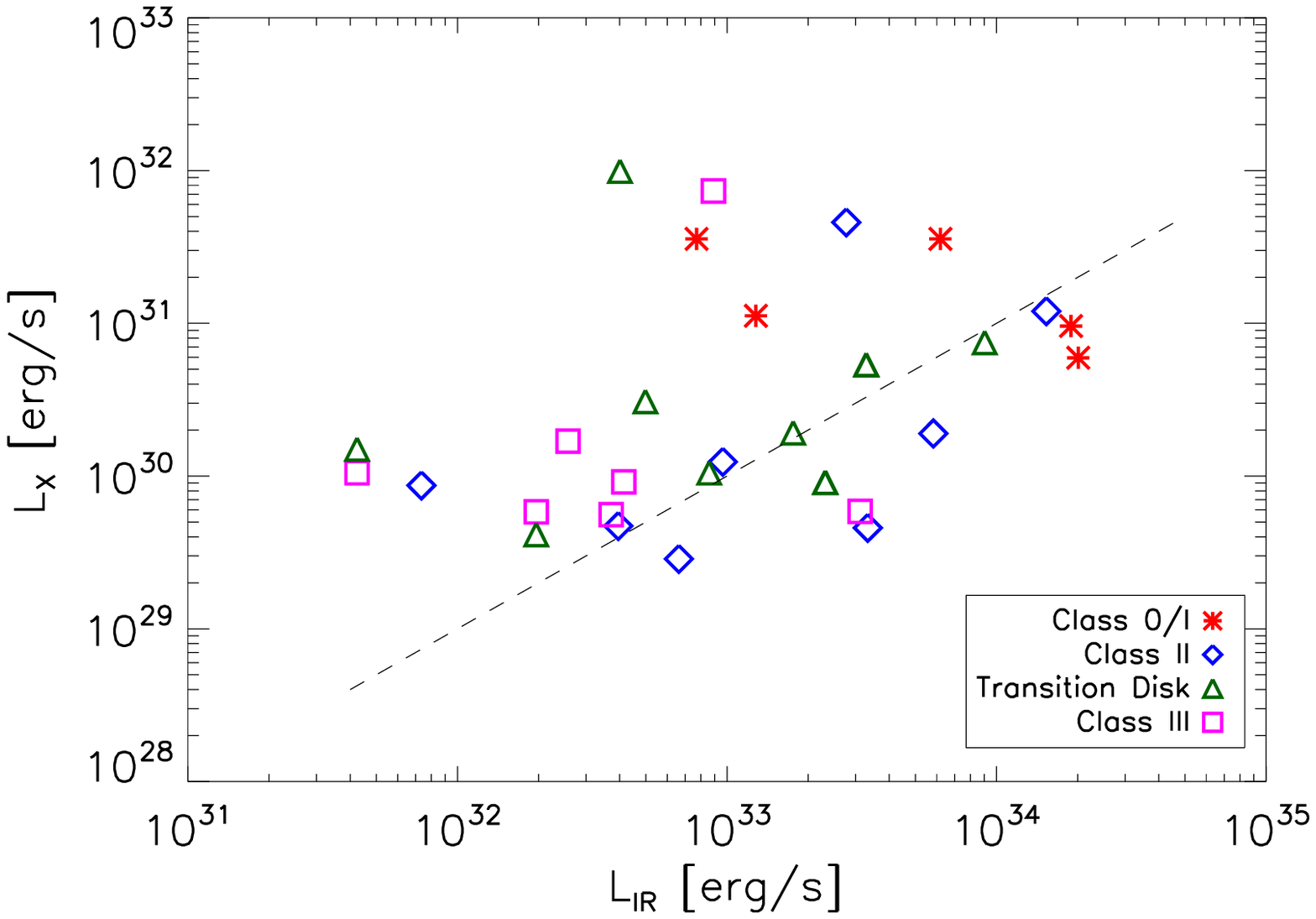}}
\subfigure[]{\includegraphics[scale=0.46]{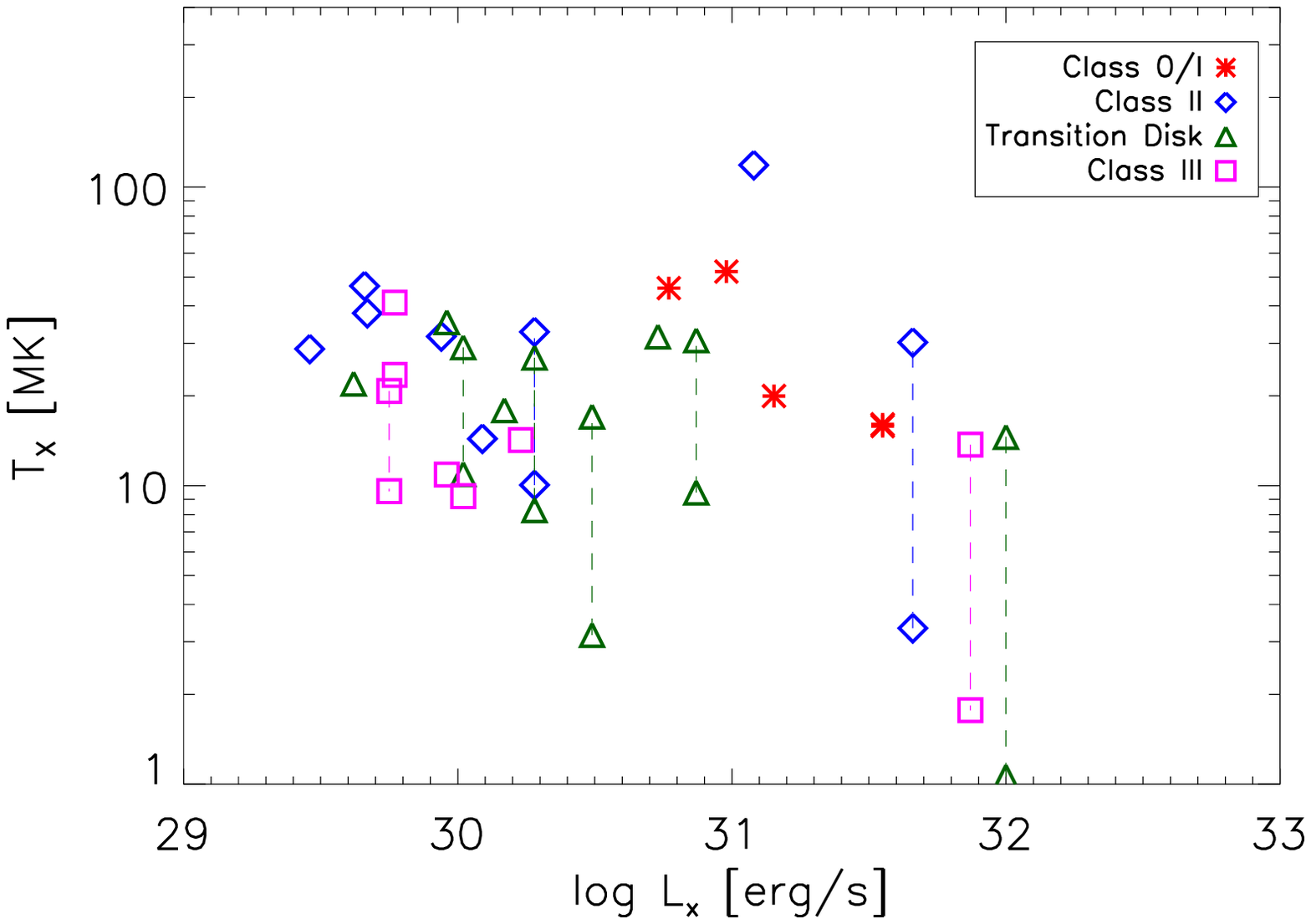}}
\subfigure[]{\includegraphics[scale=0.46]{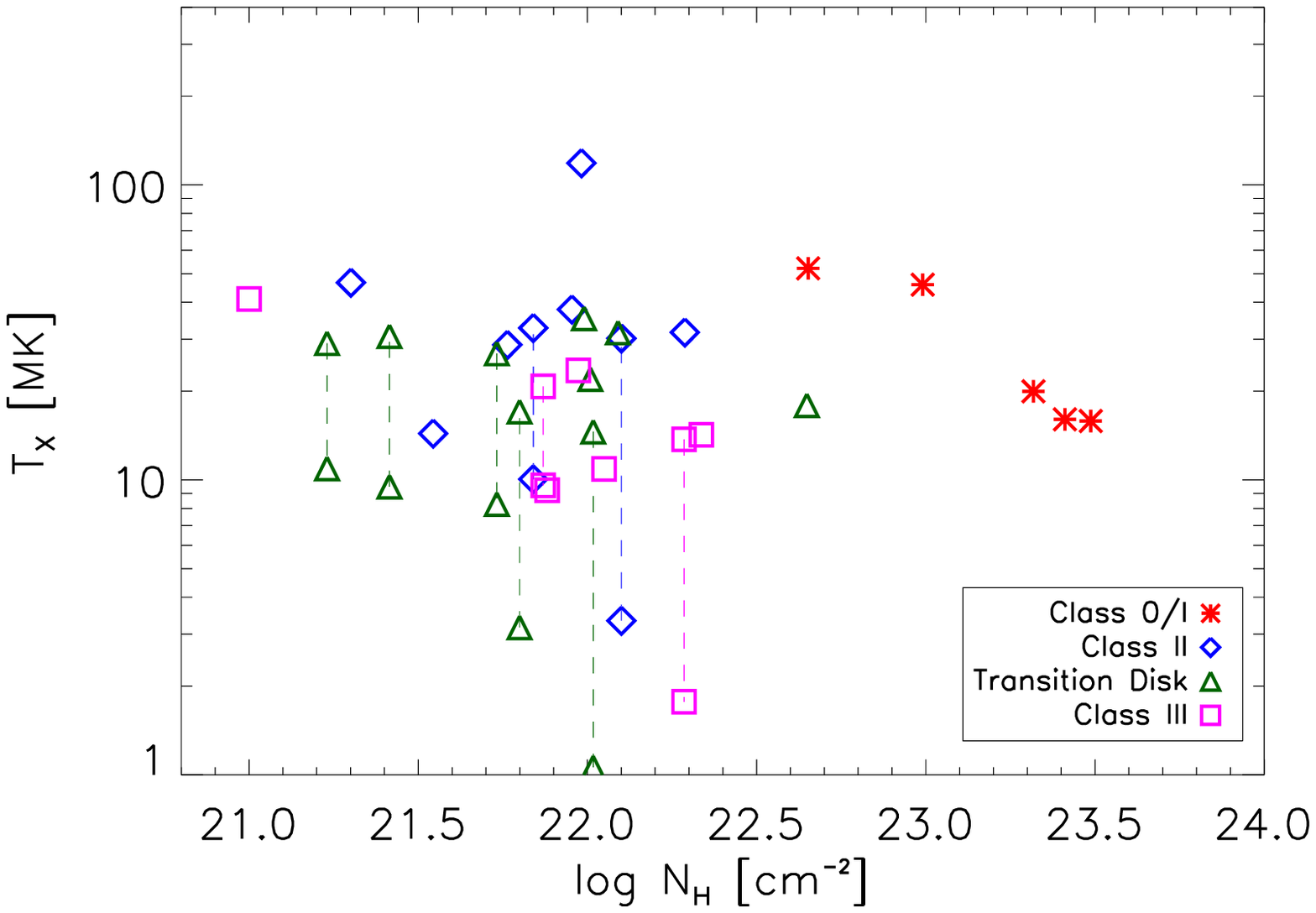}}
\subfigure[]{\includegraphics[scale=0.46]{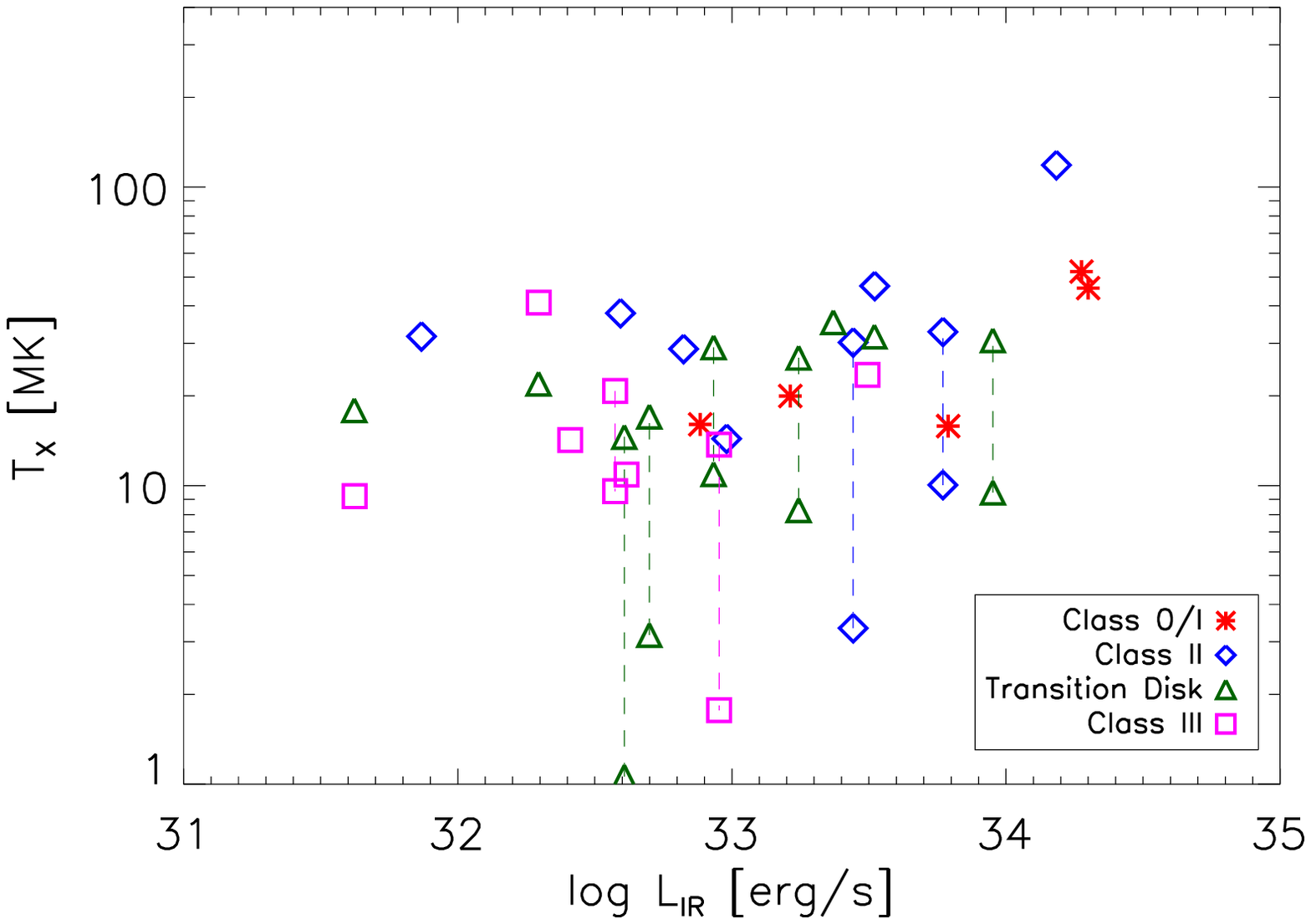}}
\subfigure[]{\includegraphics[scale=0.46]{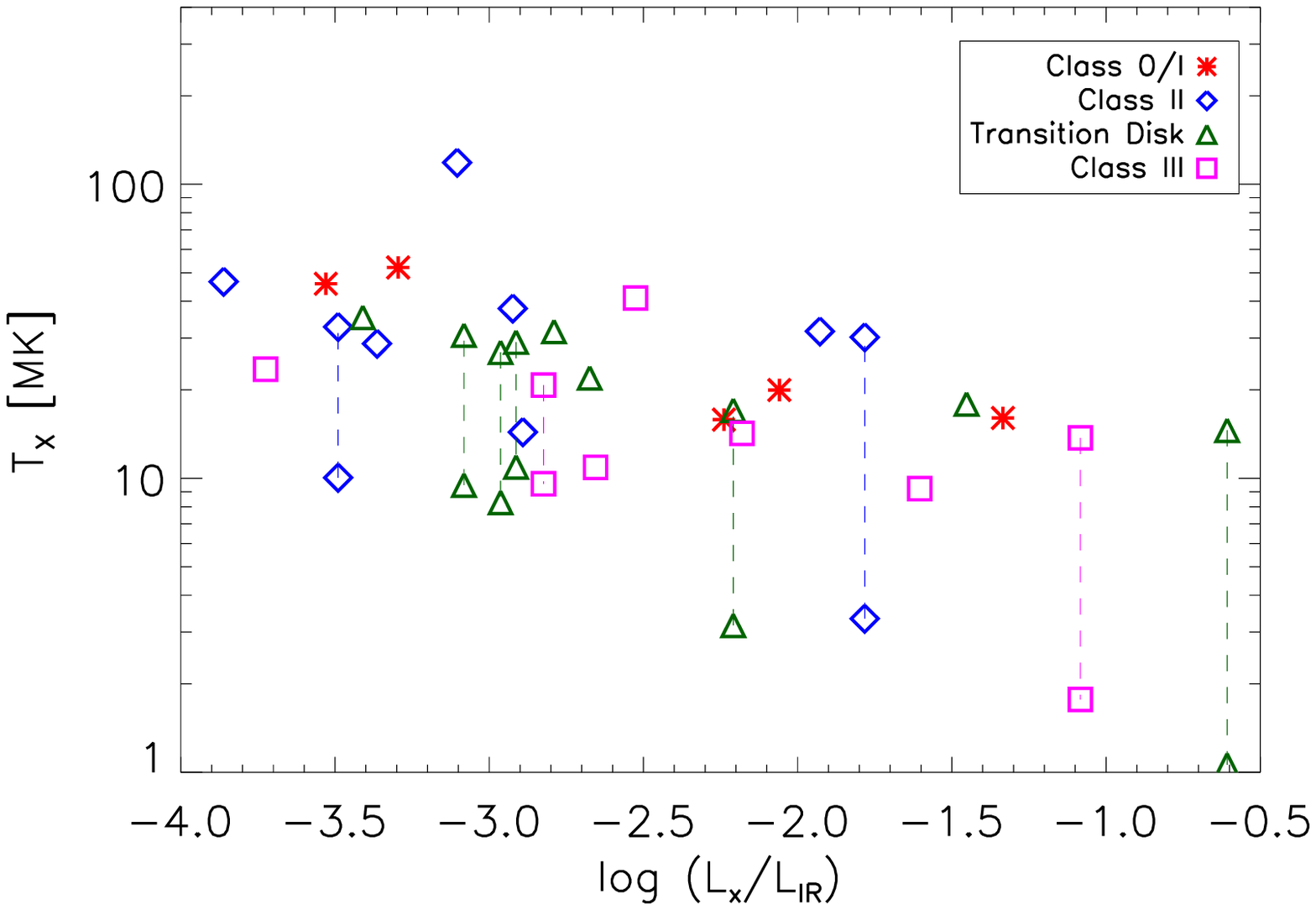}}

\caption{Spectral properties derived from model fitting of classified YSOs in L1630.  Dashed line in (a) represents $L_{X}=10^{-3}L_{IR}$.  Dashed vertical lines in (b), (c), (d) and (e) connected points determined from two-temperature model fits.  }

\label{temp_lum}

\end{figure}

 As shown in Figure \ref{nh_medE}, the line-of-sight absorption ($N_{H}$) values derived from spectral analysis of YSOs in L1630 tend to follow the same trend seen in the ONC \citep{Feigelson2005}, the Serpens Cloud \citep{Winston2007}, and the Coronet Cluster \citep{Forbrich2007}.  We find a large median X-ray absorbing column density for Class 0/I sources relative to other classifications (Table \ref{avg_prop}) and other studies \citep[e.g., $ N_H \sim $5 x $10^{22}$ cm$^{-2}$ for Class I YSOs in Rho Ophiuchi;][]{Ozawa2005}.  \citet{Winston2007} found Class 0/I sources in the Serpens Cloud core had average X-ray absorbing column densities $\sim$5 times that of Class II and Class III YSOs. We have found Class 0/I sources in L1630 have median X-ray absorbing column densities $\sim$20 times those characteristic of later YSO stages (Table \ref{avg_prop}).  However, note that three of the five X-ray detected Class 0/I sources in L1630 (ID $\#$ 12, 13 and 16; Fig. \ref{ssv}) reside within $\sim$10 arc seconds of each other and hence likely reside in the same, highly obscured region of the L1630 molecular cloud. The spatial proximity of these three YSOs with high values of N$_{H}$ could suggest that for many Class 0/I sources, much of the absorbing material resides in the molecular cloud environment immediately surrounding the protostar, rather than in the collapsing circumstellar envelope of the Class 0/I source.  Furthermore, it is interesting to note that these three YSOs have similar intrinsic X-ray luminosities and plasma temperatures (Table \ref{spec_tbl}, Figures \ref{temp_lum}, \ref{var_lum}).  These three highly embedded Class 0/I sources, which have likely formed around the same time and in a similar environment, hence represent an excellent Class 0/I YSO group for further study.   Our sample of Class II, transition disk, and Class III L1630 YSOs have very similar values of median $N_{H}$ (Table \ref{avg_prop}).  This suggests that ambient cloud material, not circumstellar, dominates $N_{H}$ for these sources and that the Class 0/I YSOs are found near the densest parts of the L1630  cores.

 The X-ray detected sample of YSOs in L1630 range over three orders of magnitude in $L_{X}$ (from $\sim 10^{29}-10^{32}$ $erg $ $ s^{-1}$), suggesting a large range of stellar masses \citep{Preibisch2005}.  In Figure \ref{temp_lum}a, many YSOs in L1630 to lie above the level $L_{X} = 10^{-3}L_{IR}$.  Unless $L_{IR}$ is significantly larger than $L_{bol}$ -- which is more likely for Class II and III YSOs than for Class I -- this suggests that many L1630 YSOs lie above the "saturation level" \citep[e.g.,][]{Preibisch2005}.  However, it is important to note that the YSO X-ray luminosities (along with other spectral properties) were calculated using merged spectra obtained from multiple observations.  Many YSOs in L1630 undergo flaring events that sometimes increase flux by an order of magnitude (Figs. \ref{lc_1}-\ref{lc_4}), possibly resulting in larger values of time-averaged $L_{X}$ than in previous studies.

  In Figure \ref{temp_lum}b and Table \ref{avg_prop}, we see that the sample of Class III YSOs in L1630 tend to have lower X-ray luminosities than Class 0/I, Class II and transition disk objects, yet these L1630 YSO classifications have similar plasma temperatures.  The similar temperatures found in the one-temperature fits and the "hot" components of the two-temperature fits for YSOs in L1630 indicate that magnetic reconnection dominates the plasma heating for all classes.  FP07 and \citet{Winston2007} also found no clear correlation between plasma temperature and YSO classification; however, FP07 found Class I and Class II objects tend to show hotter plasma temperatures than Class III YSOs.  In most cases, accretion shocks should not be contributing to $L_{X}$ or $T_{X}$ in non-accreting Class III YSOs.  Hence, either accretion does not play a significant role in increasing plasma temperature and luminosity, or Class III YSOs have another mechanism to heat plasma to similar temperatures. 
  
   Accretion shocks are expected to produce X-ray emission at characteristic temperatures of $\sim$ 3 MK and cooler\citep{Kastner2002,Huenemoerder2007,Grosso2010,Teets2012}.  FP07 concluded that Class I and II sources  in their study show no such excess soft emission indicative of accretion shocks.  As shown in Figures \ref{temp_lum}b-e, two transition disks and one Class II YSOs do display a cool plasma component indicative of accretion shocks. These sources also tend to be more X-ray luminous than other YSOs in L1630; however, this could merely reflect the fact that high count-rate sources are easier to fit with two-temperature models.  In the case of Class 0/I YSOs (which are actively accreting), non-detection of a cool plasma component does not necessarily imply that accretion shocks are not present.  Class 0/I YSOs (particularly those in L1630) are heavily absorbed (Table \ref{avg_prop}; Figs. \ref{temp_lum}c and \ref{var_lum}c) and any unabsorbed soft X-ray emission from these sources indicative of accretion shocks would be difficult to detect with X-ray CCD spectro-imaging. Furthermore, observations to include only those photons with energies $\geq$ 0.5 keV may have removed some of the soft excess detectable in L1630 YSOs by Chandra. We also note that one Class III YSO was found to have a $\sim$1 $MK$ plasma component, demonstrating that processes other than accretion (e.g., coronal activitiy) may also contribute to soft X-ray fluxes from YSOs. However, in the absence of H$\alpha$ spectroscopy, we cannot exclude the possibility that this system is actively accreting gas, despite a lack of thermal IR excess due to circumstellar dust \citep{Guenther2013}.

  A weak trend of increasing plasma temperature with infrared luminosity can be seen for YSOs with $log (L_{IR})\gtrsim 33$ in Figure \ref{temp_lum}d.  The source of infrared luminosity in Class III sources should be mostly photospheric, as they are non-accreting and have little or no disk material, whereas in earlier stages of star formation (e.g. Class 0/I, Class II, and transition disks), the disk (and for Class 0/I, the infalling envelope) contributes significantly to infrared emission.  If $T_{X}$ is an indication of magnetic activity, then Figure \ref{temp_lum}d would imply YSOs with disks tend to have more magnetic activity as $L_{IR}$ increases.  Figure \ref{temp_lum}e shows a general decrease in plasma temperature with increasing $L_{X}/L_{IR}$.  More interestingly, the Class III sources in L1630, which as a class have relatively low X-ray luminosities but a large spread of infrared luminosities, tend to decrease in plasma temperature with increasing $L_{X}/L_{IR}$.

 \section{Conclusions }

We have performed a multi-epoch study of the X-ray spectral and variability properties of young stellar objects in L1630, a star-forming region in the Orion molecular cloud.  We have used archival 2MASS and Spitzer infrared and Chandra X-ray data to classify YSOs with X-ray counterparts.   We combined 11 different Chandra observations obtained over the course of $\sim$4 years to derive variability and spectral properties of X-ray emitting YSOs.

1.  We have generated IR classifications for 87 YSOs, comprising 19 Class 0/I YSOs, 26 Class II YSOs, 21 transition disks, and 21 Class III YSOs.  We identify 52 X-ray counterparts to these YSOs; one Class 0/I, four Class I, 15 Class II, 11 transition disks and 21 Class III YSOs are detected in X-ray emission.  Our detection fractions are generally consistent with previous studies.  As expected, Class 0/I YSOs are rarely detected in soft (0.5-2.0 keV) X-rays and are more readily detected in hard (2.0-8.0 keV) X-rays.  Class II and Class III sources in our survey are more readily detected in the soft band.  The detection rate of transition disk objects is similar for both soft and hard bandpasses and is possibly indicative of stable, non-variable emission.  Based on comparison of X-ray detection fraction of YSOs of various classes with other, similarly sensitive studies, it appears that increasing the number and temporal spacing of X-ray observations may increase the detectability of X-ray emitting Class II YSOs.

2.  Consistent with previous studies, we have confirmed the strong variability of X-ray emission in many YSOs.  However, our study characterizes this variability over a timescale of years, i.e., much larger than the timescales of days or months typical of previous studies.  In our variability study, at least one YSO from each class underwent an order of magnitude increase in count rate over the $\sim$4 year duration of the L1630 Chandra observations.  Class 0/I YSOs show the highest levels of variability, with each member in our study undergoing at least one event of an order of magnitude increase in count rate.  

3.  Spectral modeling was performed on strong ($ > 100$ counts) X-ray emitting YSOs to derive X-ray plasma temperatures, line-of-sight absorbing columns, and fluxes. We find a large median absorption for Class 0/I sources in L1630.  These Class 0/I sources  display median X-ray absorbing column densities $\sim$20 times that of the more evolved YSOs in L1630.  However, these uniformly large inferred $N_{H}$ values could reflect the fact that three of the five X-ray detected Class 0/I sources likely reside in the same, highly optically thick molecular cloud core.  These three X-ray emitting YSOs (SSV 63; ID$\#$s 12, 13, and 16) represent an interesting group for further study, as such X-ray emitting Class 0/I protostars are rare and can probe high-energy phenomena during the earliest stage of star formation.  As a class, transition disks display the smallest levels of X-ray variability and Class 0/I YSOs (marginally) show the largest X-ray variability.  Class III YSOs in L1630 tend to have lower X-ray luminosities than younger classifications, but display similar plasma temperatures.  Since accretion should not be contributing to $L_{X}$ or $T_{X}$ in Class III YSOs, this could indicate a different plasma-heating mechanism is operative in Class III YSOs objects.  Evidence of cool plasma typically associated with accretion shocks was detected in three cTTSs and a Class III YSO.


\begin{figure}
\centering

\includegraphics[scale=0.46]{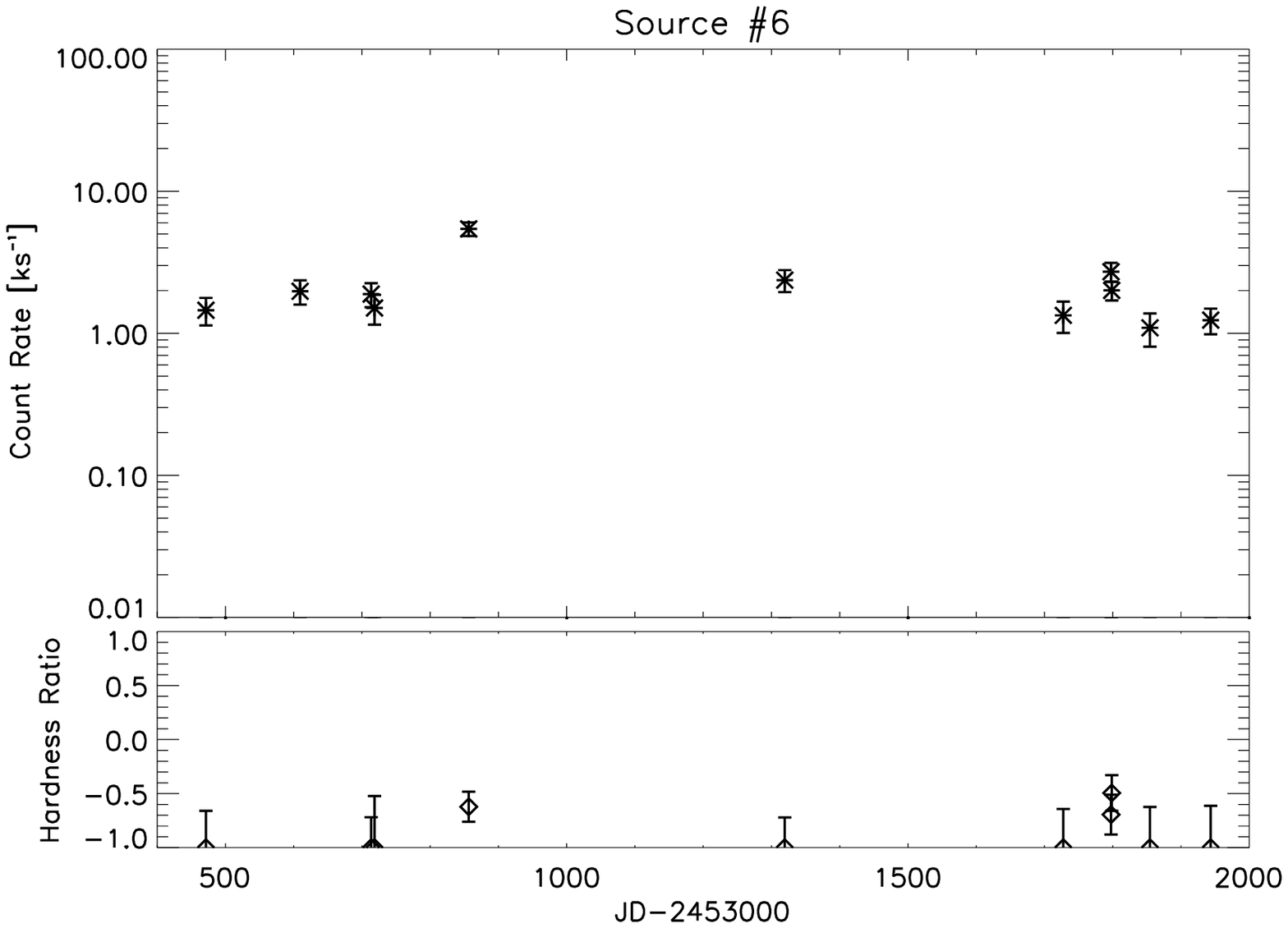} 
\includegraphics[scale=0.46]{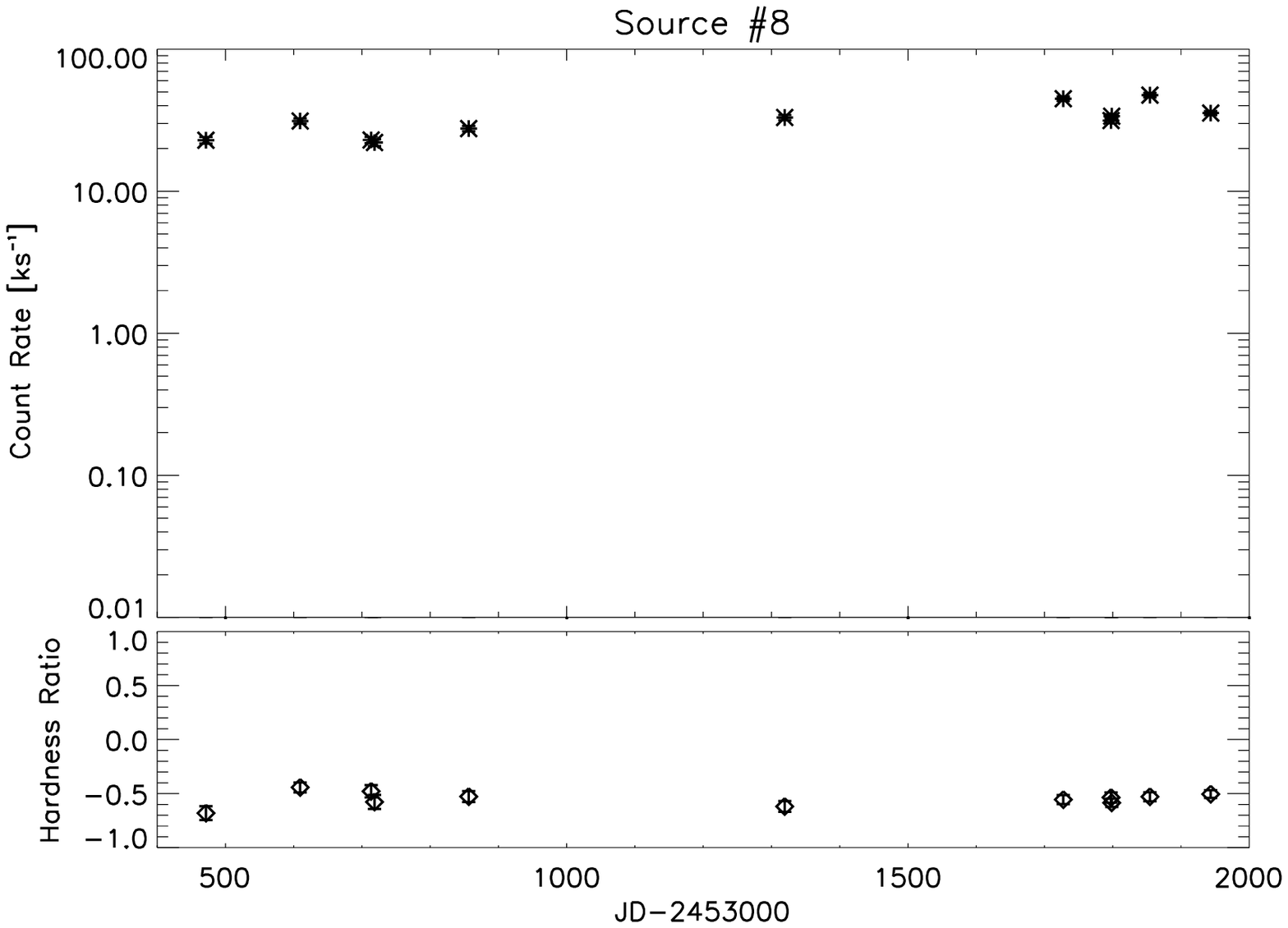} 
\includegraphics[scale=0.46]{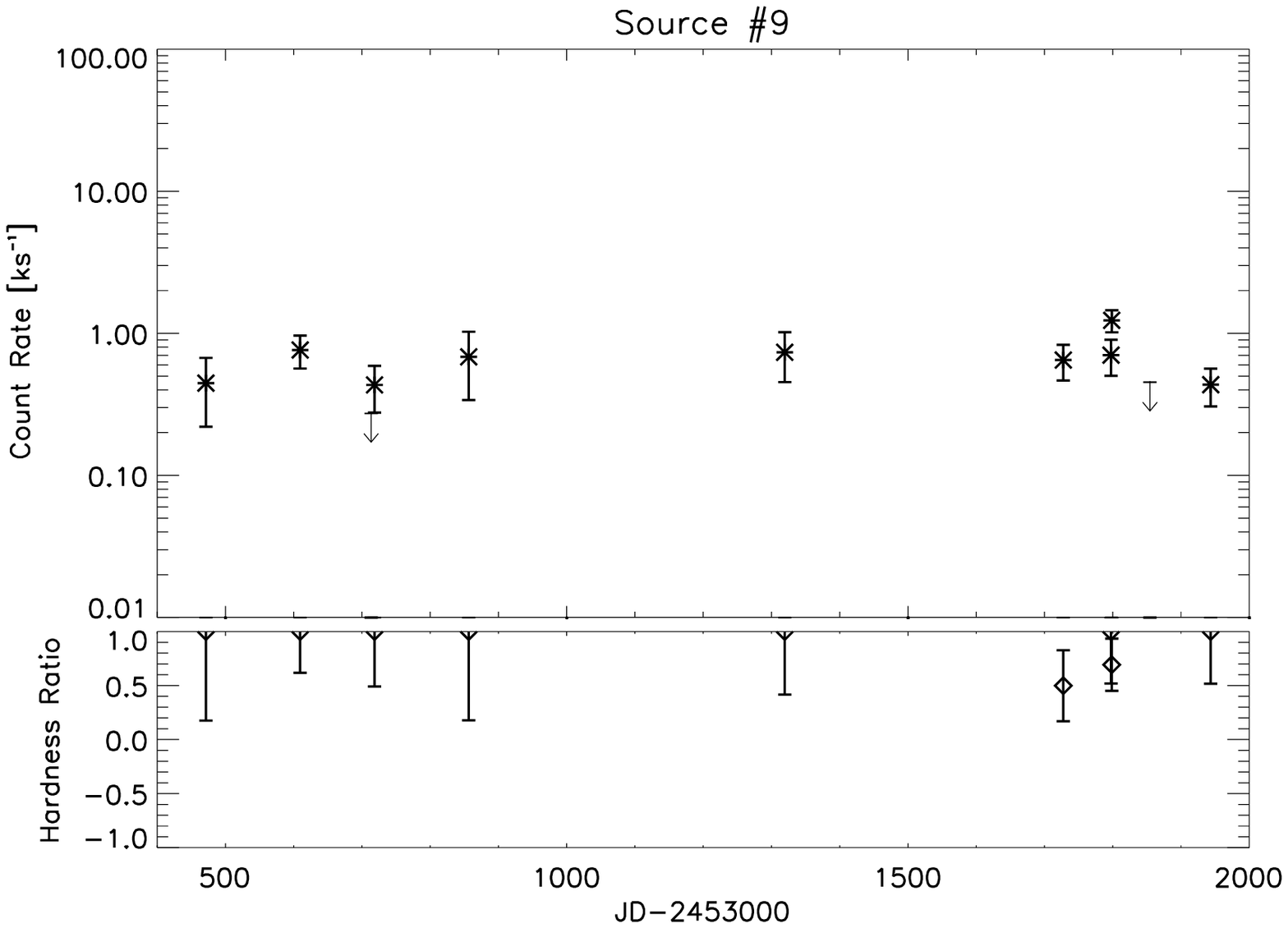} 
\includegraphics[scale=0.46]{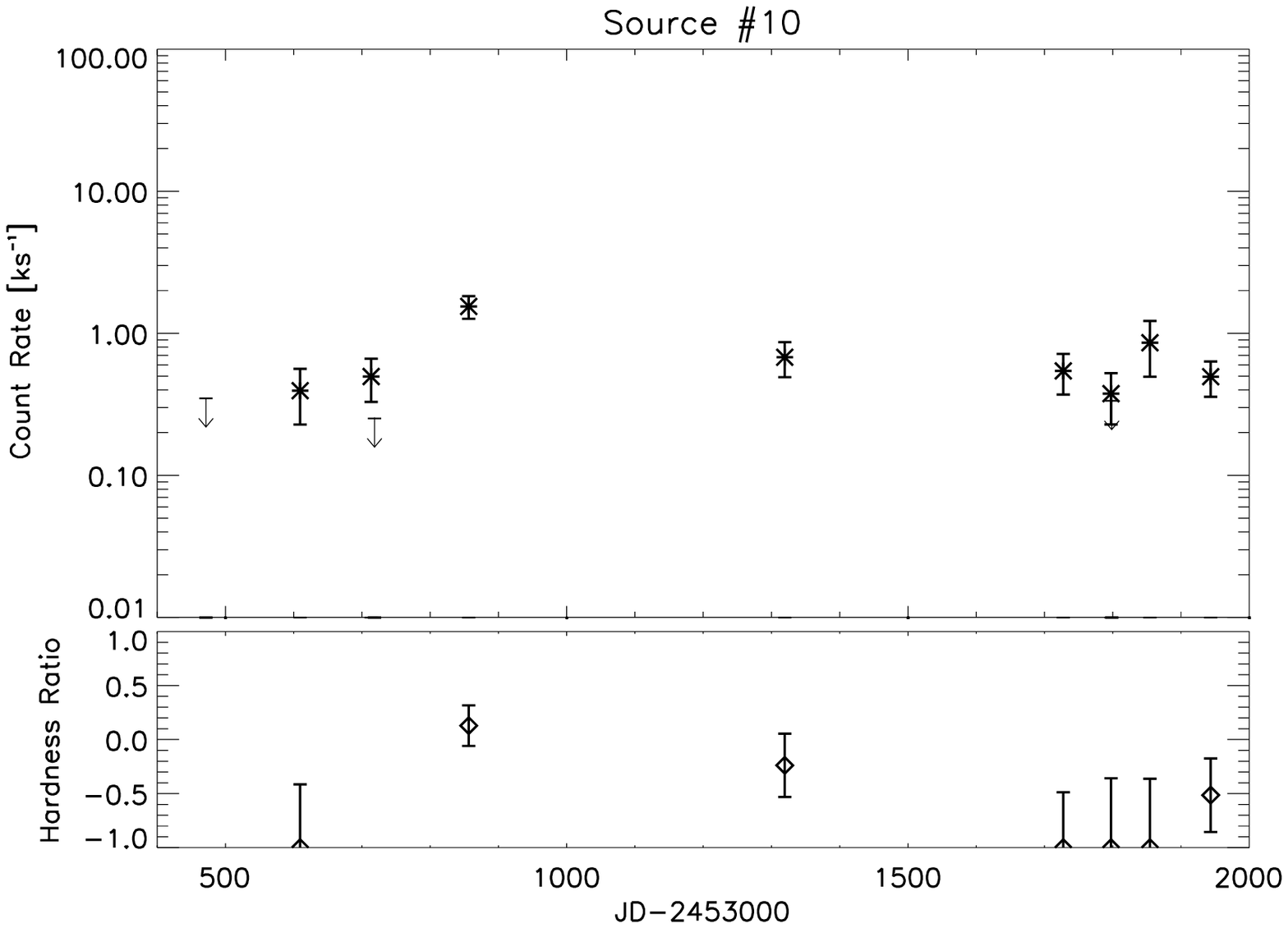} 

\caption{\label{lc_3}Count rate and hardness ratio variability over $\sim$ 4 years for transition disk YSOs in L1630. }

\end{figure}

\begin{figure}
\centering
\includegraphics[scale=0.46]{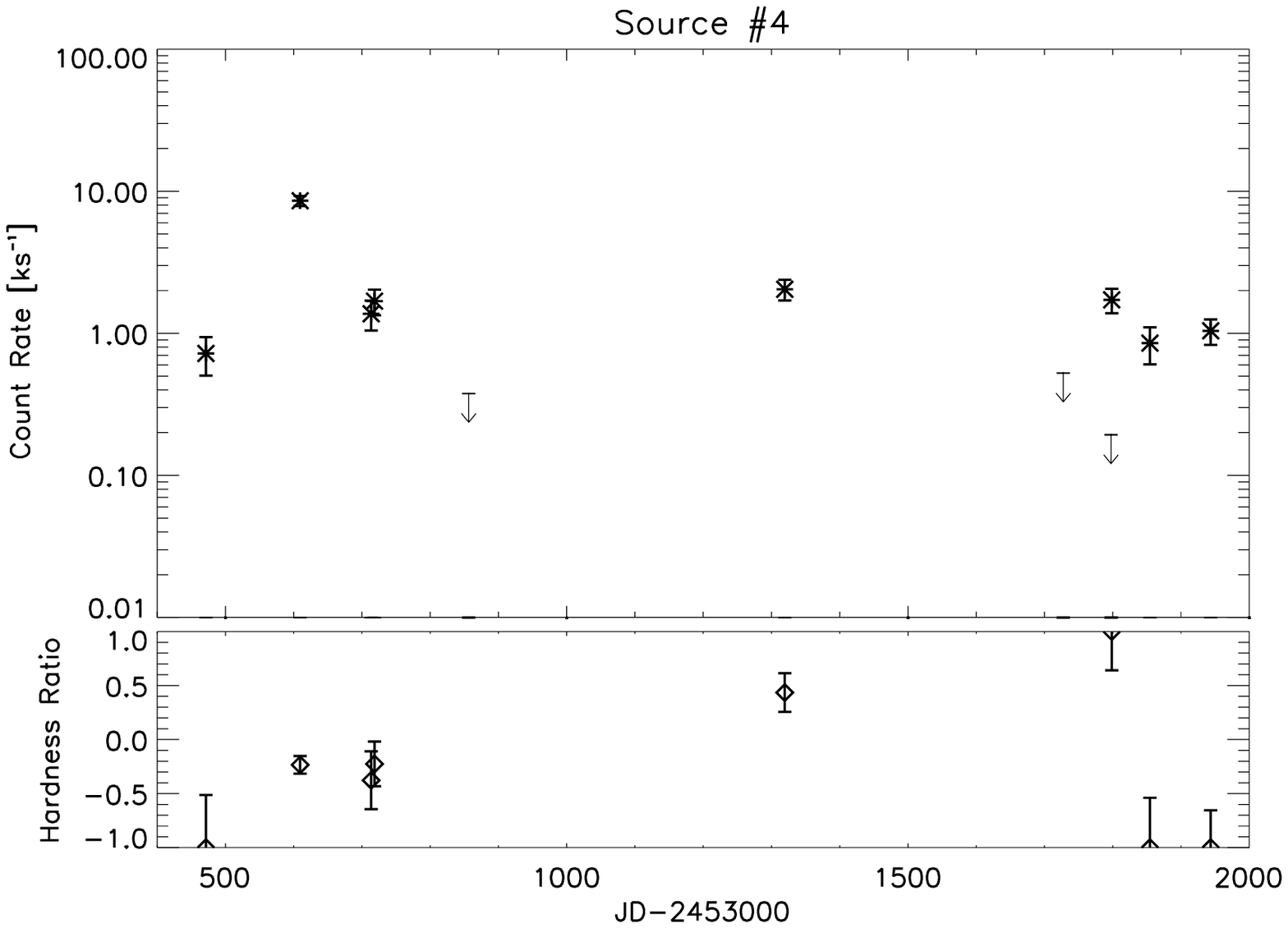} 
\includegraphics[scale=0.46]{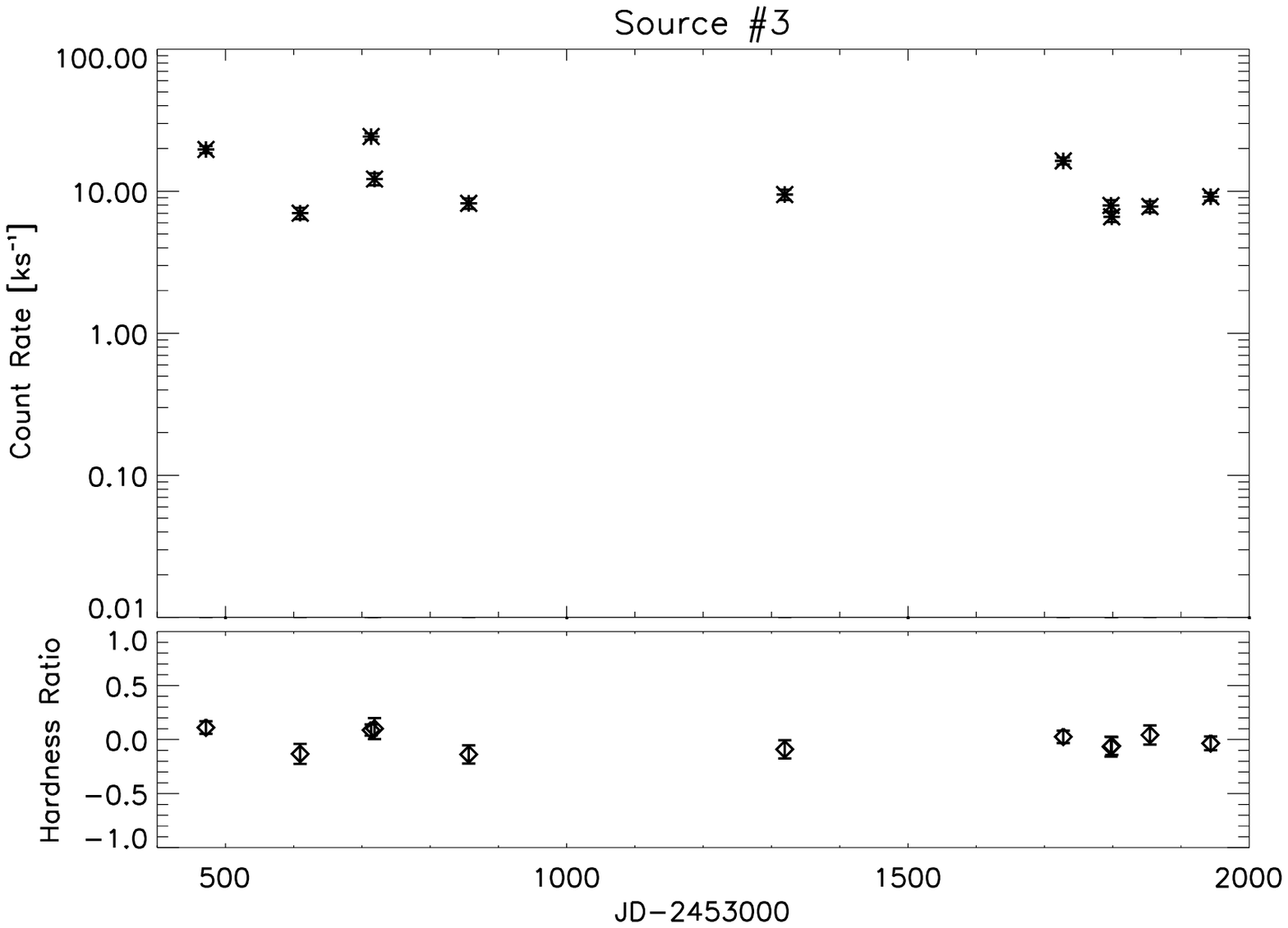} 
\includegraphics[scale=0.46]{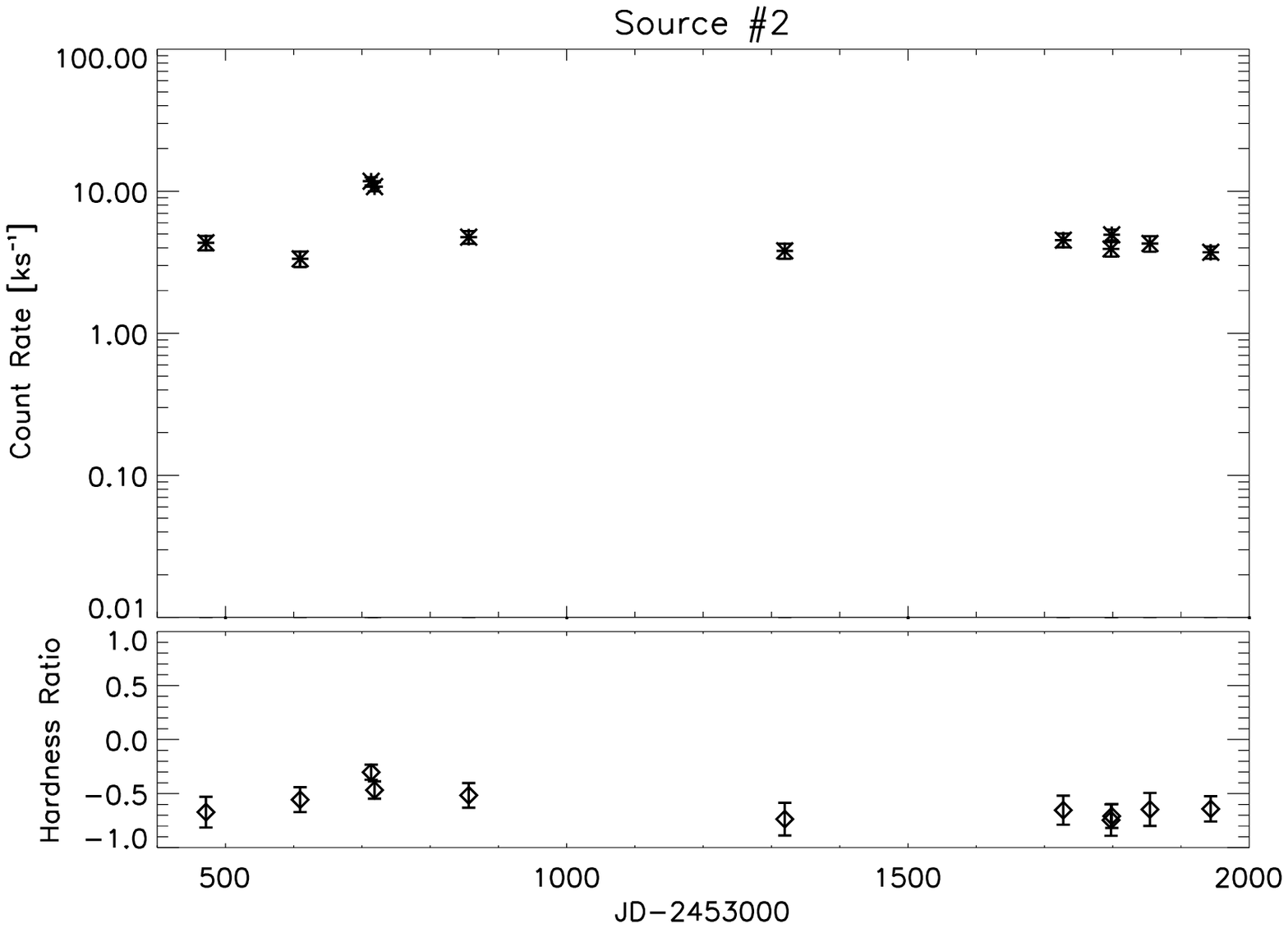} 
\includegraphics[scale=0.46]{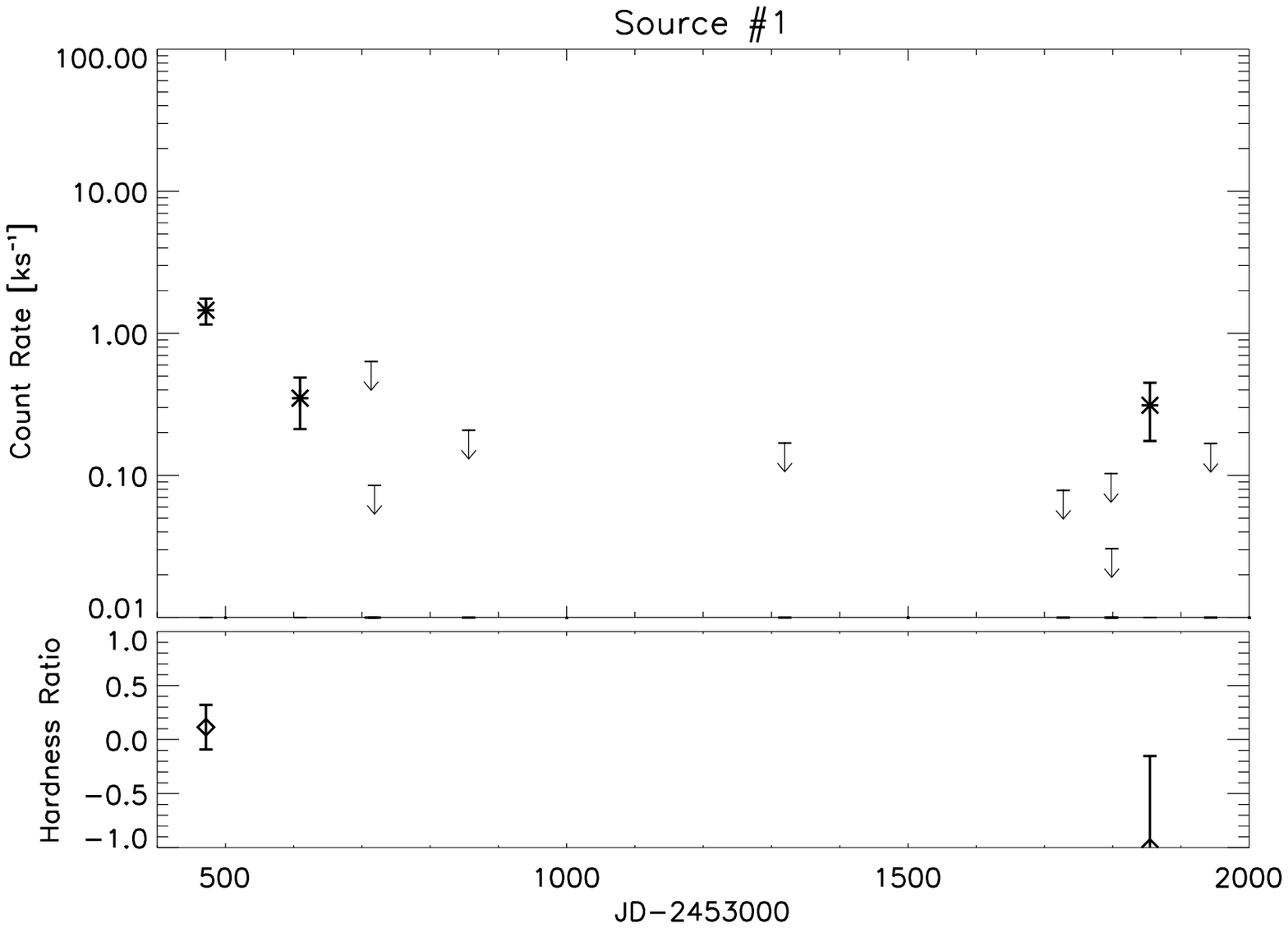} 

\setcounter{figure}{8}
\caption{(continued) Count rate and hardness ratio variability over $\sim$ 4 years for transition disk YSOs in L1630. }
\end{figure}

\begin{figure}
\centering

\includegraphics[scale=0.46]{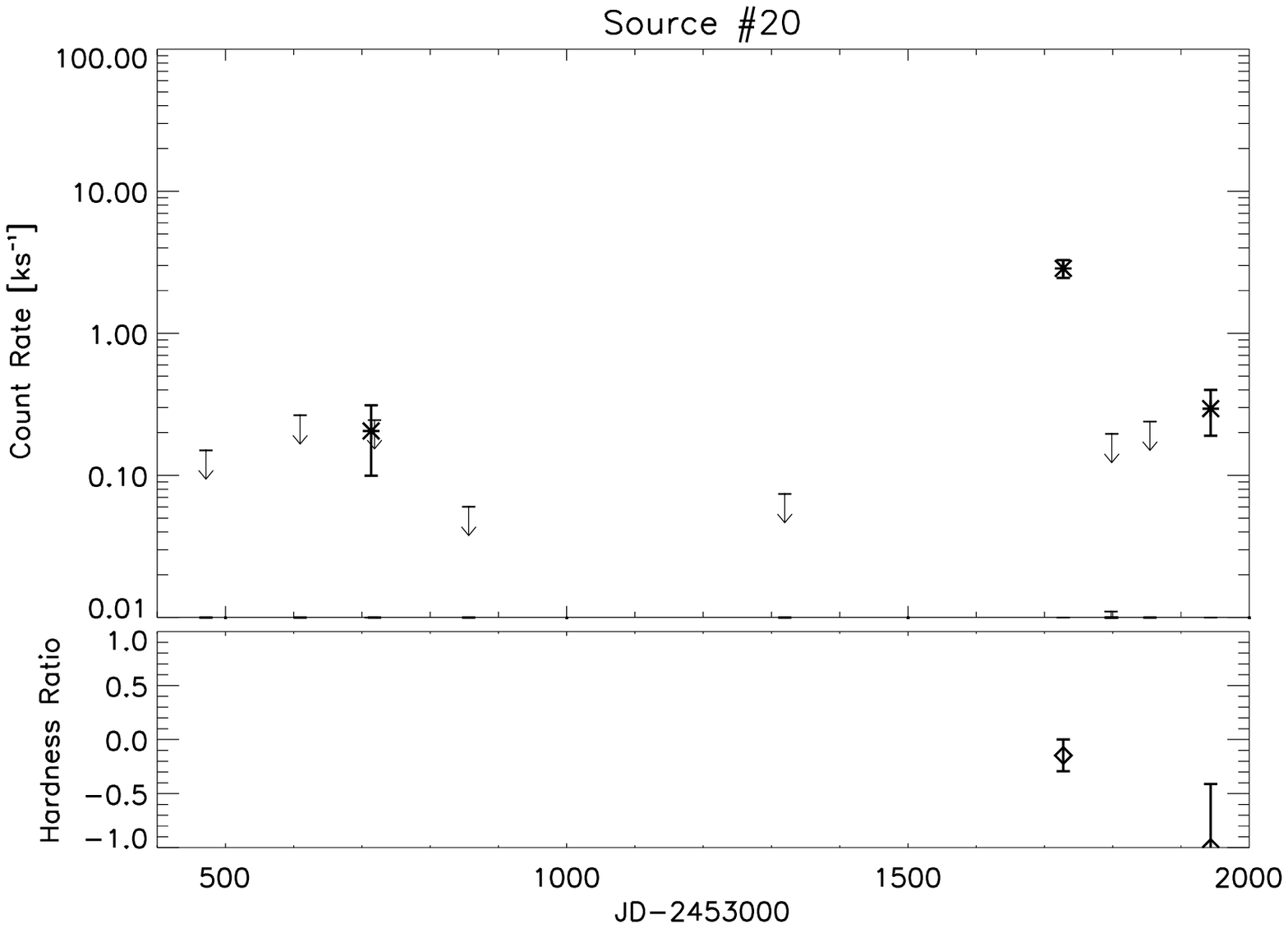} 
\includegraphics[scale=0.46]{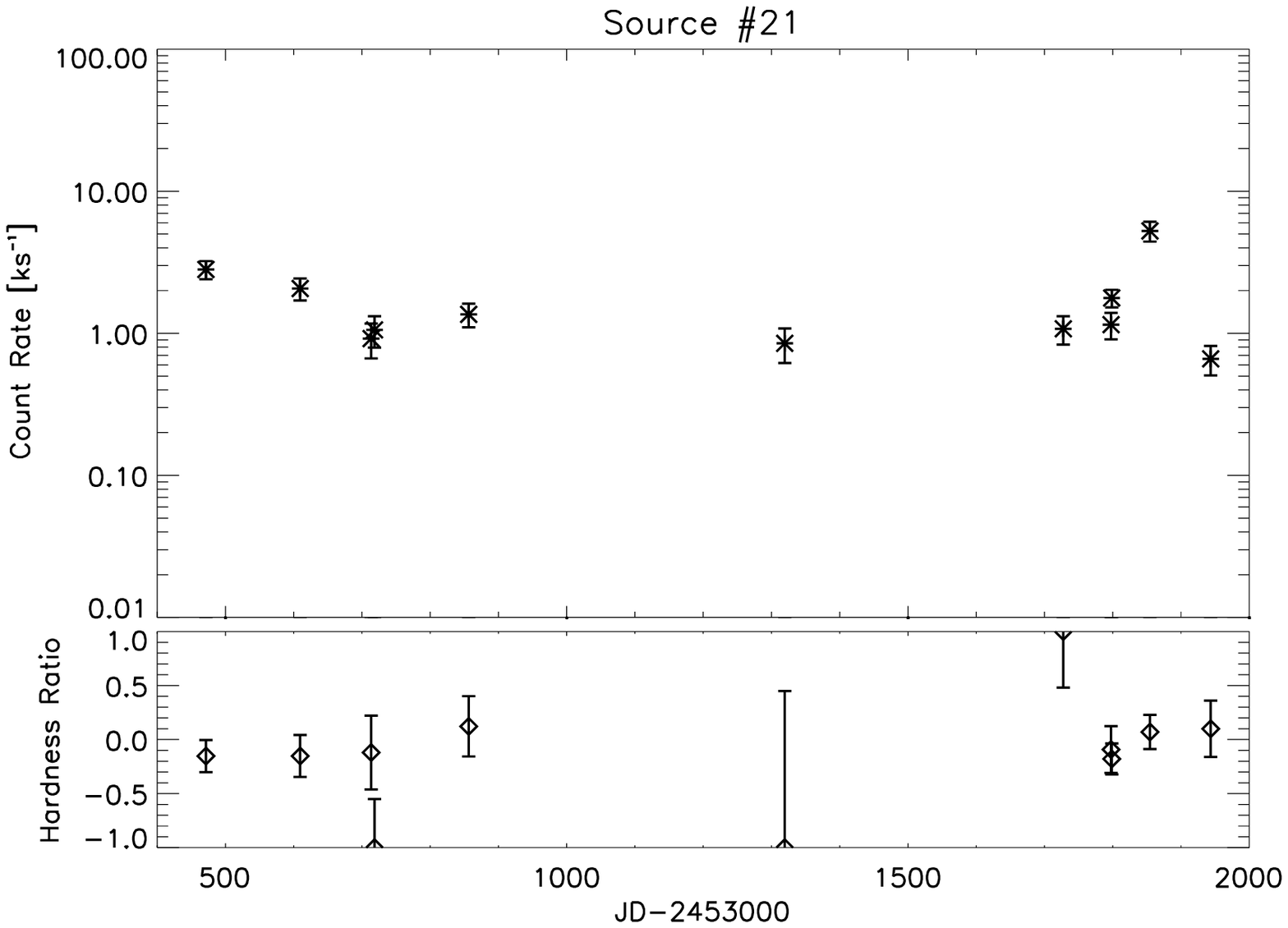} 
\includegraphics[scale=0.46]{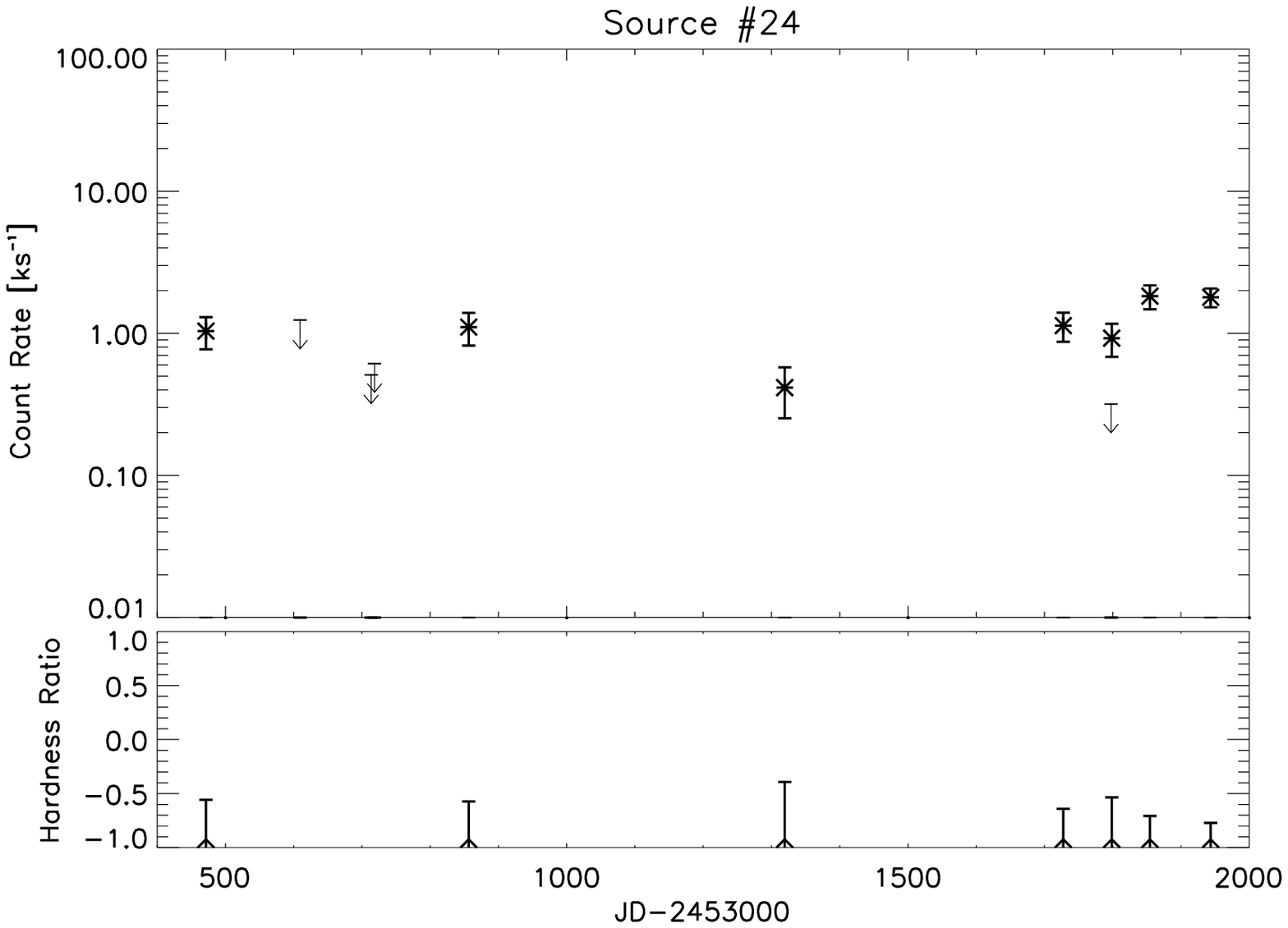} 
\includegraphics[scale=0.46]{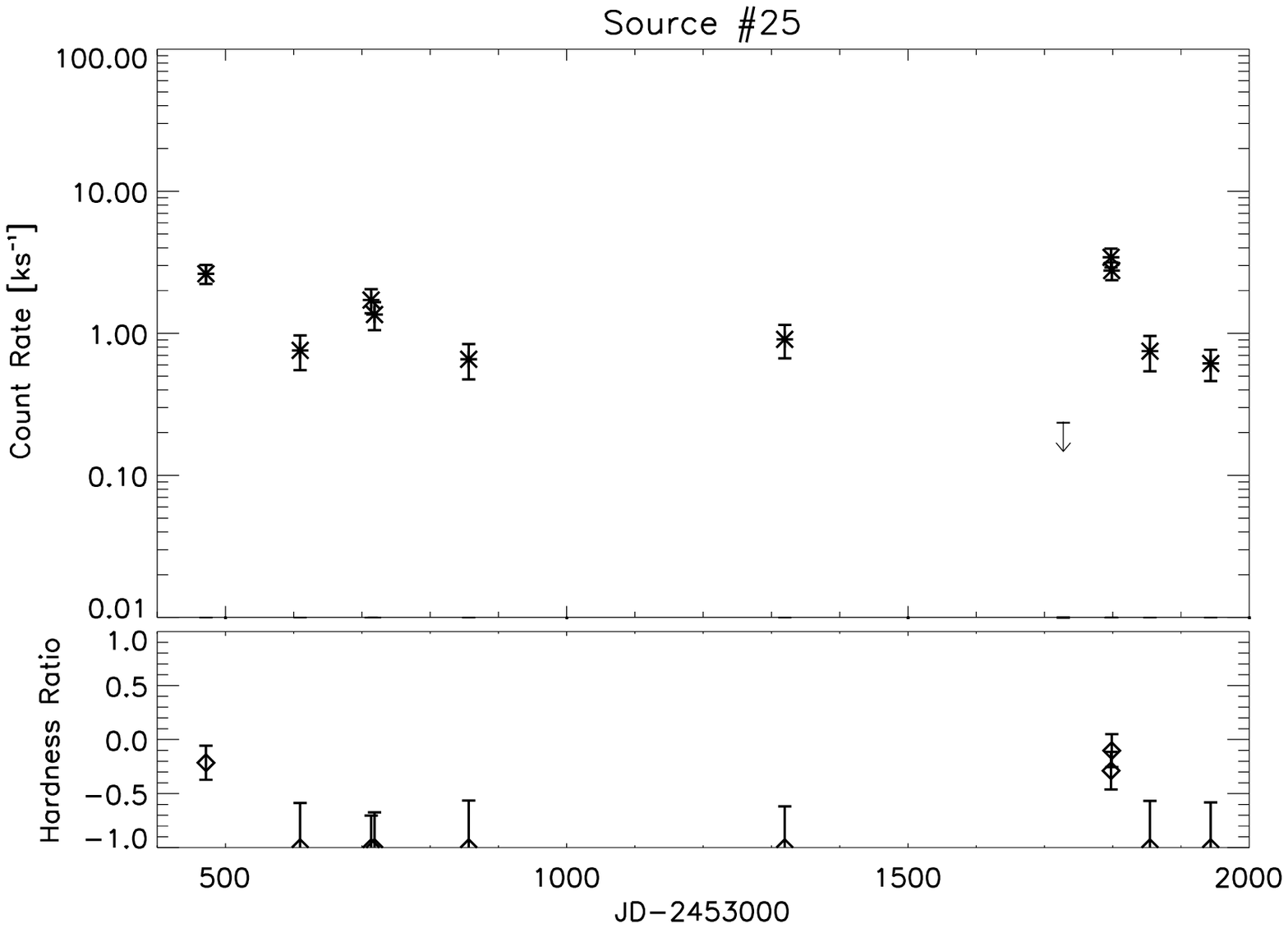} 

\caption{\label{lc_4}Count rate and hardness ratio variability over $\sim$ 4 years for Class III YSOs in L1630.}

\end{figure}

\begin{figure}
\centering
\includegraphics[scale=0.46]{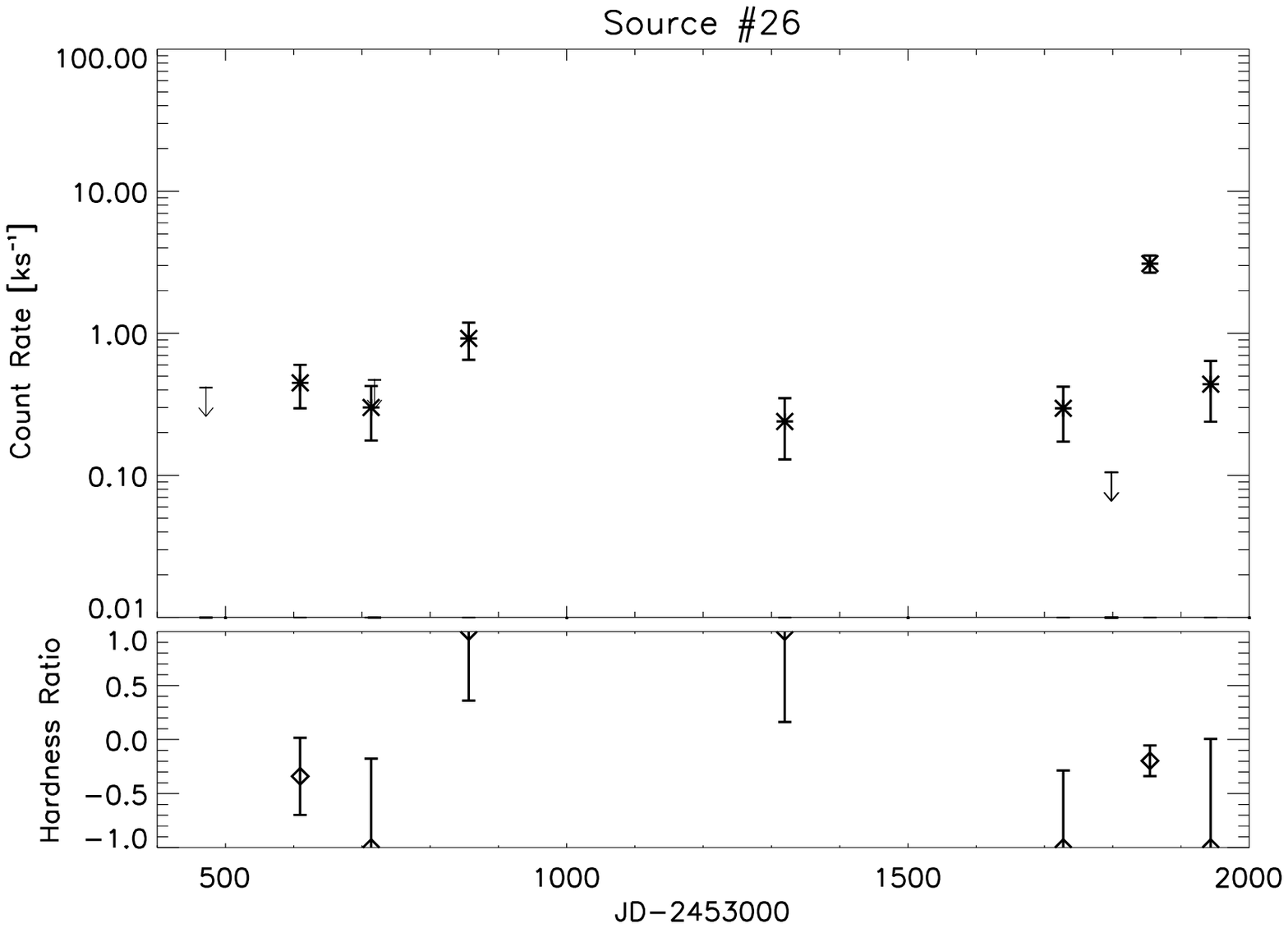} 
\includegraphics[scale=0.46]{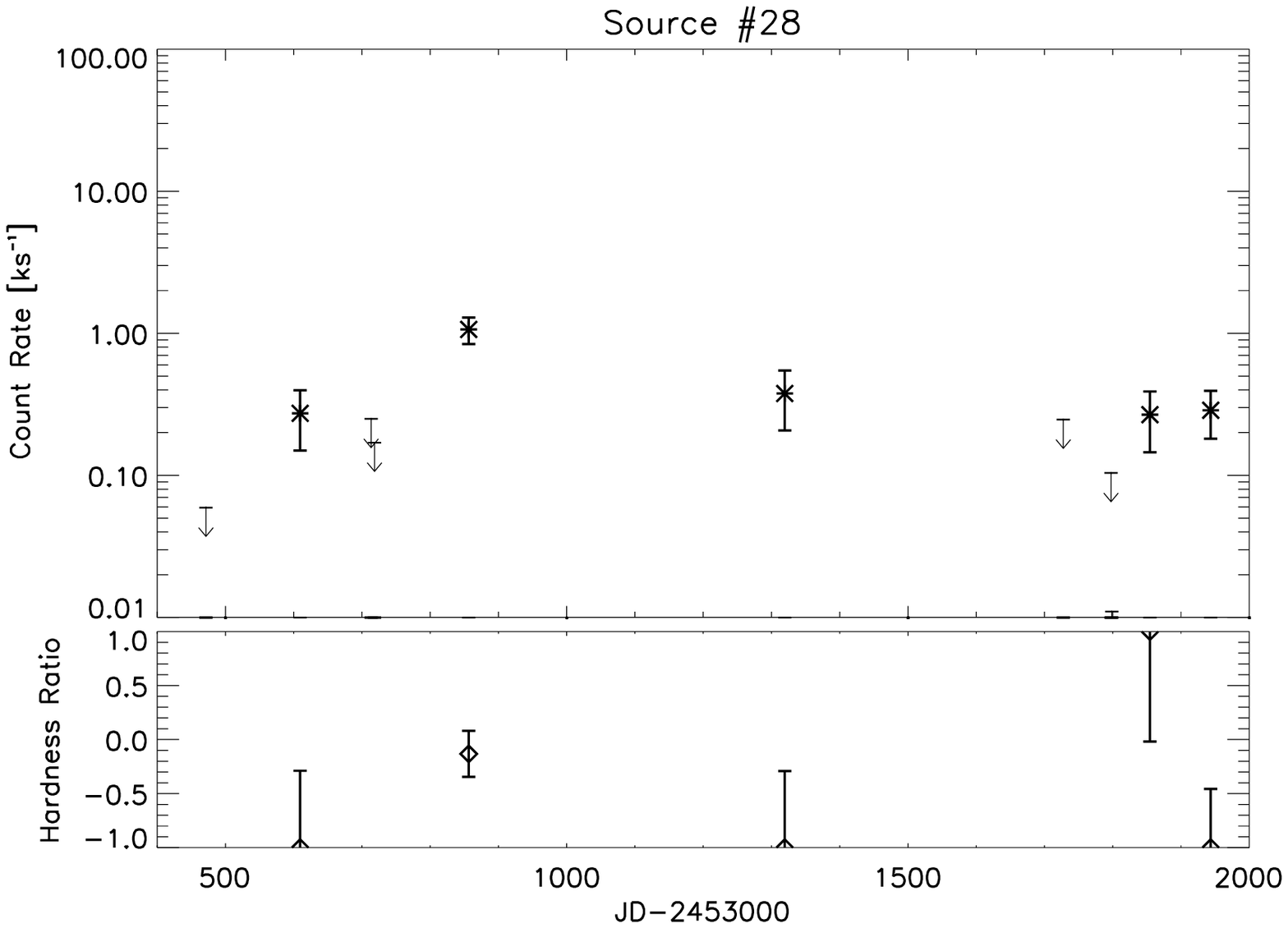} 
\includegraphics[scale=0.46]{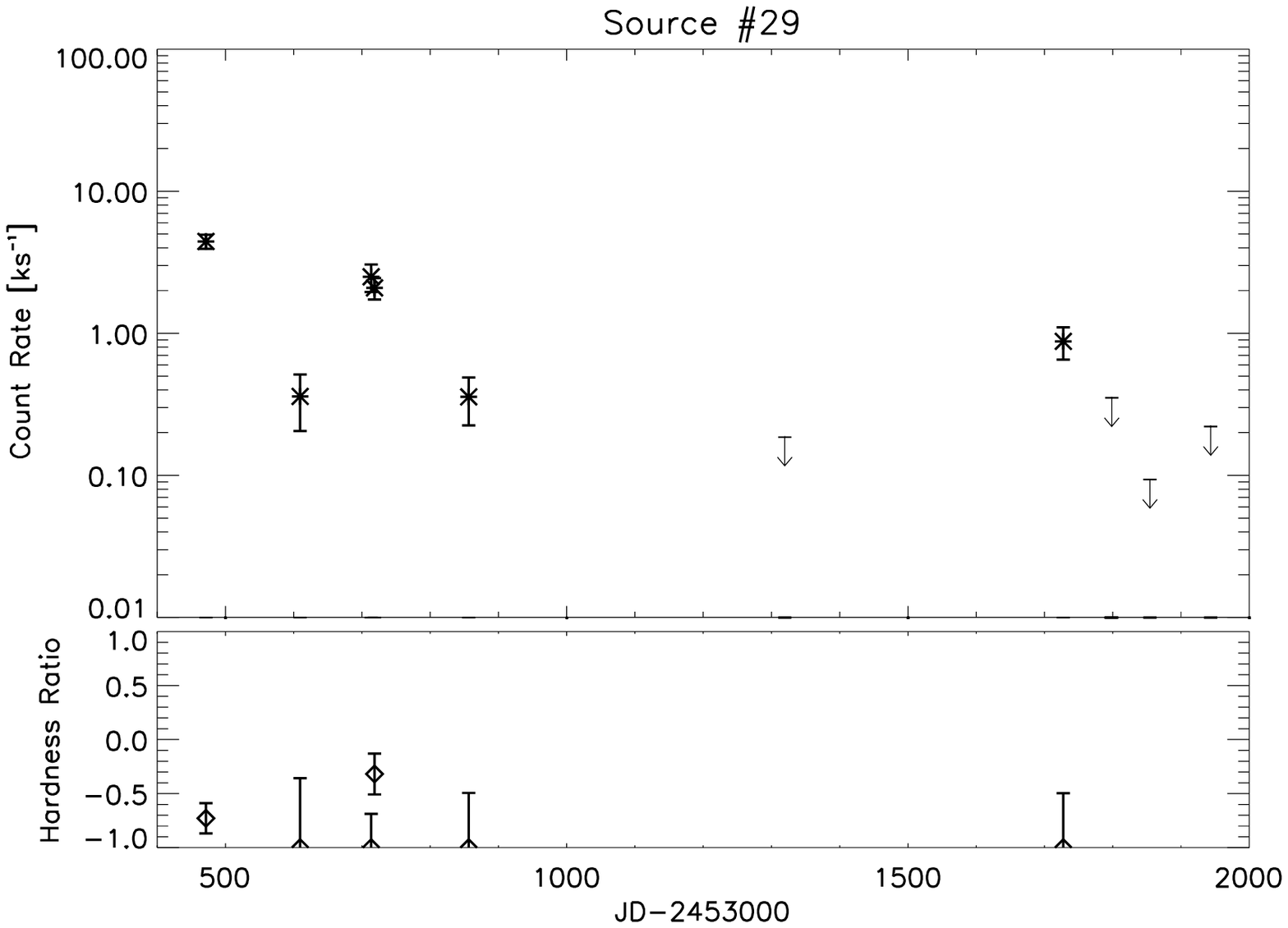} 
\includegraphics[scale=0.46]{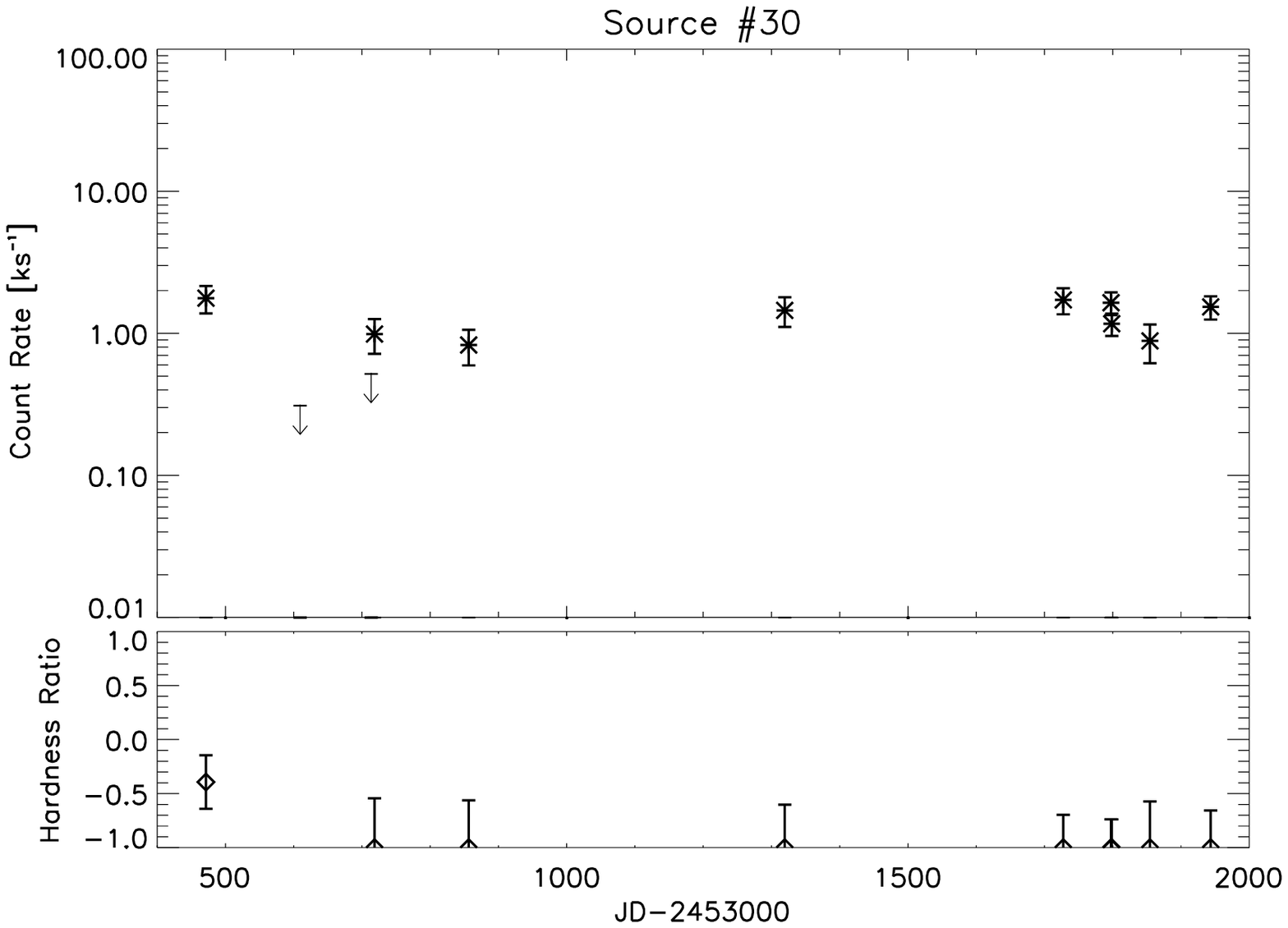} 
\includegraphics[scale=0.46]{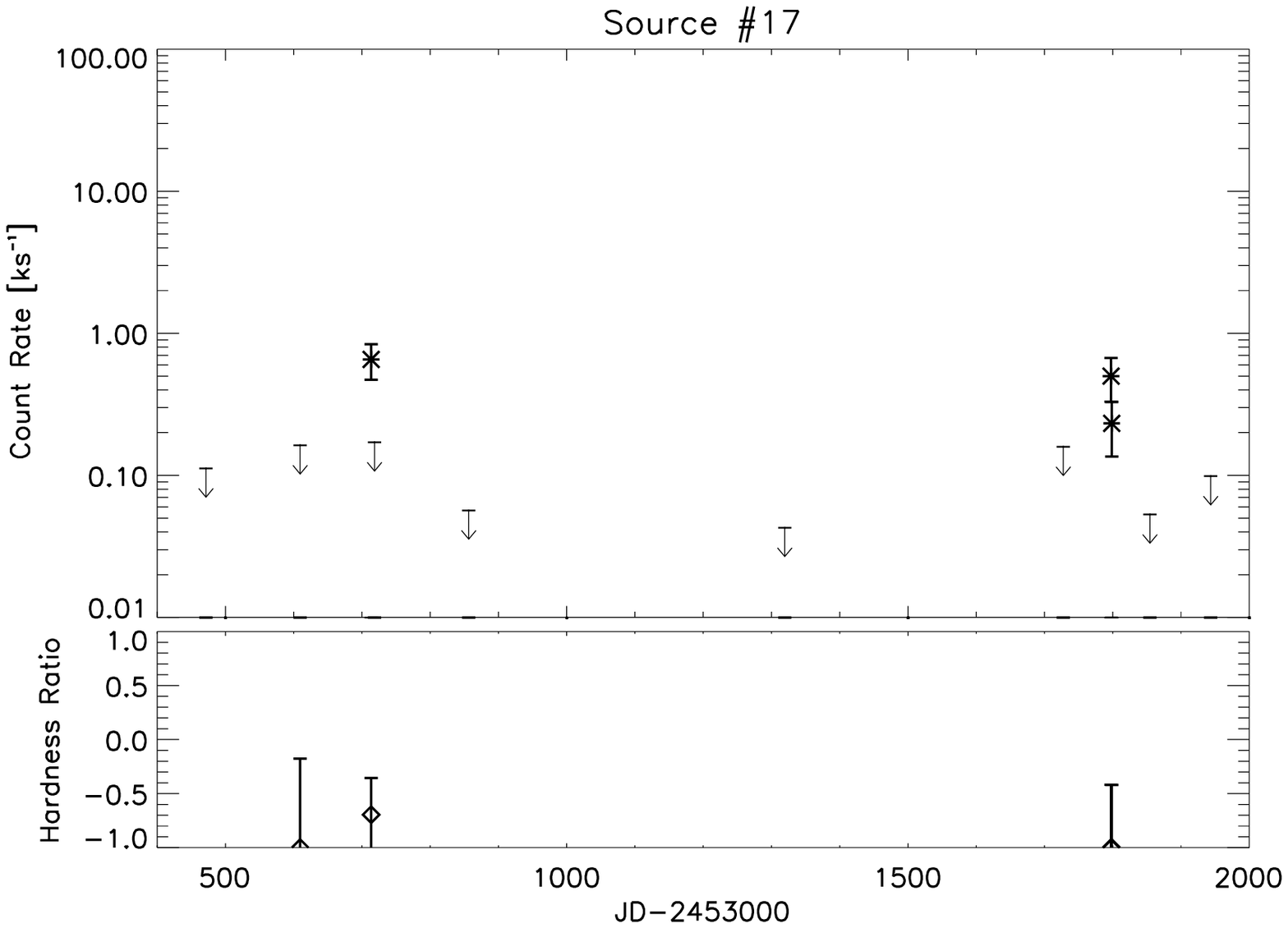}

\setcounter{figure}{9}
\caption{(continued) Count rate and hardness ratio variability over $\sim$ 4 years for Class III YSOs in L1630.}
\end{figure}

\newpage
 \begin{figure}
\centering

\includegraphics[scale=0.6,angle=90]{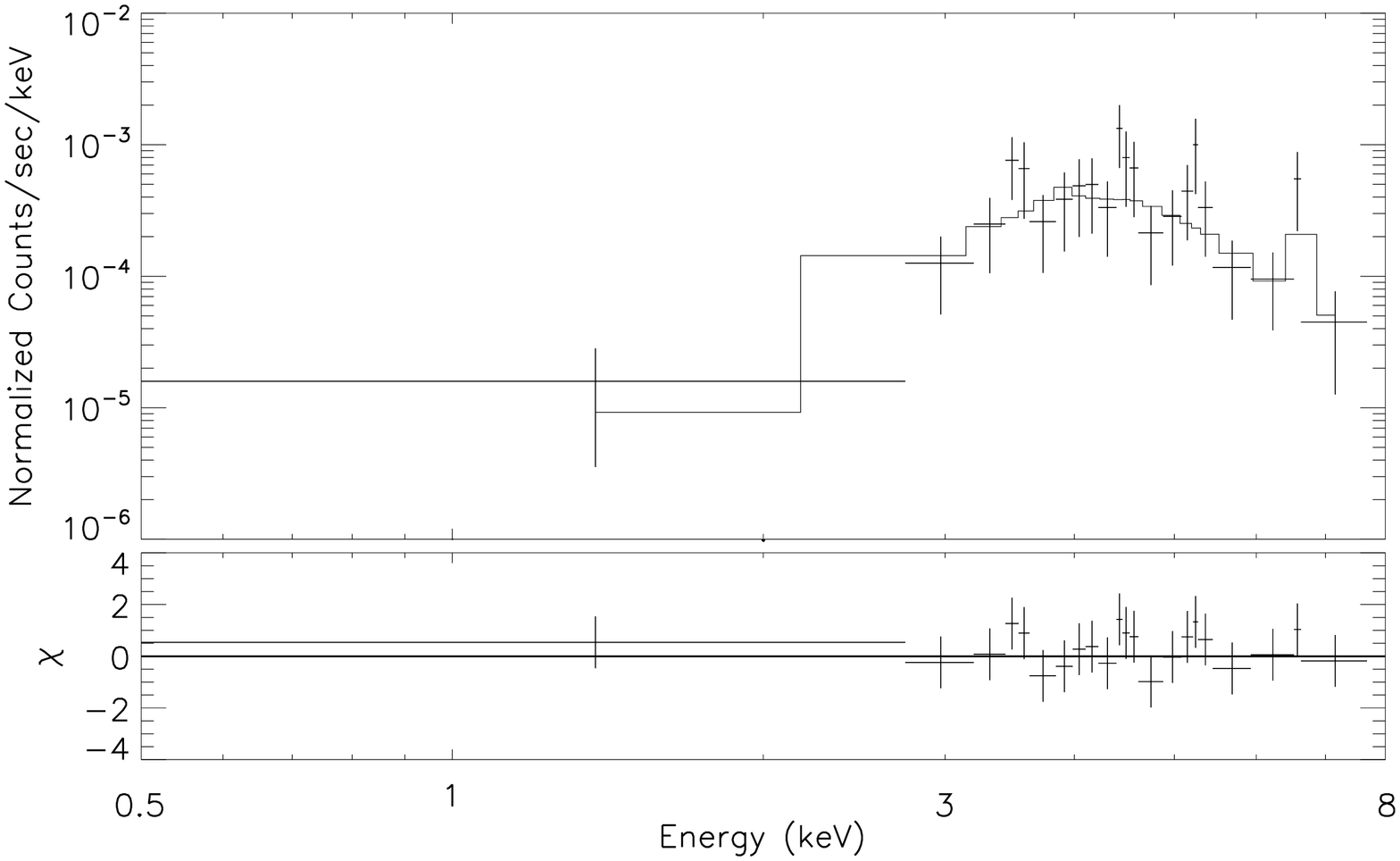}
\includegraphics[scale=0.60,angle=90]{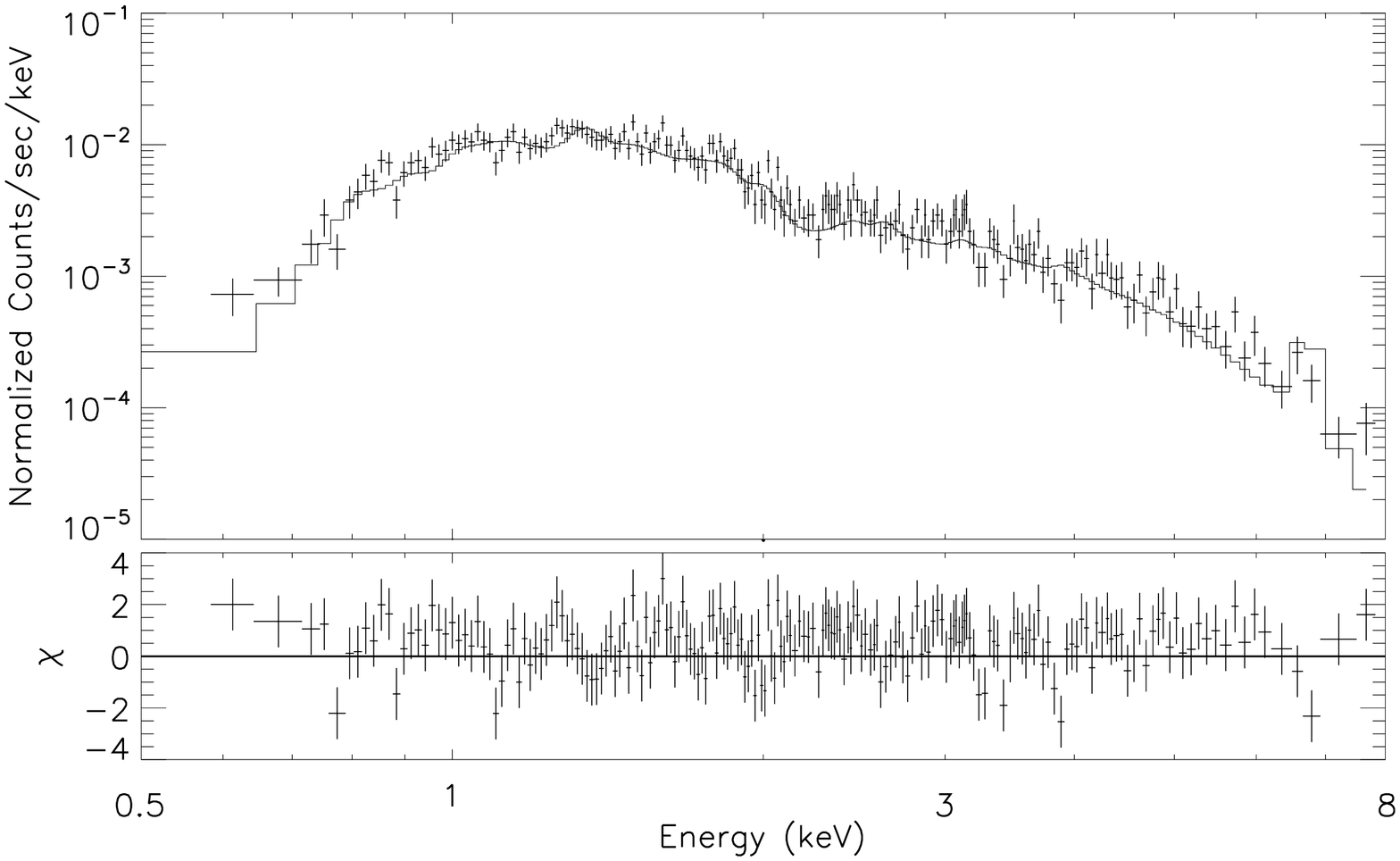}

\caption{ \label{spec} Representative X-ray spectra (crosses) fit with models consisting of one- or two-temperature plasmas suffering intervening absorption (histogram).  Examples include Class 0/I ID\# 16 (top) and Class II ID\# 46 (bottom).}

\end{figure}

\newpage
\begin{figure}

\centering
\includegraphics[scale=0.6,angle=90]{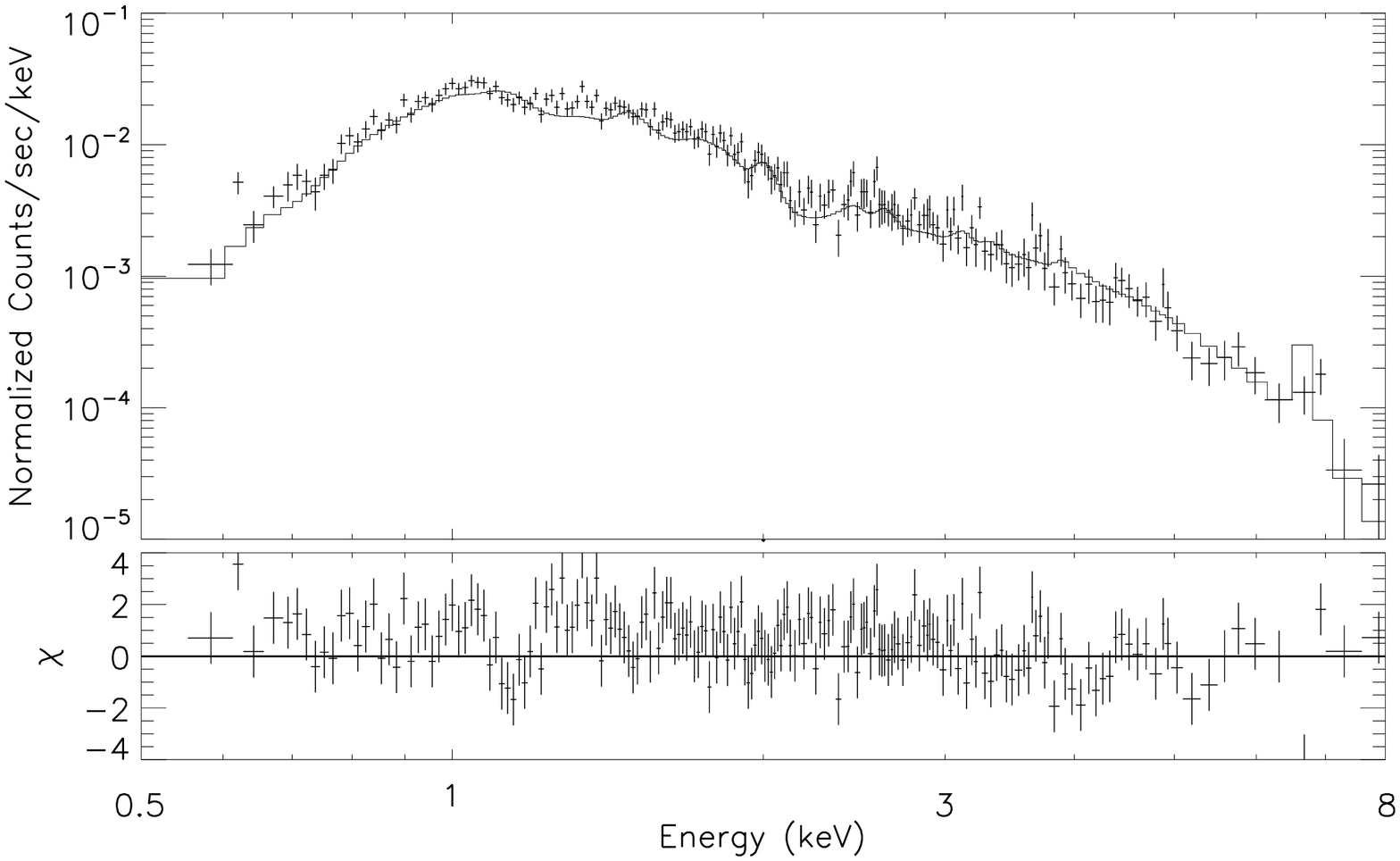}
\includegraphics[scale=0.60,angle=90]{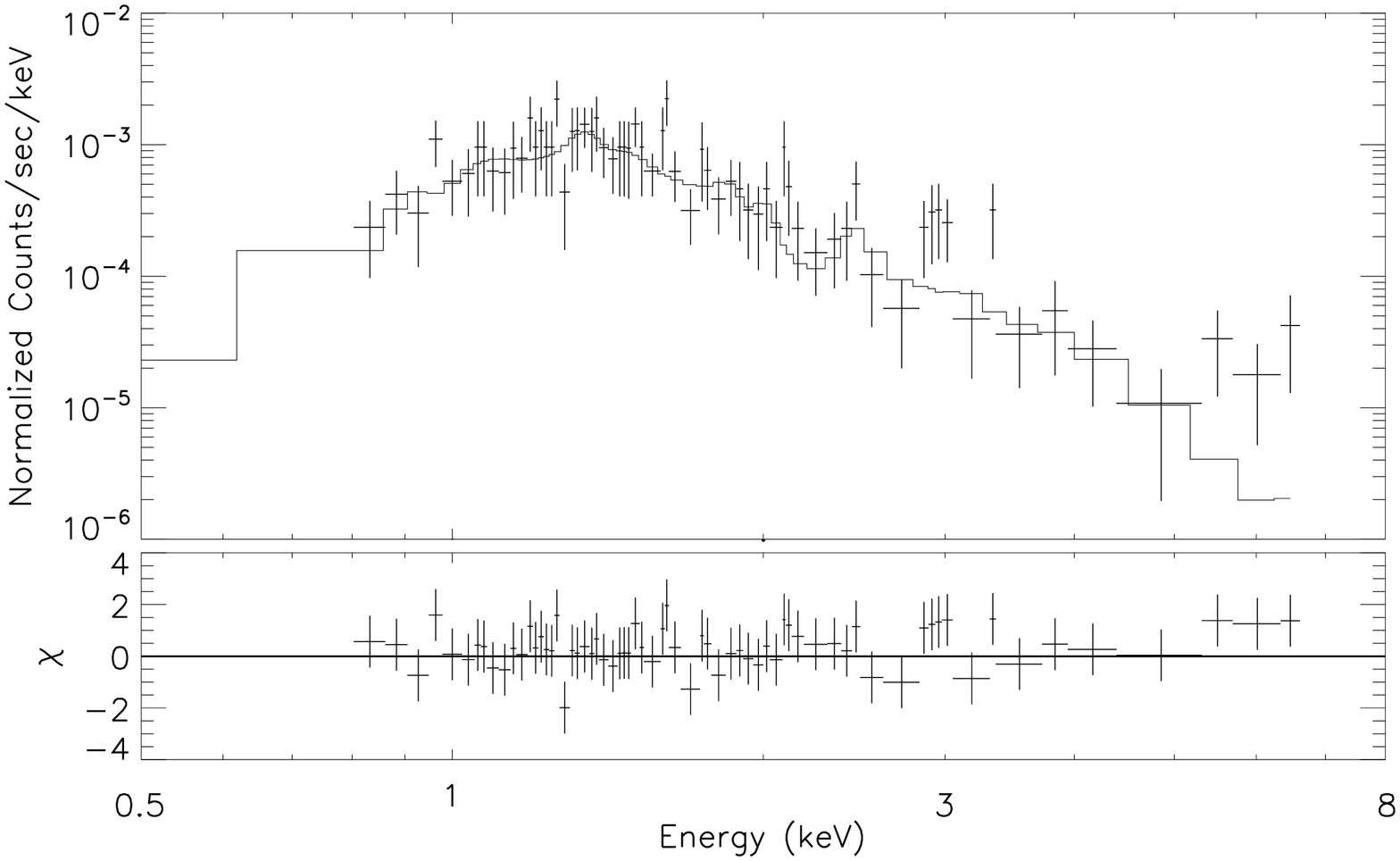}

\setcounter{figure}{12}

\caption{ (continued) Representative X-ray spectra (crosses) fit with models consisting of one- or two-temperature plasmas suffering intervening absorption (histogram).  Examples include Transition Disk ID\# 8 (top) and Class III ID\# 25 (bottom).}

\end{figure}

\newpage

\newpage
\clearpage
\begin{deluxetable}{c c c c c c}

\centering
\tablecolumns{6}
\tablewidth{400pt}
\tabletypesize{\small}

\tablecaption{Chandra X-ray Observations \label{chan_obsid}}

\tablehead{

\colhead{Obs ID} &
\colhead{Date} &
\colhead{JD-2453000} &
\colhead{Exposure Time} &
\colhead{X-ray Sources } &
\colhead{Infrared } \\

\colhead{} &
\colhead{} &
\colhead{} &
\colhead{(ks)} &
\colhead{Detected\tablenotemark{1}} &
\colhead{Counterparts} \vspace{-3mm} \\

}

\startdata

5382 & 04/11/05 & 471 &18.208 & 31 & 22\\ 
5383 & 08/27/05 & 609 & 19.879 & 41 & 27\\ 
5384 & 12/09/05 & 713 & 19.699 & 37 & 25\\ 
6413 & 12/14/05 & 718 & 18.101 & 39 & 21\\ 
6414 & 05/01/06 & 856 & 21.649 & 43 & 28\\ 
6415 & 08/07/07 & 1319 &  20.454 & 41 & 25\\ 
9915 & 09/18/08 & 1727 & 19.896 & 39 & 27\\ 
10763 & 11/27/08 & 1797 & 19.95 & 40 & 24\\ 
8585 & 11/28/08 & 1798 & 28.83 & 55 & 26\\ 
9916 & 01/23/09 & 1854 & 18.406 & 35 & 23\\ 
9917 & 04/22/09 & 1943 & 29.784 & 45 & 26\\

\tablenotetext{1}{Sources considered for detection must also have been detected in merged ($\sim$240 ks) image.}

\enddata

\end{deluxetable}

\newpage
\clearpage

\begin{deluxetable}{ccccccccccccccc}

\centering
\rotate
\tablecolumns{15}
\tablewidth{585pt}
\tabletypesize{\tiny}

\tablecaption{X-ray Sources with Infrared Counterparts \label{xray_IR_22}}

\tablehead{

\colhead{ID} &
\colhead{RA} &
\colhead{DEC} &
\colhead{Eff. Exposure Time} &
\colhead{Count Rate} &
\colhead{Med E} &
\colhead{\tablenotemark{1}$N_{H}$} &
\colhead{\tablenotemark{2}Intrinsic $L_{X}$} &
\colhead{H-K} &
\colhead{J-H} &
\colhead{[3.6-4.5]} &
\colhead{[5.8-8.0]} &
\colhead{$L_{IR}$} &
\colhead{Class} &
\colhead{2MASS ID} \vspace{0.5mm} \\

\colhead{} &
\colhead{} &
\colhead{} &
\colhead{(ks)} &
\colhead{$(ks^{-1})$} &
\colhead{(keV)} &
\colhead{$(\times 10^{22}$ $cm^{-2})$} &
\colhead{$(erg$ $s^{-1})$} &
\colhead{} &
\colhead{} &
\colhead{} &
\colhead{} &
\colhead{($erg$ $s^{-1}$)} &
\colhead{} &
\colhead{} \vspace{-1.5mm} \\

}

\startdata
1 & 05:46:08.98 & -00:02:15.4 & 199.04 & 0.23 & 1.96 & 1.21 & -- & 0.66 & 1.13 & 0.23 & 1.58 & 3.39$\times 10^{33}$ & Transition\tablenotemark{3} & 05460898-0002155\\ 
2 & 05:46:22.43 & -00:08:52.6 & 221.48 & 5.66 & 1.41 & 0.32 & 1.92$\times 10^{30}$ & 0.44 & 0.86 & 0.15 & 0.65 & 1.76$\times 10^{33}$ & Transition\tablenotemark{3} & 05462243-0008526\\ 
3 & 05:46:11.60 & -00:02:20.6 & 194.41 & 11.28 & 2.03 & 1.29 & 5.36$\times 10^{30}$ & 0.77 & 1.40 & 0.14 & 1.12 & 3.28$\times 10^{33}$ & Transition\tablenotemark{3} & 05461157-0002202\\ 
4 & 05:46:17.71 & -00:00:14.2 & 189.30 & 1.94 & 1.83 & 1.06 & 9.1$\times 10^{29}$ & 0.88 & 1.56 & 0.07 & 1.71 & 2.32$\times 10^{33}$ & Transition\tablenotemark{3} & 05461771-0000143\\ 
5 & 05:45:38.26 & -00:08:11.1 & 117.59 & 4.54 & 1.41 & 0.30 & 3.08$\times 10^{30}$ & 0.29 & 0.82 & -0.04 & 0.29 & 4.97$\times 10^{32}$ & Transition & 05453826-0008109\\ 
6 & 05:45:41.67 & -00:04:02.1 & 198.97 & 2.12 & 1.33 & 0.18 & 9.9$\times 10^{31}$ & 0.24 & 0.77 & 0.08 & 0.20 & 4.01$\times 10^{32}$ & Transition & 05454167-0004024\\ 
7 & 05:45:41.95 & -00:12:05.6 & 108.43 & 5.90 & 1.22 & 0.06 & 1.05$\times 10^{30}$ & 0.19 & 0.74 & -0.03 & 0.33 & 8.61$\times 10^{32}$ & Transition & 05454194-0012053\\ 
8 & 05:46:07.90 & -00:11:56.6 & 211.13 & 32.32 & 1.40 & 0.29 & 7.49$\times 10^{30}$ & 0.25 & 1.15 & 0.19 & 0.37 & 9.08$\times 10^{33}$ & Transition & 05460788-0011568\\ 
9 & 05:46:09.86 & -00:05:59.4 & 181.58 & 0.68 & 2.75 & 2.70 & 1.5$\times 10^{30}$ & -- & 1.50\tablenotemark{5} & 0.02 & 0.20 & 4.24$\times 10^{31}$ & Transition & 05460984-0005591\\ 
10 & 05:46:20.17 & -00:10:19.9 & 204.12 & 0.66 & 1.80 & 1.03 & 4.14$\times 10^{29}$ & 0.83 & 1.77 & 0.16 & 0.24 & 1.96$\times 10^{32}$ & Transition & 05462018-0010197\\ 
11 & 05:46:35.91 & -00:03:28.9 & 153.10 & 0.20 & 2.28 & 1.68 & -- & 0.93 & 2.00 & 0.07 & 0.27 & 2.64$\times 10^{32}$ & Transition & 05463590-0003286\\ 
12 & 05:46:07.85 & -00:10:01.2 & 215.20 & 0.90 & 4.40 & 14.33 & 1.12$\times 10^{31}$ & 1.52 & 1.73 & 1.14 & 0.98 & 1.28$\times 10^{33}$ & Class I & 05460787-0010014\\ 
13 & 05:46:08.43 & -00:10:01.0 & 203.50 & 0.84 & 4.48 & 15.46 & 3.57$\times 10^{31}$ & 2.99 & 1.71\tablenotemark{5} & 1.66 & 1.19 & 6.19$\times 10^{33}$ & Class I & 05460845-0010029\\ 
14 & 05:46:13.15 & -00:06:04.7 & 234.67 & 6.24 & 3.53 & 5.96 & 9.6$\times 10^{30}$ & 1.90 & 2.58 & 0.89 & 1.21 & 1.89$\times 10^{34}$ & Class I & \tablenotemark{4}05461313-0006048\\ 
15 & 05:46:43.12 & +00:00:52.4 & 49.94 & 2.57 & 4.12 & 10.80 & 5.95$\times 10^{30}$ & 1.68 & 2.04 & -- & 0.84 & 2.01$\times 10^{34}$ & Class I & 05464311+0000523\\ 
16 & 05:46:08.92 & -00:09:56.1 & 187.14 & 0.53 & 4.51 & 16.06 & 3.57$\times 10^{31}$ & -- & -- & 1.32 & 1.03 & 7.72$\times 10^{32}$ & Class 0/I & \\ 
17 & 05:46:12.27 & -00:08:07.6 & 223.00 & 0.19 & 1.22 & 0.07 & -- & 0.43 & 0.70 & -- & -- & 6.50$\times 10^{31}$ & Class III & 05461226-0008078\\ 
18 & 05:46:08.42 & -00:03:15.0 & 219.55 & 0.15 & 1.25 & 0.09 & -- & 0.54 & 0.92 & -- & -- & 5.50$\times 10^{31}$ & Class III & 05460840-0003150\\ 
19 & 05:45:47.56 & -00:09:33.1 & 199.96 & 0.13 & 1.46 & 0.42 & -- & 0.47 & 0.78 & -- & -- & 1.34$\times 10^{32}$ & Class III & 05454755-0009342\\ 
20 & 05:45:56.98 & -00:07:52.9 & 220.96 & 0.39 & 1.77 & 1.00 & -- & 0.54 & 1.27 & -0.25 & -0.94 & 1.03$\times 10^{32}$ & Class III & 05455695-0007510\\ 
21 & 05:46:00.22 & -00:08:26.3 & 205.28 & 1.56 & 1.76 & 0.98 & 5.87$\times 10^{29}$ & 0.58 & 1.08 & 0.12 & 0.08 & 3.12$\times 10^{33}$ & Class III & 05460024-0008262\\ 
22 & 05:46:01.38 & -00:00:13.6 & 207.37 & 0.22 & 1.79 & 1.02 & -- & 0.39 & 0.83 & 0.20 & 0.09 & 4.46$\times 10^{32}$ & Class III & 05460134-0000136\\ 
23 & 05:46:04.10 & -00:03:22.9 & 207.46 & 0.03 & 0.88 & -- & -- & 0.60 & 1.18 & -- & 1.37 & 2.67$\times 10^{31}$ & Class III \tablenotemark{6} & 05460407-0003228\\ 
24 & 05:46:04.57 & +00:00:38.1 & 194.72 & 1.07 & 1.53 & 0.59 & 9.16$\times 10^{29}$ & 0.54 & 1.08 & 0.25 & 0.21 & 4.15$\times 10^{32}$ & Class III & 05460457+0000381\\ 
25 & 05:46:05.74 & -00:02:39.7 & 200.05 & 1.33 & 1.53 & 0.57 & 7.35$\times 10^{31}$ & 0.42 & 1.15 & 0.16 & 0.12 & 8.93$\times 10^{32}$ & Class III & 05460575-0002396\\ 
26 & 05:46:09.36 & -00:06:57.6 & 178.06 & 0.64 & 1.84 & 1.07 & 1.69$\times 10^{30}$& 1.41 & 2.78 & 0.17 & -0.03 & 2.56$\times 10^{32}$ & Class III & 05460937-0006578\\ 
27 & 05:46:11.98 & -00:04:02.3 & 195.27 & 0.05 & 1.67 & 0.87 & -- & 0.80 & 1.79 & -- & - & 2.68$\times 10^{31}$ & Class III & 05461198-0004021\\ 
28 & 05:46:18.46 & -00:07:54.7 & 204.94 & 0.29 & 1.81 & 1.04 & -- & 0.73 & 1.69 & -0.48 & -0.33 & 1.46$\times 10^{32}$ & Class III & 05461846-0007551\\ 
29 & 05:46:08.00 & -00:01:52.3 & 185.89 & 0.97 & 1.22 & 0.07 & 1.05$\times 10^{30}$ & 0.28 & 0.52 & -- & -- & 4.24$\times 10^{31}$ & Class III & 05460798-0001521\\ 
30 & 05:45:53.11 & -00:13:25.0 & 206.73 & 1.24 & 1.57 & 0.68 & 5.59$\times 10^{29}$ & 0.26 & 0.82 & -- & -- & 3.73$\times 10^{32}$ & Class III & 05455311-0013248\\ 
31 & 05:46:10.95 & -00:17:39.8 & 42.99 & 3.37 & 1.62 & 0.78 & 5.83$\times 10^{29}$ & 0.37 & 1.02 & -- & -- & 1.95$\times 10^{32}$ & Class III & 05461101-0017399\\ 
32 & 05:46:47.26 & -00:08:03.6 & 137.30 & 0.46 & 1.69 & 0.89 & -- & 0.23 & 0.63 & -- & -- & 7.54$\times 10^{31}$ & Class III & 05464728-0008047\\ 
33 & 05:45:55.60 & -00:07:43.8 & 211.67 & 0.06 & 2.06 & 1.34 & -- & 0.78 & 1.78 & -- & -- & 3.59$\times 10^{31}$ & Class III & 05455557-0007435\\ 
34 & 05:45:57.04 & -00:02:02.3 & 207.00 & 0.11 & 1.03 & 0.01 & -- & 0.23 & 0.67 & -- & -- & 8.77$\times 10^{31}$ & Class III & 05455711-0002030\\ 
35 & 05:45:58.68 & -00:02:00.7 & 211.12 & 0.10 & 1.50 & 0.51 & -- & 0.34 & 0.97 & -- & -- & 1.03$\times 10^{32}$ & Class III & 05455857-0002001\\ 
36 & 05:46:03.50 & +00:02:53.4 & 27.31 & 1.94 & 1.37 & 0.24 & -- & 0.32 & 0.89 & -- & -- & 2.22$\times 10^{32}$ & Class III & 05460353+0002546\\ 
37 & 05:45:44.54 & -00:10:35.0 & 156.19 & 0.14 & 2.37 & 1.83 & -- & 0.39 & 0.96 & -- & -- & 1.14$\times 10^{32}$ & Class III & 05454455-0010355\\ 
38 & 05:46:18.62 & -00:00:18.9 & 194.40 & 0.26 & 2.51 & 2.11 & -- & 1.09 & 1.78 & 0.69 & 1.33 & 3.02$\times 10^{33}$ & Class II\tablenotemark{3} & 05461858-0000190\\ 
39 & 05:45:56.73 & -00:00:24.9 & 194.59 & 0.16 & 2.22 & 1.58 & -- & 0.47 & 1.15 & 0.39 & 1.25 & 2.19$\times 10^{32}$ & Class II\tablenotemark{3} & 05455672-0000253\\ 
40 & 05:46:19.48 & -00:05:20.1 & 191.18 & 5.71 & 1.64 & 0.83 & 1.9$\times 10^{30}$ & 0.66 & 1.04 & 0.40 & 1.14 & 5.83$\times 10^{33}$ & Class II\tablenotemark{3} & 05461946-0005199\\ 
41 & 05:46:07.75 & -00:09:37.9 & 165.76 & 0.09 & 2.37 & 1.83 & -- & -- & -- & 0.54 & 1.16 & 5.37$\times 10^{31}$ & Class II\tablenotemark{3} & \\ 
42 & 05:46:00.17 & +00:03:07.3 & 19.14 & 5.59 & 1.24 & 0.08 & 1.24$\times 10^{30}$ & 0.31 & 0.83 & 0.30 & 0.80 & 9.63$\times 10^{32}$ & Class II & 05460017+0003070\\ 
43 & 05:46:09.61 & -00:03:31.1 & 215.49 & 0.99 & 1.92 & 1.16 & 4.71$\times 10^{29}$ & 0.74 & 1.55 & 0.30 & 0.95 & 3.95$\times 10^{32}$ & Class II & 05460960-0003312\\ 
44 & 05:46:10.32 & -00:00:06.9 & 191.25 & 0.27 & 2.06 & 1.34 & -- & 0.90 & 1.32 & 0.44 & 0.92 & 3.95$\times 10^{32}$ & Class II & 05461030-0000066\\ 
45 & 05:46:11.60 & -00:06:27.8 & 183.15 & 0.97 & 1.64 & 0.82 & 2.87$\times 10^{29}$ & 0.99 & 1.75 & 0.50 & 0.82 & 6.62$\times 10^{32}$ & Class II & 05461162-0006279\\ 
46 & 05:46:18.89 & -00:05:38.2 & 194.21 & 22.87 & 1.64 & 0.82 & 4.58$\times 10^{31}$ & 0.44 & 1.03 & 0.38 & 0.65 & 2.78$\times 10^{33}$ & Class II & 05461889-0005381\\ 
47 & 05:46:19.08 & +00:03:29.7 & 32.71 & 26.90 & 2.58 & 2.26 & 1.20$\times 10^{31}$ & 1.16 & 1.60 & 0.61 & 0.77 & 1.53$\times 10^{34}$ & Class II & 05461906+0003295\\ 
48 & 05:46:29.14 & +00:02:59.2 & 69.91 & 0.35 & 2.61 & 2.33 & -- & 1.21 & 2.35 & 0.67 & 0.70 & 1.06$\times 10^{33}$ & Class II & 05462913+0002590\\ 
49 & 05:46:33.26 & +00:02:51.5 & 69.10 & 0.89 & 3.47 & 5.59 & -- & 0.96 & 1.54 & 0.48 & 0.76 & 1.93$\times 10^{33}$ & Class II & 05463328+0002518\\ 
50 & 05:46:35.50 & +00:01:39.7 & 80.15 & 0.78 & 2.07 & 1.35 & -- & 0.62 & 1.04 & 0.23 & 1.04 & 4.67$\times 10^{32}$ & Class II & 05463549+0001389\\ 
51 & 05:46:35.61 & -00:01:37.0 & 163.72 & 0.74 & 2.53 & 2.15 & 8.68$\times 10^{29}$ & 1.38 & 2.44\tablenotemark{5} & 0.73 & 0.58 & 7.35$\times 10^{31}$ & Class II & 05463560-0001371\\ 
52 & 05:46:37.06 & +00:01:22.0 & 81.41 & 2.00 & 1.68 & 0.88 & 4.58$\times 10^{29}$ & 0.90 & 1.25 & 0.72 & 0.95 & 3.33$\times 10^{33}$ & Class II & 05463705+0001217\\ 
53 & 05:45:51.58 & -00:16:34.1 & 45.29 & 1.69 & 2.46 & 2.00 & -- & -- & -- & -- & 0.47 & 2.29$\times 10^{30}$ & -- & \\ 
54 & 05:45:57.79 & -00:09:28.9 & 219.62 & 1.17 & 2.62 & 2.36 & -- & -- & -- & -- & 1.00 & 2.01$\times 10^{30}$ & -- & \\ 
55 & 05:46:05.77 & -00:19:49.0 & 40.79 & 4.59 & 2.81 & 2.86 & -- & -- & -- & -- & 0.19 & 1.86$\times 10^{30}$ & -- & \\ 
56 & 05:46:10.79 & -00:07:44.5 & 195.59 & 0.03 & 1.72 & 0.95 & -- & -- & -- & -- & - & 1.35$\times 10^{29}$ & -- & \\ 
57 & 05:46:13.19 & -00:09:10.3 & 214.81 & 0.07 & 0.95 & -- & -- & 0.05 & 0.95 & -- & -- & 1.45$\times 10^{31}$ & -- & 05461317-0009100\\ 
58 & 05:46:44.40 & -00:08:36.0 & 138.91 & 1.63 & 2.83 & 2.92 & -- & 0.84 & 0.69\tablenotemark{5} & -- & -- & 1.21$\times 10^{31}$ & -- & 05464432-0008361\\ 
59 & 05:46:32.47 & -00:02:38.2 & 162.98 & 0.18 & 1.08 & 0.01 & -- & 0.78\tablenotemark{5}  & 0.75 & -- & -- & 9.00$\times 10^{30}$ & -- & 05463246-0002385\\


\thispagestyle{empty}
\tablenotetext{1}{$N_{H}$ derived from median energy}
\tablenotetext{2}{Spectrally derived from high count rate sources. }
\tablenotetext{3}{Tentative classification due to excess 8 $\mu$m extended emission.  }
\tablenotetext{4}{V1647 Ori}
\tablenotetext{5} {Upper limit measurement in J, H, or K}
\tablenotetext{6}{Likely cosmic ray afterglow detection within 2.5''  of a Class III YSO}

\enddata

\end{deluxetable}



\newpage
\clearpage

\begin{deluxetable}{c c c c c c c | c c c c c c | c c c c c c | c c c c c }

\centering
\rotate
\tablecolumns{24}
\tablewidth{570pt}
\tabletypesize{\scriptsize}

\tablecaption{\label{yso_fov} X-ray Detection of YSOs by Observation}

\tablehead{

\colhead{} &
  \multicolumn{6} {c}{Class 0/I} &
  \multicolumn{6} {c}{Class II} &
  \multicolumn{6} {c}{Transition Disk} & 
  \multicolumn{5} {c}{Class III\tablenotemark{1}}  \\

  \hline
  
\colhead{} &
\colhead{} &
\colhead{} &
\colhead{} &
    \multicolumn{2}{c}{YSOs in FOV\tablenotemark{2}} &
\colhead{} &    
\colhead{} &
\colhead{} &
\colhead{} &
    \multicolumn{2}{c}{YSOs in FOV\tablenotemark{2}} &
\colhead{} &    
\colhead{} &
\colhead{} &
\colhead{} &
    \multicolumn{2}{c}{YSOs in FOV\tablenotemark{2}} &
\colhead{} &    
\colhead{} &
\colhead{} &
\colhead{} &
    \multicolumn{2}{c}{YSOs in FOV\tablenotemark{2}}  \\

\colhead{JD-2453000} &    
\colhead{S} &
\colhead{H} &
\colhead{B} &
\colhead{X-ray} &
\colhead{IR} &
\colhead{\%} &
\colhead{S} &
\colhead{H} &
\colhead{B} &
\colhead{X-ray} &
\colhead{IR} &
\colhead{\%} &
\colhead{S} &
\colhead{H} &
\colhead{B} &
\colhead{X-ray} &
\colhead{IR} &
\colhead{\%} &
\colhead{S} &
\colhead{H} &
\colhead{B} &
\colhead{X-ray} &
\colhead{IR} \\

}

\startdata

471 & 1 & 3 & 3 & 5 & 17 & 29 & 4 & 4 & 4 & 9 & 17 & 53 & 8 & 6 & 9 & 11 & 15 & 73 & 5 & 4 & 5 & 18 &  18\\ 
609 & 1 & 4 & 4 & 5 & 15 & 33 & 5 & 4 & 7 & 14 & 21 & 67 & 6 & 5 & 8 & 9 & 14 & 64 & 8 & 3 & 5 & 19 &  19\\ 
713 & 1 & 2 & 2 & 4 & 14 & 29 & 7 & 3 & 7 & 13 & 19 & 68 & 5 & 5 & 6 & 9 & 13 & 69 & 9 & 2 & 8 & 18 &  18\\ 
718 & 1 & 1 & 1 & 5 & 15 & 33 & 7 & 4 & 7 & 12 & 18 & 67 & 5 & 5 & 6 & 9 & 13 & 69 & 7 & 1 & 4 & 19 &  19\\ 
856 & 0 & 3 & 2 & 4 & 14 & 29 & 7 & 4 & 7 & 10 & 16 & 63 & 7 & 6 & 8 & 11 & 17 & 65 & 9 & 3 & 9 & 19 &  19\\ 
1319 & 0 & 3 & 3 & 4 & 14 & 29 & 7 & 5 & 7 & 14 & 23 & 61 & 6 & 6 & 7 & 9 & 16 & 56 & 8 & 1 & 6 & 18 &  18\\ 
1727 & 1 & 3 & 3 & 5 & 15 & 33 & 5 & 5 & 6 & 13 & 20 & 65 & 7 & 4 & 8 & 10 & 16 & 63 & 8 & 2 & 8 & 20 &  20\\ 
1797 & 1 & 3 & 3 & 4 & 13 & 31 & 4 & 4 & 4 & 8 & 14 & 57 & 8 & 7 & 8 & 10 & 16 & 63 & 5 & 2 & 5 & 20 &  20\\ 
1798 & 1 & 3 & 3 & 4 & 13 & 31 & 4 & 3 & 4 & 8 & 14 & 57 & 7 & 8 & 8 & 10 & 16 & 63 & 7 & 3 & 6 & 20 &  20\\ 
1854 & 1 & 2 & 2 & 4 & 14 & 29 & 4 & 5 & 5 & 11 & 18 & 61 & 8 & 3 & 8 & 10 & 16 & 63 & 8 & 3 & 7 & 19 &  19\\ 
1943 & 2 & 4 & 4 & 4 & 13 & 31 & 5 & 4 & 5 & 9 & 16 & 56 & 9 & 6 & 9 & 11 & 16 & 69 & 8 & 1 & 7 & 18 &  18\\
\enddata

\tablenotetext{1}{X-ray detection is a requirement for Class III classification.}
\tablenotetext{2}{X-ray detected YSOs from the merged $\sim$240 ks observation.}

\end{deluxetable}



\newpage
\clearpage

\begin{deluxetable}{c c c c l l c c c c c c}
\thispagestyle{empty}

\centering

\rotate
\tablecolumns{12}
\tablewidth{650pt}
\tabletypesize{\scriptsize}

\tablecaption{\vspace{-2mm}L1630 YSO X-ray Spectral Analysis \label{spec_tbl}}

\tablehead{

\colhead{ID} &
\colhead{RA} &
\colhead{DEC} &
\colhead{Count Rate} &
\colhead{$N_{H}$} &
\colhead{$T_{1}$ } &
\colhead{$T_{2}$} &
\colhead{Norm. 1} &
\colhead{Norm. 2} &
\colhead{Absorbed Flux\tablenotemark{1}} &
\colhead{Red. $\chi^{2}$ } &
\colhead{Class} \vspace{1mm} \\

\colhead{} &
\colhead{} &
\colhead{} &
\colhead{($ks^{-1}$)} &
\colhead{$(\times 10^{22}$ $cm^{-2})$} &
\colhead{(MK) } &
\colhead{(MK)} &
\colhead{} &
\colhead{} &
\colhead{($erg$ $s^{-1}cm^{-2}$)} &
\colhead{ } &
\colhead{} \vspace{-1.5mm} \\

}

\startdata

2 & 05:46:22.43 & -00:08:52.6 & 5.66 & 0.54$_{-0.1}^{+0.1}$ & 26.82$_{-3.4}^{+4.9}$ & 8.28$_{-0.8}^{+1.0}$ & 3.97E-5$_{-7.0E-6}^{+5.8E-6}$ & 2.57E-5$_{-1.2E-5}^{+1.7E-5}$ & 4.33E-14$_{-1.0E-14}^{+5.1E-15}$ & 1.25 & Transition/ClassII\vspace{1mm}\\
3 & 05:46:11.60 & -00:02:20.6 & 11.28 & 1.23$_{-0.1}^{+0.1}$ & 31.57$_{-3.3}^{+3.7}$ & -- & 2.15E-4$_{-1.8E-5}^{+2.5E-5}$ & -- & 1.39E-13$_{-1.9E-14}^{+1.5E-14}$ & 1.22 & Transition\tablenotemark{2}\vspace{1mm}\\
4 & 05:46:17.71 & -00:00:14.2 & 1.94 & 0.98$_{-0.3}^{+0.3}$ & 35.28$_{-10}^{+19.2}$ & -- & 3.53E-5$_{-9.7E-6}^{+1.3E-5}$ & -- & 2.65E-14$_{-1.1E-14}^{+7.3E-15}$ & 0.94 & Transition\tablenotemark{2}\vspace{1mm}\\
5 & 05:45:38.26 & -00:08:11.1 & 4.54 & 0.63$_{-0.2}^{+0.3}$ & 3.16$_{-1.1}^{+3.8}$ & 17.04$_{-2.9}^{+2.9}$ & 9.25E-5$_{-5.5E-5}^{+5.9E-4}$ & 3.75E-5$_{-7.2E-6}^{+1.0E-5}$ & 3.03E-14$_{-1.5E-14}^{+6.3E-15}$ & 0.88 & Transition\vspace{1mm}\\
6 & 05:45:41.67 & -00:04:02.1 & 2.12 & 1.04$_{-0.2}^{+0.5}$ & 14.56$_{-3.2}^{+3.9}$ & 1.06$_{-0.1}^{+0.5}$ & 2.08E-5$_{-5.1E-6}^{+5.4E-6}$ & 5.40E-2$_{-9.8E-3}^{+3.0E-1}$ & 1.45E-14$_{-9.9E-15}^{+1.3E-15}$ & 0.74 & Transition\vspace{1mm}\\
7 & 05:45:41.95 & -00:12:05.6 & 5.90 & 0.17$_{-0.1}^{+0.1}$ & 29.05$_{-10.9}^{+24.2}$ & 10.93$_{-1.8}^{+1.0}$ & 1.76E-5$_{-7.4E-6}^{+7.9E-6}$ & 1.81E-5$_{-7.4E-6}^{+1.0E-5}$ & 4.00E-14$_{-1.0E-14}^{+6.1E-15}$ & 1.07 & Transition\vspace{1mm}\\
8 & 05:46:07.90 & -00:11:56.6 & 32.32 & 0.26$_{-0.0}^{+0.0}$ & 9.48$_{-0.6}^{+1.3}$ & 30.67$_{-1.6}^{+2.1}$ & 3.94E-5$_{-1.4E-5}^{+1.4E-5}$ & 2.44E-4$_{-1.5E-5}^{+1.1E-5}$ & 2.76E-13$_{-2.0E-14}^{+1.5E-14}$ & 1.1 & Transition\vspace{1mm}\\
9 & 05:46:09.86 & -00:05:59.4 & 0.68 & 4.45$_{-1.5}^{+2.1}$ & 17.9$_{-6.1}^{+11.2}$ & -- & 6.18E-5$_{-6.2E-5}^{+1.1E-4}$ & -- & 1.09E-14$_{-1.1E-14}^{+2.0E-15}$ & 0.66 & Transition\vspace{1mm}\\
10 & 05:46:20.17 & -00:10:19.9 & 0.66 & 1.02$_{-0.5}^{+0.7}$ & 21.95$_{-8.0}^{+11.5}$ & -- & 1.78E-5$_{-8.7E-6}^{+1.5E-5}$ & -- & 9.24E-15$_{-9.2E-15}^{+2.9E-15}$ & 0.51 & Transition\vspace{1mm}\\
12 & 05:46:07.85 & -00:10:01.2 & 0.90 & 20.8$_{-13.39}^{+13.49}$ & 1.72$_{-0.63}^{+4.65}$ & - & 4.82E-4$_{-4.82E-4}^{+2.65E-3}$ & - & 3.06E-14$_{-3.06E-14}^{+1.00E-14}$ & 0.7 & Class 0/I\vspace{1mm}\\
13 & 05:46:08.43 & -00:10:01.0 & 0.84 & 30.72$_{-11.5}^{+12.5}$ & 15.83$_{-5.2}^{+19}$ & -- & 1.41E-3$_{-1.4E-3}^{+7.4E-3}$ & -- & 3.03E-14$_{-3.0E-14}^{+7.0E-15}$ & 0.92 & Class 0/I\vspace{1mm}\\
14 & 05:46:13.15 & -00:06:04.7 & 6.24 & 4.49$_{-0.6}^{+0.5}$ & 52.13$_{-11.1}^{+27.5}$ & -- & 3.30E-4$_{-7.3E-5}^{+7.6E-5}$ & -- & 2.01E-13$_{-4.4E-14}^{+2.3E-14}$ & 0.84 & Class 0/I\vspace{1mm}\\
15 & 05:46:43.12 & +00:00:52.4 & 2.57 & 9.79$_{-3.3}^{+6.8}$ & 45.89$_{-25.8}^{+71.0}$ & -- & 2.13E-4$_{-1.0E-4}^{+6.6E-4}$ & -- & 8.32E-14$_{-8.3E-14}^{+3.5E-14}$ & 0.67 & Class 0/I\vspace{1mm}\\
16 & 05:46:08.92 & -00:09:56.1 & 0.53 & 25.76$_{-12.6}^{+15.4}$ & 16.04$_{-6.9}^{+27.3}$ & -- & 1.42E-3$_{-1.4E-3}^{+1.8E-2}$ & -- & 3.91E-14$_{-3.9E-14}^{+9.2E-15}$ & 0.64 & Class 0/I\vspace{1mm}\\
21 & 05:46:00.22 & -00:08:26.3 & 1.56 & 0.94$_{-0.3}^{+0.3}$ & 23.49$_{-5.1}^{+6.7}$ & -- & 2.51E-5$_{-7.0E-6}^{+9.8E-6}$ & -- & 1.42E-14$_{-5.8E-15}^{+3.6E-15}$ & 0.55 & Class III\vspace{1mm}\\
24 & 05:46:04.57 & +00:00:38.1 & 1.07 & 1.12$_{-0.2}^{+0.2}$ & 10.89$_{-2.3}^{+2.6}$ & -- & 2.67E-5$_{-8.6E-6}^{+1.4E-5}$ & -- & 8.11E-15$_{-5.2E-15}^{+3.2E-15}$ & 0.88 & Class III\vspace{1mm}\\
25 & 05:46:05.74 & -00:02:39.7 & 1.33 & 1.93$_{-0.4}^{+0.7}$ & 13.72$_{-3.6}^{+8.1}$ & 1.77$_{-0.8}^{+2.0}$ & 2.83E-5$_{-1.5E-5}^{+2.4E-5}$ & 5.89E-3$_{-2.5E-3}^{+2.5E-1}$ & 9.32E-15$_{-9.3E-15}^{+1.4E-15}$ & 0.76 & Class III\vspace{1mm}\\
26 & 05:46:09.36 & -00:06:57.6 & 0.64 & 2.17$_{-1.4}^{+1.4}$ & 14.22$_{-5.0}^{+10.0}$ & -- & 6.06E-5$_{-6.1E-5}^{+1.1E-4}$ & -- & 1.33E-14$_{-1.3E-14}^{+6.4E-15}$ & 0.86 & Class III\vspace{1mm}\\
29 & 05:46:08.00 & -00:01:52.3 & 0.97 & 0.76$_{-0.2}^{+0.2}$ & 9.22$_{-1.6}^{+6.2}$ & -- & 2.83E-5$_{-1.2E-5}^{+1.6E-5}$ & -- & 1.13E-14$_{-1.0E-14}^{+3.7E-15}$ & 1.29 & Class III\vspace{1mm}\\
30 & 05:45:53.11 & -00:13:25.0 & 1.24 & 0.74$_{-0.4}^{+0.3}$ & 20.74$_{-7.4}^{+21.5}$ & 9.58$_{-2.1}^{+2.2}$ & 7.89E-6$_{-6.0E-6}^{+6.9E-6}$ & 1.02E-5$_{-8.3E-6}^{+1.4E-5}$ & 8.88E-15$_{-5.2E-15}^{+3.6E-15}$ & 0.62 & Class III\vspace{1mm}\\
31 & 05:46:10.95 & -00:17:39.8 & 3.37 & 0.1$_{-0.1}^{+0.2}$ & 41.03$_{-16.2}^{+63.2}$ & -- & 2.15E-5$_{-5.7E-6}^{+7.2E-6}$ & -- & 2.73E-14$_{-1.6E-14}^{+1.2E-14}$ & 1.02 & Class III\vspace{1mm}\\
40 & 05:46:19.48 & -00:05:20.1 & 5.71 & 0.688$_{-0.2}^{+0.3}$ & 10.05$_{-2.5}^{+2.7}$ & 32.73$_{-6.2}^{+10.6}$ & 1.50E-5$_{-1.0E-5}^{+2.2E-5}$ & 5.38E-5$_{-1.3E-5}^{+6.2E-6}$ & 4.96E-14$_{-9.6E-15}^{+8.2E-15}$ & 1.13 & Class II\tablenotemark{2}\vspace{1mm}\\
42 & 05:46:00.17 & +00:03:07.3 & - & 0.35$_{-0.4}^{+0.4}$ & 14.36$_{-3.9}^{+4.6}$ & -- & 4.48E-5$_{-1.7E-5}^{+3.7E-5}$ & -- & 3.53E-14$_{-3.5E-14}^{+1.5E-14}$ & 0.64 & Class II\vspace{1mm}\\
43 & 05:46:09.61 & -00:03:31.1 & 0.99 & 0.9$_{-0.4}^{+0.4}$ & 37.81$_{-12.5}^{+33.8}$ & -- & 1.79E-5$_{-5.6E-6}^{+9.9E-6}$ & -- & 1.44E-14$_{-7.9E-15}^{+4.3E-15}$ & 0.54 & Class II\vspace{1mm}\\
45 & 05:46:11.60 & -00:06:27.8 & 0.97 & 0.58$_{-0.3}^{+1.1}$ & 28.7$_{-16.9}^{+13.5}$ & -- & 1.18E-5$_{-3.5E-6}^{+2.5E-5}$ & -- & 9.06E-15$_{-3.9E-15}^{+3.0E-15}$ & 0.75 & Class II\vspace{1mm}\\
46 & 05:46:18.89 & -00:05:38.2 & 22.87 & 1.26$_{-0.1}^{+0.1}$ & 3.33$_{-0.7}^{+0.8}$ & 30.18$_{-2.6}^{+3.0}$ & 1.64E-3$_{-8.9E-4}^{+2.5E-3}$ & 2.73E-4$_{-2.5E-5}^{+2.7E-5}$ & 2.10E-13$_{-4.4E-14}^{+2.0E-14}$ & 1.56 & Class II\vspace{1mm}\\
47 & 05:46:19.08 & +00:03:29.7 & 26.90 & 0.96$_{-0.2}^{+0.2}$ & 118.51$_{-48.2}^{+234.7}$ & -- & 3.65E-4$_{-3.3E-5}^{+5.2E-5}$ & -- & 4.43E-13$_{-7.8E-14}^{+5.6E-14}$ & 0.88 & Class II\vspace{1mm}\\
51 & 05:46:35.61 & -00:01:37.0 & 0.74 & 1.94$_{-0.9}^{+0.8}$ & 31.62$_{-15.1}^{+36.8}$ & -- & 3.48E-5$_{-1.7E-5}^{+3.1E-5}$ & -- & 1.91E-14$_{-1.9E-14}^{+9.2E-15}$ & 0.62 & Class II\vspace{1mm}\\
52 & 05:46:37.06 & +00:01:22.0 & 2.00 & 0.2$_{-0.2}^{+0.3}$ & 46.65$_{-19.7}^{+53.7}$ & -- & 1.63E-5$_{-4.1E-6}^{+6.5E-6}$ & -- & 2.00E-14$_{-7.1E-15}^{+7.1E-15}$ & 0.97 & Class II\vspace{1mm}\\

\tablenotetext{1}{Model flux including absorption (i.e., observed flux).}
\tablenotetext{2}{Tentative classification due to excess 8 $\mu$m extended emission. }

\enddata

\end{deluxetable}



\newpage
\clearpage

\begin{deluxetable}{c c c c}

\centering
\tablecolumns{4}
\tablewidth{300pt}
\tabletypesize{\small}

\tablecaption{Spectral Properties of X-ray Bright YSOs \label{avg_prop} }

\tablehead{

\colhead{Class} &
\colhead{Median N$_{H}$} &
\colhead{\tablenotemark{1}Median $L_{X}$ } &
\colhead{\tablenotemark{1}Median ($L_{X}/L_{IR}$)}   \\

\colhead{} &
\colhead{$ \times 10^{22}$ $[cm^{-2}]$} &
\colhead{$[erg$  $s^{-1}]$ } &
\colhead{}   \\

}

\startdata

0/I & 20.8 & 9.6 $ \times$ $10^{30}$ & 5.8 $ \times$ $10^{-3}$\\ 
II & 0.90 & 1.2 $ \times$ $10^{30}$ & 9.9 $ \times$ $10^{-4}$\\ 
Transition & 0.98 & 1.9 $ \times$ $10^{30}$ & 1.6 $ \times$ $10^{-3}$\\ 
III & 0.94 & 9.2 $ \times$ $10^{29}$ & 3.0 $ \times$ $10^{-3}$\\

\tablenotetext{1}{Intrinsic $L_{X}$}

\enddata

\end{deluxetable}



\begin{deluxetable}{c c c c c c c c c c}

\centering
\tablecolumns{10}
\tabletypesize{\small}

\tablecaption{YSO Infrared Detections with X-ray Counterparts \label{yso_compare}}

\tablehead{

\colhead{} &
  \multicolumn{3} {c}{Serpens (W07)} &
  \multicolumn{3} {c}{Coronet (FP07)} &
  \multicolumn{3} {c}{L1630 (this study)}  \\
\hline
\colhead{Class} &
\colhead{Infrared} &
\colhead{X-ray} &
\colhead{\%} &
\colhead{Infrared} &
\colhead{X-ray} &
\colhead{\%} &
\colhead{Infrared} &
\colhead{X-ray} &
\colhead{\%} \vspace{-3mm}\\

}

\startdata

Class 0/I & 22 & 9 & 40.9 & 10 & 9& 90 & 19 & 5 & 26.3\\ 
Flat Spectrum & 16 & 8 & 50 & - & - & - & - & - & -\\ 
Class II & 62 & 20 & 32.3 & 17 & 14 & 82.5 &  26 & 15 & 57.7\\ 
Transition Disk & 17 & 2 & 11.8 & - & - & - & 21 & 11& 52.4\\  
Class III & 21 & (21) & ... &  12 & (12) & ... & 21 & (21) & ...\\ 
\hline
Total & 138 & 60 & 43.5 & 27 & 23 & 85.2 &  87 & 52 & 59.77\\ 

\enddata

\end{deluxetable}







\clearpage
\newpage

\begin{appendices}

\begin{figure}[H]
\centering

\includegraphics[scale=0.45]{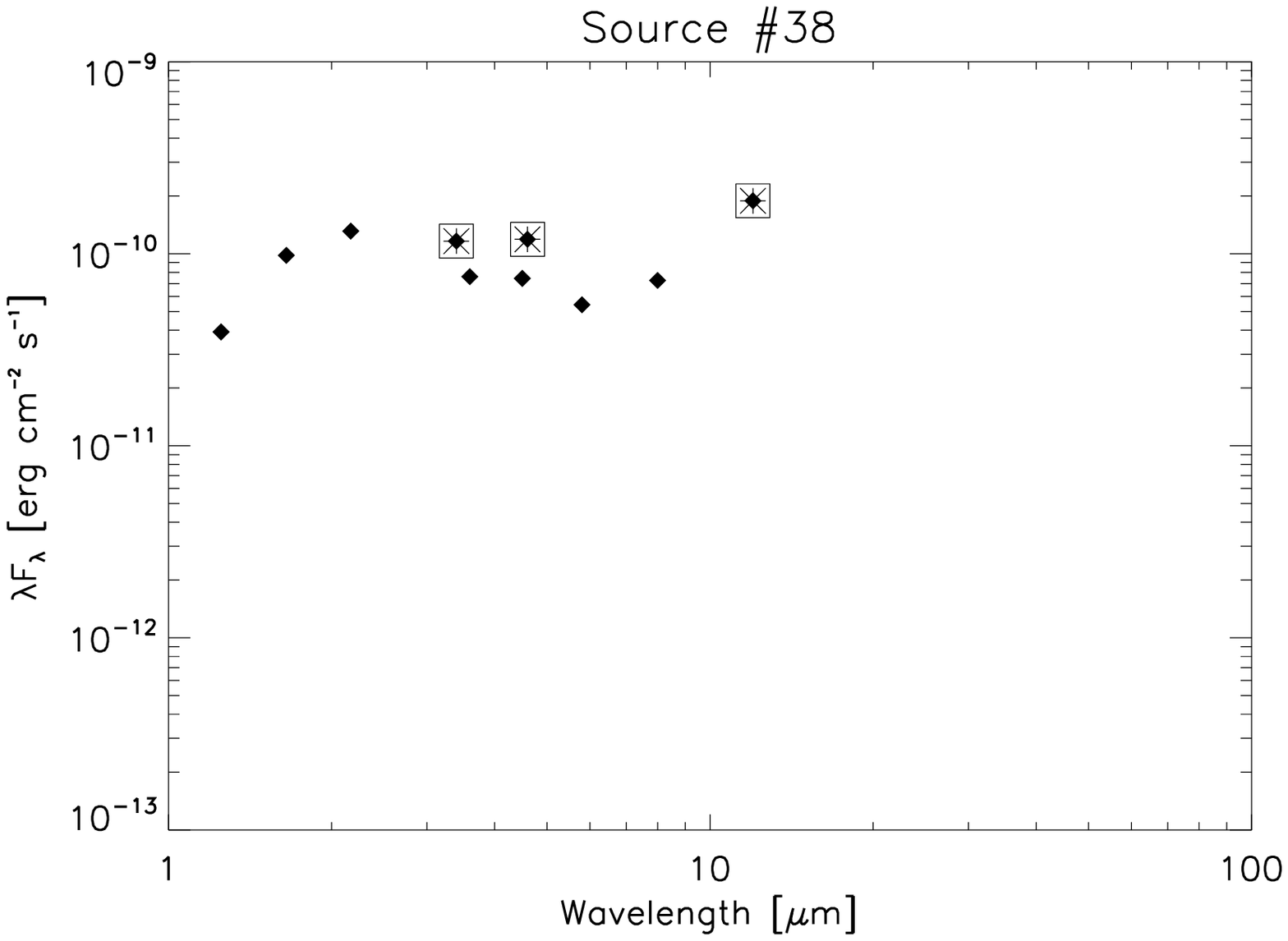}
\includegraphics[scale=0.45]{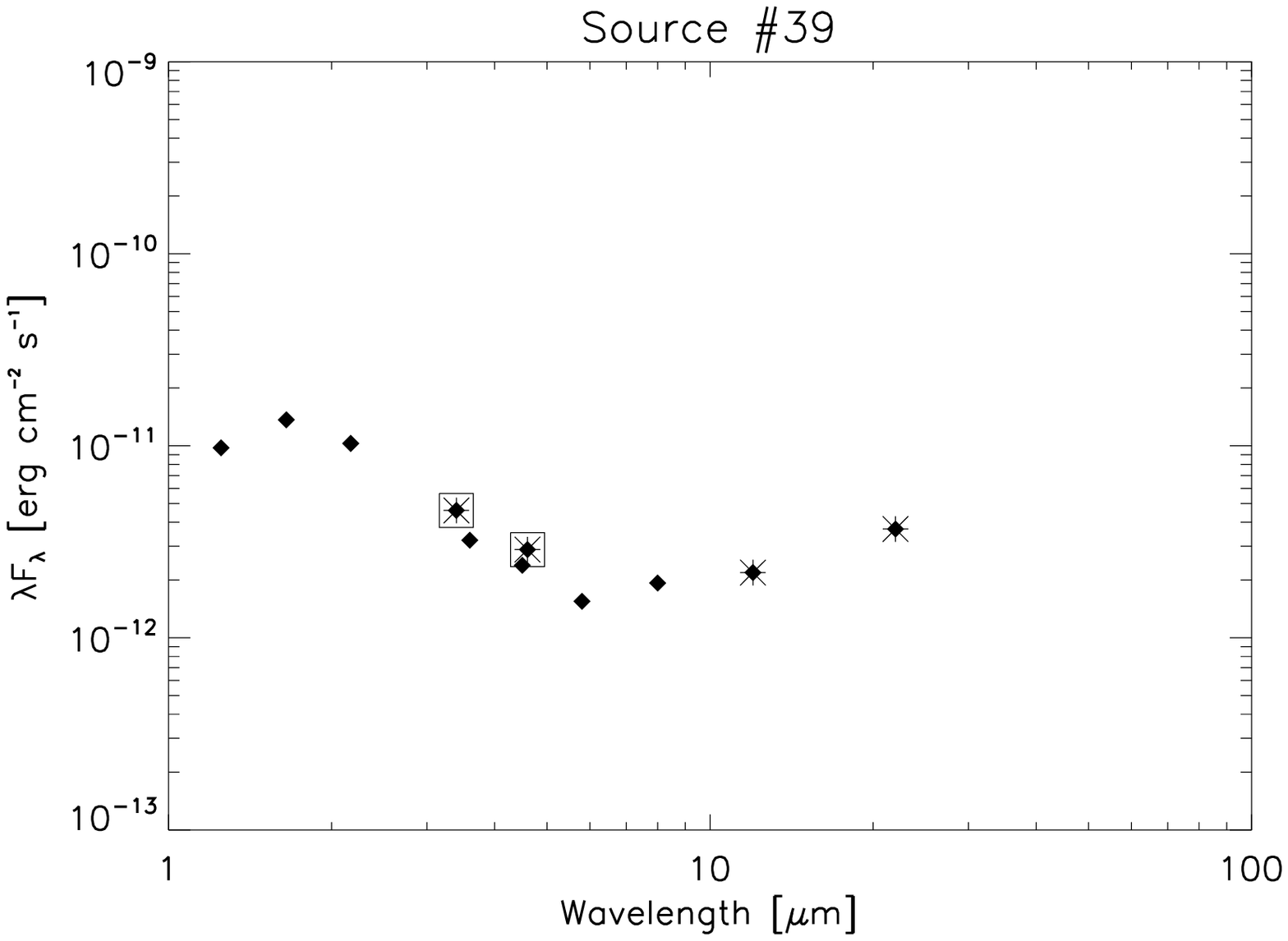}
\includegraphics[scale=0.45]{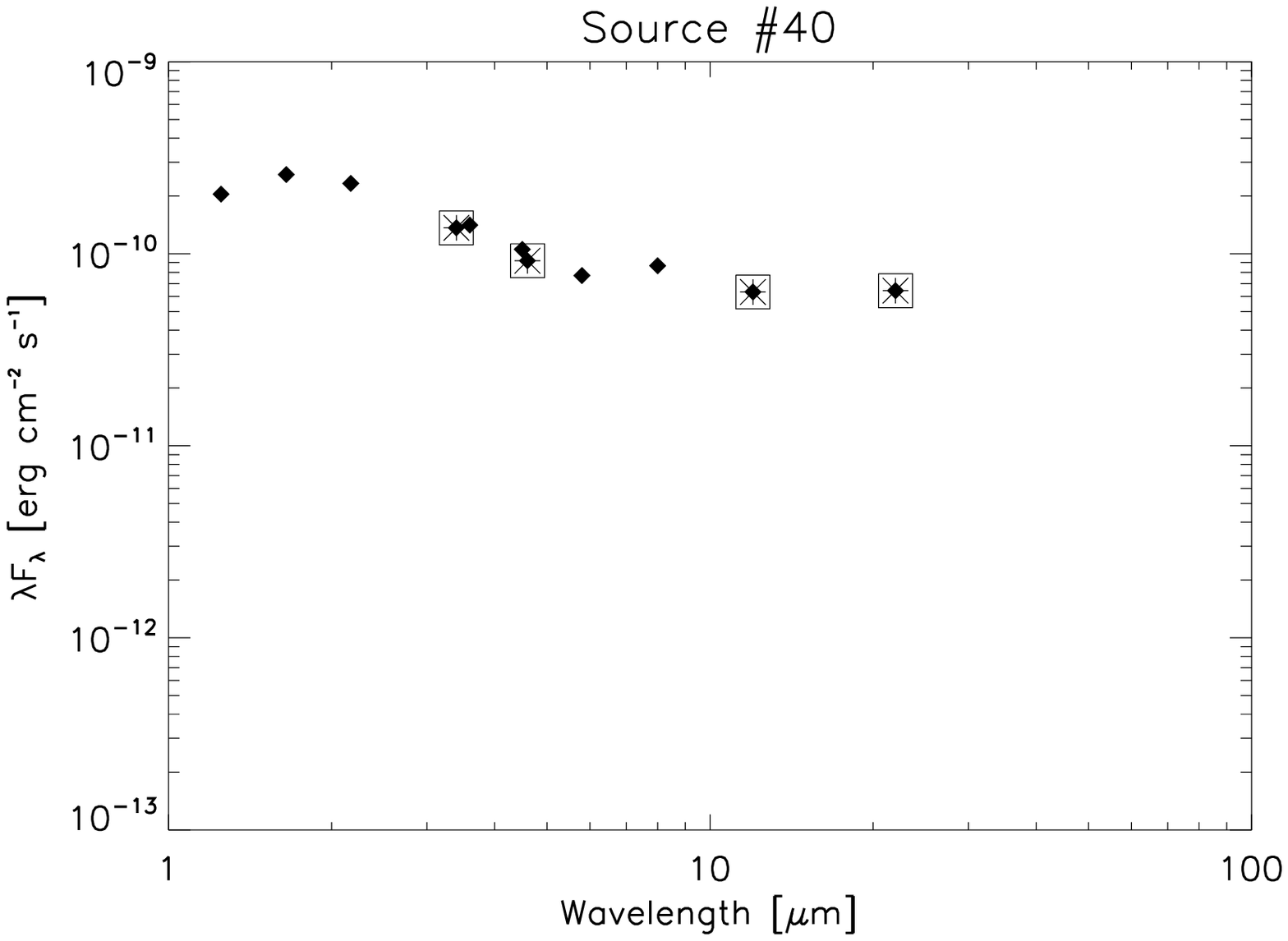}
\includegraphics[scale=0.45]{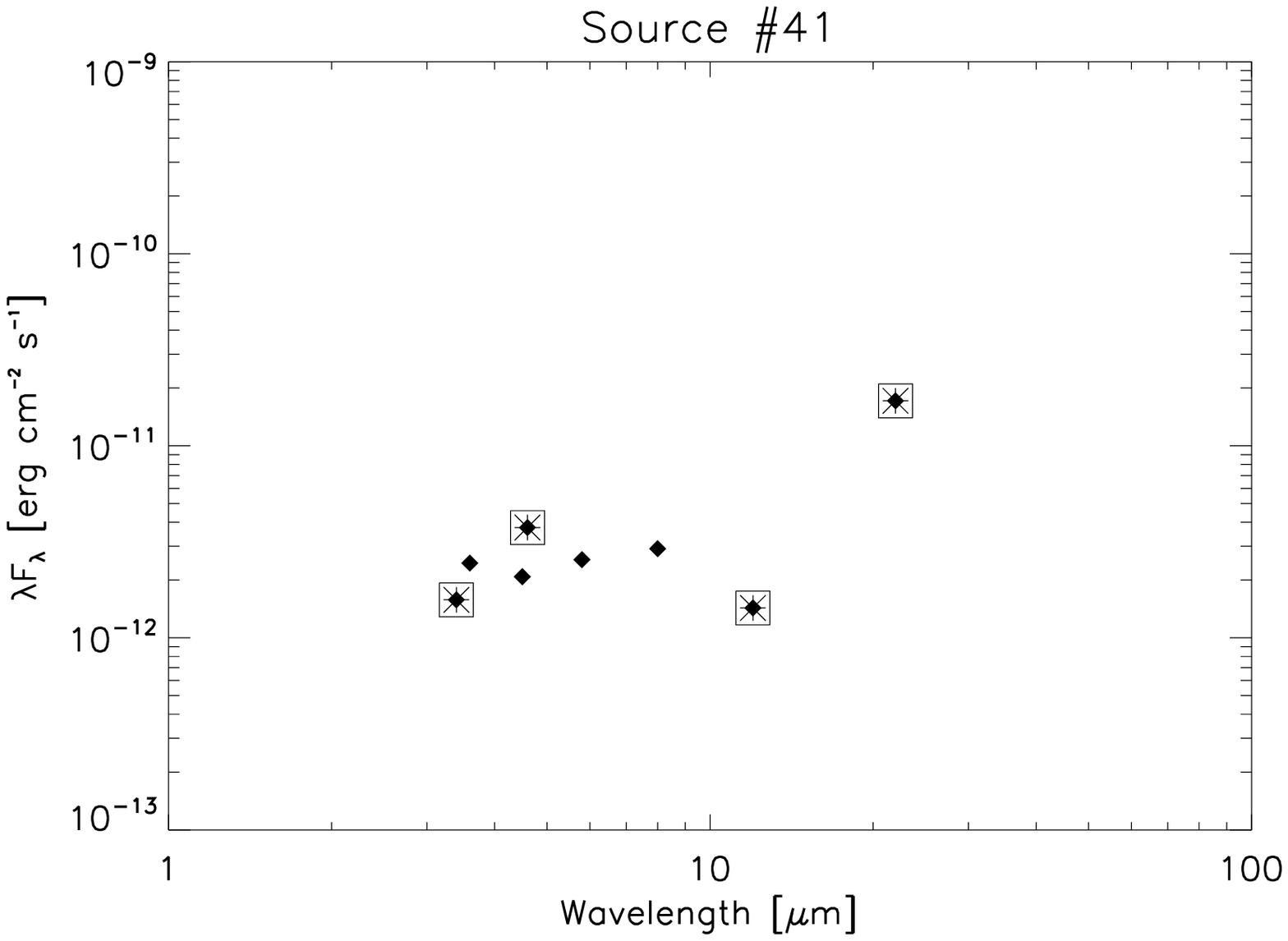}
\includegraphics[scale=0.45]{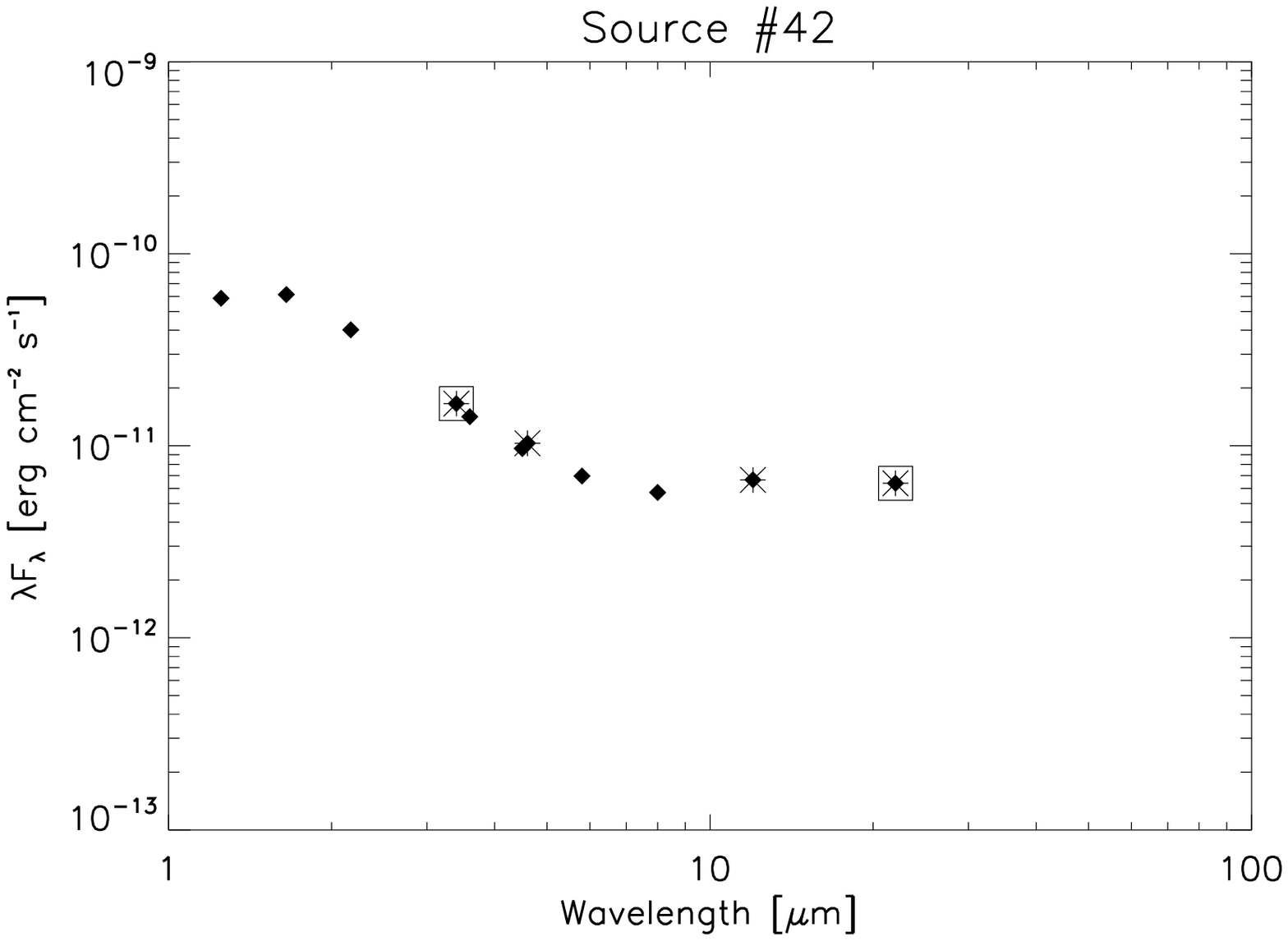}
\includegraphics[scale=0.45]{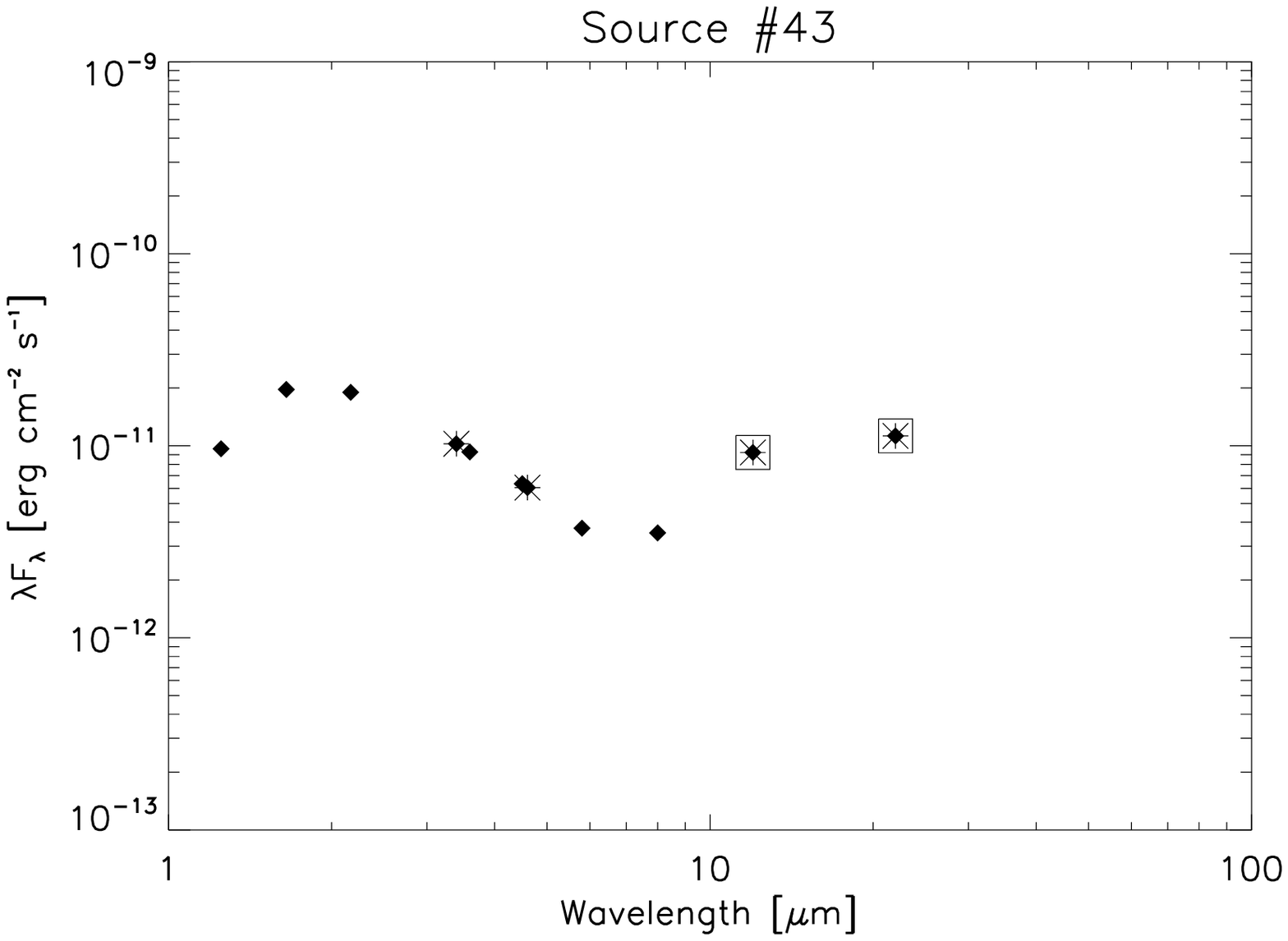}

\caption{\label{sed_2} SEDs of Class II X-ray detected YSOs from 2MASS (diamond), Spitzer (diamond), and WISE (asterisk) data when available.  Boxes plotted over WISE data signify high variability and/or possible contamination or confusion of photometry.}

\end{figure}

\begin{figure}
\centering
\includegraphics[scale=0.45]{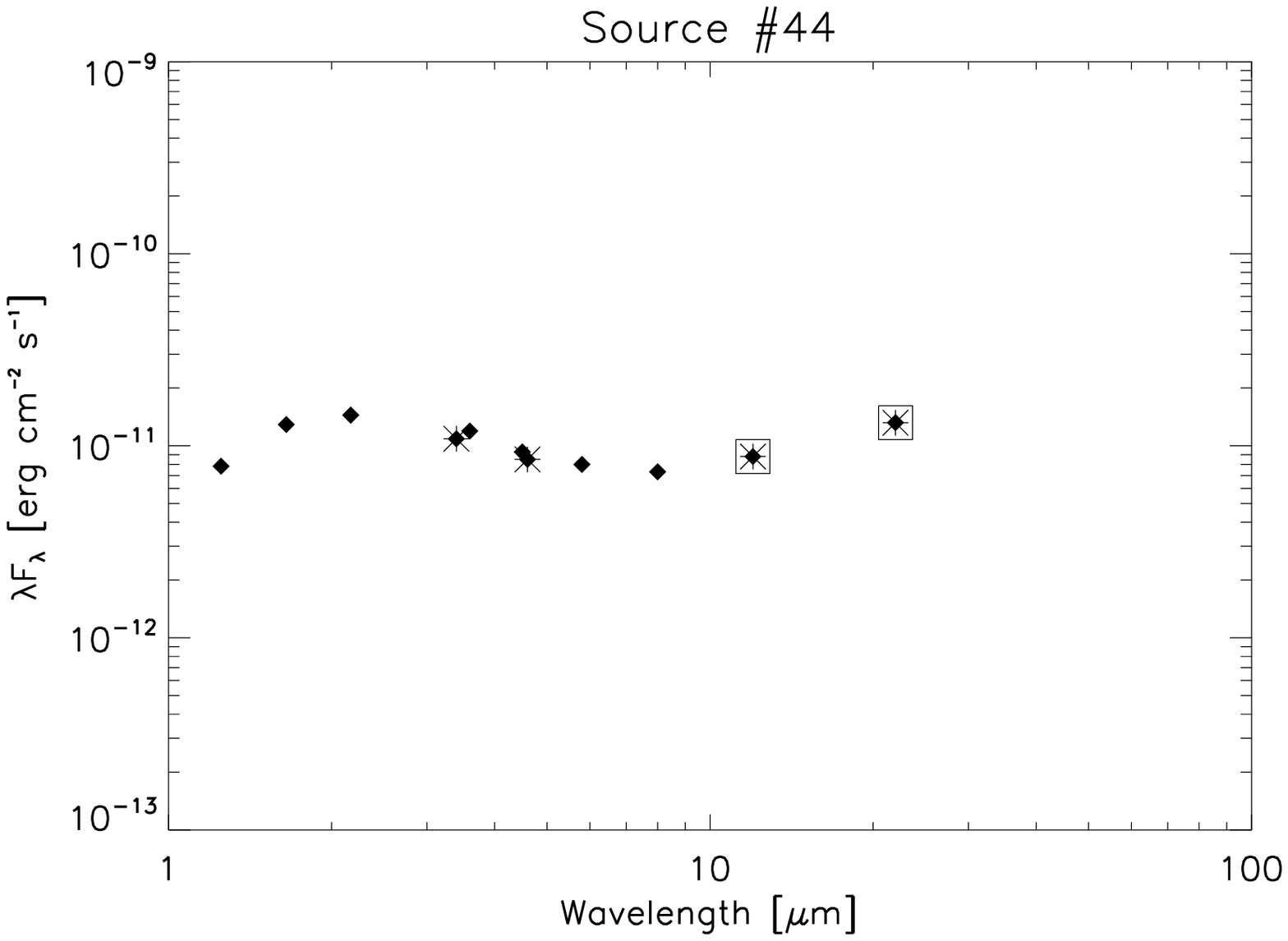}
\includegraphics[scale=0.45]{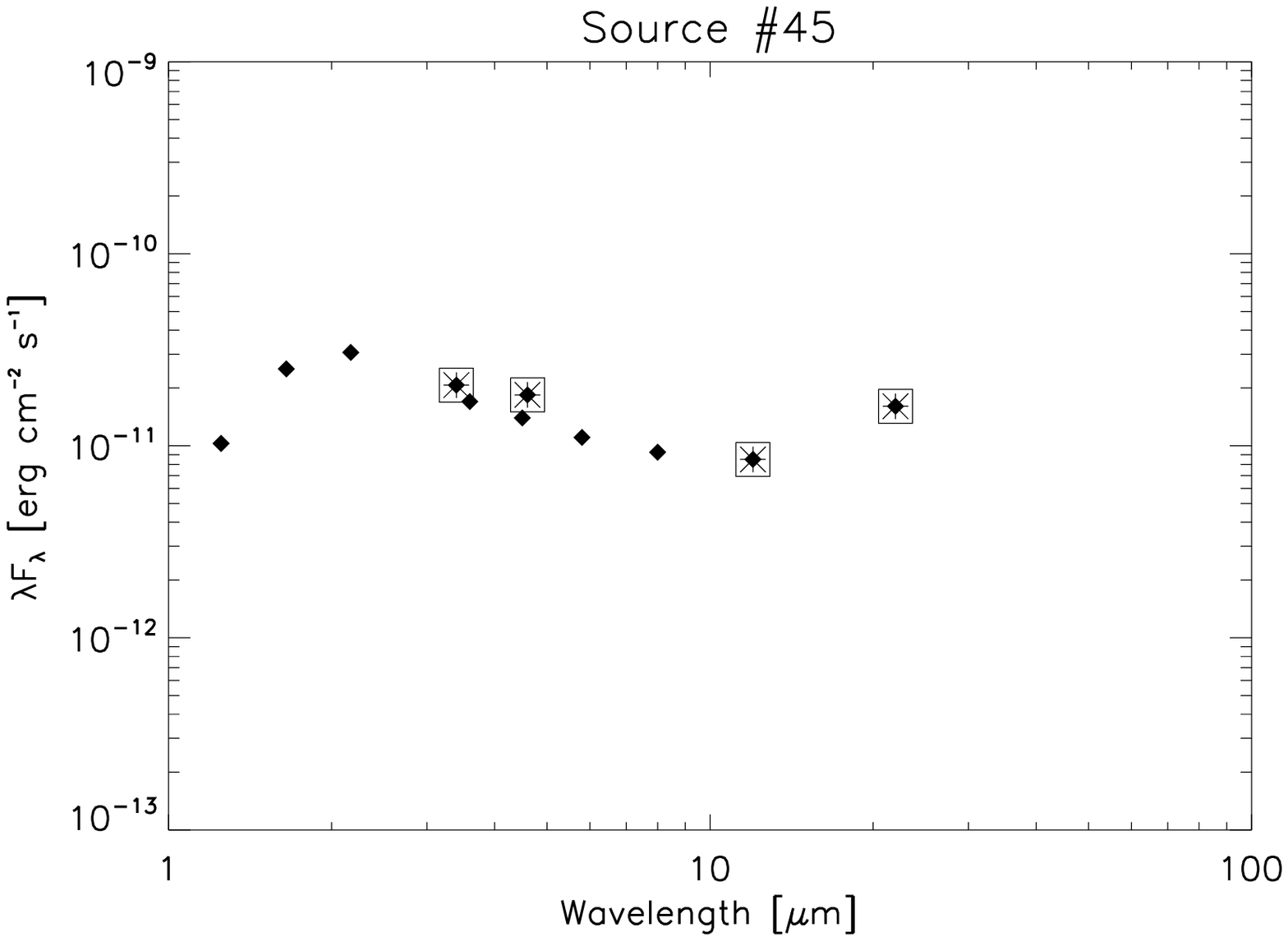}
\includegraphics[scale=0.45]{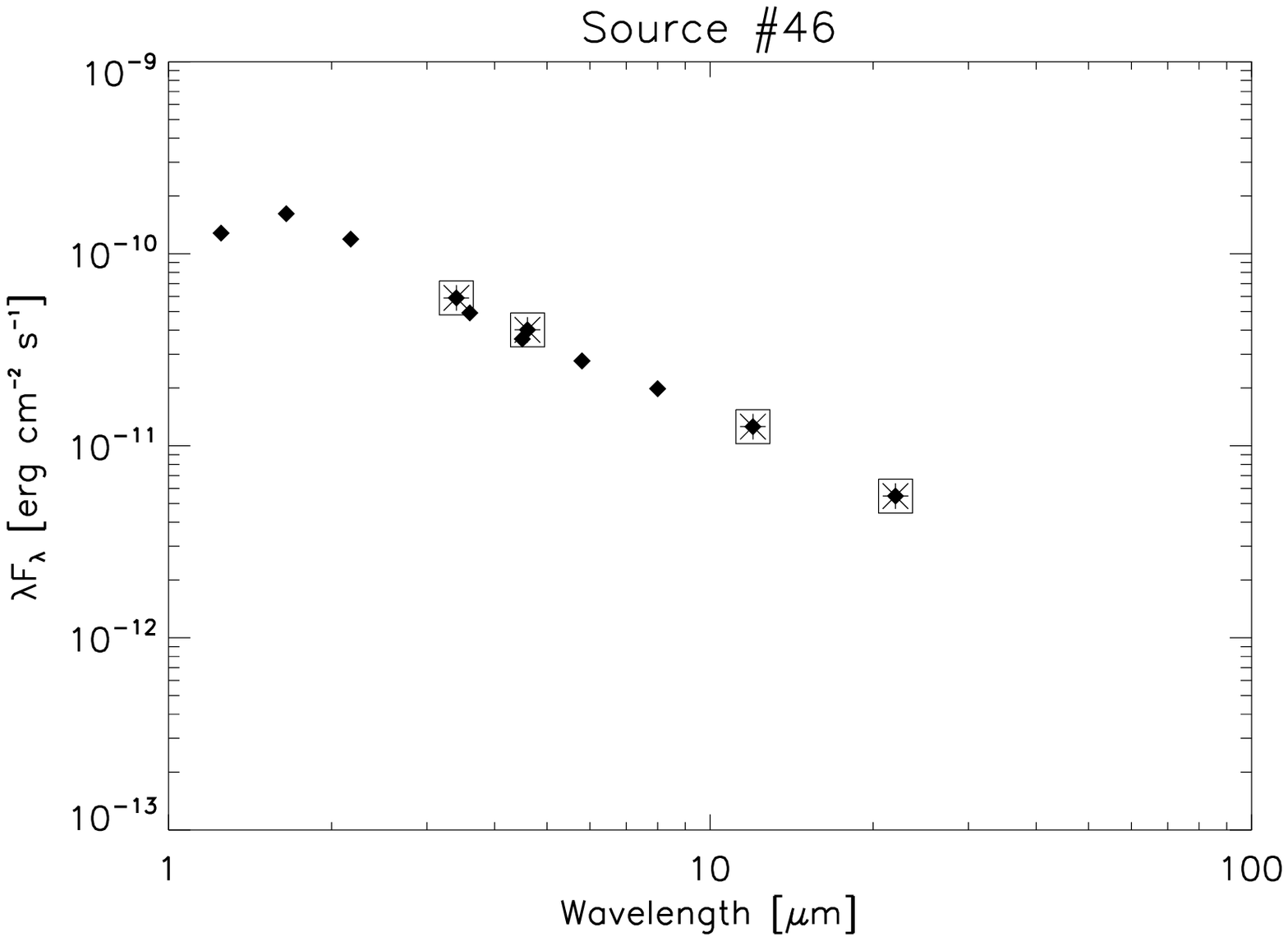}
\includegraphics[scale=0.45]{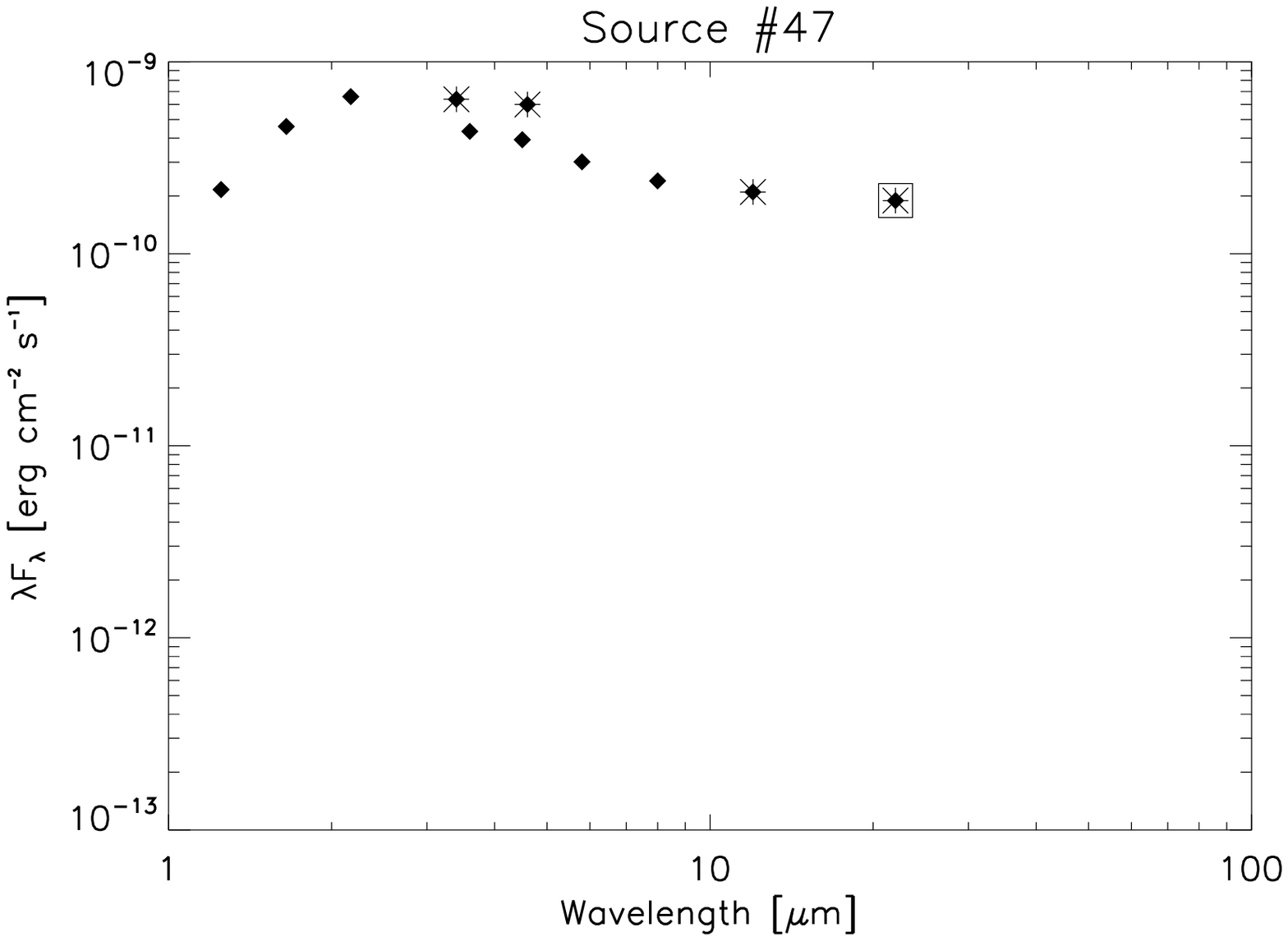}
\includegraphics[scale=0.45]{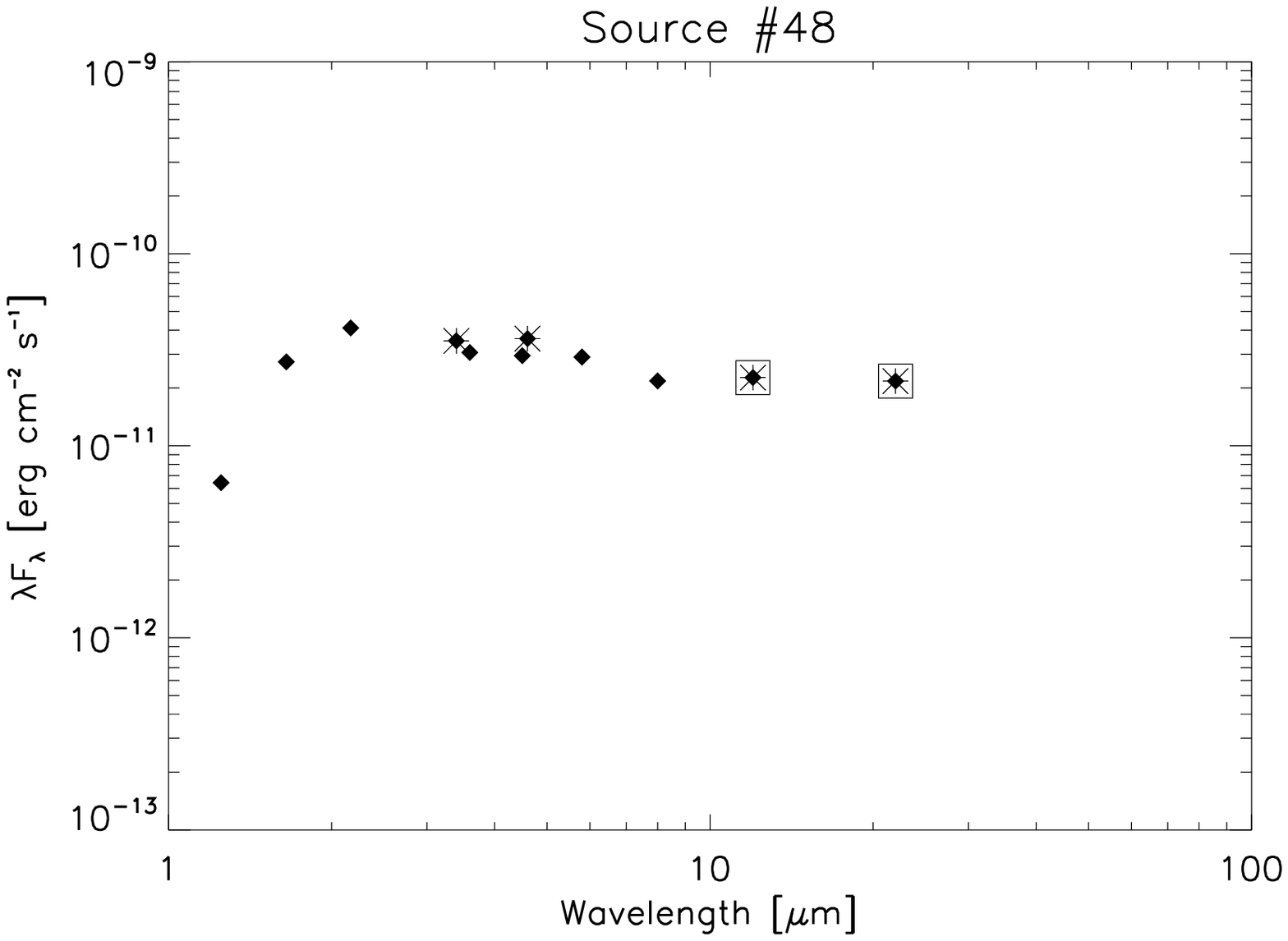}
\includegraphics[scale=0.45]{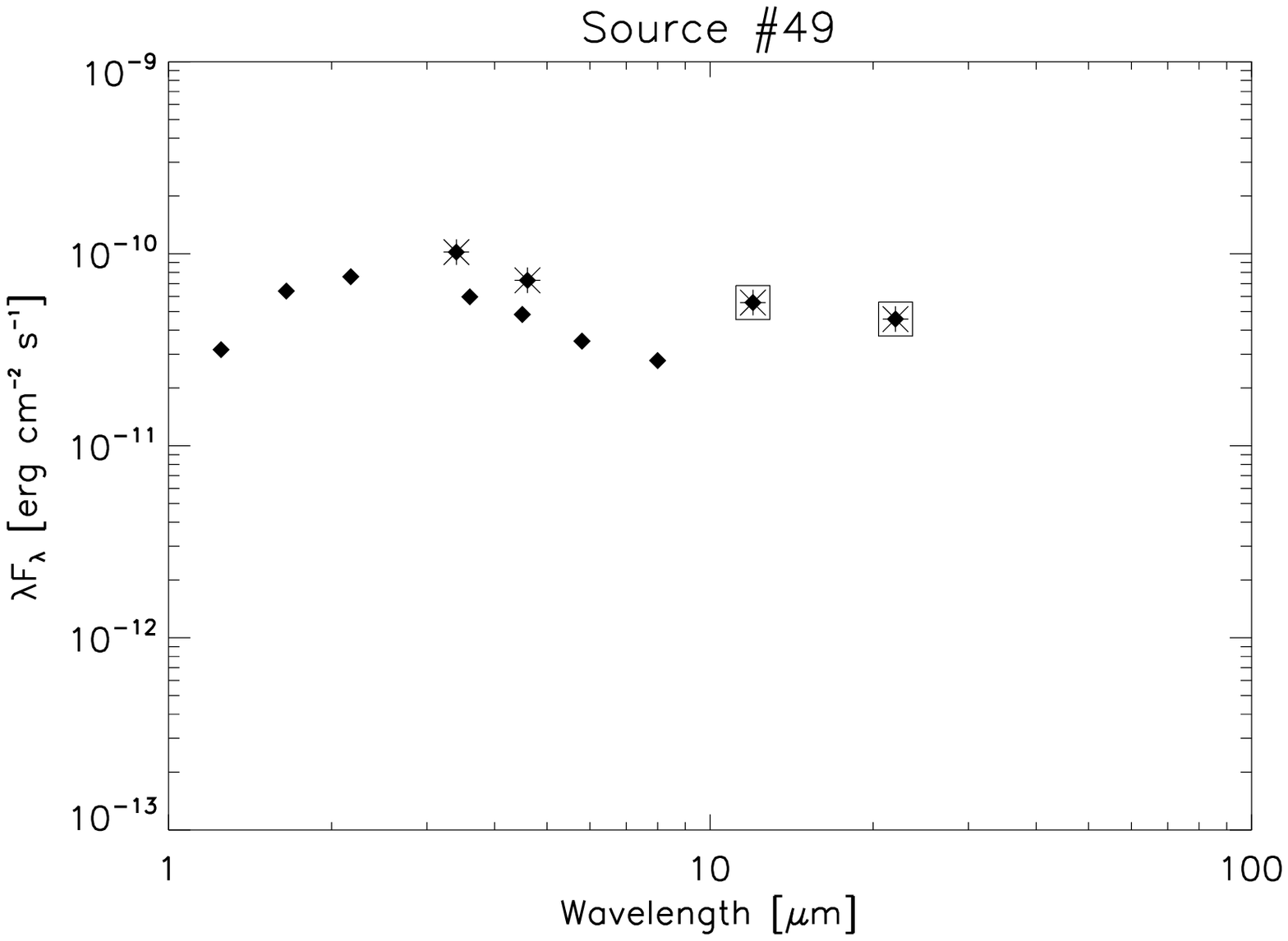}

\setcounter{figure}{17}

\caption{(continued) SEDs of Class II X-ray detected YSOs from 2MASS (diamond), Spitzer (diamond), and WISE (asterisk) data when available.  Boxes plotted over WISE data signify high variability and/or possible contamination or confusion of photometry.}

\end{figure}

\begin{figure}
\centering
\includegraphics[scale=0.45]{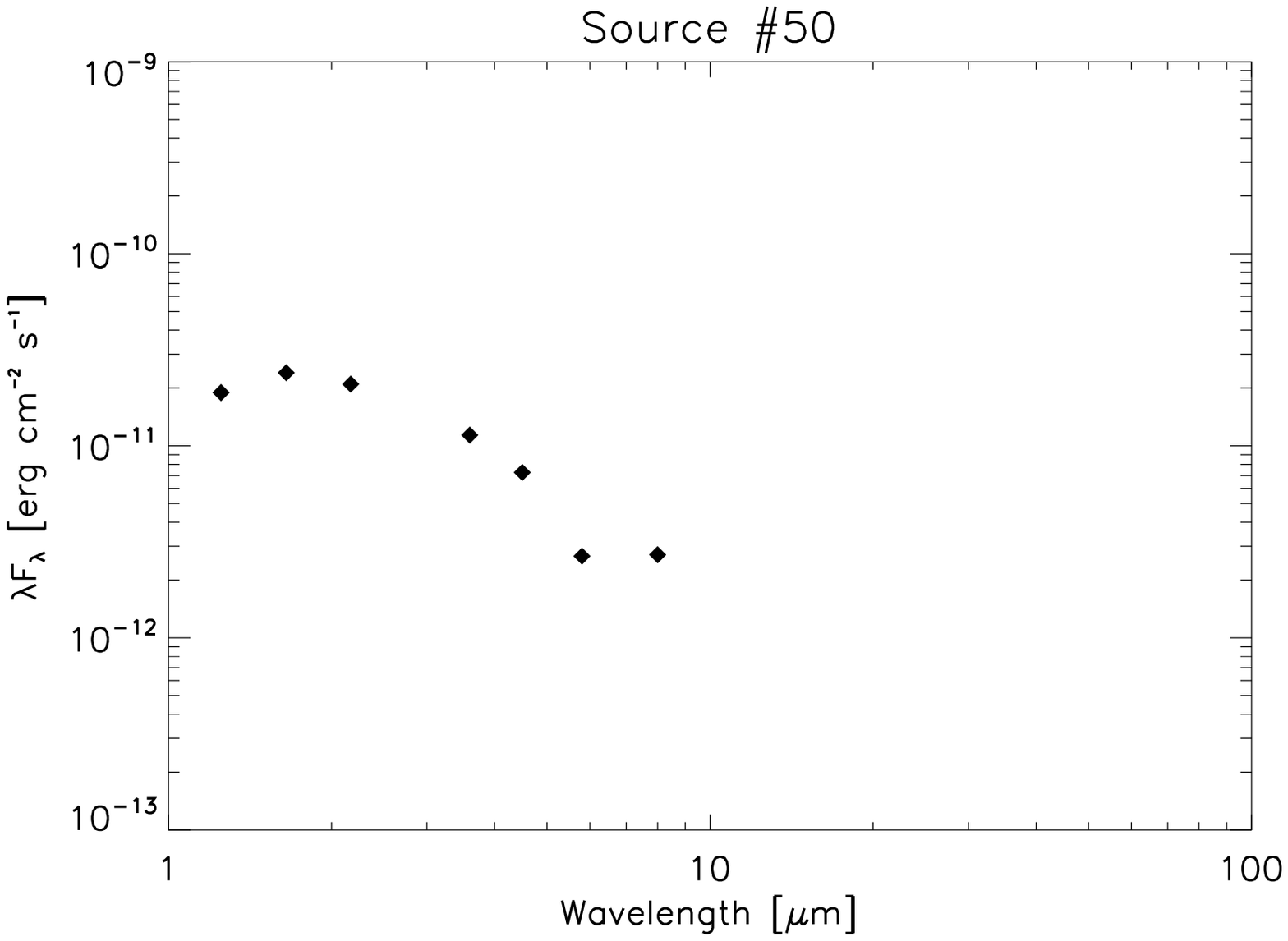}
\includegraphics[scale=0.45]{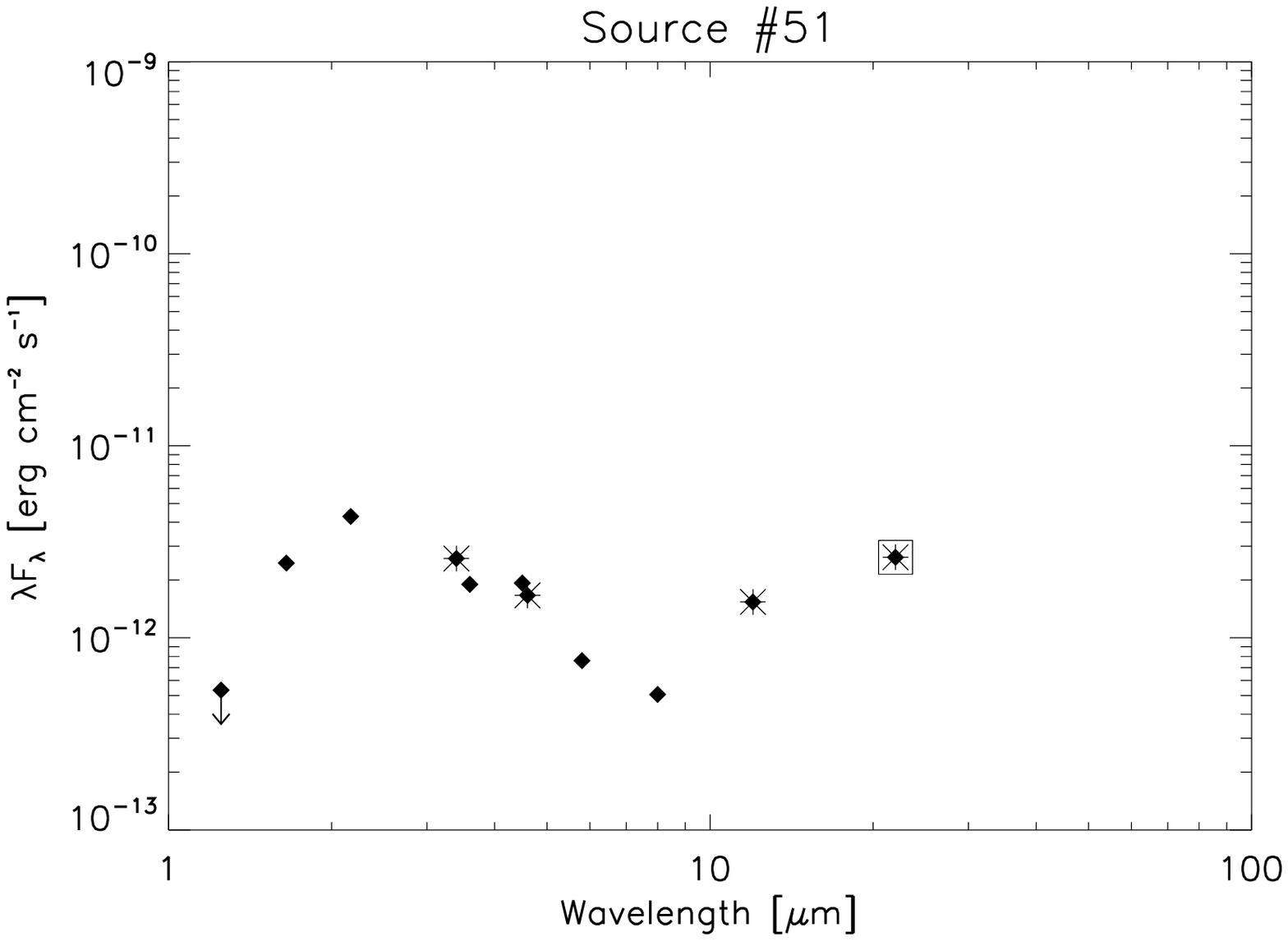}
\includegraphics[scale=0.45]{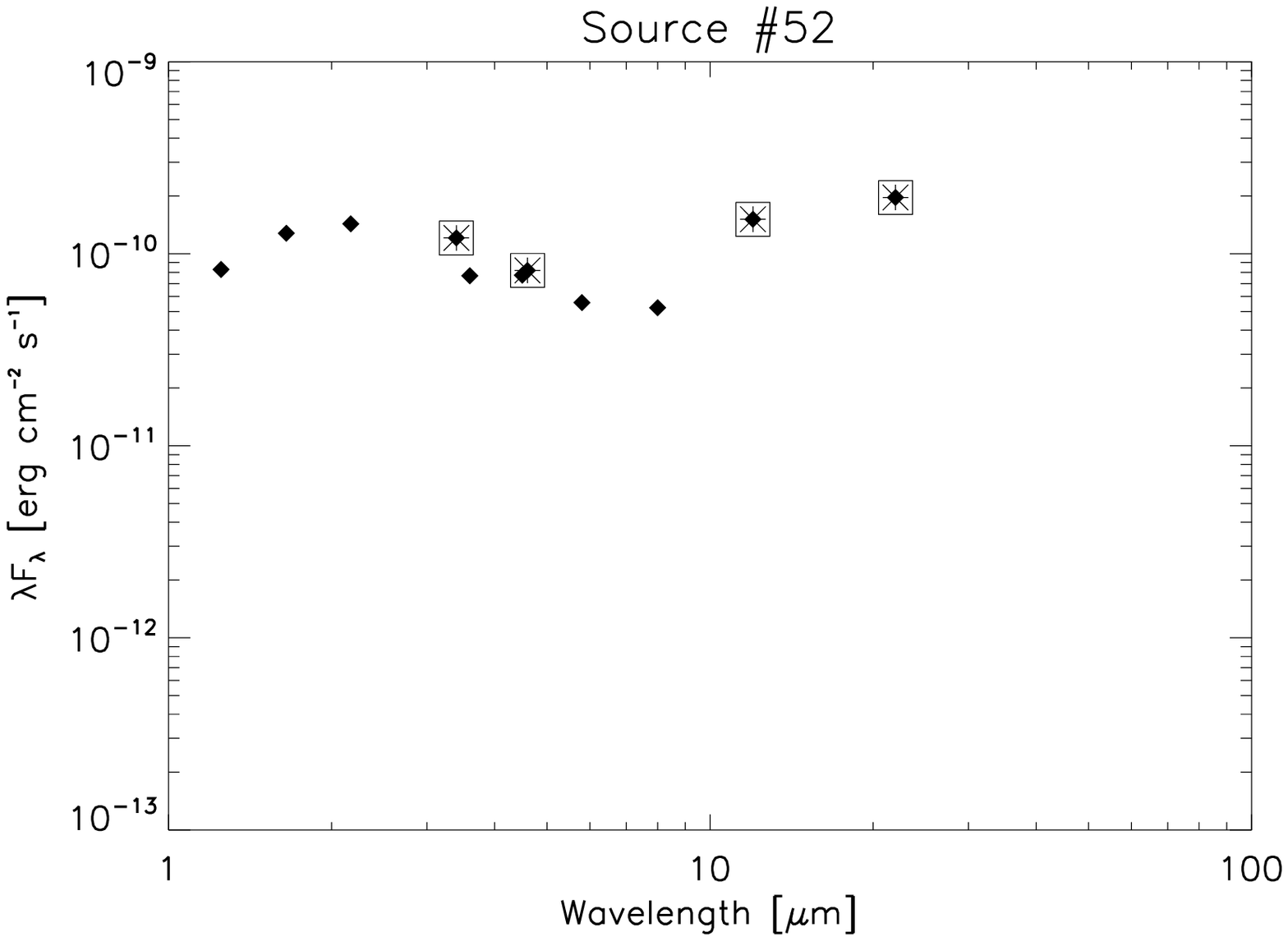}

\setcounter{figure}{17}

\caption{(continued) SEDs of Class II X-ray detected YSOs from 2MASS (diamond), Spitzer (diamond), and WISE (asterisk) data when available.  Boxes plotted over WISE data signify high variability and/or possible contamination or confusion of photometry.}
\end{figure}

\begin{figure}[H]
\centering

\includegraphics[scale=0.45]{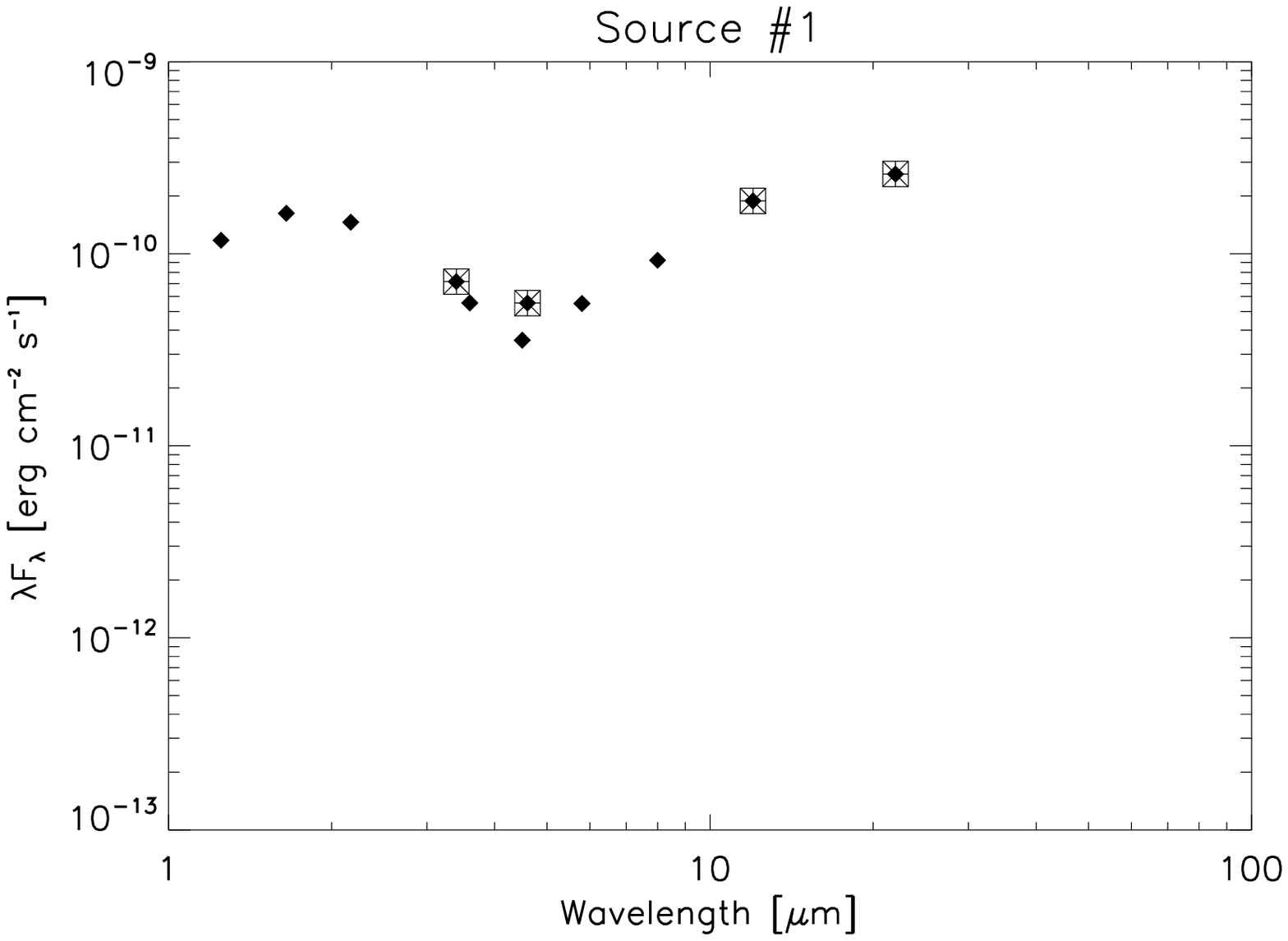}
\includegraphics[scale=0.45]{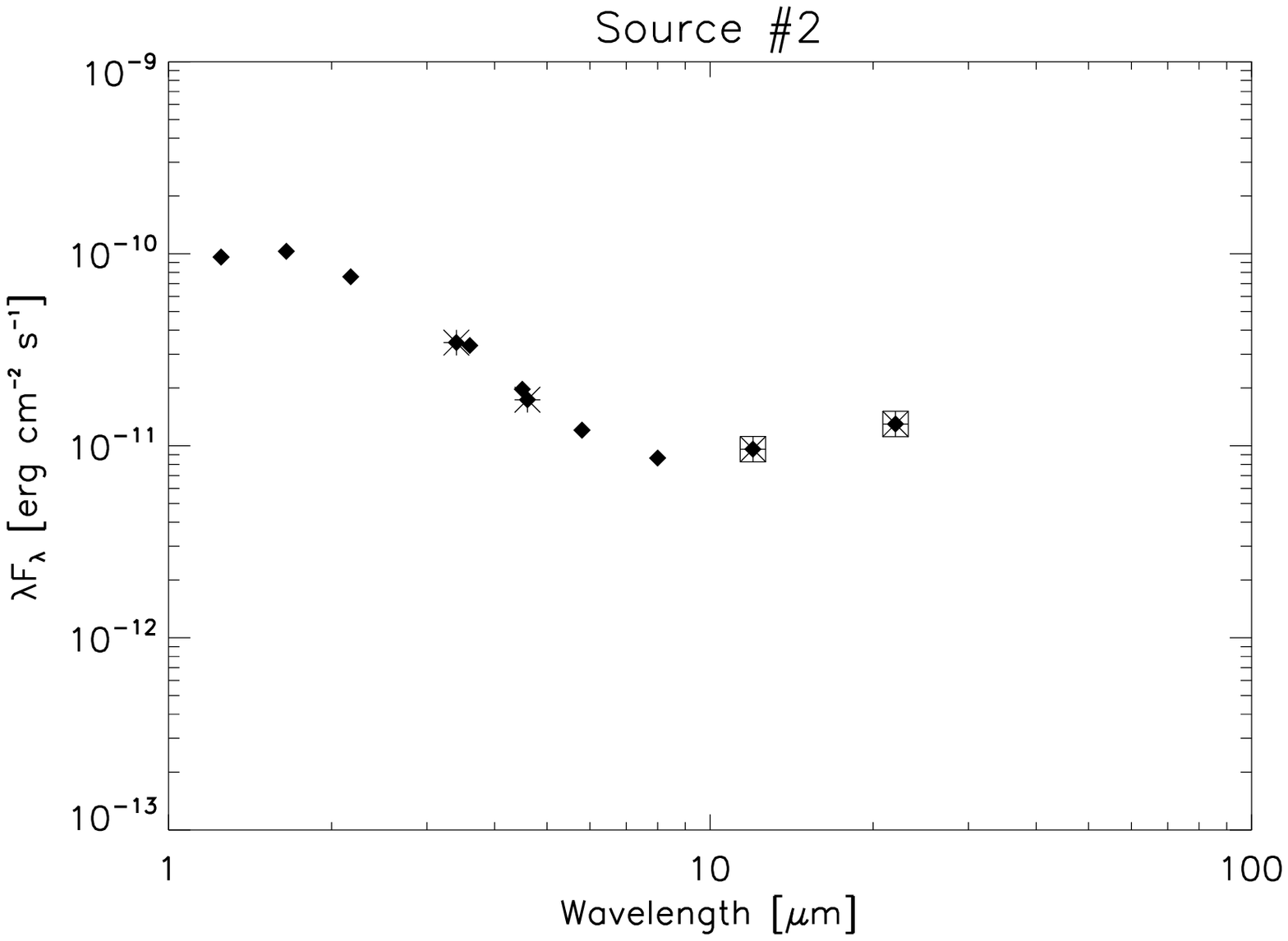}
\includegraphics[scale=0.45]{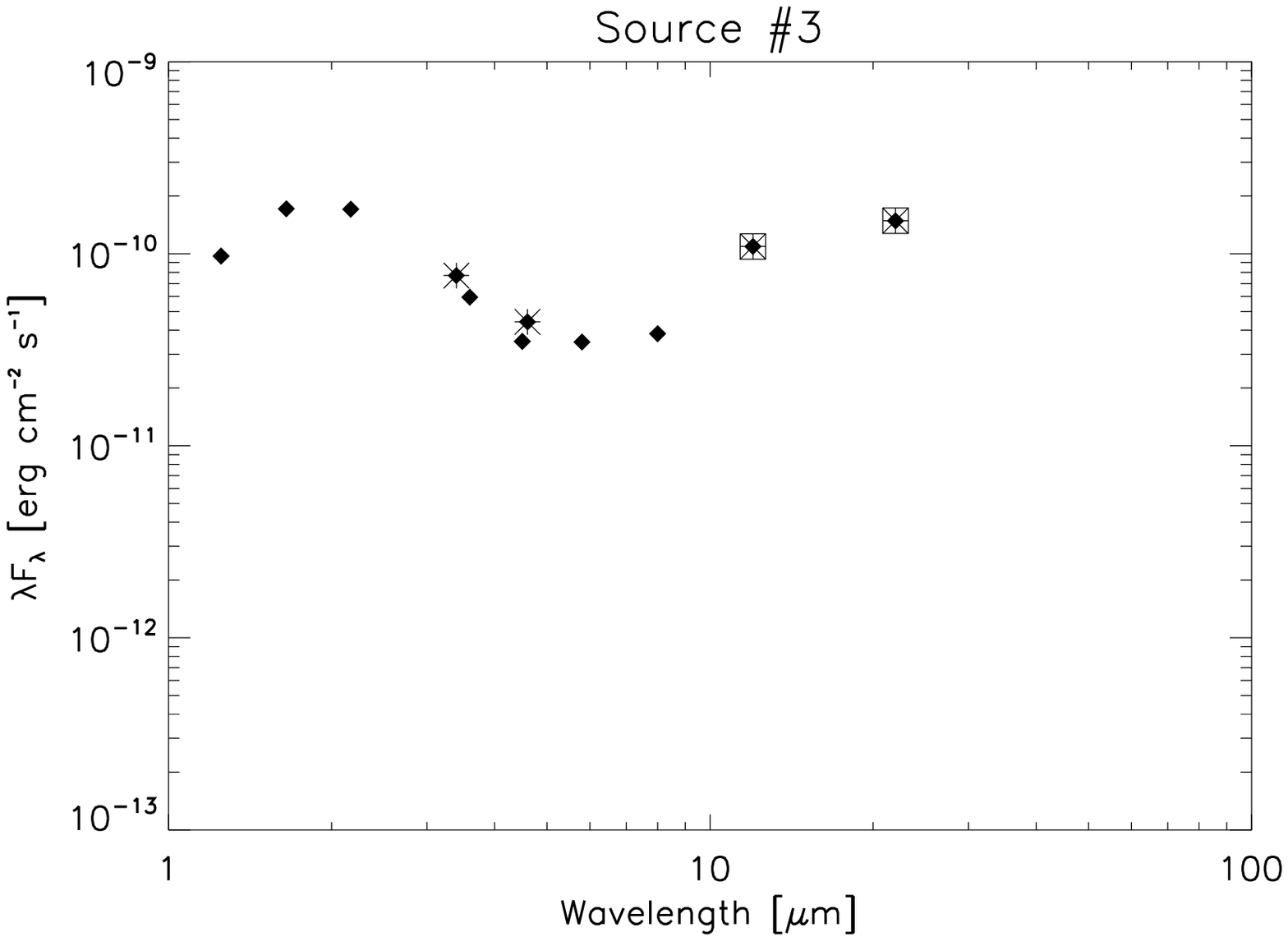}
\includegraphics[scale=0.45]{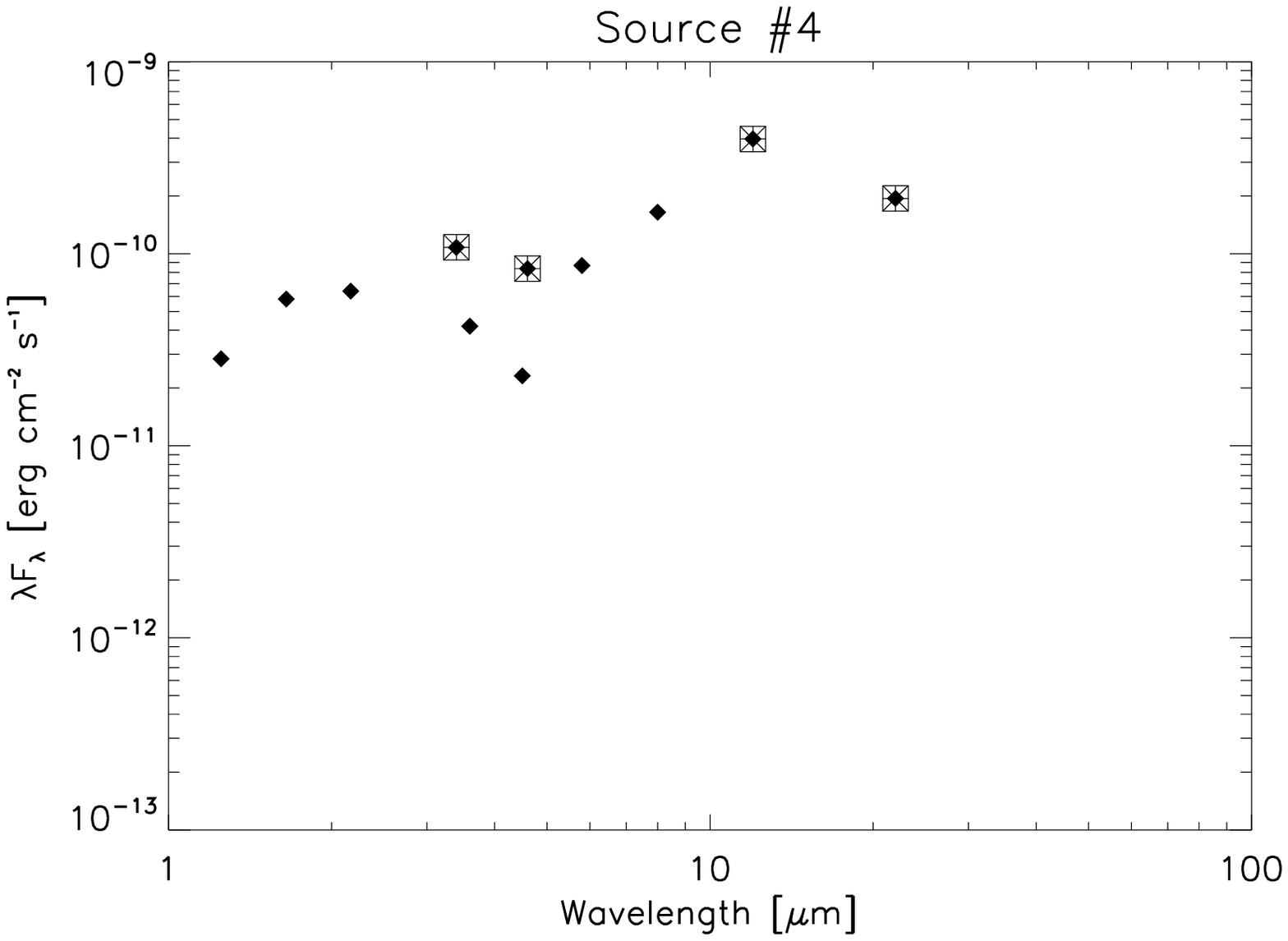}
\includegraphics[scale=0.45]{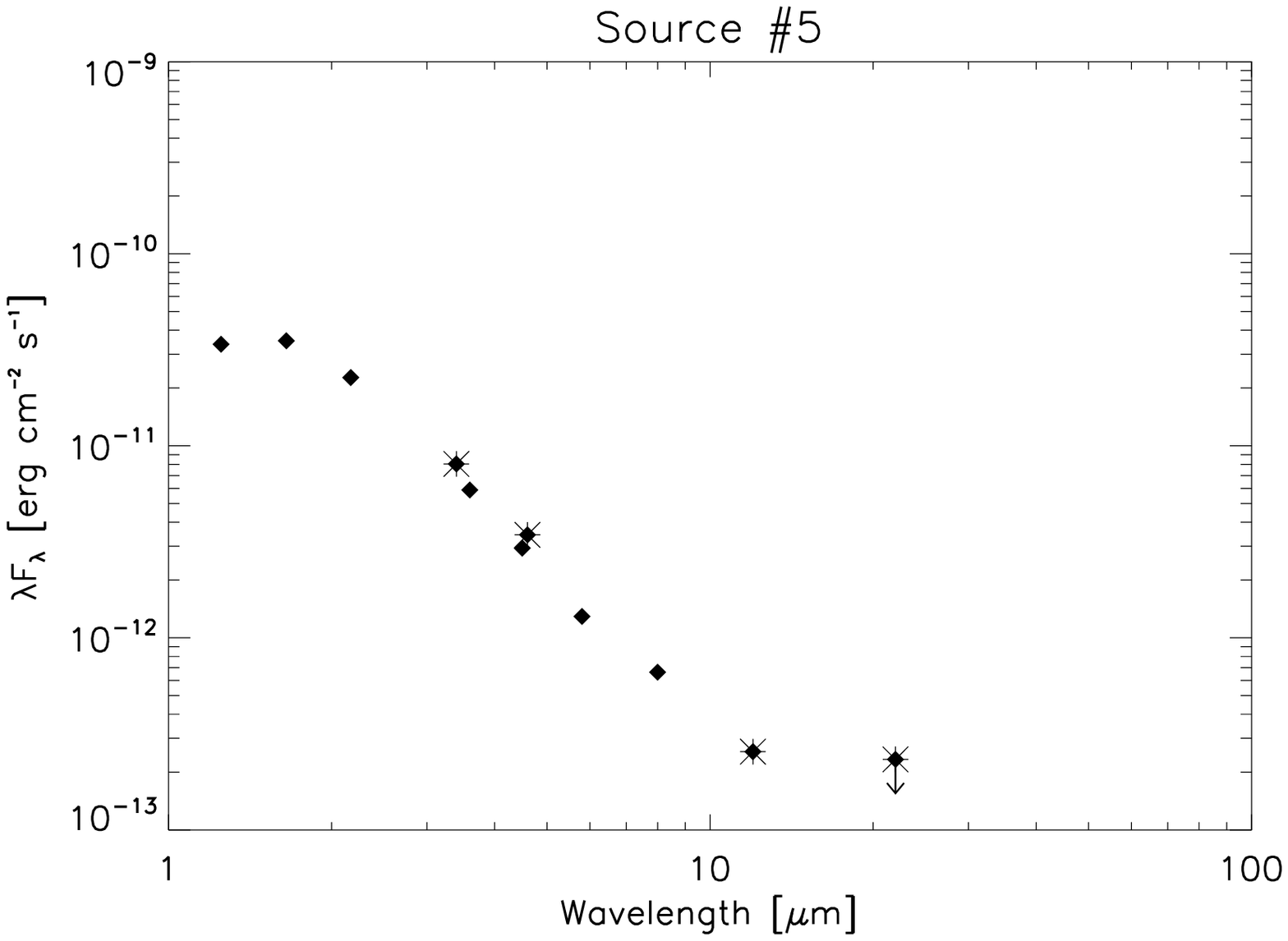}
\includegraphics[scale=0.45]{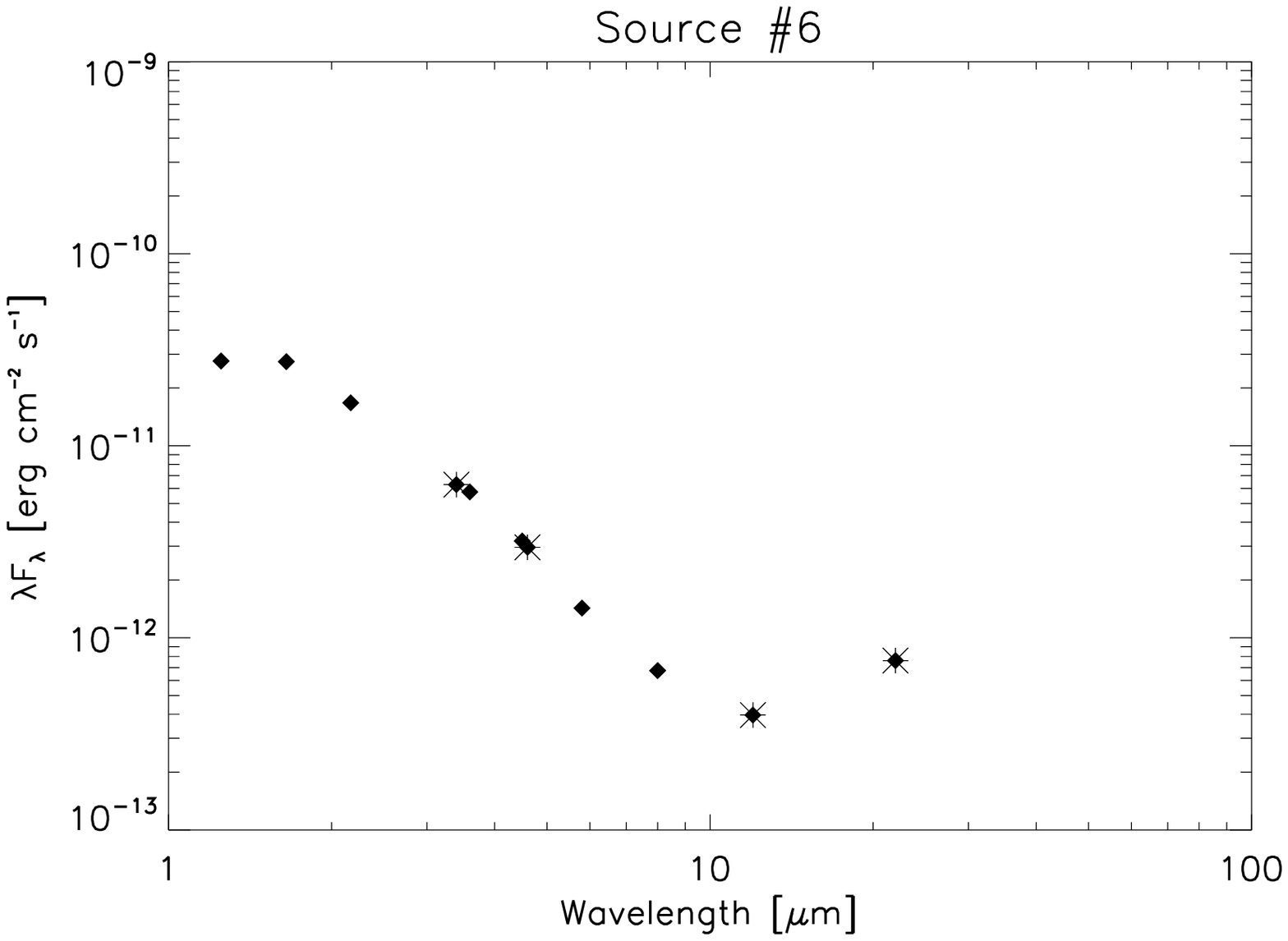}

\caption{\label{sed_tran} SEDs of X-ray detected Transition Disk YSOs from 2MASS (diamond), Spitzer (diamond), and WISE (asterisk) data when available. Boxes plotted over WISE data signify high variability and/or possible contamination or confusion of photometry.}

\end{figure}

\begin{figure}
\centering
\includegraphics[scale=0.45]{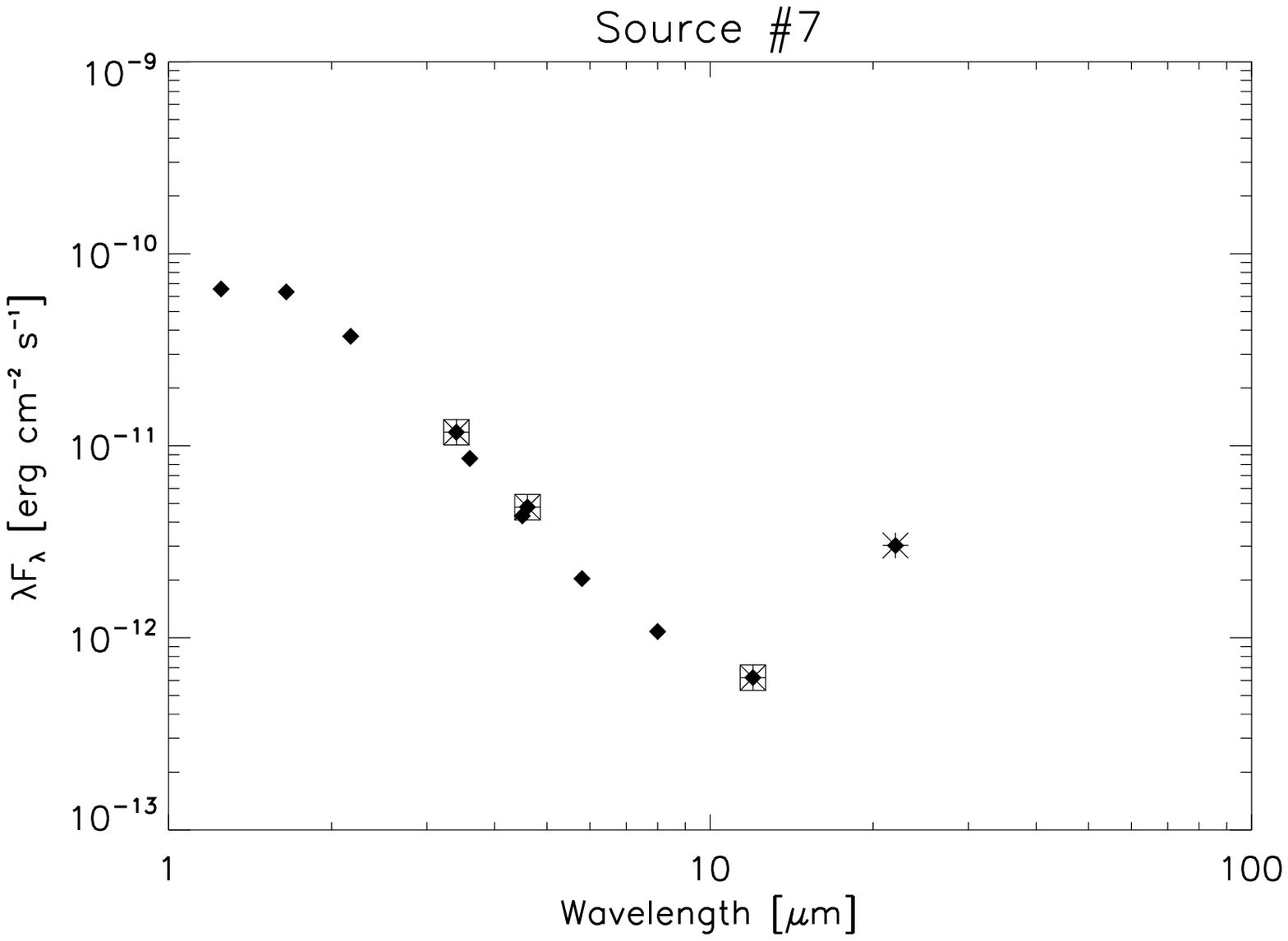}
\includegraphics[scale=0.45]{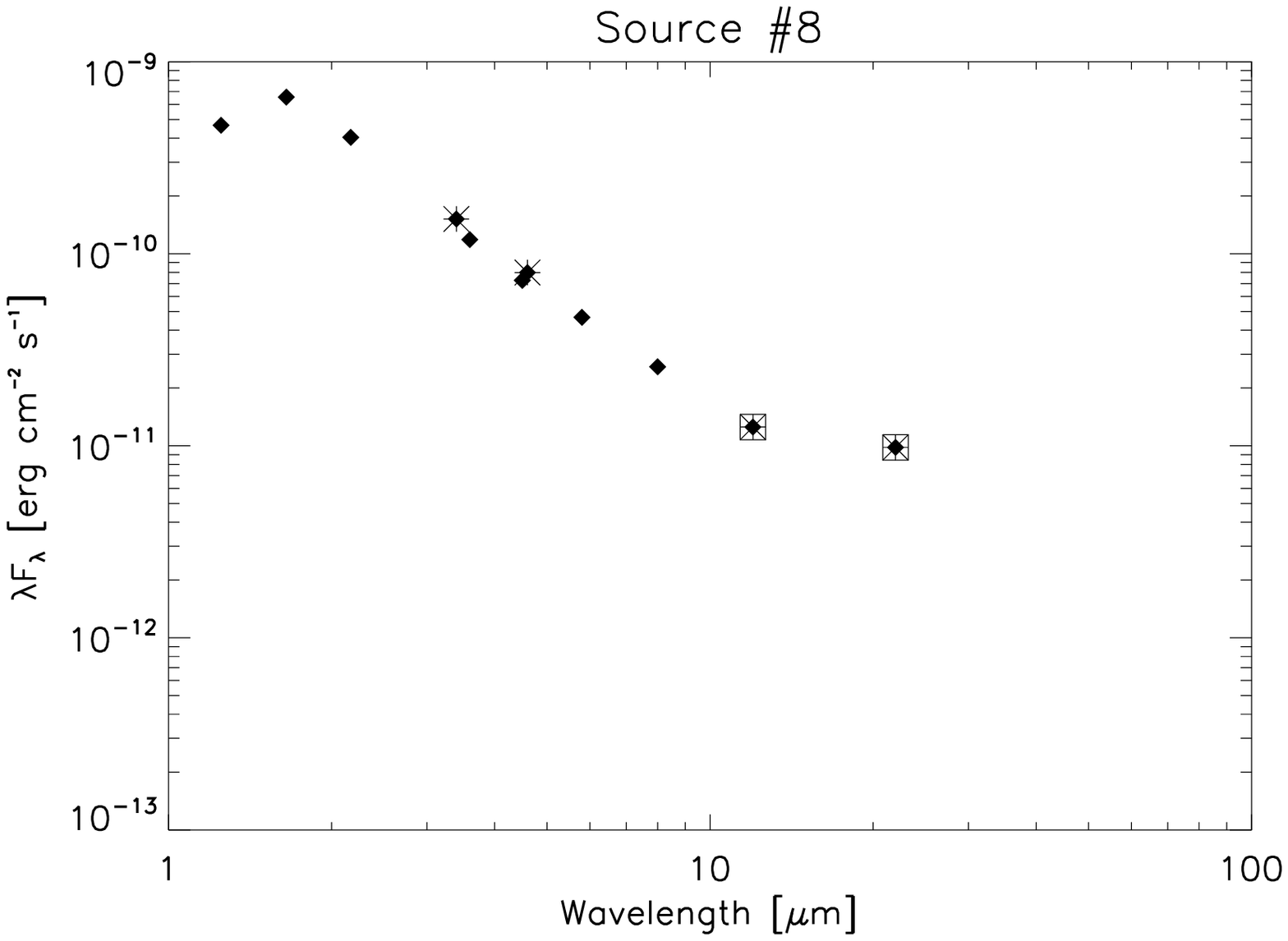}
\includegraphics[scale=0.45]{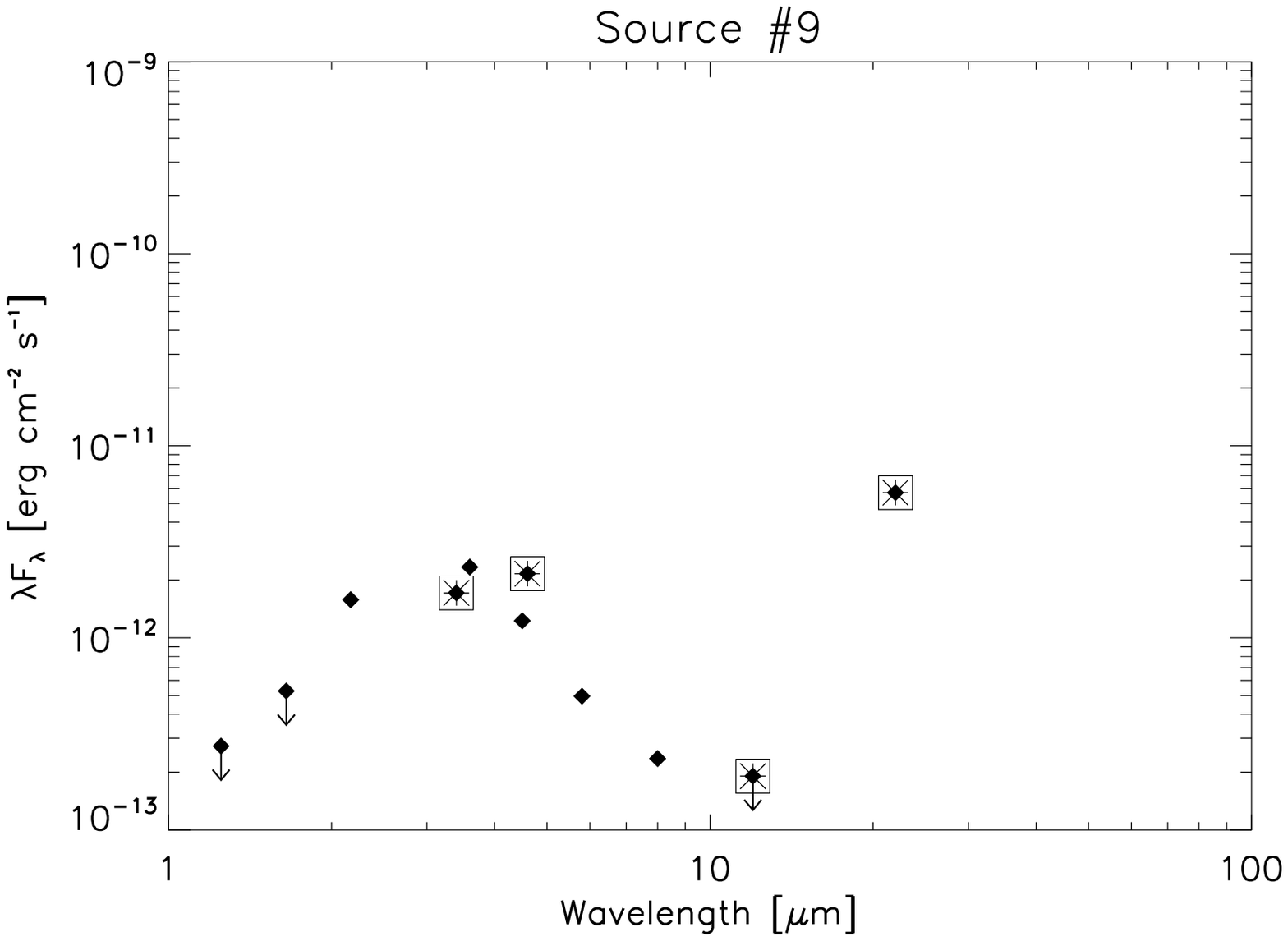}
\includegraphics[scale=0.45]{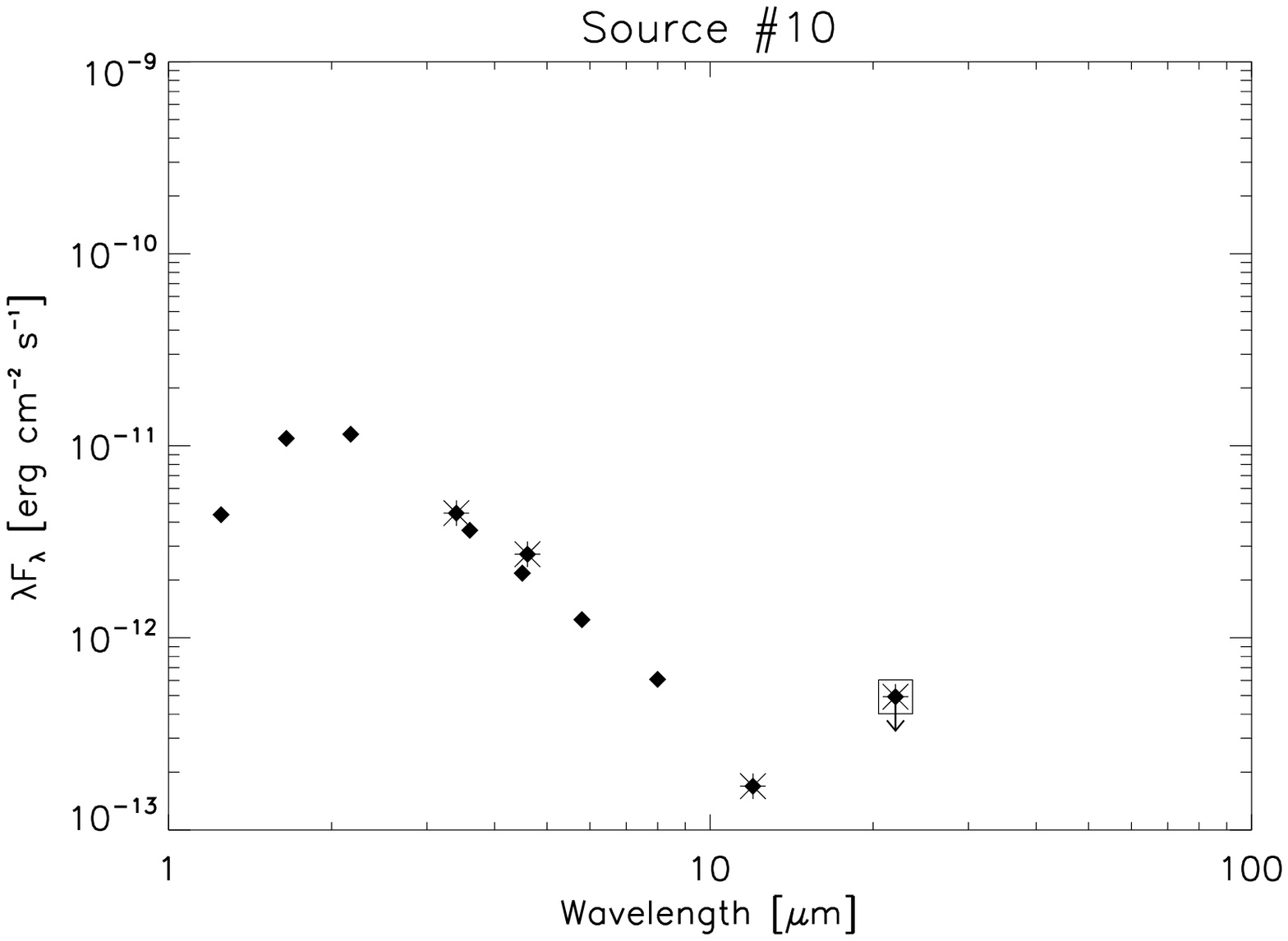}
\includegraphics[scale=0.45]{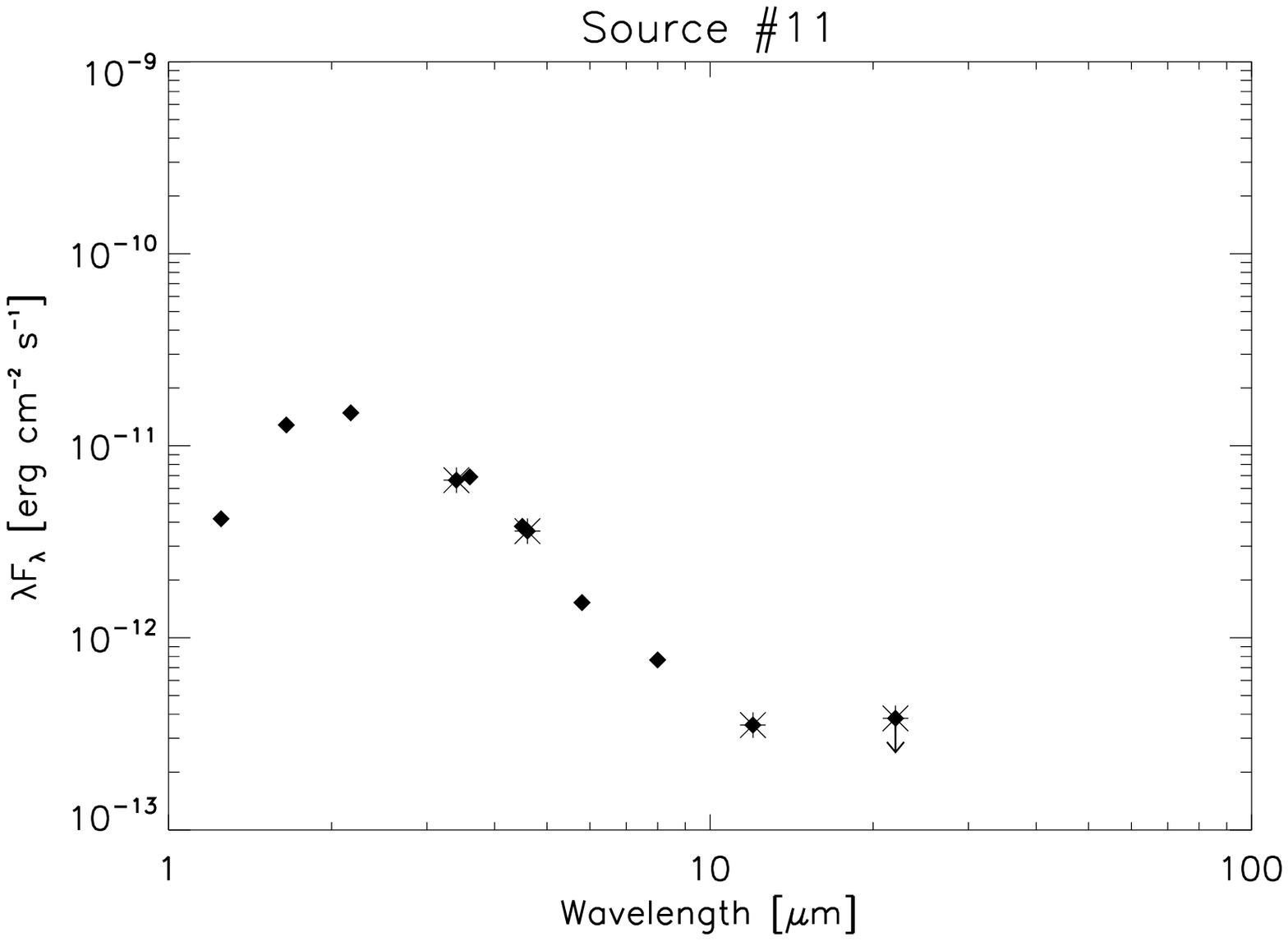}

\setcounter{figure}{18}
\caption{(continued) SEDs of X-ray detected Transition Disk YSOs from 2MASS (diamond), Spitzer (diamond), and WISE (asterisk) data when available. Boxes plotted over WISE data signify high variability and/or possible contamination or confusion of photometry.}
\end{figure}

\begin{figure}[H]
\centering

\includegraphics[scale=0.45]{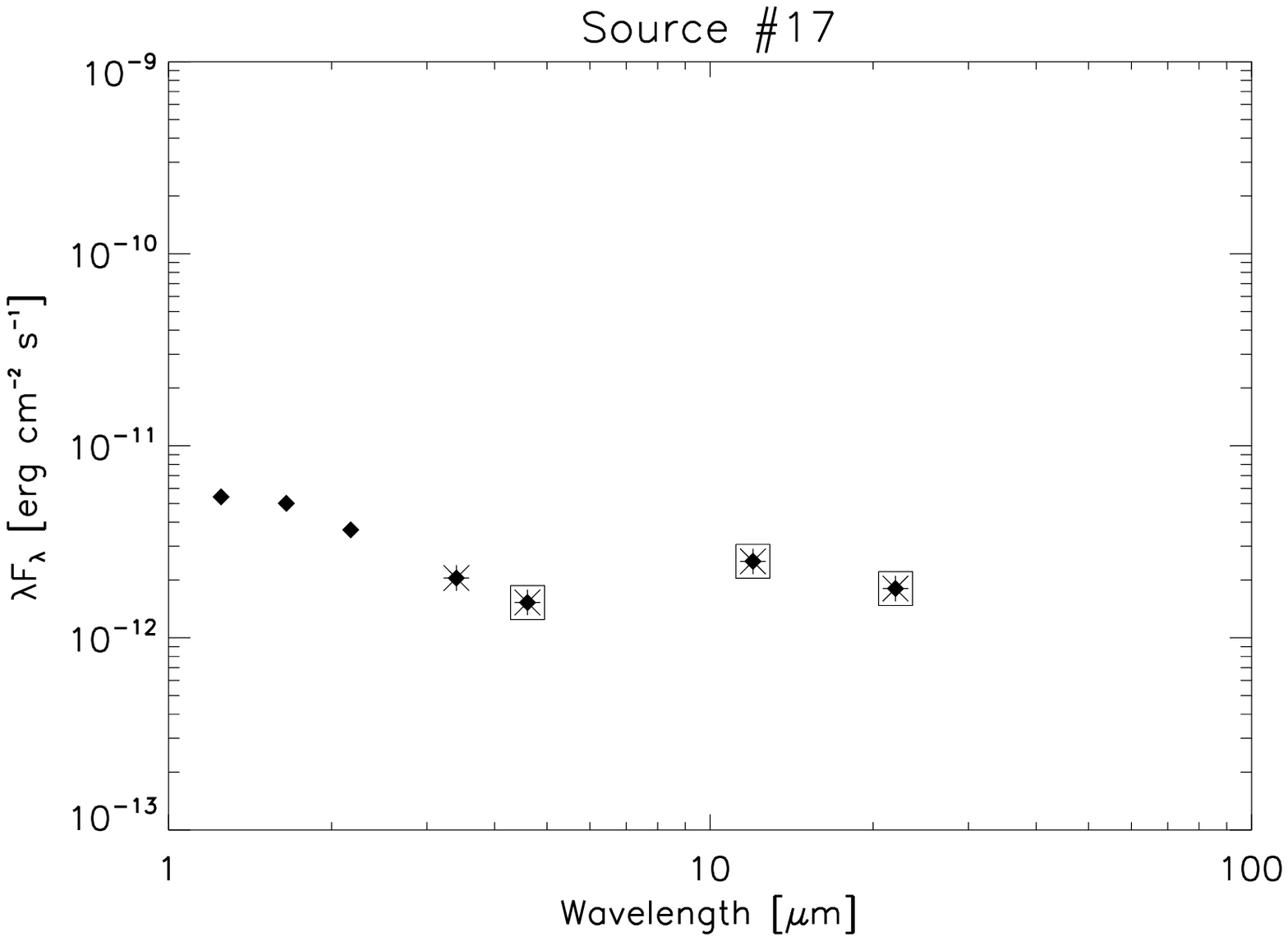}
\includegraphics[scale=0.45]{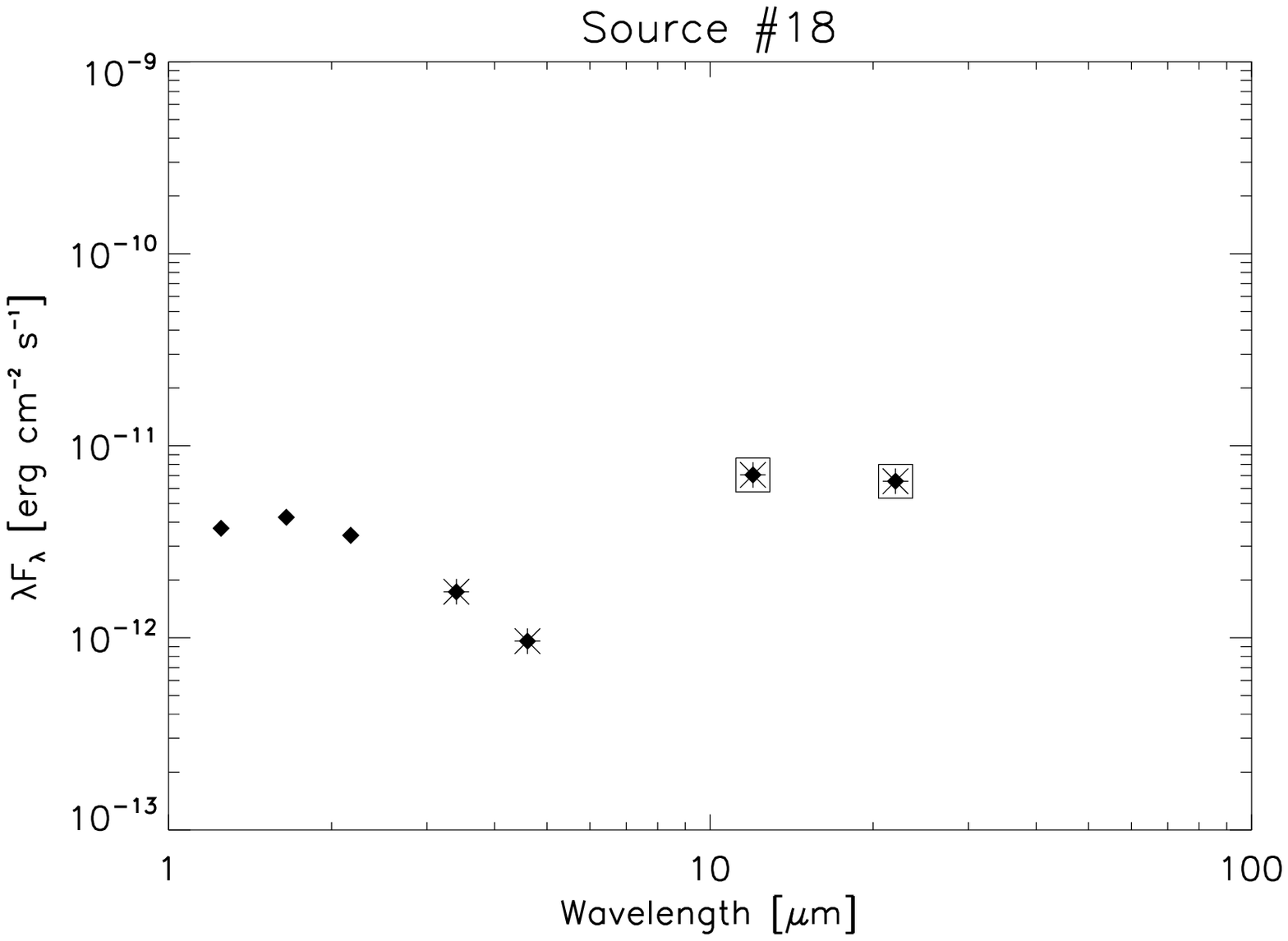}
\includegraphics[scale=0.45]{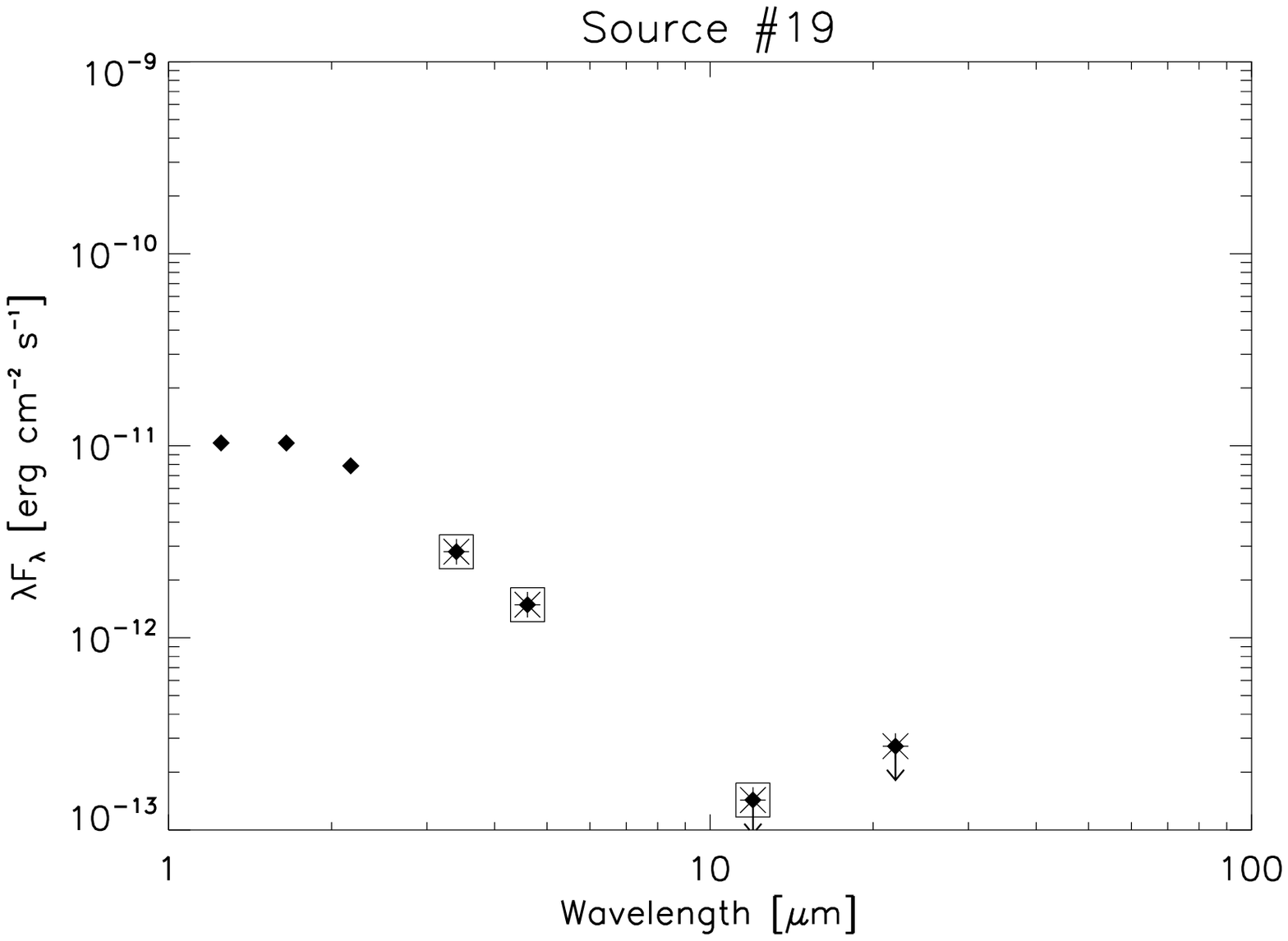}
\includegraphics[scale=0.45]{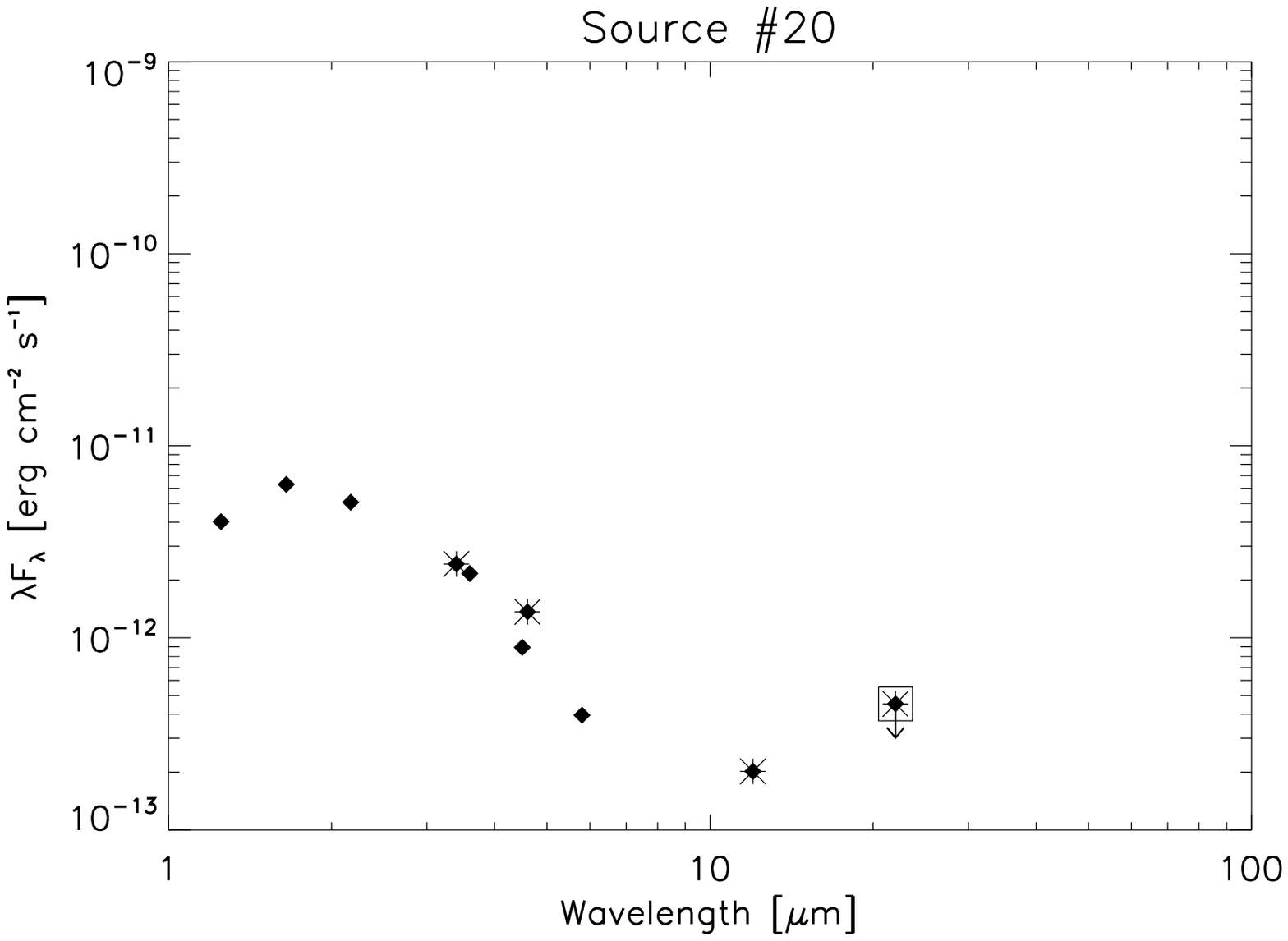}
\includegraphics[scale=0.45]{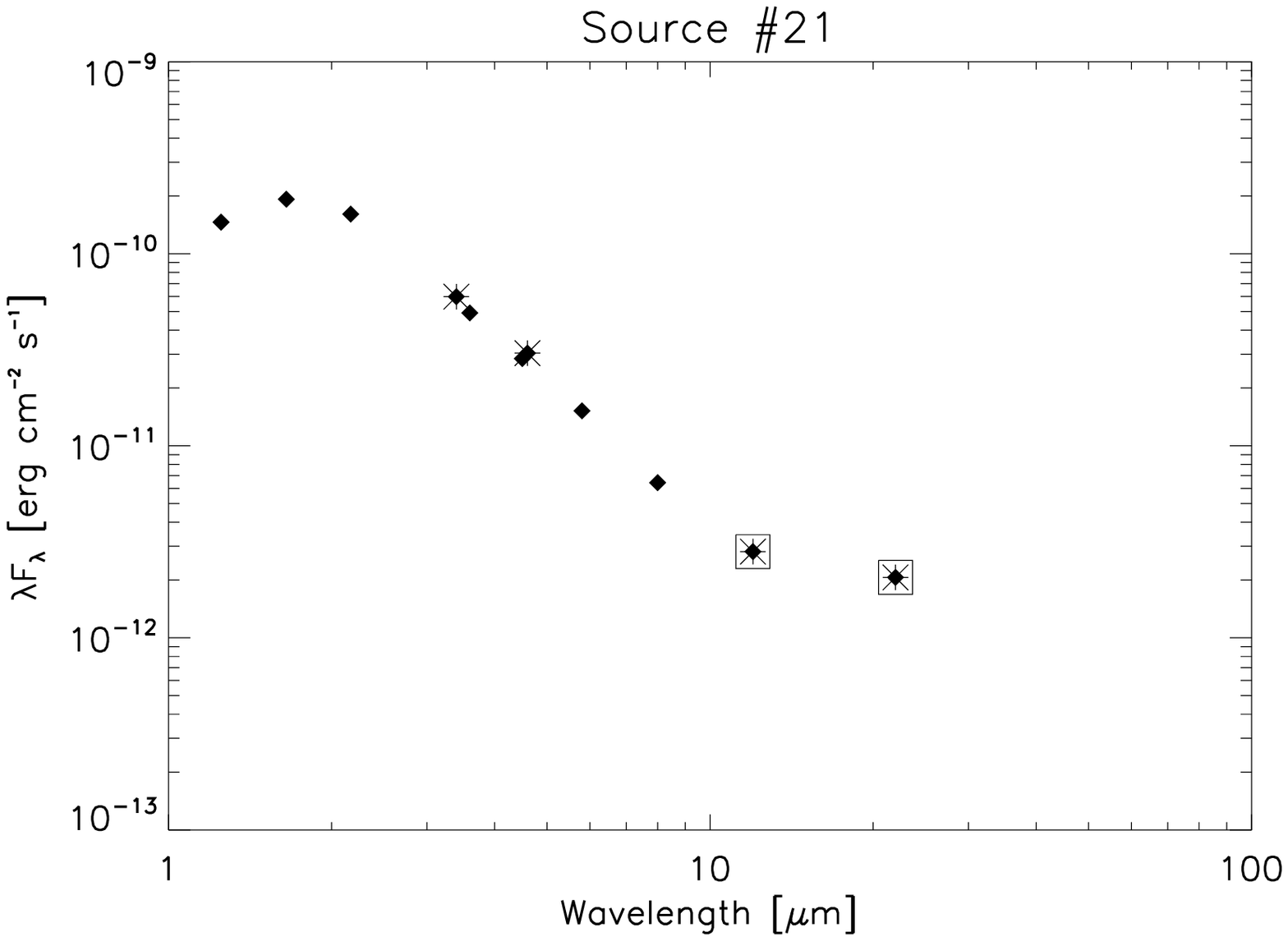}
\includegraphics[scale=0.45]{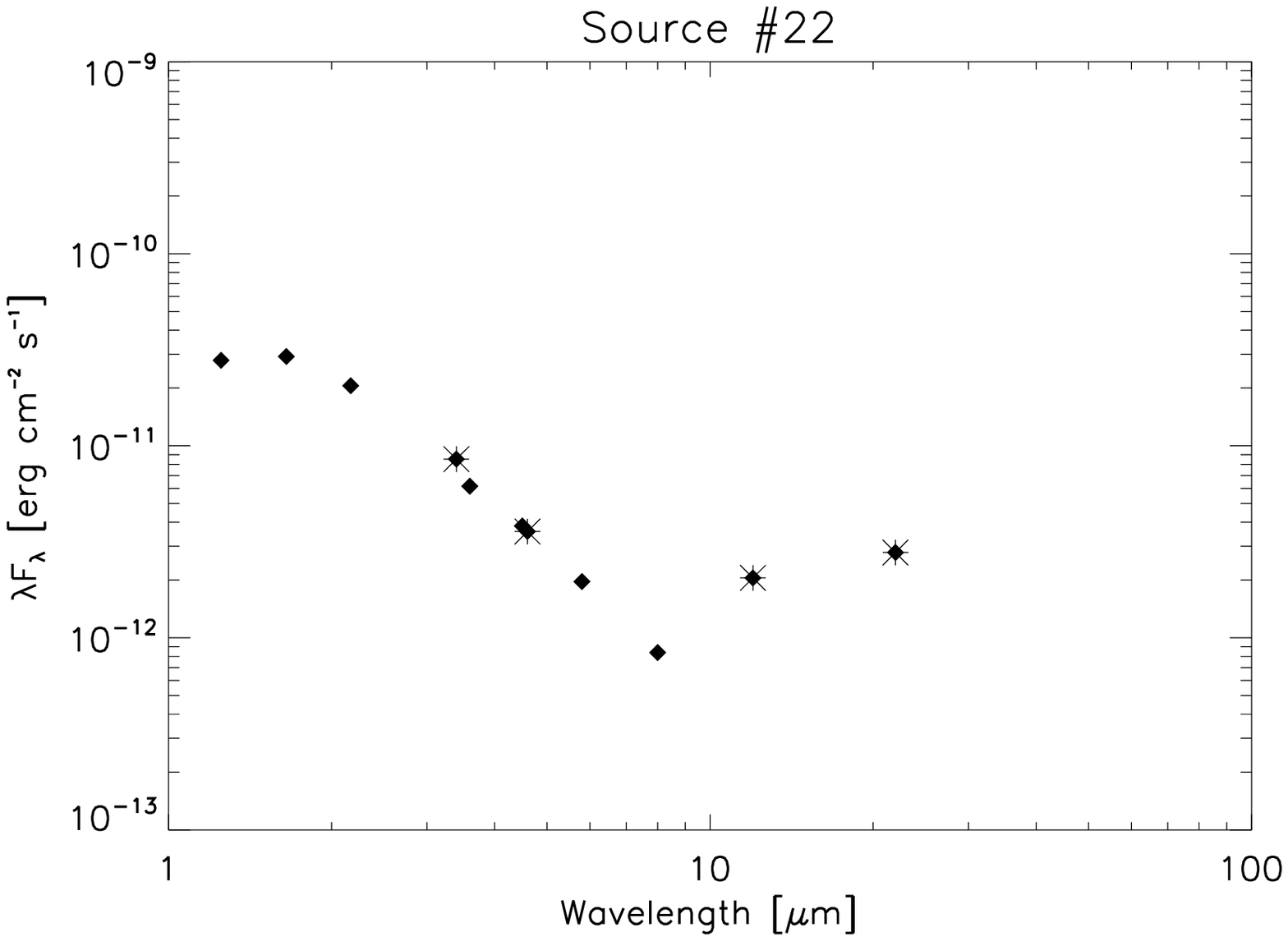}

\caption{\label{sed_3} SEDs of Class III X-ray detected YSOs from 2MASS (diamond), Spitzer (diamond), and WISE (asterisk) data when available.  Boxes plotted over WISE data signify high variability and/or possible contamination or confusion of photometry.}

\end{figure}

\begin{figure}
\centering
\includegraphics[scale=0.45]{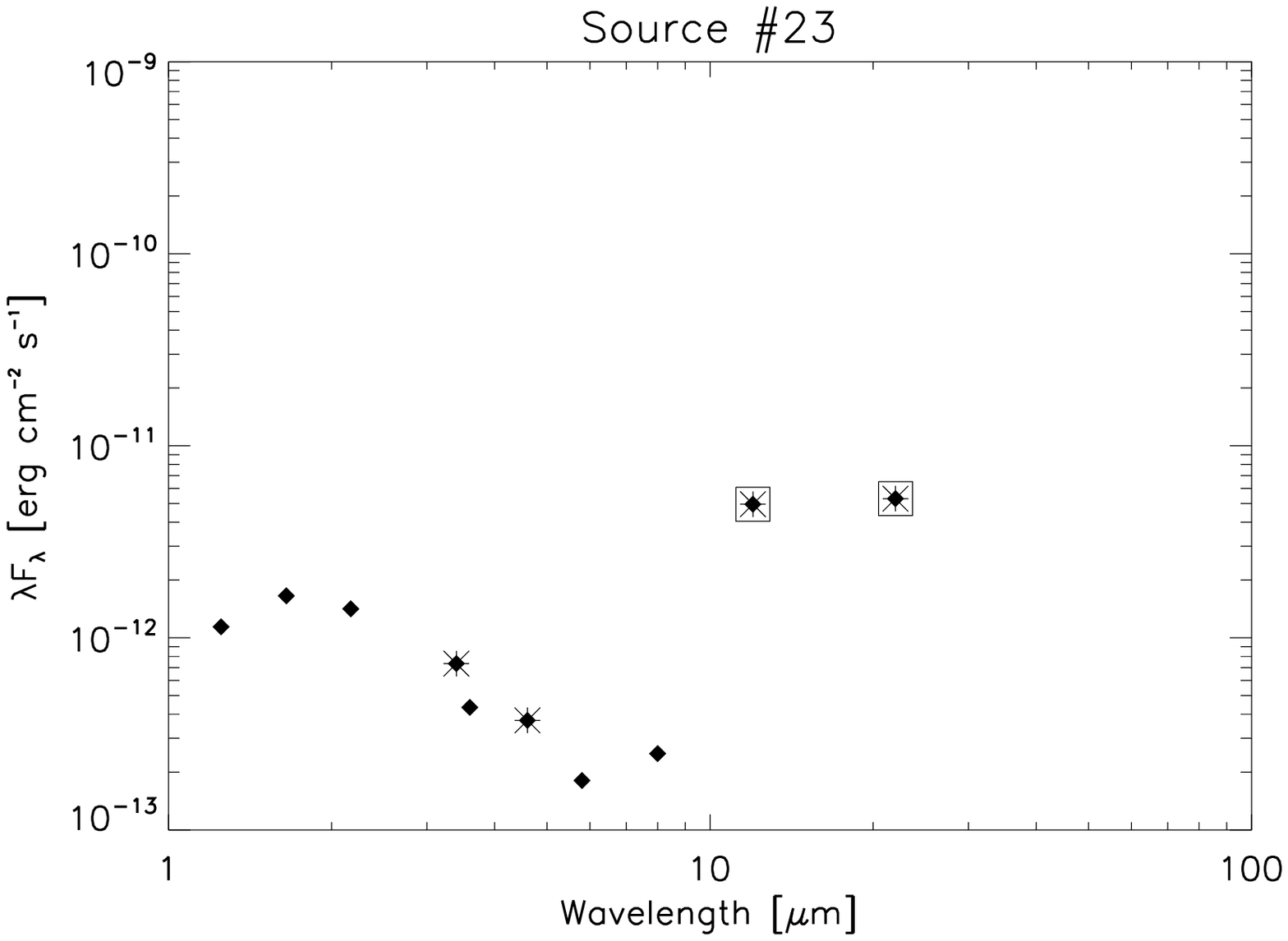}
\includegraphics[scale=0.45]{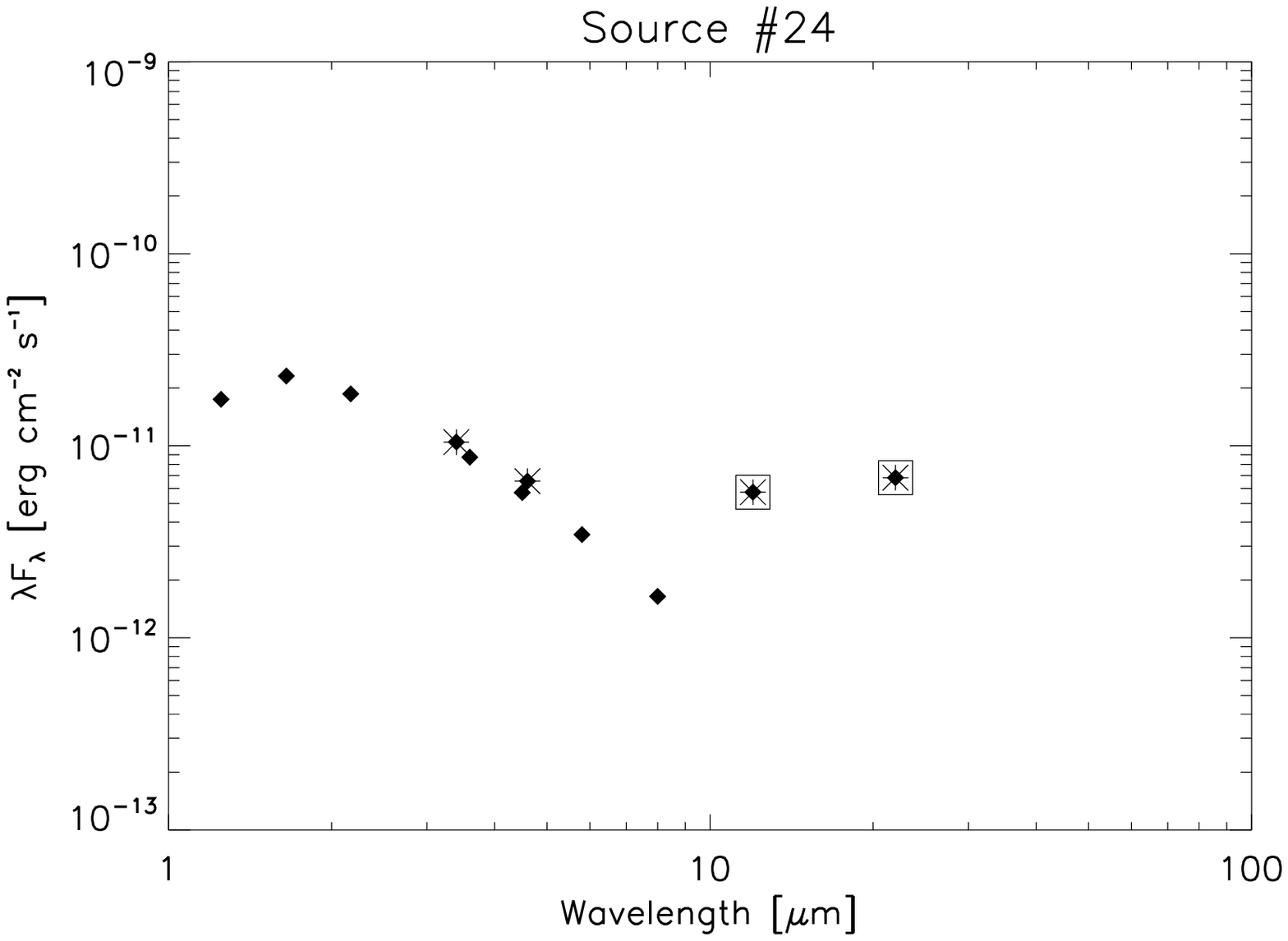}
\includegraphics[scale=0.45]{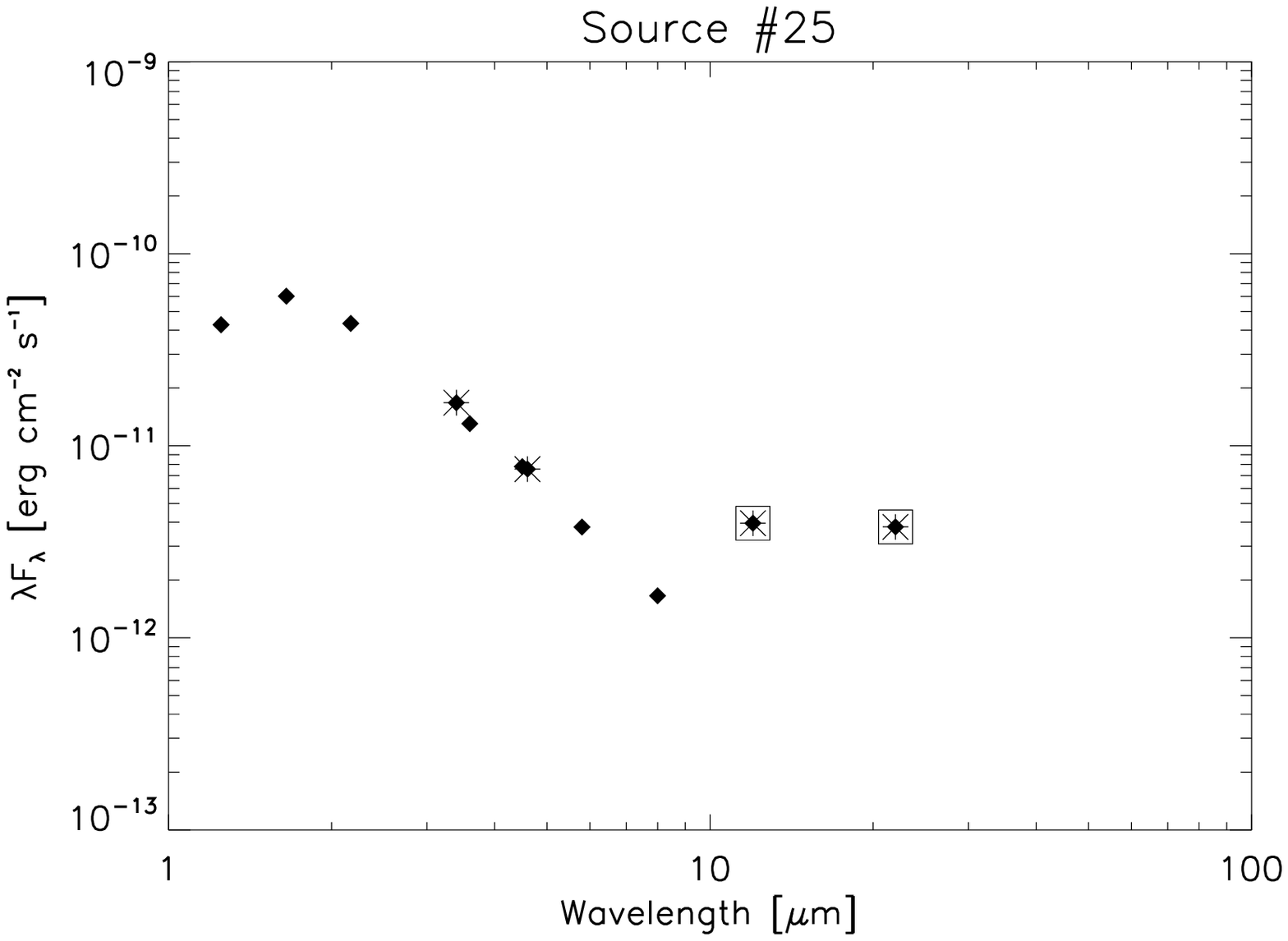}
\includegraphics[scale=0.45]{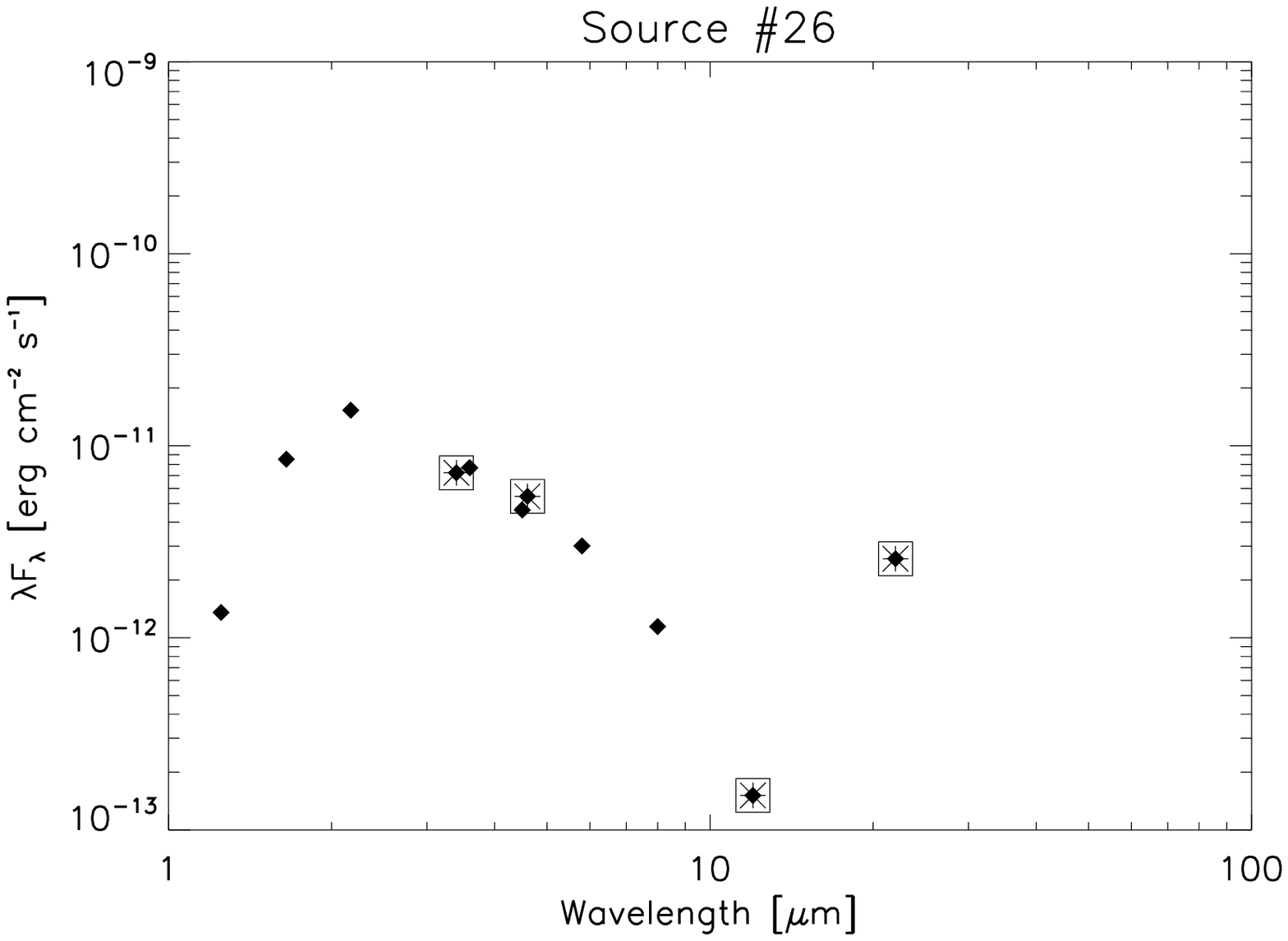}
\includegraphics[scale=0.45]{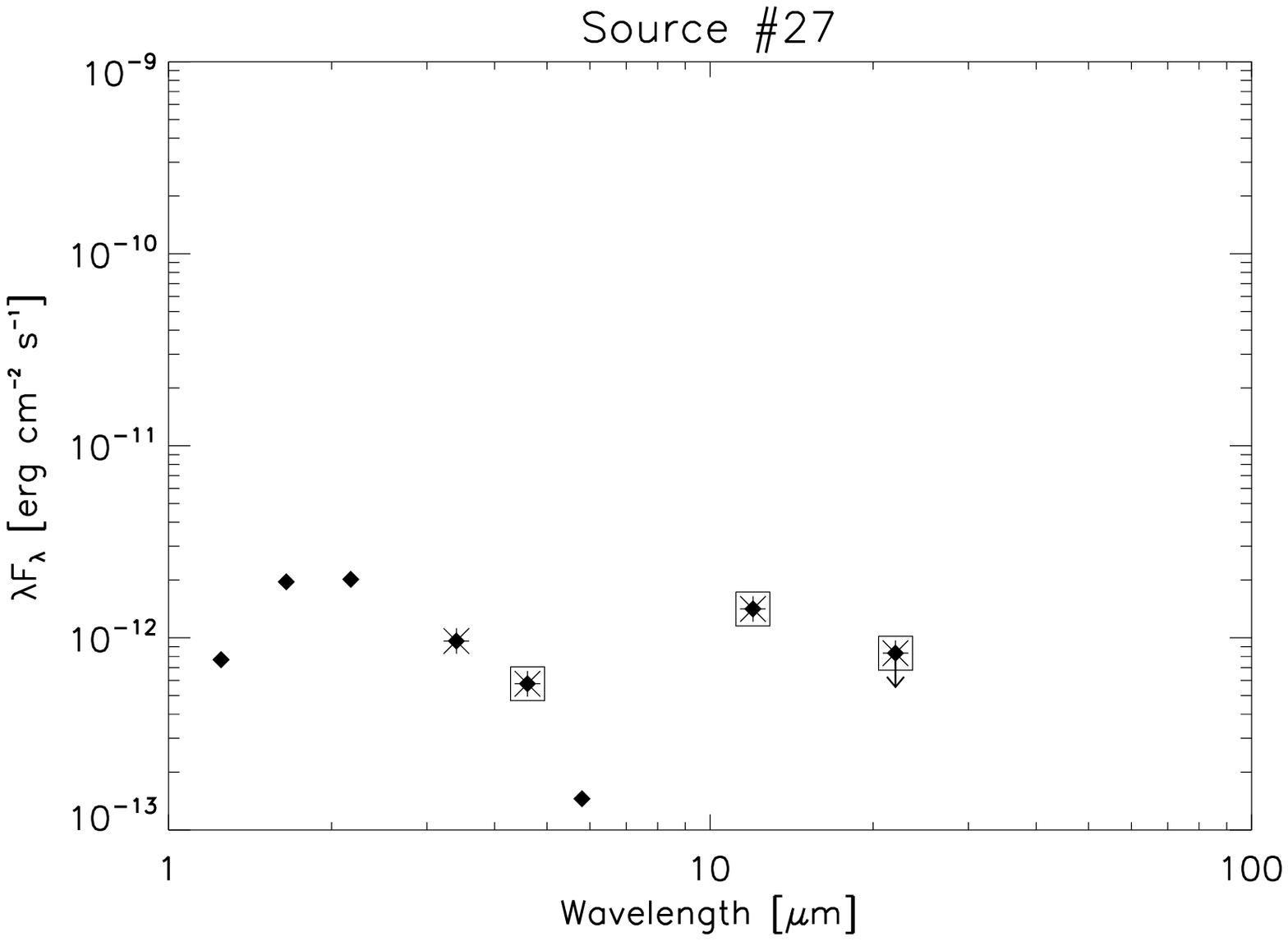}
\includegraphics[scale=0.45]{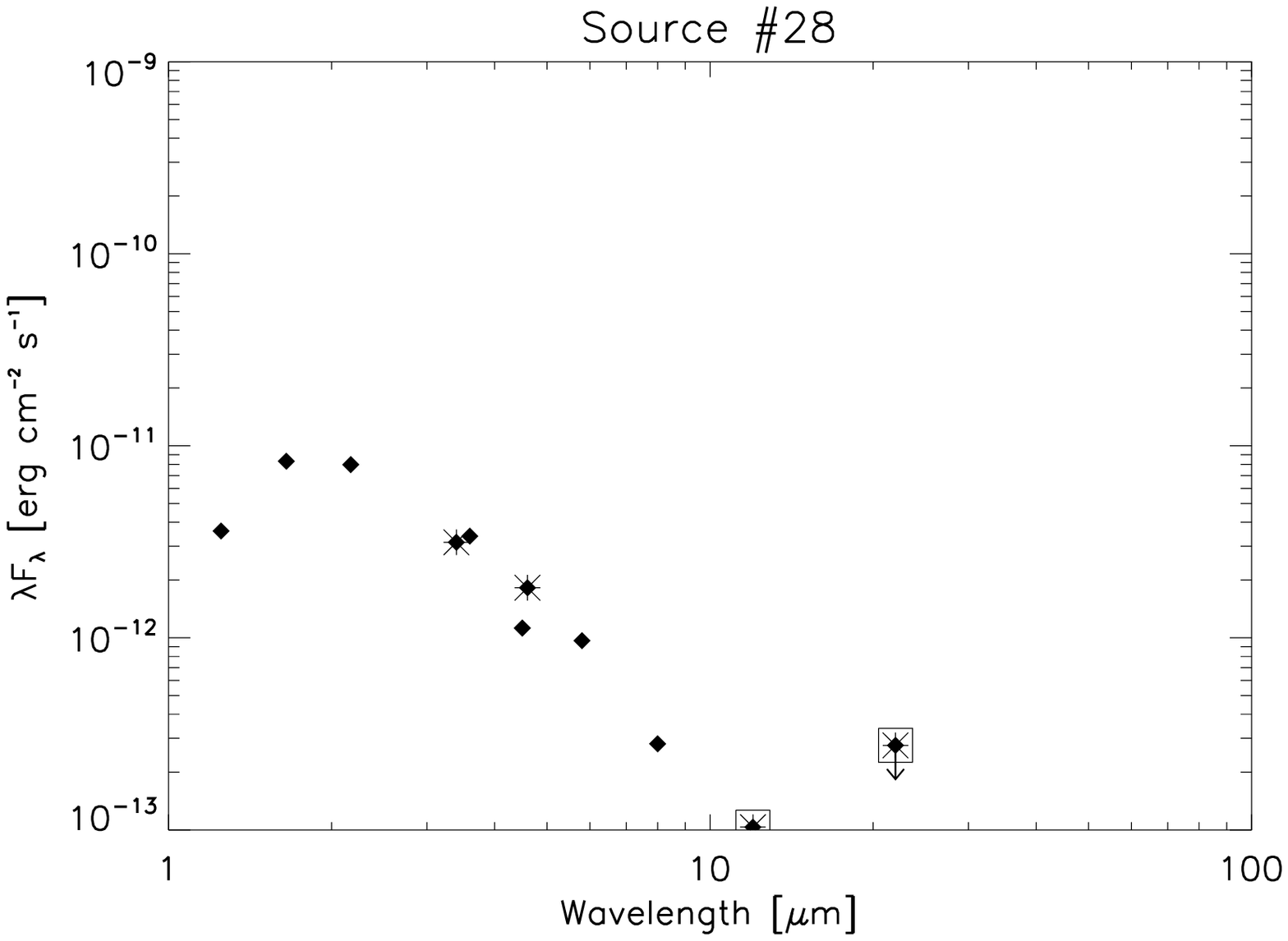}

\setcounter{figure}{19}
\caption{(continued) SEDs of Class III X-ray detected YSOs from 2MASS (diamond), Spitzer (diamond), and WISE (asterisk) data when available. Boxes plotted over WISE data signify high variability and/or possible contamination or confusion of photometry.}

\end{figure}

\begin{figure}
\centering
\includegraphics[scale=0.45]{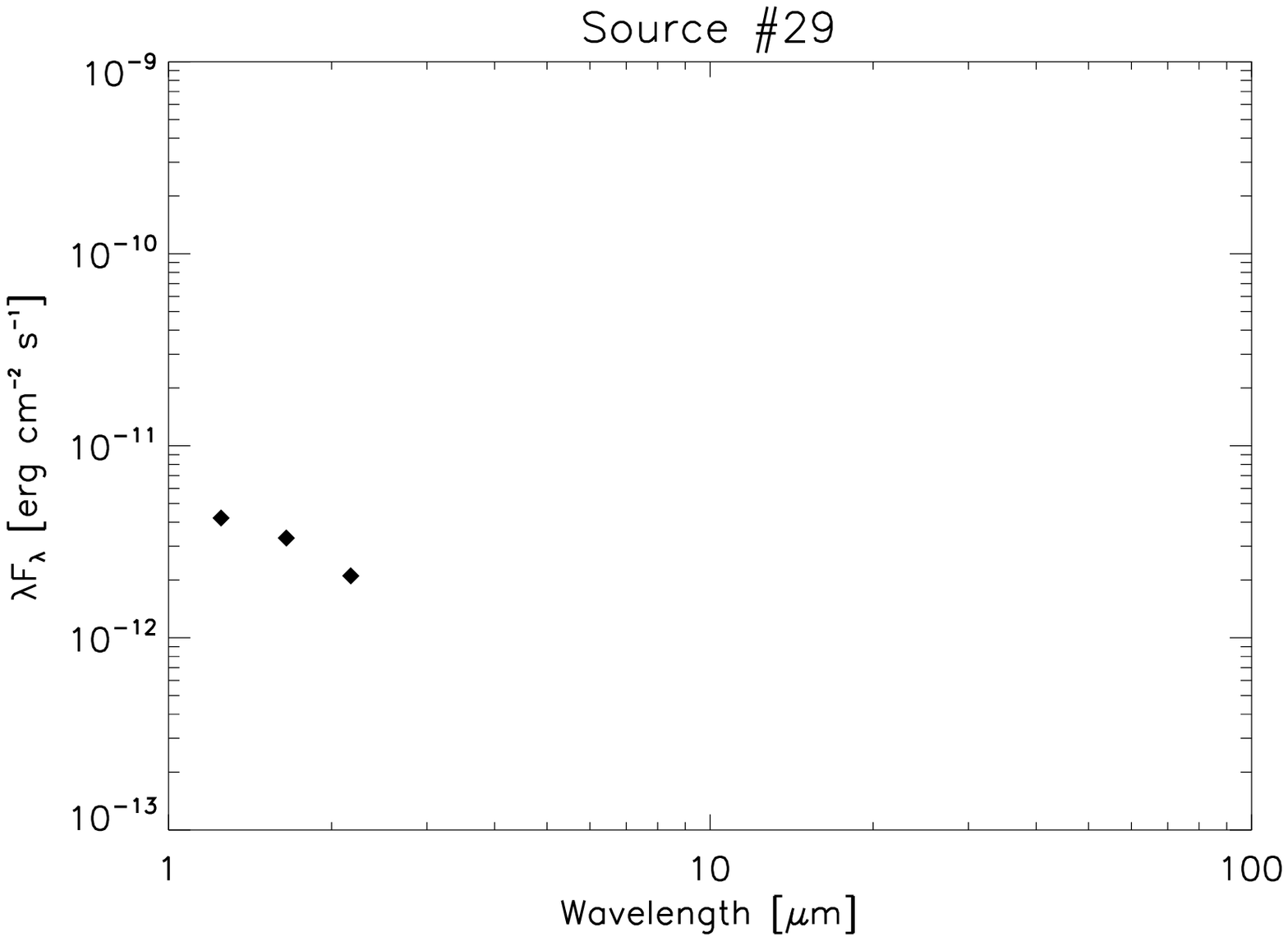}
\includegraphics[scale=0.45]{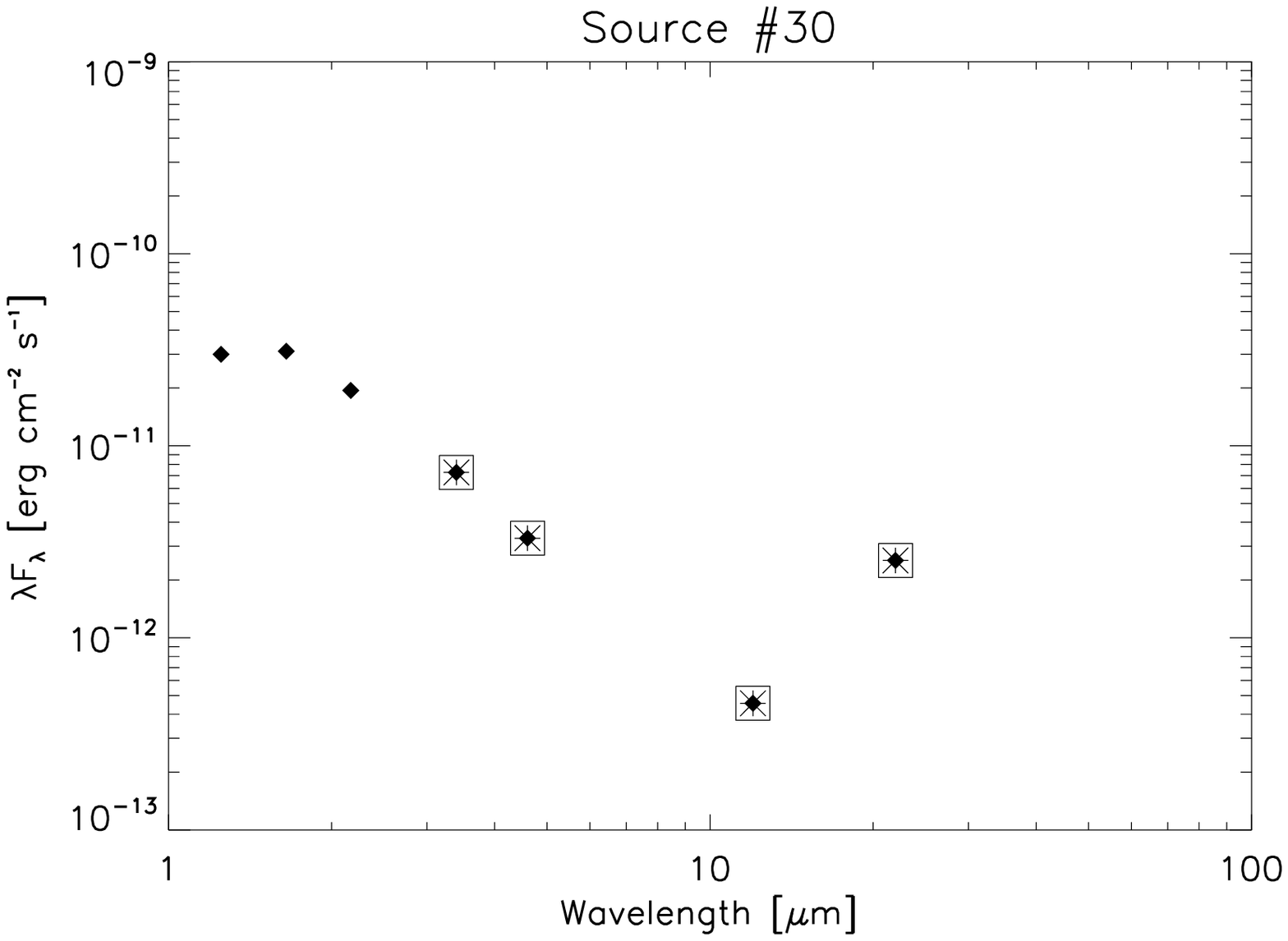}
\includegraphics[scale=0.45]{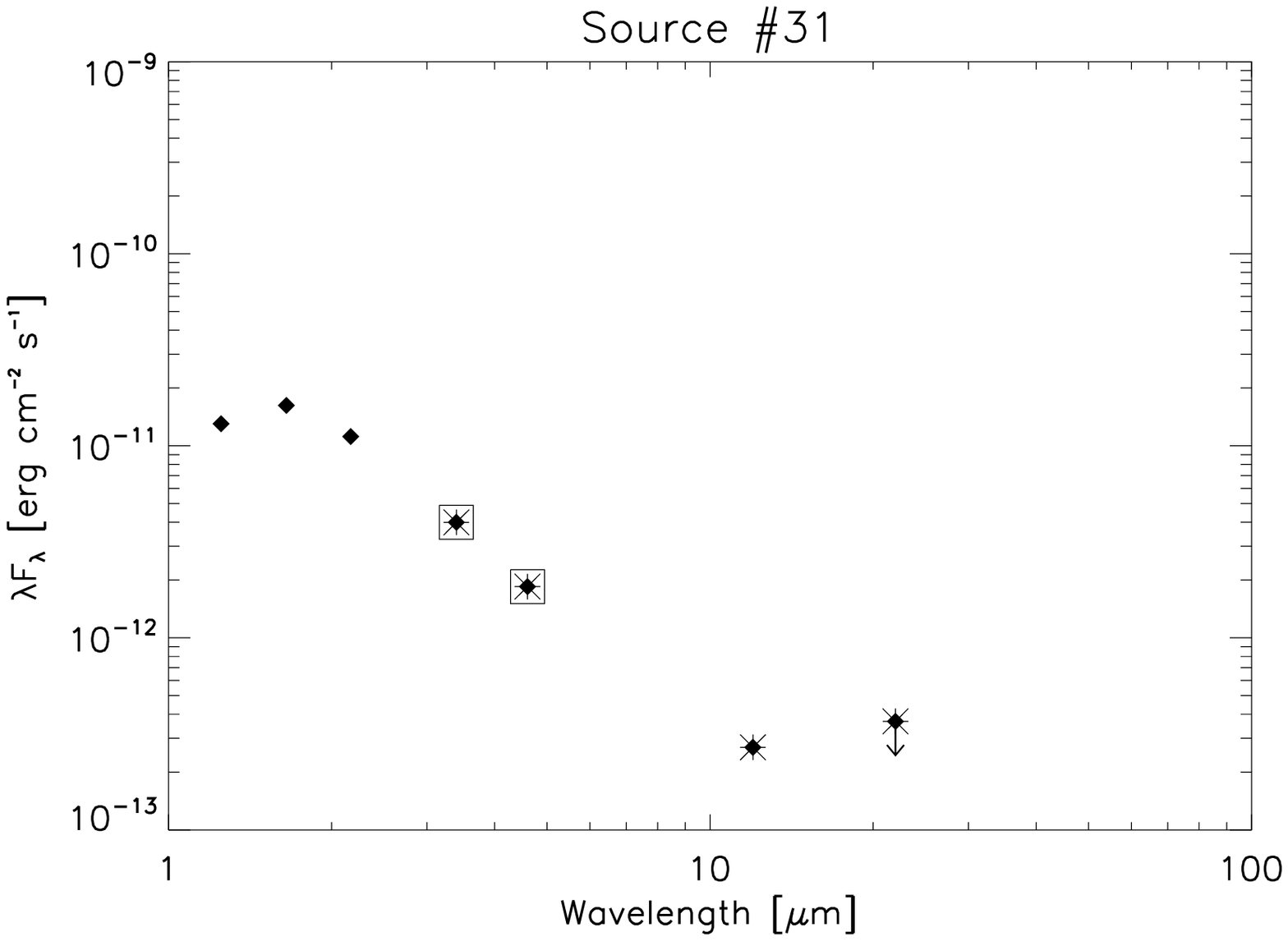}
\includegraphics[scale=0.45]{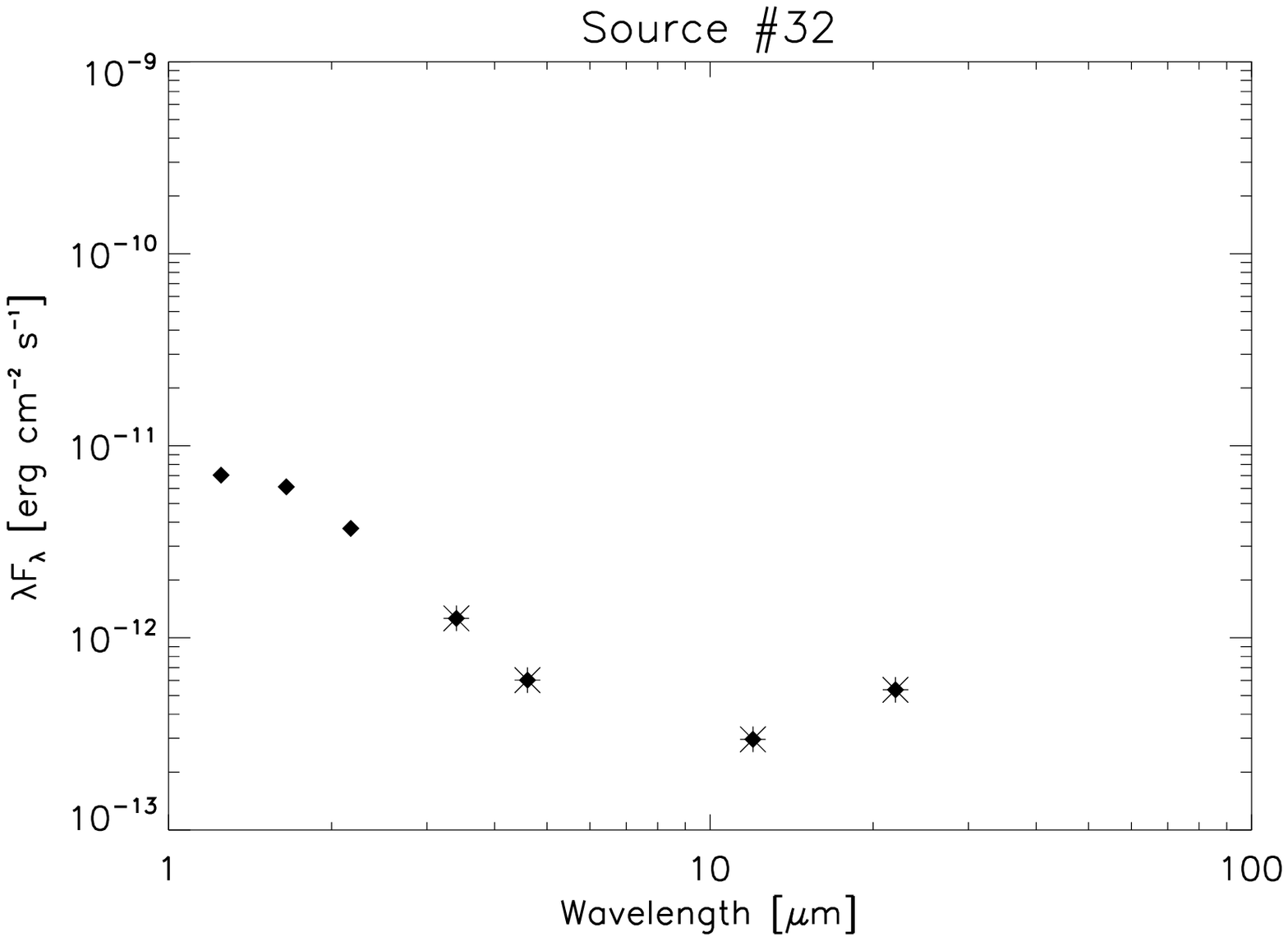}
\includegraphics[scale=0.45]{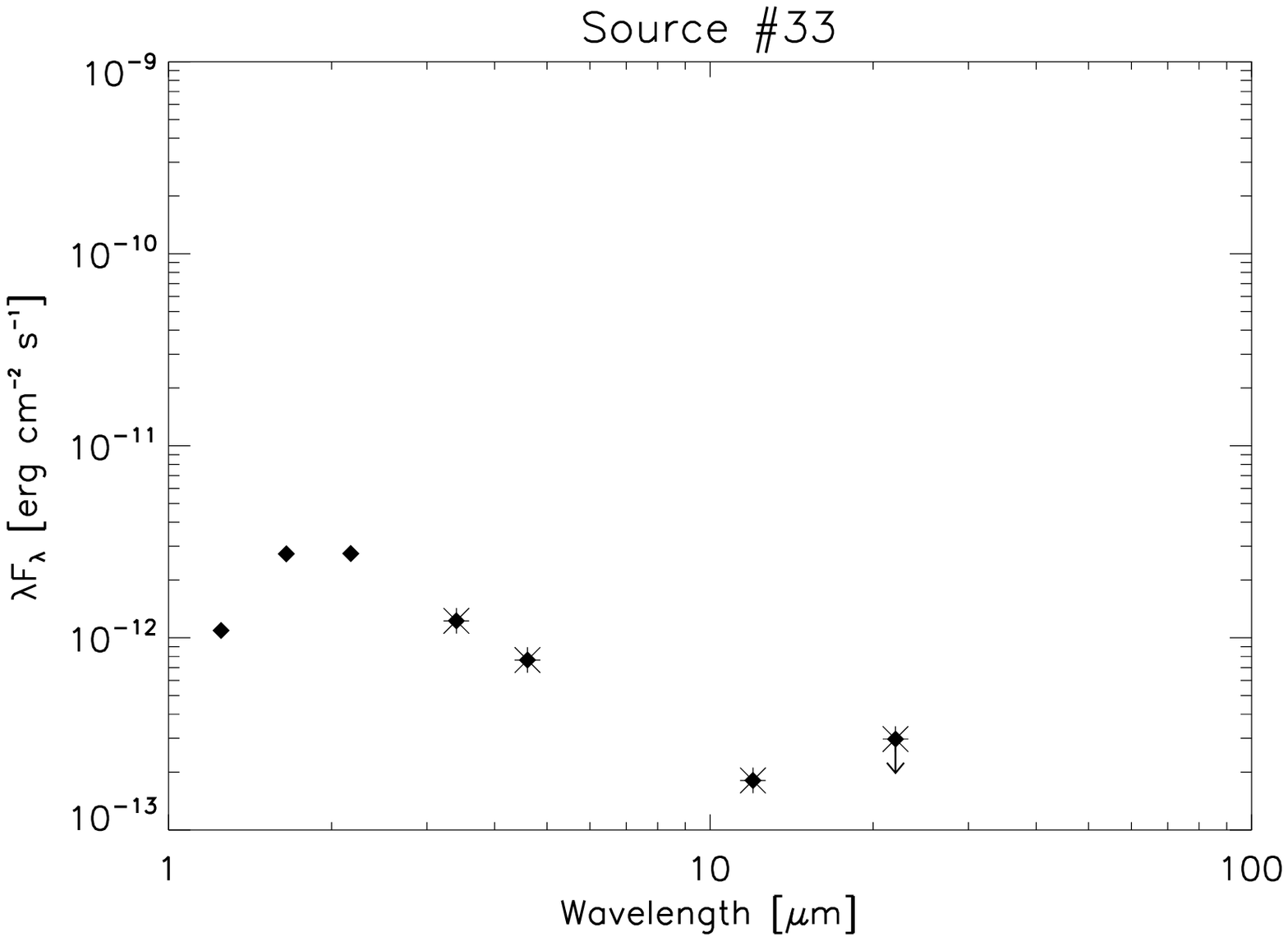}
\includegraphics[scale=0.45]{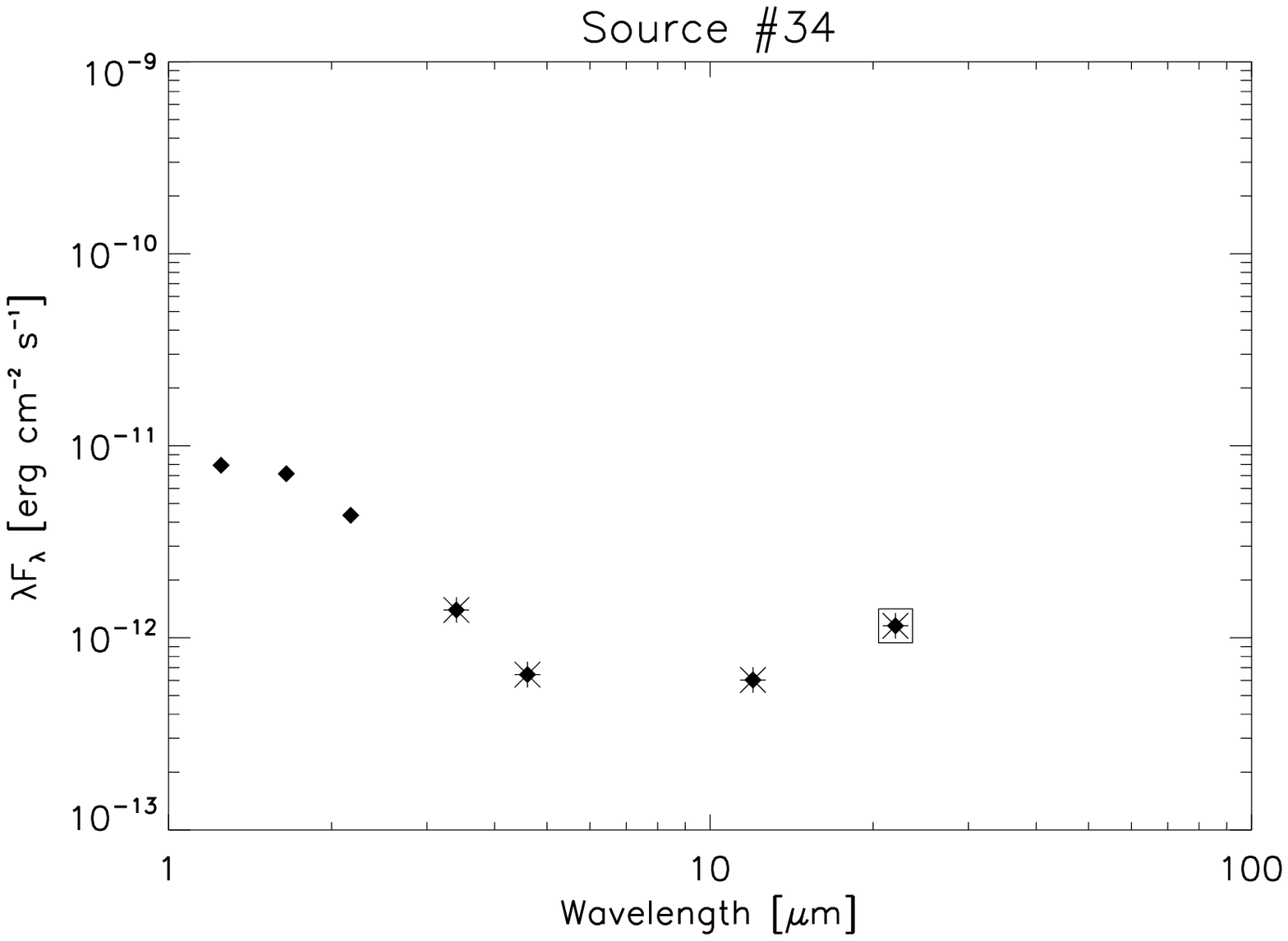}

\setcounter{figure}{19}
\caption{(continued) SEDs of Class III X-ray detected YSOs from 2MASS (diamond), Spitzer (diamond), and WISE (asterisk) data when available.  Boxes plotted over WISE data signify high variability and/or possible contamination or confusion of photometry.}

\end{figure}

\begin{figure}
\centering

\includegraphics[scale=0.45]{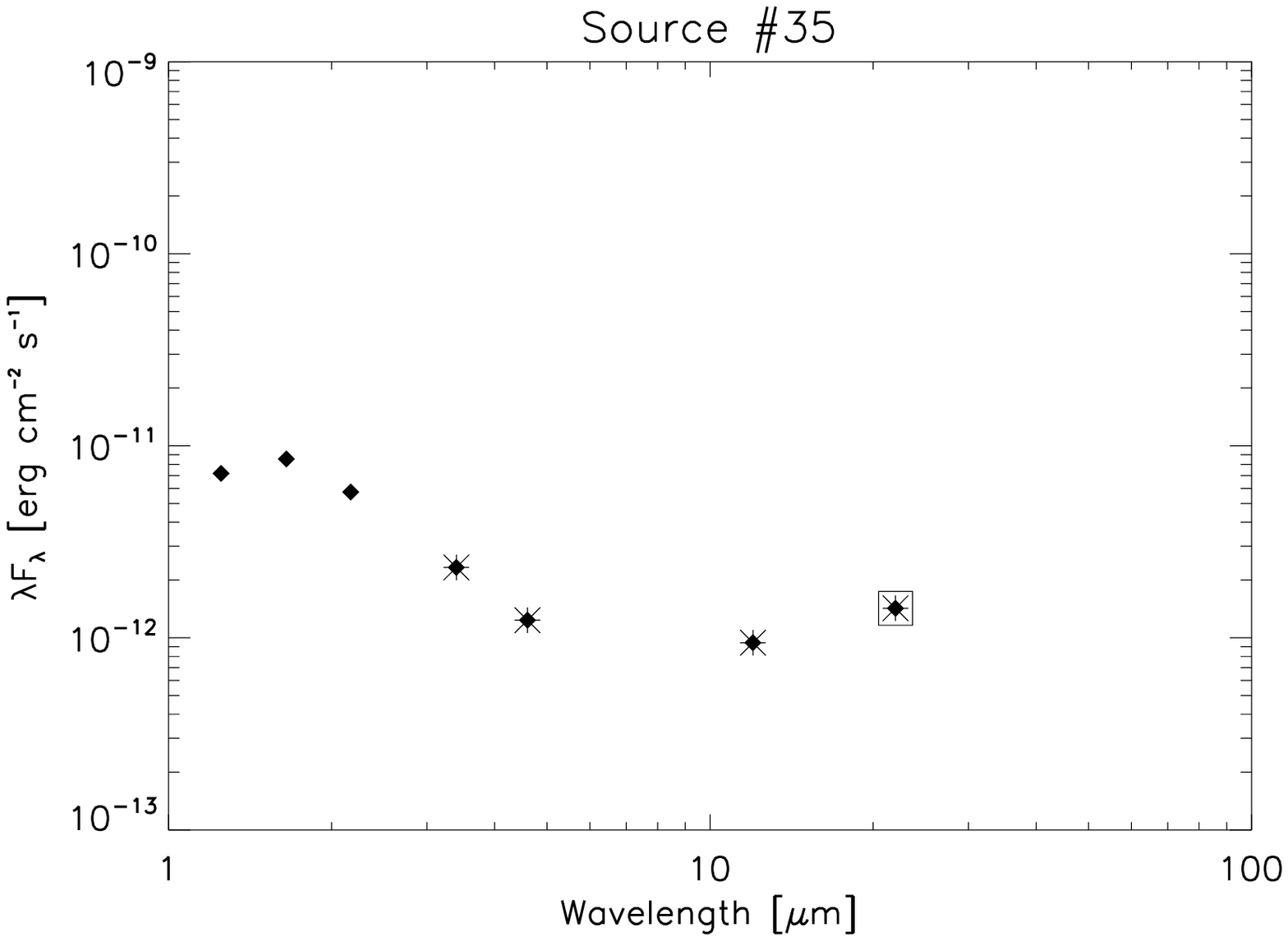}
\includegraphics[scale=0.45]{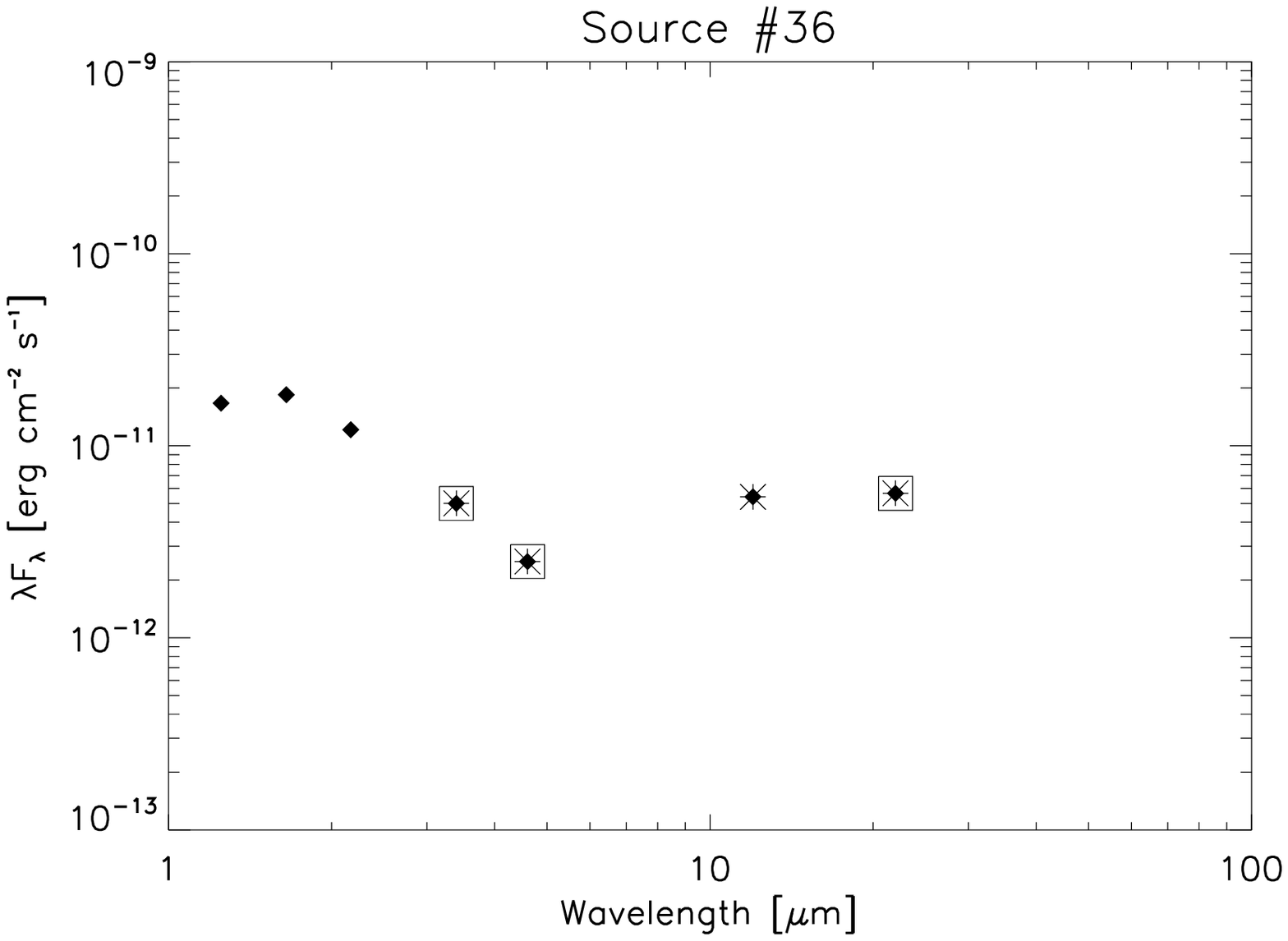}
\includegraphics[scale=0.45]{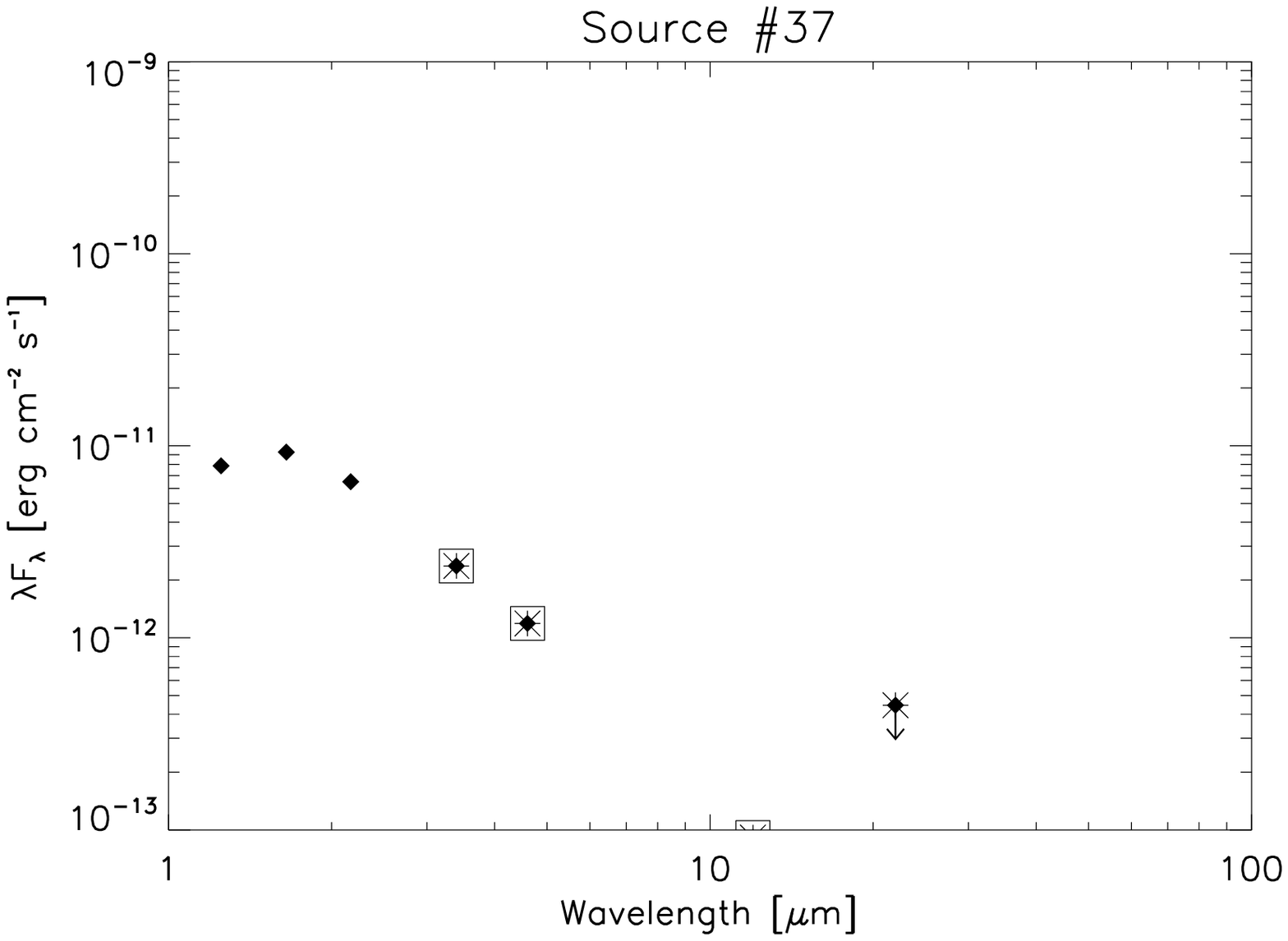}

\setcounter{figure}{19}
\caption{(continued) SEDs of Class III X-ray detected YSOs from 2MASS (diamond), Spitzer (diamond), and WISE (asterisk) data when available.  Boxes plotted over WISE data signify high variability and/or possible contamination or confusion of photometry.}

\end{figure}

\end{appendices}

\end{document}